\documentclass[12pt,a4,openright,nohyper]{book}
\usepackage{a4wide}
\usepackage{fancyhdr,graphicx}
\usepackage{latexsym,amssymb,epsfig}
\usepackage{amsmath,amssymb}
\usepackage{mathbarr}

\usepackage{axodraw,color}

\def\epp{\varepsilon^{\prime}}
\newcommand{\dis}{\displaystyle}
\newcommand{\tr}{\mbox{tr}}
\newcommand{\re}{\mbox{Re }}
\newcommand{\im}{\mbox{Im }}
\newcommand{\be}{\begin{equation}}
\newcommand{\ee}{\end{equation}}
\newcommand{\ba}{\begin{eqnarray}}
\newcommand{\ea}{\end{eqnarray}}
\newcommand{\mkd}{m_K^2}
\newcommand{\mpd}{m_{\pi}^2}

\newcommand{\order}{{\cal O}}
\newcommand{\dsp}{\displaystyle}
\newcommand{\SMN}{\overline \Pi_{Sij}^M}

\newcommand{\vl}{\overline \Pi_{Vij}^{(0)}}

\newcommand{\vlt}{\overline \Pi_{Vij}^{(0+1)}} 
\newcommand{\vltc}{\overline \Pi_V^{(0+1)\chi}}
\newcommand{\vlti}{\overline \Pi_V^{(0+1)I}}
\newcommand{\vt}{\overline \Pi_{Vij}^{(1)}}

\newcommand{\s}{\overline \Pi_{Sij}}

\newcommand{\vlf}{\Pi_{Vij}^{(0)}}
\newcommand{\vtf}{\Pi_{Vij}^{(1)}}
\newcommand{\smnf}{\Pi_{Sij}^M}      
\newcommand{\ssf}{\Pi_{Sij}}
\newcommand{\PMN}{\overline \Pi_{Pij}^M}
\newcommand{\al}{\overline \Pi_{Aij} ^{(0)}}
\newcommand{\alt}{\overline \Pi_{Aij}^{(0+1)}} 
\newcommand{\altc}{\overline \Pi_A^{(0+1)\chi}}
\newcommand{\alti}{\overline \Pi_A^{(0+1)I}}
\newcommand{\at}{\overline \Pi_{Aij}^{(1)}}

\newcommand{\PMNc}{\overline \Pi_{P}^{M\,\chi}}
\newcommand{\PMNi}{\overline \Pi_{P}^{M\,I}}
\newcommand{\p}{\overline \Pi_{Pij}}

\newcommand{\atf}{\Pi_{Aij}^{(1)}}
\newcommand{\alf}{\Pi_{Aij}^{(0)}}
\newcommand{\ppf}{\Pi_{Pij}}
\newcommand{\pmnf}{\Pi_{Pij}^M}
\newcommand{\Plr} {\Pi_{LR}}
\newcommand{\cond}{\langle\overline q q\rangle}
\newcommand{\condc}{\langle\overline q q\rangle_{\chi}}

\newcommand{\eq}{\begin{eqnarray}}
\newcommand{\en}{\end{eqnarray}}

\newcommand{\et}{\!\!\!&&\!\!\!}
\newcommand{\fr}{\frac}
\newcommand{\alp}{\alpha}
\newcommand{\la}{\lambda}
\newcommand{\de}{\delta}
\newcommand{\sig}{\sigma}
\newcommand{\nn}{\nonumber}

\newcommand{\mev}{\mbox{\rm MeV}}
\newcommand{\gev}{\mbox{\rm GeV}}
\newcommand{\ksls}{\not \! {\bf k}}

\begin{document}

%\include{portadaes}
%%%%%%%%%%%%%%%%%%%%%%%%%%%%%%%%%%%%%%%%%%%%%%%%%%%%%%%%%%%%%%%%%%%%%%%%%%%
% This is a sample header for a sample dissertation. Fill in the name,
% and the other information. LaTeX will work out the table of
% content, the list of figures and of tables for you. 
%%%%%%%%%%%%%%%%%%%%%%%%%%%%%%%%%%%%%%%%%%%%%%%%%%%%%%%%%%%%%%%%%%%%%%%%%%%

%\newpage
%\thispagestyle{empty}

% ******* Title page ******* 
% ************************** 

%\vspace*{1cm}
%\begin{center}
%{\Large{\bf Kaon Physics: CP Violation and Hadronic Matrix Elements}\\[.5cm] 
%}
%\vspace{2cm}
%{\large 
%Mar\'{\i}a Elvira G\'amiz S\'anchez\\
%\vspace{0.5cm}
%Departamento de F\'{\i}sica 
%Te\'orica y del Cosmos\\
%Universidad de Granada}
%\end{center}

%\vspace{1cm}
%\begin{figure}[!h]
%\begin{center}
%\epsfig{file=escudo_color.ps,height=5cm}
%\epsfig{file=plots/escudo.eps,height=7cm}
%\end{center}
%\end{figure}
%}
%%%%%%%%%%%%%%%%%%%%%%%%%%%%%%%%%%%%%%%%%%%%%%%%%%5
%\vspace{1cm}

%\begin{center}
%{\large  
%Thesis submitted for the Degree of Doctor
%in the\\
% University of Granada\\
%\vspace{0.3cm}
%$\cdot$ October 31, 2003 $\cdot$}
%\end{center}

%\newpage
%\thispagestyle{empty}
%\mbox{ }
%\newpage

\thispagestyle{empty}

% ******* Title page ******* 
% ************************** 

\vspace*{1cm}
\begin{center}
{\fontsize{24.pt}{9pt}\selectfont{
{\bf Kaon Physics: CP Violation and}\\
\vspace*{0.5cm}
{\bf Hadronic Matrix Elements}\\
}}
\vspace{2.5cm}
{\fontsize{18.pt}{9pt}\selectfont{
Mar\'{\i}a Elvira G\'amiz S\'anchez\\
\vspace{0.5cm}
\emph{Departamento de F\'{\i}sica 
Te\'orica y del Cosmos}}}
\end{center}

\vspace{2.cm}
\begin{figure}[!h]
\begin{center}
\epsfig{file=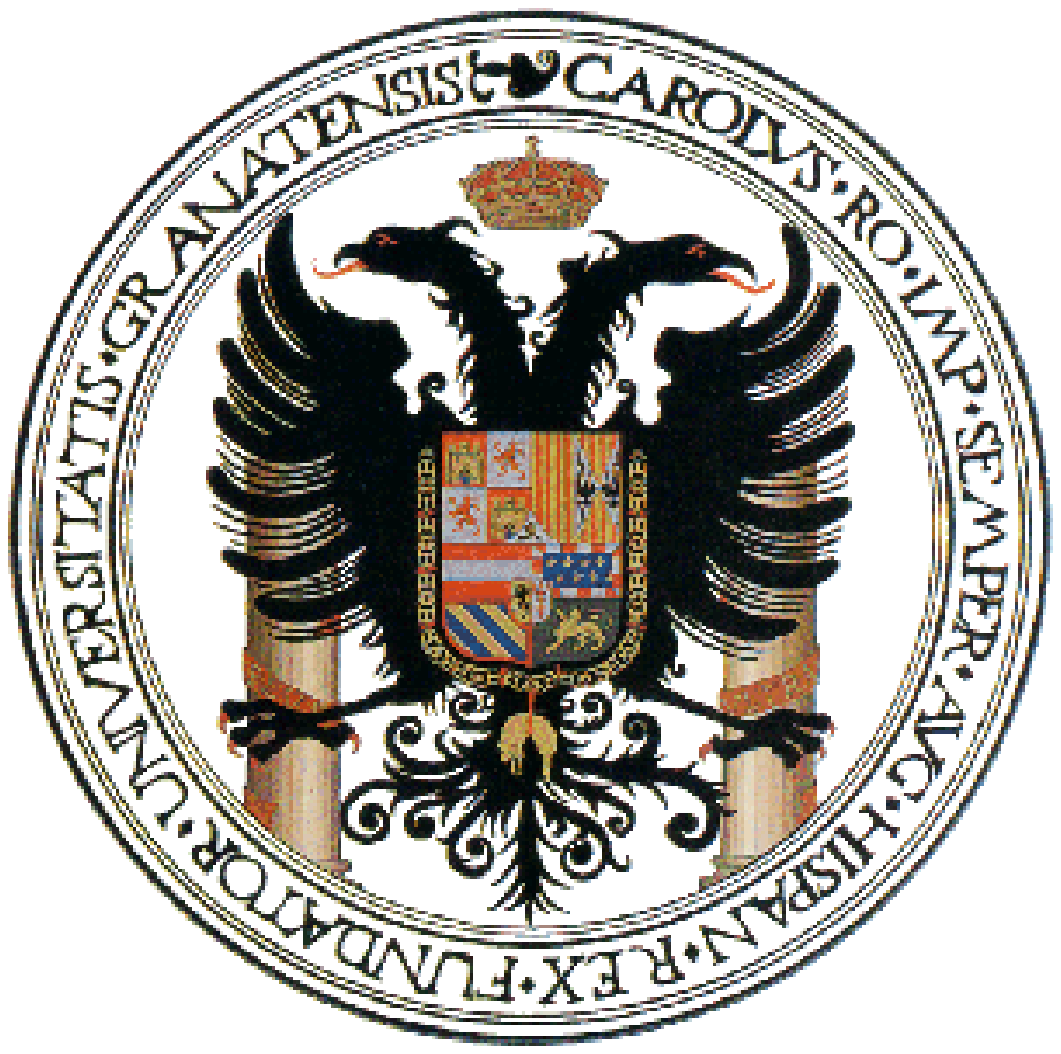,height=4cm}
\end{center}
\end{figure}
%}
%%%%%%%%%%%%%%%%%%%%%%%%%%%%%%%%%%%%%%%%%%%%%%%%%%5
\vspace{2.cm}

\begin{center}
{\fontsize{18.pt}{9pt}\selectfont{
%Thesis submitted for the Degree of Doctor
%in the\\
\emph{Universidad de Granada}\\
\vspace{0.5cm}
$\cdot$ October 2003 $\cdot$}}
\end{center}

\newpage
\thispagestyle{empty}
\mbox{ }
\newpage

\pagenumbering{arabic}
\setcounter{page}{1} \pagestyle{fancy} %\pagestyle{plain}

\renewcommand{\chaptermark}[1]{\markboth{\chaptername%
\ \thechapter:\,\ #1}{}}
\renewcommand{\sectionmark}[1]{\markright{\thesection\,\ #1}}

\newpage

%{\setlength{\baselineskip}{24pt} 
%\pagestyle{fancy}
\addtolength{\headheight}{3pt}    % more space for rule under running head
\fancyhead{}
%\fancyhead[LE]{\sl\leftmark}
%\fancyhead[LO,RE]{\rm\thepage}
%\fancyhead[RO]{\sl\rightmark}
\fancyhead[RE]{\sl\leftmark}
\fancyhead[RO,LE]{\rm\thepage}
\fancyhead[LO]{\sl\rightmark}
\fancyfoot[C,L,E]{}

\tableofcontents      

%\chapter{Introduction}
\chapter{Introduction}
\label{introd0}

CP symmetry violation was discovered several decades ago in 
neutral kaon decays \cite{CCFT64}. Effects of  CP symmetry breaking 
have been also recently observed in $B$ meson decays 
\cite{BaBar,Belle}, but kaon physic continues being an exceptional 
ground to study this kind of phenomena. 
The analysis of the parameters that describe CP violation 
constitutes a great source of information about flavour 
changing processes and, in the Standard Model, they can  
provide us with information about the worst known part 
of the Lagrangian, the scalar sector, where CP violation has its origin. 

The main CP-violating parameter in the kaon decays is 
$\varepsilon_K'/\varepsilon_K$, 
for which there exist very precise experimental measures 
$\re(\varepsilon_K'/\varepsilon_K)=(1.66 \pm 0.16)\times 10^{-3}$ 
-see Chapter \ref{CPviolation} for definitions and discussions. 
Experimental data for this quantity and 
another CP--violating parameters --as well as other quantities related with 
different fundamental aspects in particle physics as flavour changing 
neutral currents or quark masses-- are based on the analysis of observables 
involving hadrons, that interact through strong interactions. 
In particular, these observables are governed in a 
decisive way by the non-perturbative regime of 
Quantum Chromodynamics (QCD), the theory describing strong interactions.  
At low energies --in the non-perturbative regime-- QCD is not completely 
understood, so it is not easy to get theoretical predictions for 
the hadronic matrix elements involved in these processes 
without large uncertainties. Any improvement in the calculation of 
these hadronic matrix elements would thus been fundamental in the 
understanding of experimental CP-violating results. The goal is twofold,  
on one hand we pretend to analyze theoretically with great precision 
some observables in the Standard Model, specially those related 
to the CP-violating sector which is embedded in the scalar sector, the worst 
known part of the Standard Model lagrangian. On the other hand, 
the precise knowledge of the Standard Model predictions for CP-violating 
observables can serve to unveil the existence of new physics and 
check the validity of certain models beyond the Standard Model. 

The main problem is dealing with strong interactions at 
intermediate energies. At very low and high energies we can use 
Chiral Perturbation Theory (CHPT) \cite{ChPT1,ChPT2} and perturbative 
QCD respectively, that are well established theories. They let us 
do reliable calculations using next orders in the expansion as an 
estimate of the error associated to them. Several methods and 
approximations have been developed in order to try to suitably describe
the intermediate region. Any reliable calculation give a 
good matching between the long- and 
short- distance regions. The most important of these methods are listed in 
Chapter \ref{matrixelements} and their predictions for several 
CP-violating parameters are commented along the different chapters 
of this Thesis. 

The basic objects that are needed for the description of the low energy
physics are the two-quark currents and densities Green's functions.
The couplings of the strong CHPT lagrangian are coefficients of the 
Taylor expansion in powers of masses and external momenta 
of some of these Green's functions. 
While other order parameters needed in kaon 
physics such as $B_K$, $G_8$, $\im G_E$ or $G_{27}$ are obtained 
by also doing the appropriate identification of the coefficients in the 
Taylor expansion of integrals of this kind of Green's functions 
over all the range of energies. 
See Section \ref{secq6} of Chapter \ref{CPviolation} for more details.  
A good description of Green's functions would thus 
provide predictions for the parameters we are interested in. A description 
of a general method to perform a program of this kind is given 
in Chapter \ref{matrixelements} and some possible applications of it are 
discussed in the conclusions.

This Thesis is organized in two parts. In the first part, Chapters 
\ref{introd0}-\ref{chCHPT}, we describe the framework in which 
the Thesis has been developed and establish the definitions and notation   
necessary in the second part. 

In the next section of this chapter we give an overview of 
the Standard Model, explaining the different sectors in which 
it can divided and the grade of knowledge we have about each of them.  
We define the symmetry CP and discuss in which conditions it can be 
measured experimentally. Discussions about the value of the SM parameters 
$m_s$ and $|V_{us}|$ calculated in \cite{CDGHPP01,GJPPS03} 
are also provided in this chapter. The Chapter \ref{CPviolation} is a more 
detailed description of the theory involved in the 
CP violation in $K$ decays, in which we define the main CP-violating 
observables and outline the theoretical calculation of these parameters. 
The present experimental and theoretical status of direct CP violation 
in the kaon decays is also given. 
In Chapter \ref{chCHPT} we introduce CHPT, 
collect the Lagrangians at leading and next-to-leading order in this 
theory and discuss the values of the couplings of these Lagrangians. 
In addition, leading order in the chiral expansion predictions for 
some CP-violating observables are given.

The second part of the Thesis, Chapters \ref{chK3pi}-\ref{secepsilonprime}, 
is composed by the calculations of some of the CP-violating 
observables defined before. In Chapter \ref{chK3pi} we study 
CP-violating asymmetries in the decays of charged kaons 
$K\to 3\pi$ \cite{GPS03}. The work in \cite{BGP01}, where the $\Delta I=3/2$ 
contribution to $\varepsilon_K'$ was calculated in the chiral limit,  
is reported in Chapter \ref{chq7q8}. In Chapter \ref{secepsilonprime}, 
we perform a calculation of the direct CP-violating parameter 
$\varepsilon_K'$, as discussed in \cite{BGP03}, using some of the 
results obtained in the other chapters. The different approaches 
that can be used to calculate hadronic matrix elements are listed in 
Chapter \ref{matrixelements}. In this chapter, 
there is also a new approach that can be used to 
systematically determine hadronic matrix elements from 
the calculation of a set of Green's functions compatible with all 
the QCD and phenomenology constraints, which was developed in \cite{BGLP03}. 

The results will be summarized in Chapter \ref{conclusions0}. Applications 
of the ladder resummation approach described in Chapter \ref{matrixelements}, 
in which we are working at the moment or are planning to do in the future 
are also summarized in Chapter \ref{conclusions0}.

We give the Operator Product Expansion of the two-point 
functions relevant in the calculation of the matrix elements of the 
electroweak penguin $Q_7$ and $Q_8$ in Appendix \ref{appendixq7q8} and the  
analytical formulas we got for CP-conserving and CP-violating 
observables in $K\to 3\pi$ decays, as well as the notation 
used in writing these formulas in Appendix \ref{appendixK3pi}.

\section{Overview of the Standard Model}

The Standard Model (SM) is a non-abelian gauge theory based on the 
$SU(3)_C\times SU(2)_L\times U(1)_Y$ 
symmetry group, which describes strong,  
weak and electromagnetic interactions 
\cite{Glashow61,Weinberg67,Salam69,GIM70,FGL73}. 
The Standard Model Lagrangian can be divided into four parts 
\ba \label{SMlagrangian}
{\cal L}_{SM} &=&
\dsp\underbrace{{\cal L}_H(\phi)}_{\mbox{Higgs}}
+
\underbrace{{\cal L}_G(W,Z,G)}_{\mbox{Gauge}}
\dsp+\underbrace{\sum_{\psi=\mbox{fermions}}
\bar\psi iD\hskip-0.7em/\hskip0.4em \psi}_{\mbox{gauge-fermion}}
\nonumber\\&&
+
\underbrace{\sum_{\psi,\psi^\prime=\mbox{fermions}}
g_{\psi\psi'}\bar\psi\phi\psi'}_{\mbox{Yukawa}} \hspace{3cm}\,.
\ea

Fermionic-matter content is described by leptons and quarks 
which are organized in three generation with two quark flavours 
($u$ and $d$ like) and two leptons (neutrino and electron-like) each one:
\be \label{fermionfam}
\left[\begin{array}{cc} \nu_e&u\\e^-& d\end{array}\right], \quad
\left[\begin{array}{cc} \nu_\mu&c\\\mu^-& s\end{array}\right], \quad
\left[\begin{array}{cc} \nu_\tau&t\\\tau^-& b\end{array}\right],
\ee
where each quark appears in three different colours and each family 
is composed by
\be
\left[\begin{array}{cc} \nu_l&q_u\\l^-& q_d\end{array}\right] \equiv
\left(\begin{array}{c} \nu_l \\ l^- \end{array}\right),\,
\left(\begin{array}{c} q_u\\q_d\end{array}\right),\,
l_R^-,\,(q_u)_R,\,(q_d)_R.
\ee
All these particles are accompanied by their corresponding antiparticle 
with the same mass and opposite charge. The masses and flavour quantum 
numbers of the three families in (\ref{fermionfam}) are different, but 
they have the same properties under gauge interactions. The left-handed 
fields are $SU(2)_L$ doublets and their right-handed partners transform 
as $SU(2)_R$ singlets. 

The gauge sector ${\cal L}_G(W,Z,G)$ 
collect the purely kinetic terms of the spin-1 gauge 
fields which are exchanged between the fermion fields to generate the 
interactions, as well as the self-interactions of these gauge fields 
due to the non-abelian structure of the $SU(2)$ and $SU(3)$ groups. 
There are 8 massless gluons and 1 massless photon 
for the strong and electromagnetic interactions, respectively, and 3 
massive bosons, $W^\pm$ and $Z$, for the weak interaction. 
The interaction terms between fermions and gauge bosons are encoded in 
the gauge-fermion sector together with the kinetic terms (those 
corresponding to free massless particles) for the fermions. 
These two part of the SM are well tested at LEP (at CERN) and SLC (at SLAC). 
An overview of the present experimental status of the SM tests 
and the determination of its parameters can be found in \cite{SMtests}.

%???? poner cosas que se han medido y referencias. Buscar en el review 
%de Toni.

The gauge symmetry in which is based the Standard Model is spontaneously 
broken by the vacuum that is not invariant under the whole group 
but only under the electromagnetic and the $SU(3)_C$ symmetries that remain 
exact
\be
SU(3)_C\times SU(2)_L\times U(1)_Y \xrightarrow{SSB}
SU(3)_C\times U(1)_{QED}\,.
\ee
The spontaneous symmetry breaking (SSB) 
of electroweak interactions is responsible for the 
generation of masses for the weak gauge bosons, quarks and leptons.   
It is also at the origin of fermion mixing and CP violation.  
The other important 
consequence of SSB is the appearance of physical scalars particles 
in the model, the so-called Higgs. The simplest realization of 
SSB --the minimal SM-- is made by the appearance of 
one scalar $\phi$. 
The Higgs and Yukawa parts of the Lagrangian in 
(\ref{SMlagrangian}) constitutes the scalar sector of the Standard Model 
and are associated to SSB. The Higgs Lagrangian contains the kinetic 
terms of the scalar particle/s that appear due to SSB mechanism and 
the interaction terms of Higgs and gauge particles. These interaction 
terms generate the mass of the massive gauge bosons $W^{\pm}$ and $Z$. 
The Yukawa terms, which describe the interactions between fermions and the 
scalar particles after SSB, originate the fermion masses 
and CP violation (see bellow). 

The scalar sector is the worst tested part of the Standard Model, 
LEP at CERN and SLC at SLAC have started to test its basic features. 
It is expected that Tevatron and LHC can give more information 
about it in the future. Since the scalar sector is the one in which 
we are more interested in this Thesis, we will treat it more 
extensively in the next section.

There are three discrete symmetries specially relevant, namely, C (Charge 
Conjugation), P (Parity) and T (Time Reversal). Local Field Theory 
by itself implies the conservation of CPT. The asymmetries 
C, P and T hold separately for strong and electromagnetic interactions. 
The fermion and Higgs part of the Lagrangian in 
(\ref{SMlagrangian}) conserve CP and T, 
so the only source of CP violation can be the Yukawa part. We will see 
in section \ref{SMCPViolation} how this can be carried out.

Finally, we must remark that the 
Standard Model depend on a number of parameters that are not 
fixed by the model itself and are left free. 
The Higgs part is responsible for 
two parameters in the minimal SM 
and the gauge part for three. Neglecting Yukawa couplings 
to neutrinos, the Higgs-Fermion part contains 54 real (27 complex) 
parameters, however most of them are unobservables 
since they can be removed by field transformations. 
With neutrino mixing recently observed \cite{neutrmix} this last number 
increases to a total number of parameters that depend on the nature 
of the neutrino fields, being Dirac or Majorana.

The SM provide a theoretical framework in which one can accommodate 
all the experimental facts in particle physics up to date with 
great precision, with the exception of a few parameters that differ 
from the SM predictions in (2-3)$\sigma$. For detailed 
descriptions of the Standard Model and the phenomenology associated 
to it, see \cite{reviewsSM}.

\subsection{The Scalar Sector}

The scalar sector is the main source of unknown SM parameters 
and the best ground to get information about the electroweak 
symmetry breaking and its phenomenology. 
Due to SSB, the Yukawa couplings and the Higgs vacuum expectation value 
give rise to a mass matrix for quarks, electron-like leptons and neutrinos  
that is not diagonal in the family space. After diagonalizing the mass 
matrix only the three charged lepton masses and the six quark masses 
survive if we don't consider masses for the neutrinos.

In the rest of the Chapter we remark and comment the aspects 
related to this part of the SM in which we are interested.

\subsubsection{Quark masses}

As we shall see in Chapter \ref{CPviolation}, two  
parameters of the SM that are relevant in the calculation of 
direct CP violation in kaon decays  
within some approaches \cite{ChPT,PPS01} are the top quark mass $m_t$ 
and the strange quark mass $m_s$. While $m_t$ is already very 
well known and its precise value is less important, the exact  value 
of the strange quark mass as well as the uncertainty associated to it 
is more relevant within the $N_C$ approach in that calculation. The running 
top quark mass can be obtained from converting the corresponding pole mass 
value \cite{PDG02} with an error of about 3\%, however, $m_s$ 
is a more controversial parameter an its value have decreased in the last 
years since 1999 by 15\% \cite{BJ03}. This decreasing of $m_s$ 
has enhanced the theoretical value of 
the parameter of direct CP violation in kaon decays that use its value 
by a factor around 1.3 \cite{BJ03}.

Several methods have been used to determine the value of the strange 
quark mass. Sum rule determinations of $m_s$ 
have been performed on
the basis of the divergence of the vector or axial-vector spectral functions
alone \cite{SRms,JOP02,KM02}, with results that agreed very well between them. 
The status of the extraction of $m_s$ from the hadronic $e^+e^-$ cross
section is less clear.  
Recent reviews of determinations of the strange quark mass from lattice QCD
have been presented in \cite{latticems,WIT02,Kaneko02}, with the
conclusions $m_s(2\,\gev)= 108\pm 15\,\mev$ and $m_s(2\,\gev)= 90\pm 20\;\mev$
in the quenched and unquenched cases respectively.

Other analysis of the strange quark mass 
\cite{CDGHPP01,GJPPS03,taums,ALEPH,PPms99,KM00} 
are based on the available data for hadronic $\tau$ decays 
and the experimental separation of the Cabibbo-allowed decays and
Cabibbo-suppressed modes into strange particles. 
.
Some of these strange mass determinations suffer from sizable
uncertainties due to higher order perturbative corrections. In the sum rule
involving $SU(3)$ breaking effects in the $\tau$ hadronic width on which they 
are based, scalar and pseudoscalar correlation functions contribute,
which are known to be inflicted with large higher order QCD corrections, 
and these corrections are additionally
amplified by the particular weight functions which appear in the $\tau$ sum
rule. As a natural continuation, it was realized that one remedy of the
problem would be to replace the QCD expressions of scalar and pseudoscalar
correlators by corresponding phenomenological hadronic parametrizations
\cite{ALEPH,PPms99,KM00}, which are expected to be more precise than
their QCD counterparts. In the work in \cite{GJPPS03}, 
we presented a complete analysis of
this approach, and it was shown that the determination of the strange quark
mass can indeed be significantly improved.

By far the dominant contributions to the pseudoscalar correlators come from
the kaon and the pion, which are very well known. The corresponding parameters
for the next two higher excited states have been recently estimated
\cite{KM02}. Though much less precise, the corresponding contributions to
the $\tau$ sum rule are suppressed, and thus under good theoretical control.
The remaining strangeness-changing scalar spectral function has been extracted
very recently from a study of S-wave $K\pi$ scattering 
\cite{JOP0001,JOP01} in
the framework of chiral perturbation theory with 
explicit inclusion of resonances \cite{EGPR89,Eckeretal}. The resulting scalar
spectral function was then employed to directly determine the strange quark
mass from a purely scalar QCD sum rule \cite{JOP02}. In \cite{GJPPS03} 
we incorporated this contribution into the $\tau$ sum
rule. The scalar $ud$ spectral function is still only very poorly determined
phenomenologically, but it is well suppressed by the small factor $(m_u-m_d)^2$
and can be safely neglected \cite{GJPPS03}.

An average over the most recent of these determinations gives the value 
\cite{BJ03}
\be
m_s(2\gev)=100\pm17 MeV \,.
\ee
The $m_s$ dependence of the hadronic matrix elements necessary to 
calculate the parameters of direct CP violation in kaon decays appears 
when one use GMOR relations \cite{GMOR68} to relate the quark condensate 
in the chiral limit with $m_s$. These GMOR relations \cite{GMOR68} have 
chiral corrections that usually are not taken into account. 
This dependence doesn't appear in the $\Delta I=3/2$ contribution 
to $\varepsilon_K'$ since it can be related to integrals over 
experimental spectral functions.

\subsubsection{The Cabibbo--Kobayashi--Maskawa matrix}

The mass-eigenstates found by 
the diagonalization process are not the same as the weak interaction 
eigenstates. This fact generates extra terms that are conventionally put 
in the couplings of the $W$ gauge boson through the 
Cabibbo--Kobayashi--Maskawa (CKM) matrix ($V_{CKM}=\left(V_{ij}\right)$) 
as follows
\ba
&\dsp
-\frac{g}{2\sqrt{2}}W^-_\mu 
\left(\overline u^\alpha~\overline c^\alpha~\overline t^\alpha\right)
\gamma^\mu\left(1-\gamma_5\right)
\left(\begin{array}{ccc}V_{ud}&V_{us}&V_{ub}\\
                        V_{cd}&V_{cs}&V_{cb}\\
                        V_{td}&V_{ts}&V_{td}\end{array}\right)
\left(\begin{array}{c}d_\alpha\\s_\alpha\\b_\alpha\end{array}\right)
&\nonumber\\ \nonumber&
\dsp-\frac{g}{2\sqrt{2}}W^-_\mu \sum_{\ell=e,\mu,\tau}
\bar \nu_\ell\gamma^\mu\left(1-\gamma_5\right)\ell \,,
\ea
that is, the CKM-matrix connects the weak eigenstates and the mass 
eigenstates. C and P symmetries are broken by the factor 
$(1-\gamma_5)$ in the weak interactions. CP can be broken in this 
sector if $V_{CKM}$ is irreducibly {\emph complex}. With non-zero neutrino 
masses there are analogous mixing effect in the lepton sector. We don't 
consider this possibility in this Thesis, for a review on this fact  
see \cite{GGN02}.

The CKM-matrix $V_{CKM}$ \cite{Cabibbo63,KM73}
is a general complex unitary matrix so, 
in principle, it should depend on 9 real parameters. 
However, part of these independent parameters 
can be eliminated by performing a redefinition of the phases of the 
quark fields. The matrix $V_{CKM}$ thus contains four independent 
parameters, which are usually parametrized as three angles 
($\theta_{12}$, $\theta_{13}$ and $\theta_{23}$) and one phase $\delta_{13}$. 
The Particle Data Group preferred parametrization, the standard 
parametrization, is \cite{PDG02}
\be \label{CKMmatrix}
V_{CKM}=\left(\begin{array}{ccc}
c_{12}c_{13} & s_{12}c_{13} & s_{13}e^{-i\delta_{13}}\\
-s_{12}c_{23}-c_{12}s_{23}s_{13}e^{i\delta_{13}} &
c_{12}c_{23}-s_{12}s_{23}s_{13}e^{i\delta_{13}} & s_{23}c_{13}\\
s_{12}s_{23}-c_{12}c_{23}s_{13}e^{i\delta_{13}} &
-c_{12}s_{23}-s_{12}c_{23}s_{13}e^{i\delta_{13}} & c_{23}c_{13}\,.
\end{array}\right)
\ee
where $c_{ij}=\cos{\theta_{ij}}$ and $s_{ij}=\sin{\theta_{ij}}$, and 
the indices $i,j=1,2,3$ label the three families.

Taking into account the experimental fact that $s_{12}\ll s_{23}\ll
s_{13}$, an approximate parametrization of the CKM-matrix, known as the 
Wolfenstein parametrization \cite{Wolfenstein83}, can be given via the 
change of variables
\be
s_{12}=\lambda,\quad s_{23}=A\lambda^2,\quad s_{13}=A\lambda^2(\varrho -i\eta) 
\ee
To order $\lambda^4$ the CKM-matrix in the Wolfenstein parametrization is 
\be \label{Wolfpar}
\left(\begin{array}{ccc}
1-\frac{\lambda^2}{2} & \lambda & A\lambda^3(\rho-i\eta)\\
-\lambda & 1-\frac{\lambda^2}{2} & A\lambda^2\\
A\lambda^3(1-\rho-i\eta) & -A\lambda^2 & 1
\end{array}\right)\,.
\ee

The classification of different parametrizations can be found 
in \cite{FX00}.

\begin{figure}[hbt]
\begin{center}
\vspace{0.10in}
\centerline{
\epsfysize=2.0in
\epsffile{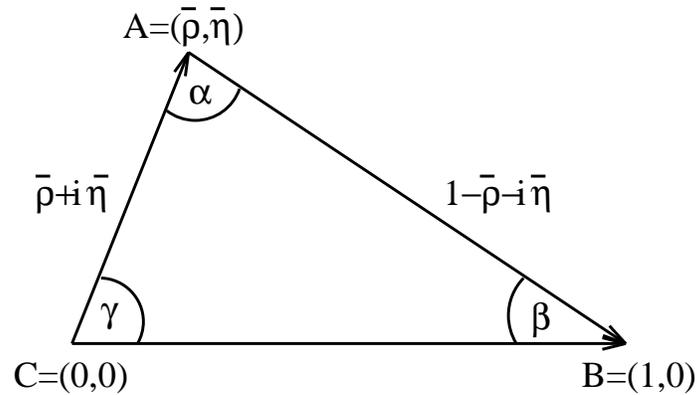}
}
\vspace{0.08in}
\end{center}
\caption{Unitarity Triangle.}\label{Utriangle}
\end{figure}

Using the parametrization in (\ref{Wolfpar}), the CKM-matrix can 
be fully described by $|V_{us}|$, $|V_{cb}|$ and the triangle 
show in Figure \ref{Utriangle}. 
The relation between the parameters depicted in the figure and those 
in  (\ref{Wolfpar}) are
\be
\bar \varrho \equiv \varrho \left(1-\frac{\lambda^2}{2}\right)\,,\quad
\bar \eta \equiv \eta \left(1-\frac{\lambda^2}{2}\right)\,.
\ee
Using trigonometry one can relate the angles $\phi_i=\alpha,\beta,\gamma$ 
with $(\bar\varrho,\bar\eta)$ as follows
\ba \label{anglesdef}
\sin{(2\alpha)}&=&\frac{2\bar\eta(\bar\eta^2+\bar\varrho^2-\bar\varrho)}
{(\bar\varrho^2+\bar\eta^2)((1-\bar\varrho)^2+\bar\eta^2)}\,,\nonumber\\
\sin{(2\beta)}&=&\frac{2\bar\eta(1-\bar\varrho)}
{(1-\bar\varrho)^2+\bar\eta^2}\,,\nonumber\\
\sin{(2\gamma)}&=&\frac{2\bar\eta\bar\varrho}
{\bar\varrho^2+\bar\eta^2}=\frac{2\eta\varrho}{\varrho^2+\eta^2}\,.
\ea
CP violation is given by a non-vanishing value of $\bar\eta$ or 
$\gamma\ne 0,\pi$, what can be predicted within the SM by the 
measurements of CP conserving decays sensitive to $|V_{us}|$,  
$|V_{ub}|$, $|V_{cb}|$ and $|V_{td}|$. The semi-leptonic $K$ and $B$ 
decays are very important in the determination of these CKM-matrix elements. 
For a detailed discussion of the present knowledge of these quantities, 
see \cite{BBG03}. The results can be summarized by 
\be
|V_{us}| = \lambda =  0.2240 \pm 0.0036\,
\quad\quad
|V_{cb}|=(41.5\pm0.8)\cdot 10^{-3},
\ee
\be
\frac{|V_{ub}|}{|V_{cb}|}=0.086\pm0.008, \quad\quad
|V_{ub}|=(3.57\pm0.31)\cdot 10^{-3}.
\ee

Before ending this section we would like to remark some points about 
the calculation of $|V_{us}|$. Recently it has been proposed a new  
route to determine this parameter using hadronic $\tau$ decays 
experimental data \cite{GJPPS03}. This requires the value of the 
strange quark mass, that can be obtained from other sources like 
QCD sum rules or the lattice --see last subsection. The result 
got from this calculation is $V_{us}=0.2179\pm0.0045$, 
where the uncertainty is dominated by the experimental error and 
can thus be improved through and improved measurement of the 
hadronic $\tau$ decay rate into strange particles. A reduction of 
this uncertainty by a factor of two, would let us obtain a value for 
$V_{us}$ more precise than the one given by the current PDG average and, 
eventually, determine both $m_s$ and $V_{us}$ simultaneously \cite{GJPPS03}. 
Such improvement of the precision of the measurements can 
hopefully be achieved by the BaBar and Belle $\tau$ data samples.

\subsection{CP Symmetry Violation in the Standard Model}
\label{SMCPViolation}

CP violation requires the presence of a complex phase and, as we have 
discussed in the last section, the only possible origin of this phase in 
the 3-generation SM is the Yukawa sector. More specifically, if one writes 
the CKM-matrix as in (\ref{CKMmatrix}), CP violation is present if 
$\delta \ne 0,\pi$ or, equivalently, if $\eta \ne 0$ in (\ref{Wolfpar}). 
The CKM mechanism for CP violation requires several necessary conditions. 
All the CKM-matrix elements must be different from zero and the quarks of 
a given charge and different families can not have the same mass. In addition, 
CP can be violated only in processes where the 
three generations are involved. All these conditions can be summarize as 
\cite{Jarlskog85}
\be
{\rm CP\,violation} \,\Longleftrightarrow \im\left(\det
[ \bf{mm^\dagger},\bf{\widetilde m \widetilde m^\dagger}]
\right)\ne 0 \,,
\ee
where $\bf{m}$ and $ \bf{\widetilde m}$ are the original quark-mass matrices. 

From these necessary conditions one can deduce several implications 
of the CKM mechanism of CP violation without performing  
any calculation \cite{PichCPV}. For example, one knows that 
the violations of the CP 
symmetry must be small since the CP-violating 
observables must be proportional to   
a given combination of CKM-matrix elements that is itself small 
\cite{Jarlskog85} 
\be
J_{CP}=\im(V_{ai}V_{bj}V_{aj}^*V_{bi}^*)=\eta A^2\lambda^6
+\order(\lambda^8)\le 10^{-4}\,.
\ee
The transitions most suitable to detect CP violation in them 
are those where the CP-conserving amplitude already suppressed 
by small CKM-matrix elements as $|V_{ub}|$ and $|V_{td}|$. 
The condition that processes violating CP must 
involve the three fermion generations makes the 
$B$ system a better place to look for such kind of effects 
than the kaon system, since in the first one the 3 generations 
enter at tree level and in the second case only at one loop level. 

In fact, CP-violating effects in $B$ decays in a large number of channels 
are expected to be observable in the near future, what constitutes one 
of the main motivation of $B$-factories and another $B$ experiments. 
The most interesting processes are the decays of neutral $B$ into final 
states $f_{CP}$ that are CP-eigenstates. For this kind 
of decays one can define the time dependent asymmetry
\be \label{acpdef}
a_{CP}(t,f)=\frac{\Gamma[B^0(t)\to f_{CP}]-\Gamma[\bar B^0(t)\to f_{CP}]}
{\Gamma[B^0(t)\to f_{CP}]+\Gamma[\bar B^0(t)\to f_{CP}]}\,.
\ee
These asymmetries are generated via the interference of mixing and 
decays (see Chapter \ref{CPviolation} for definitions). 
In the case when a single mechanism dominates the decay amplitude 
or the different mechanisms have the same weak phases, the hadronic 
matrix elements and strong phases drop out and the asymmetry is given by 
\cite{Krawczyk88}
\be \label{acpB}
a_{CP}(t,f)=-\sin{(2\beta)}\sin{(\Delta Mt)}\,,
\ee
where $\Delta M\equiv M_{B_H}-M_{B_L}$ is the difference of masses 
between the mass eigenstates --see for example \cite{Buras03} for 
a definition of such states--
and $\beta$ is one the angles in the unitarity triangle in Figure 
\ref{Utriangle}. The CP-violating asymmetry (\ref{acpB}) thus 
provides a direct and clean measurement of the angle $\beta$ and also of 
the CKM-matrix elements through their relation with $\beta$ 
--see equation (\ref{anglesdef}).

The experimental determination of $\sin{(2\beta)}$ have been 
improved considerably in the 
last four years by the measurements of the time dependent asymmetry 
in (\ref{acpdef}) for the decay $B_d\to\psi K_S$ 
\be
a_{\psi K_s}(t)\equiv=-a_{\psi K_s}\sin{(\Delta M_d t)}
=-\sin{(2\beta)}\sin{(\Delta M_d t)}\,
\ee
in the B-factories. Previously, The $\beta$ parameter had been measured 
by LEP and CLEO.

The last data from BaBar \cite{BaBar} and Belle \cite{Belle} collaborations 
give the results
\be
(\sin{( 2\beta)})_{\psi K_S}=\left\{
\begin{array}{ll}
0.741\pm 0.067 \, \mbox{(stat)} \pm0.033 \, \mbox{(syst)} & \mbox{(BaBar)}\\
0.719\pm 0.074 \, \mbox{(stat)} \pm0.035  \, \mbox{(syst)} & \mbox{(Belle).}
\end{array}
\right.
\ee
that combined with the earlier measurements by CDF, ALEPH and OPAL give 
the average \cite{Nir03}
\be \label{sinbetaexp}
(\sin{( 2\beta)})_{\psi K_S}=0.734\pm0.054 \,.
\ee

The experimental measurement of $\sin{( 2\beta)}$ in (\ref{sinbetaexp}) 
agrees very well with the result obtained within the SM 
$\sin{( 2\beta)}=0.705^{+0.042}_{-0.032}$
\cite{Buras03}, this indicates that the CKM mechanism 
is suitable to be the dominant source of CP violation in flavour violating 
decays. Notice, however, that recently have been found the first 
discrepancies  with the CKM mechanism in the measure of $\sin(2\beta)$ in the 
penguin loop dominated decay modes $B^0 \to \phi K_S,\eta' K_S$ 
\cite{Belle2,BaBar2}. The deviation from the Standard 
Model is about 2.5$\sigma$ \cite{abe03}.

For a more detailed discussion about CP violation in $B$ decays within 
the SM see \cite{BJ03} and references therein, 
and for analysis in models beyond SM 
see, for example, \cite{Nir03}. The violation of the CP symmetry in kaons, 
that is the main subject of this Thesis, is treated more extensively 
in the next chapters.

%\chapter{CP Violation in the Kaon System} \label{CPviolation}
\chapter{CP Violation in the Kaon System} \label{CPviolation}

The interference between various amplitudes that carry complex phases 
contributing to the same physical transition is always needed to generate the 
CP-violation observable effects. These effects can be classified into three 
types
\begin{itemize}
\item CP Violation in Mixing
\item CP Violation in Decay
\item CP Violation in the interference of Mixing and Decay
\end{itemize}

In this chapter we are going to review the main definitions 
and CP-violating 
observables in the kaon system we will discuss later, 
giving useful formulas for them.

\section{CP Violation in $K^0-\bar K^0$ Mixing: Indirect CP Violation}

The $K^0$ and $\bar K^0$ states have strangeness equal to -1 and 1 
respectively, 
as their quark content is $\bar s d$ and $\bar d s$. These states have no a 
definite value of the CP parity, 
but they transform one into another under the action of this transformation 
in the next way \footnote{Since the flavour quantum number is conserved by 
strong interactions, there is some freedom in defining the phases of the 
flavour eigenstates. In general, one could use 
\be
|K_\phi^0\rangle\equiv e^{-i\phi}|K^0\rangle,\quad \quad
|\bar K_\phi^0\rangle\equiv e^{i\phi}|\bar K^0\rangle \,,
\ee
which under CP symmetry are related by $CP|K_\phi^0\rangle 
=-e^{-2i\phi}|\bar K_\phi^0\rangle$ . Here we use the phase convention 
implicit in (\ref{phaseconv}).
}
\be \label{phaseconv}
CP|K^0\rangle =-|\bar K^0\rangle\,.
\ee
We can construct eigenstates with a definite CP transformation by 
combining 
$K^0$ and $\bar K^0$
\ba
&&K_1\,=\,\frac{1}{\sqrt 2}\left(K^0-\bar K^0\right)\,,\quad \,
K_2\,=\,\frac{1}{\sqrt 2}\left(K^0+\bar K^0\right)\, ,\nonumber\\
&&\hspace{2 cm}CP|K_{1(2)}\rangle =+(-)|\bar K_{1(2)}\,\rangle \, .
\ea

The strange particles can decay only via weak interactions as 
strong and electromagnetic interactions preserve the strangeness quantum 
numbers. If we assume that weak interactions are symmetric under CP violation 
as strong and electromagnetic interactions are, then the $K_{1(2)}$ states 
must decay into an state with even(odd) CP parity. Taking into account that 
the main decay mode of $K^0$-like states is $\pi\pi$ and the fact that a two 
pion state with charge zero in spin zero is always CP even, the decay 
$K_1\rightarrow\pi\pi$ is possible (as well as $K_2\rightarrow\pi\pi\pi$) but 
$K_2\rightarrow\pi\pi$ is impossible. However, in 1969 it was observed 
the decay of $K_L$ mesons, that were identified with $K_2$, 
in states of two pions \cite{CCFT64}. 
This meant that or the physical $K_L$ were not purely CP eigenstates but 
the result of a mixing between both CP odd $K_2$ and CP even $K_1$, 
or that these transitions directly violated CP since an odd state 
decayed into an even state. 

Assuming CPT symmetry to hold, the $K^0\bar K^0$ system, seen as a two state 
system, can be described by the Hamiltonian
\be
i\frac{d}{dt}\left(\begin{array}{c}K^0 \\ \bar K^0 \end{array}\right )=
\left(\begin{array}{cc}M_{11}-\frac{i}{2}\Gamma_{11}&
M_{12}-\frac{i}{2}\Gamma_{12}\\ M_{21}-\frac{i}{2}\Gamma_{21}& 
M_{22}-\frac{i}{2}\Gamma_{22}\end{array}\right ) \left(\begin{array}{c}
K^0\\ \bar K^0\end{array}\right )
\ee
where $M\,=\,M_{ij}$ and $\Gamma\,=\,\Gamma_{ij}$ are hermitian matrices. 
However, the Hamiltonian itself is allowed to have a non-hermitian part since
the probability is not conserved. The kaons can decay and the anti-hermitian 
part $\Gamma$ describes the decays of the kaons into states out of this 
system. 

If one impose CPT, not all the components in the mixing matrix are free 
(see Ref. \cite{rafaelTASI} for a derivation)
\ba
M_{11}=M_{22} \,,\quad && \quad \,\Gamma_{11}=\Gamma_{22}\, ,\nonumber\\
M_{12}=M_{21}^* \,,\quad && \quad \,\Gamma_{12}=\Gamma_{21}^*\,.
\ea

The physical propagating eigenstates of the Hamiltonian, obtained by 
diagonalizing the mixing matrix, are
\be
|K_{S(L)}\rangle \,=\,\frac{1}{\sqrt{1+|\bar \varepsilon_K|^2}}\,
\left(|K_{1(2)}\rangle+\bar \varepsilon_K |K_{2(1)}\rangle\right)
\ee
with the parameter $\bar \varepsilon_K$ defined by
\be
\frac{1-\bar \varepsilon_K}{1+\bar \varepsilon_K} \,=\,\left(\frac{M_{12}^*-
\frac{i}{2}\Gamma_{12}^*}{M_{12}-\frac{i}{2}\Gamma_{12}}\right)\,.
\ee

If $M_{12}$ and $\Gamma_{12}$ were real, $\bar \varepsilon_K$ would vanish 
and the states $|K_{S(L)}\rangle$ would correspond to the CP-even(odd) 
$|K_{1(2)}\rangle$ states. If this is not true and CP is violated, both states 
are no longer orthogonal 
\be
\langle K_L|K_S\rangle \, \approx \, 2\rm{Re}(\bar \varepsilon_K)\,.
\ee
The parameter $\bar \varepsilon_K$ depends on the phase convention chosen for 
$K^0$ and $\bar K^0$. Therefore it may not be taken as a physical measure of 
CP violation. On the other hand, $\re (\bar \varepsilon_K)$ is independent on 
phase conventions and can be measured in semi-leptonic decays via the ratio
\be
\delta \equiv \frac{\Gamma[K_L\rightarrow \pi^-l^+\nu_L]-
\Gamma[K_L\rightarrow \pi^+l^-\nu_L]}{\Gamma[K_L\rightarrow \pi^-l^+\nu_L]+
\Gamma[K_L\rightarrow \pi^+l^-\nu_L]}\,=\,\frac{2\rm{Re}(\bar \varepsilon_K)}
{(1+|\bar\varepsilon_K|^2)}\,.
\ee

$\delta$ is determined purely by the quantities related to $K^0-\bar K^0$ 
mixing. Specifically, it measures the difference between the phases of 
$\Gamma_{12}$ and $M_{12}$.

\section{$\varepsilon_K$ in the Standard Model}
\label{scepsilonk}

Between the various components relevant for the determination of 
$\varepsilon_K$ -see (\ref{epsilonfin})-, the most accurately known is 
the $K_L-K_S$ mass difference $\Delta m$. Its experimental value is 
\cite{PDG02}
\be
\Delta m \equiv m_{K_L}-m_{K_S}=(3.490\pm0.006)\times10^{-12}\mev
\ee
assuming CPT to hold.

The term proportional to the ratio $\im a_0/\re a_0$ in (\ref{epsilonfin}) 
constitutes a small correction to $\varepsilon_K$. Its value is 
analyzed in Section \ref{secq6}.

The off-diagonal element $M_{12}$ in the neutral kaon mass matrix represents 
the $K^0-\bar K^0$ mixing and its short-distance contributions comes from the 
effective $\Delta S=2$ Hamiltonian \cite{BJW90,UKJS98} obtained once the 
heaviest particles (top, W, bottom and charm) have been integrated out
\be \label{hamiltonianS2}
H_{eff}^{\Delta S=2} = C_{\Delta S=2}(\mu)\int d^4x \, Q_{\Delta S=2}(x)\, .
\ee
The four-quark operator $Q_{\Delta S=2}$ is the product of two left handed 
currents
\be \label{operatorDS2}
Q_{\Delta S=2}(x) = 4\,\left[\bar s_{\alpha}\gamma_{\mu}
d_{\alpha}\right]_L(x)
\left[\bar s_{\beta}\gamma^{\mu}d_{\beta}\right]_L(x)
\ee
with $(\bar q\gamma_\mu q')_{L}=\frac{1}{2}
\bar q \gamma_\mu(1- \gamma_5)q'$.  The function 
$C_{\Delta S=2}(\mu)$ depend on 
the Cabibbo-Kobayashi-Maskawa (CKM) matrix elements, top and charm masses, W, 
boson mass, and some QCD factor collecting the running of the Wilson 
coefficients between each threshold appearing in the process of integrating 
out the heaviest particles. It is scale and scheme dependent and can be 
written as \cite{Herrlich96}
\ba
 C_{\Delta S=2}(\mu) & = & \frac{G_F^2M_W^2}
{16\pi^2}\left[\lambda_c^2\eta_1
S_0(x_c)+\lambda_t^2\eta_2S_0(x_t)+2\lambda_c\lambda_t\eta_3
S_0(x_c,x_t)\right]
\nonumber\\
&&\times \alpha_S^{-2/9}(\mu)\left(1+\frac {\alpha_S(\mu)}{4\pi}J_3\right)
\ea
with
\be
x_c=\frac{m_c^2}{M_W^2} \, ,\quad
x_t=\frac{m_t^2}{M_W^2} \, , \quad
\lambda_q=-V_{qd}V_{qs}^* \,.
\ee
The scale dependence is encoded in the $\mu$ dependence of the strong 
coupling $\alpha_S(\mu)$ and it is given by the 
value of the parameter $J_3$. $J_3$ also depend on the 
renormalization scheme and it is known in NDR and 
HV schemes \cite{Herrlich96}. 
The parameters $\eta_i$ are functions of the heavy quark masses and 
are independent on 
the renormalization scheme and scale of the operator $Q_{\Delta S=2}$. 
The first calculation of $\eta_1$ and $\eta_2$ at NLO 
are in \cite{HN94} and \cite{BJW90} respectively. $\eta_3$ and updated 
values of $\eta_1$ and $\eta_2$ can be found in \cite{Herrlich96} and 
the first explicit expressions for the functions $S_0(x_q)$ 
and $S_0(x_c,x_t)$ in \cite{Inami81}. 

The matrix element between $\bar K^0$ and $K^0$ of the 
hamiltonian in (\ref{hamiltonianS2}) is parametrized by the 
so called $\hat B_K$ parameter, that is defined by 
\be \label{bkdef}
C_{\Delta S=2}(\mu)\langle \bar K^0|\int d^4x \,Q_{\Delta S=2}(x)|K^0
\rangle\equiv\frac{16}{3}f_K^2 m_K^2 \hat B_K\,.
\ee
Here, $f_K$ denotes the $K^+\to \mu^+\nu$ coupling 
($f_K=113\,MeV$) and $m_K$ 
is the $K^0$ mass. 
The quantity $\hat B_K$ defined in (\ref{bkdef}) is scale and scheme 
independent, what means that the scale and scheme dependences of both 
the coefficient $C_{\Delta S=2}(\mu)$ and the matrix element 
$\langle \bar K^0|\int d^4x \,Q_{\Delta S=2}(x)|K^0
\rangle$ must cancel against each other at a given order. 

This matrix element enters in the \emph{indirect} 
CP violating parameter $\varepsilon_K$. It is an important input 
for the analysis of one of the Cabibbo-Kobayashi-Maskawa (CKM) 
unitarity triangles -see \cite{BBG03,ckmandbk} for more information. 
More details about the calculation of $\hat B_K$ 
are given in Section \ref{scbk}.

\section{CP Violation in the Decay: Direct CP Violation} \label{DirectCPV}

Any observed difference between a decay rate $\Gamma(P\rightarrow f)$ and 
the CP conjugate $\Gamma(\bar P\rightarrow \bar f)$ would indicate that CP is 
directly violated in the decay amplitude.
% lo siguiente es prácticamente copiado

We are going to suppose that the amplitudes of the transitions $P\rightarrow f$
 and  $\bar P\rightarrow \bar f$ have two interfering amplitudes
\ba 
A(P\rightarrow f)\, = \quad M_1 e^{i\phi_1} e^{i\alpha_1}&
+&M_2 e^{i\phi_2} e^{i\alpha_2}\,,\nonumber\\
A(\bar P\rightarrow \bar f)\,  =\,\,  M_1 e^{-i\phi_1} e^{i\alpha_1}
&+& M_2 e^{-i\phi_2} e^{i\alpha_2}\,,
\ea
where $\phi_i$ are weak phases, $\alpha_i$ strong final-state phases and 
$M_i$ real moduli of the matrix elements. The asymmetry can be written as
\be
\frac{ \Gamma(P\rightarrow f)-\Gamma(\bar P\rightarrow \bar f)}
{\Gamma(P\rightarrow f)+\Gamma(\bar P\rightarrow \bar f)}\, = \,
\frac{-2M_1 M_2 \sin(\phi_1-\phi_2)\sin(\alpha_1-\alpha_2)}
{|M_1|^2+|M_2|^2+2M_1M_2\cos(\phi_1-\phi_2)\cos(\alpha_1-\alpha_2)}.
\ee
From this equation one can deduced that to have a non-zero value of the 
asymmetry the next requirements are necessary
\begin{itemize}
\item At least two interfering amplitudes
\item Two different weak phases
\item Two different strong phases
\end{itemize}
In addition, in order to get a sizable asymmetry, the two amplitudes $M_1$ 
and $M_2$ should be of comparable size. Note that the value of the asymmetry 
are related only to differences of phases not to the phases themselves as they 
are convention dependent.

When direct CP violation is studied in the decay of neutral kaons 
where also 
$K^0-\bar K^0$ mixing is involved, both direct and indirect CP violation
effects need to be considered. 

We can define the following observables
\ba \label{CPobservables}
\eta_{+-} &\equiv& \frac {A[K_L\rightarrow \pi^+ \pi^-]} 
{A[K_S\rightarrow \pi^+ \pi^-]}\, ,\nonumber\\
\eta_{00} &\equiv& \frac {A[K_L\rightarrow \pi^0 \pi^0]} 
{A[K_S\rightarrow \pi^0 \pi^0]}\, ,\nonumber\\
\varepsilon_K & \equiv & \frac {A[K_L\rightarrow (\pi \pi)_{I=0}]} 
{A[K_S\rightarrow (\pi \pi)_{I=0}]}\nonumber\\
\ea
and
\ba \label{epsilonprime1def}
\frac{\sqrt{2}\varepsilon_K'}{\varepsilon_K}=
\frac{A\left[K_L\!\to\!(\pi\pi)_{I=2}\right]}
{A\left[K_L\!\to\!(\pi\pi)_{I=0}\right]}-
\frac{A\left[K_S\!\to\!(\pi\pi)_{I=2}\right]}
{A\left[K_S\!\to\!(\pi\pi)_{I=0}\right]}\, .
\ea
In the latter definition the transition $K^0-\bar K^0$ has been removed, so 
the parameter $\varepsilon_K'$ is related to direct CP violation only. 
% as one 
%can deduce from equation (\ref{})????. 
The parameter 
$\varepsilon_K$, in the other hand, is related to indirect CP violation. 
Since they are ratios of decay rates $|\varepsilon_K|$, $|\eta_{+-}|$ and 
$|\eta_{00}|$ are directly measurable.

The decay amplitudes of a kaon into a system of two pions can be expressed 
in the isospin symmetry limit in terms of amplitudes with definite isospin 
[$A\equiv -iT$]; 
\ba \label{A0A2def}
i \, A[K^0\to \pi^0\pi^0] &\equiv& {\frac{a_0}{\sqrt 3}} \, 
 e^{i\delta_0}
-\frac{ 2 \, a_2}{\sqrt 6} \, e^{i\delta_2} \, , \nonumber \\
i \, A[K^0\to \pi^+\pi^-] &\equiv& {\frac{a_0}{\sqrt 3}} \, 
 e^{i\delta_0}
+\frac{a_2}{\sqrt 6} \, e^{i\delta_2}\, , \nonumber\\
i \, A\left[K^+ \rightarrow \pi^+\pi^0\right] &\equiv& 
\frac{\sqrt{3}}{2} a_2\, e^{i\delta_2}\, ,
\ea
with $\delta_I$ the final state interaction
(FSI) phases that can be used together with  the amplitudes $a_I$ to 
rewrite the parameters in (\ref{CPobservables}). 
In doing it we make the next 
approximations, experimentally valid,
\ba
|\im a_0|,|\im a_2| & << & |\re a_2|\,<<\,|\re a_0| \, ,\nonumber\\
|\varepsilon_K|,|\bar \varepsilon_K| & << & 1 \, , \nonumber\\
|\varepsilon_K'| & << & |\varepsilon_K|\, ;
\ea
and obtain the next expression for the two main parameters of CP Violation
\be \label{epsilonprimedef}
\varepsilon_K' \, \simeq\, \frac{i}{\sqrt 2}
\frac{\re  a_2 }{\re  a_0}\left[\frac{\im  a_2}{\re  a_2}
-\frac{\im  a_0}{\re  a_0} \right]e^{i(\delta_2-\delta_0)}
\ee
and
\be \label{epsilondef}
\varepsilon_K \,= \, \bar \varepsilon_K + i\frac{\im  a_0}{\re  a_0}
\ee
%% lo siguiente es muy parecido al original de  Bijnens about QCD and Weak 
%% Interactions of Light Quarks.
The latter can be rewritten using the facts that $\Delta m=m_L-m_S\approx 
\frac{\Delta \Gamma}{2}$, $\Gamma_L<<\Gamma_S$ and $\Gamma_{12}$ is dominated 
by $\pi\pi$ states
\be \label{epsilonfin}
\varepsilon_K \,=\,\frac{1}{\sqrt{2}} e^{i\pi/4}\left(\frac {\im(M_{12})}
{\Delta m}+\frac{\im a_0}{\re a_0}\right)\,.
\ee
All the above information let us relate the observables defined in 
(\ref{CPobservables}) and (\ref{epsilonprime1def}) in the next way
\be
\eta_{+-}\,=\,\varepsilon_K\,+\, \varepsilon_K'\,\quad \rm{and} \quad \,
\eta_{00}\,=\,\varepsilon_K\,-\,2 \varepsilon_K'\,.
\ee

%% poner ahora cómo se ve lo que son epsilon y epsilon'. Mirar lo del israeli

The ratio  $\varepsilon_K '/\varepsilon_K$ is measured experimentally via the 
double ratio (see equations (\ref{CPobservables}) and (\ref{epsilonprime1def}))
\be
\rm{Re}\left(\frac{\varepsilon_K'}{\varepsilon_K}\right)=\frac{1}{6}
\left\lbrace 1-\frac{\Gamma\left[K_L\rightarrow\pi^+\pi^-\right]/
\Gamma\left[K_S\rightarrow\pi^+\pi^-\right]}
{\Gamma\left[K_L\rightarrow\pi^0\pi^0\right]/
\Gamma\left[K_S\rightarrow\pi^0\pi^0\right]}\right\rbrace \, .
\ee

\section{$\varepsilon_K'$ in the Standard Model}
 \label{imai}

The first measure of a non-vanishing value of the parameter $Re\left(\frac
{\varepsilon_K '}{\varepsilon_K}\right)$ 
defined in (\ref{epsilonprimedef}) and (\ref{epsilondef}) was performed in 
1988 \cite{directCPV}. For a long time the experimental situation was unclear 
since two different experiments, NA31 at CERN and E731 at FNAL, obtained 
conflicting results at the end of the 1980's. This situation has been 
clarified by improved versions of these two experiments, NA48 at CERN and KTeV 
in FNAL. The last values measured by both of them are

\be
\rm{Re}\left.\left(\frac{\varepsilon_K'}{\varepsilon_K}\right)\right|_{exp}
=\left\lbrace \begin{array}{c}
(1.47\pm0.22)\times 10^{-3}\quad NA48 \,\cite{NA4802}\,\\
(2.07\pm0.28)\times 10^{-3}\quad KTeV \,\cite{KTEV03}
\end{array}\right.
\ee
In combination with previous results \cite{oldexperiments}, the present world 
average is
\be
\rm{Re}\left.\left(\frac{\varepsilon_K'}{\varepsilon_K}\right)\right|_{exp}
=(1.66 \pm 0.16)
\cdot 10^{-3} \,.
\ee

Let us analyze the values of the various terms in (\ref{epsilonprimedef}). 
The ratio $\rm{Re} a_0/\rm{Re} a_2= 21.8$ \cite{BDP03} 
is an experimentally 
well known quantity and reflects the $\Delta I=1/2$ rule. The smallness of the 
ratio suppresses $\varepsilon'_K/\varepsilon_K$.

The phase of $\varepsilon_K'$ is also a model independent 
quantity that can be determined from hadronic parameters and its value is
$\rm{arg}(\varepsilon_K')=\pi/2+\delta_2-\delta_0\approx \pi/4$. The relation 
between the phases of the two main parameters of CP violation is the kaon 
system is, accidentally, $arg(\varepsilon_K')\approx arg(\varepsilon_K)$ what 
means that
\be\label{realepsilon}
\re \left(\varepsilon_K'/\varepsilon_K\right)\,\approx\,
\varepsilon_K'/\varepsilon_K \,.
\ee

In a theoretical calculation of the direct CP-violation parameter 
$\varepsilon_K'$ the ratio of the amplitudes  
$\re  a_0/\re  a_2$ and $\varepsilon_K$ are usually set to their 
experimental values, so the only quantities we need to evaluate 
theoretically are $\im (a_I)/\re(a_I)$.

The calculation of $K\to\pi\pi$ amplitudes is a several 
step process in the Standard Model. Above the electroweak scale, the 
usual gauge-coupling perturbative expansion let one analyze the 
flavour-changing process in terms of quarks, leptons and gauge bosons in 
a well established way. The first step to calculate $K\to\pi\pi$ amplitudes 
consists in integrating out 
the heavy particles, top, Z and W, 
replacing the effects of their exchanges by an effective Hamiltonian given by
\be \label{hamiltonians1}
H_{eff}^{\Delta S =1}=\frac{G_F}{\sqrt 2}V_{ud}V_{us}^*\sum_{i=1}^{10}
C_i(\mu)Q_i(\mu) \, ,
\ee
where the $C_i(\mu)$ are Wilson coefficients containing information on the 
heavy fields that have been integrated out and the 10 four-quark operators 
constructed with the light degrees of freedom are 
\ba \label{operators}
 Q_1 &=& (\bar s_\alpha\gamma_\mu u_\beta)_L
(\bar u_\beta\gamma^\mu d_\alpha)_L \, ,\nonumber\\
 Q_2 &=& (\bar s_\alpha\gamma_\mu u_\alpha)_L
(\bar u_\beta\gamma^\mu d_\beta)_L \, ,\nonumber\\
 Q_3 &=& (\bar s_\alpha\gamma_\mu d_\alpha)_L
\sum_{q=u,d,s}(\bar q_\beta\gamma^\mu q_\beta)_L \, ,\nonumber\\
 Q_4 &=& (\bar s_\alpha\gamma_\mu d_\beta)_L
\sum_{q=u,d,s}(\bar q_\beta\gamma^\mu q_\alpha)_L \, ,\nonumber\\
 Q_5 &=& (\bar s_\alpha\gamma_\mu d_\alpha)_L
\sum_{q=u,d,s}(\bar q_\beta\gamma^\mu q_\beta)_R \, ,\nonumber\\
 Q_6 &=& (\bar s_\alpha\gamma_\mu d_\beta)_L
\sum_{q=u,d,s}(\bar q_\beta\gamma^\mu q_\alpha)_R \, ,\nonumber\\
 Q_7 &=& (\bar s_\alpha\gamma_\mu d_\alpha)_L
\sum_{q=u,d,s}\frac{3}{2}e_q(\bar q_\beta\gamma^\mu q_\beta)_R \, ,\nonumber\\
 Q_8 &=& (\bar s_\alpha\gamma_\mu d_\beta)_L
\sum_{q=u,d,s}\frac{3}{2}(\bar q_\beta\gamma^\mu q_\alpha)_R \, ,\nonumber\\
 Q_9 &=& (\bar s_\alpha\gamma_\mu d_\alpha)_L
\sum_{q=u,d,s}\frac{3}{2}e_q(\bar q_\beta\gamma^\mu q_\beta)_L \, ,\nonumber\\
 Q_{10} &=& (\bar s_\alpha\gamma_\mu d_\beta)_L
\sum_{q=u,d,s}\frac{3}{2}(\bar q_\beta\gamma^\mu q_\alpha)_L \, ,
\ea
with $(\bar q\gamma_\mu q')_{(L,R)}=\bar q \gamma_\mu(1\mp \gamma_5)q'$; 
$\alpha$ and $\beta$ are colour indices and $e_q$ are the quark charges 
($e_u=2/3$,$e_d=e_s=-1/3$).

The unitarity of the CKM matrix allows to write the Wilson coefficients in 
terms of real functions $z_i(\mu)$ and $y_i(\mu)$ and the CKM
 matrix elements $\tau=-V_{ts}^*V_{td}/(V_{us}^*V_{ud})$ in the 
next way
\be
C_i(\mu)=z_i(\mu)+\tau y_i(\mu) \, .
\ee
The CP-violating decay amplitudes are proportional to the $y_i$ components 
and $\im(\tau)$. For three fermion generations the Cabibbo-Kobayashi-Maskawa 
(CKM) matrix is described by 3 angles an 1 phase. This is the only complex 
phase in the Standard Model, thus, it is a unique source 
for violations of the CP symmetry. 

One of the advantages of having a formulation like this is that one can 
separate trough the scale $\mu$ the perturbative effects enclosed in the 
Wilson coefficients $C_i(\mu)$ and the 
non-perturbative effects contained in the matrix elements 
of the operators $Q_i$. 
The coefficients $z_i$ and $y_i$ are calculated 
by equating the matrix elements between quarks and gluons 
of the effective hamiltonian in (\ref{hamiltonians1}) and the same 
matrix elements evaluated in the Standard Model . 
The Wilson coefficients are known at NLO \cite{NLOWilscoef,CFMR94}.

From this hamiltonian we can go down in energy until an 
hadronic scale using 
the renormalization group evolution equations to change the scale of the 
Wilson coefficients. In this second step one resumes large 
logarithms containing heavy masses. An introductory review of this method 
is \cite{Buras98,HansQCDweak} and a review with numerical 
results for all the Wilson coefficients is \cite{BBL96}.

The last step is to take the wanted hadronic 
matrix elements of the operators in 
(\ref{operators}) at an scale $\mu$ low enough to avoid 
large logarithms of the type ${\rm ln}(m_K^2/\mu^2)$. 
The Wilson coefficients $y_i$ and $z_i$ depend on 
this scale $\mu$ and on the definitions of the $Q_i$, so this 
dependences must be consistently accounted for in the evaluation 
of the matrix elements in 
order to have physical quantities without any scale or scheme dependence. 
This is not a trivial goal and we will speak about the different ways of 
doing it in Section \ref{matrixelements}.

All the operators in (\ref{operators}) enter in the evaluation of  
$\varepsilon_K'/\varepsilon_K$, but numerically the contribution 
with $\Delta I=3/2$ (the ratio $\im a_2/\re a_2$ 
in (\ref{epsilonprimedef}))
is dominated by the matrix element of the electroweak penguin $Q_8$ 
and the $\Delta I=1/2$ contribution (the ratio $\im a_0/\re a_0$ 
in (\ref{epsilonprimedef})) by the matrix 
element of  the QCD penguin $Q_6$. The former contribution 
is suppressed by isosping breaking, i.e., strong isospin violation 
($m_u\ne m_d$) and electromagnetic effects. The strong isospin violation 
was traditionally parametrized by 
\be
\Omega _{IB}=\left(\frac{\re a_0}{\im a_0}\right)^{(0)}
\frac{\im a_2^{IB}}{\re a_2^+}\,,
\ee
where the superscript $(0)$ means that these amplitudes are in the isospin 
limit. For the definition of the amplitude $a_2^+$, see \cite{isosbreak}. 
The 
imaginary amplitude $\im a_2$ is first order in isospin breaking and 
we can split it (in an scheme dependent way)
 in the electroweak penguin contribution -coming from 
the operators $Q_{7-10}$ in (\ref{operators})- and the 
isospin breaking contribution generated by other four quark operators. 
The last one, that corresponds to $\im a_2^{IB}$, is 
dominated by the fact that $\pi^0$, 
$\eta$ and $\eta'$ mix. Originally this effect was estimated to be 
$\Omega_{IB}^{\pi^0-\eta}=0.25$ \cite{IBold}. The most recent 
calculation lower the total value of the isospin breaking effects
 to be \cite{isosbreak}
\be
\Omega_{{\rm eff}}=0.060\pm0.077 \,;
\ee
which increases the estimate of $\varepsilon_K'/\varepsilon_K$. 
The quantity $\Omega_{{\rm eff}}$ includes all effects to leading order in 
isospin breaking and it generalizes the parameter $\Omega_{IB}$ 
\cite{isosbreak}.
 
The main uncertainty in the theoretical calculation 
of $\varepsilon_K'$ comes from this isospin breaking parameter 
and, further away, from the calculation of the hadronic matrix elements 
$\langle  Q_i\rangle_I(\mu)$ defined as $\langle(\pi\pi)_0|Q_i|K\rangle$ 
with $I=0,2$, to which the ratios $\im(a_I)/\re(a_I)$ are 
proportional. The contributions to $\varepsilon_K'$ from these two ratios, 
i.e., with $I=0,2$ tend to cancel each other, so an accurate determination 
of both of them turn to be necessary.

One can write another effective field theory bellow the resonance region using 
global symmetry considerations only. This is the Chiral Perturbation Theory 
of the Standard Model (CHPT) that is discussed in Chapter 
\ref{chCHPT}. The theory is defined in terms of the Goldstone bosons 
$(\pi,K,\eta)$ and is organized in powers of momenta and masses of the light 
mesons according to chiral symmetry. It can be used to make predictions for 
the CP--violating observables described in Section \ref{DirectCPV}. 
The operators appearing in the chiral 
lagrangians at each order in momenta are fixed only by symmetry requirements, 
but the chiral couplings modulating each of these operators are not. The 
calculation of such couplings can be done using short-distance effective 
hamiltonians as the one in (\ref{hamiltonians1}) by performing the matching 
between the two effective field theories taking the same hadronic matrix 
elements of both groups of hamiltonians, so it turns to be very important 
to have accurate determinations of these matrix elements.

\subsection{$\Delta I =\frac{1}{2}$ Contribution} \label{secq6}

The LO chiral lagrangian in (\ref{lagdS1}), which is explained in 
Chapter \ref{chCHPT}, let us make the next prediction 
for the $\Delta I =\frac{1}{2}$ contribution to the ratio 
$\varepsilon_K'/\varepsilon_K$
\be \label{delta12}
\left(\frac{\im a_0}{\re a_0}\right)^{LO} 
\simeq \frac{\im G_8}{\re G_8 + G_{27}/9} \, , 
\ee
where we disregard the corrections proportional to $\re(e^2G_E)$ and  
$\im(e^2G_E)$, and take the isospin limit $m_u=m_d$; in order to 
be able to deal with all the first order isospin 
breaking corrections by using the parameter $\Omega_{eff}$ 
\cite{isosbreak} in (\ref{ratio32}).

The chiral corrections to (\ref{delta12}) can be introduced as follows
\be \label{ratio12}
\frac{\im a_0}{\re a_0}=
\left(\frac{\im a_0}{\re a_0}\right)^{LO} 
{\cal C}_0 \, .
\ee
The value of the $C_0$ factor is given in Section \ref{secepsilonprime}. 

The theoretical calculation of the ratio ${\im a_2}/
{\re a_2}$ has its main source of error in the value of $\im G_8$. It 
has been seen,  
%\cite{}???bertolini et al, 
even using only vacuum insertion approximation 
(VIA), that this imaginary coupling 
is dominated by the hadronic matrix element of the operator $Q_6$, 
although all the operators in (\ref{operators}) contribute to it in 
a less determinant way. 

In Section \ref{LOcounterterms} we discuss the 
different results for $\im G_8$ that exist in the literature, 
however, we would like to pay more attention to the determination in 
\cite{epsprime} whose results we update in \cite{BGP03}. 
The two basic technical ingredients in that calculation, namely, the 
X-boson method and the short-distance matching, are the same as those 
used in the determination of the $\Delta I =\frac{3}{2}$ contribution 
in \cite{BGP01} and are described in Section \ref{chq7q8}. 
In general in \cite{BGP01,epsprime,BP95bk,BP99,scheme}, the two-point function
\ba
\label{twopointcal}
{\bf \Pi}_{ij}(q^2) &\equiv & i \int {\rm d}^4 x \, e^{iq \cdot x}
\, \langle 0 | T \left( P_{i}^\dagger(0) P_j(x) e^{i {\bf \Gamma_{LD}}}
 \right)| 0 \rangle
\ea
is computed in the presence of the long-distance effective
action of the Standard Model ${\bf \Gamma_{\rm LD}}$. 
The pseudo-scalar sources $P_i(x)$ have the appropriate quantum
numbers to describe $K\to\pi$ transitions. 
The effective action ${\bf \Gamma_{LD}}$
reproduces the physics of the SM at low energies by the exchange of 
colorless heavy  X-bosons.  To obtain it one must make  a
short-distance matching analytically, which takes into account exactly the
short-distance scale and scheme dependence.
 We are left with the couplings 
of the  X-boson long-distance effective action
completely fixed in terms of the Standard Model ones.
 This action is regularized with a four-dimensional
 cut-off, $\mu_C$. The X-boson
effective action has the technical advantage to separate the short-distance
of the two-quark currents or densities from the purely four-quark
short-distance which is always only logarithmically divergent and 
regularized by the X boson mass in our approach.
 The cut-off $\mu_C$ only appears in the short-distance of  the
two-quark currents or densities and can be thus taken into account 
exactly.

 Taylor expanding the two-point function  (\ref{twopointcal})
in $q^2$ and quark masses one can extract  the CHPT couplings
$G_{\Delta S=2}$, $G_8$, $G_{27}$, $\cdots$ -see definitions in 
Chapter \ref{chCHPT}- and make the
predictions of the physical quantities at lowest order. One can also
go further and extract the NLO CHPT weak counterterms needed
for instance in the isospin breaking corrections or in the
rest of NLO CHPT corrections.
 
After following the procedure sketched above one is able to write
$\im G_8$ (as well as $\hat B_K$, $\re G_8$, and $G_{27}$) 
as some known effective coupling  \cite{epsprime,scheme}
$|g_i|^2(M_X,\mu_C,\cdots)$ times
\be
\label{PPAB}
\int^\infty_0 {\rm d} Q^2 \,  \frac{Q^{2}}{Q^2+M_X^2} \, 
{\bf \Pi}_{PPAB}(Q^2,q^2)
\ee
where ${\bf \Pi}_{PPAB}(Q^2,q^2)$ is a four-point function
with $AB$  being either $L^\mu L_\mu$ or $L^\mu R_\mu$;
$L^\mu$ and $R^\mu$ are  left and right currents, respectively, 
and $Q=|p_X^E|$
is the X-boson momentum in Euclidean space. 

Similar way can be covered to get an analytical expression for 
$\im (e^2G_E)$ as described in Chapter \ref{chq7q8}.

\subsection{$\Delta I =\frac{3}{2}$ Contribution} \label{secq7q8}

In the limit $m_u=m_d$ and $\alpha_{QED}^2=0$, and neglecting
the tiny electroweak corrections to Re(a$_2$) proportional to 
$\re\left(e^2G_E\right)$, one gets
\be \label{delta32}
\left(\frac{\im a_2}{\re a_2}\right)^{LO} \simeq 
 - \frac{3}{5} \, \frac{F_0^2}{m_K^2-m_\pi^2} 
\, \frac{\im(e^2 G_E)}{G_{27}} \, ,
\ee 
including FSI to {\emph all} orders in CHPT and up to $\order(p^4)$ in the 
non-FSI corrections \cite{PPS01,PalPich00,BPP98,CG00}.  
The coupling $G_{27}$
modulates the 27-plet operator describing $K\to \pi\pi$
at $O(p^2)$ in CHPT. The coupling $G_E$ appears in CHPT to $O(e^2 p^0)$
\cite{BW84}. See Chapter \ref{chCHPT} for definitions of both couplings. 

The chiral corrections to (\ref{delta32}) can be parametrized in the next way
\be \label{ratio32}
\frac{\im a_2}{\re a_2}=
\left(\frac{\im a_2}{\re a_2}\right)^{LO} 
{\cal C}_2+\Omega_{{\rm eff}}\frac{
\im a_0}{\re a_0}\, .
\ee
The full isospin breaking corrections are include here through the 
effective parameter $\Omega_{{\rm eff}}=(0.060\pm 0.077)$ recently calculated 
in \cite{isosbreak}.

In the Standard Model, there are just two operators contributing
to $\im (e^2 G_E)$; namely, the so-called electroweak penguins,
$Q_7$ and $Q_8$ (see definition in (\ref{operators})), 
being the $Q_8$ contribution the dominant one. 
In the chiral limit, 
these operators form a closed system under QCD corrections.
Its anomalous dimensions mixing matrix is known to NLO in  the
NDR and HV schemes \cite{NLOWilscoef,CFMR94}.

There has been recently a lot of work devoted to calculate
$\im (e^2 G_E)$, both analytically \cite{BGP01,DG00,NAR01,KPR01,CDGM01,Atalks,
CDGM02} and using lattice QCD \cite{CPPACS01,RBC01,BFKGLS03,papinutto}. 
Lattice and analytical methods are in good agreement what shows that 
the calculation of the matrix elements entering in $\im (e^2 G_E)$ is 
quite robust. However, there is tendency of the lattice results to be
lower that the analytical one. Some discrepancies also exist 
between different lattice approaches: the results using Wilson fermions 
are lower than those using domain wall fermions.

In Chapter \ref{chq7q8} we report one of this calculations \cite{BGP01} 
in which analytical expressions for the $\Delta S=1$ coupling 
$\im (e^2 G_E)$ in the chiral limit in 
terms of observable spectral functions are given. This is done 
at NLO in $\alpha_S$ and, since we use experimental data, in a model 
independent way. Further discussions and 
comparison with other results are also contained in that chapter.

\subsection{Discussion on the Theoretical Determinations of 
$\varepsilon_K'$} 

There exist in the literature many calculations of $\varepsilon_K'$ within 
the SM using different methods and approximations
\cite{PPS01,epsprime,HPR03,Martinelli,BJL93,BEF01,HKPS00}. All these 
analysis use the NLO Wilson coefficients  calculated by the Rome 
and Munich groups \cite{NLOWilscoef,CFMR94} so the disagreement between  
the results got by them are due to the way of calculating the hadronic 
matrix elements -see Chapter \ref{matrixelements}. 
The different results obtained are listed in the tables 
of the Section \ref{comparison} in Chapter \ref{chq7q8}.

Many of these calculation are based on the large $N_c$ limit \cite{largeN}
-briefly discussed in Chapter \ref{matrixelements}- with $N_c$ 
the number of colours. This method was first applied to the 
calculation of weak hadronic matrix elements in \cite{ChPT}, 
where they simply identified the cut-off in meson loops with the scale 
in the renormalization group. A more sophisticated way of performing 
the identification of scales was given in \cite{BP95bk,bosonX} 
by using colour-singlet bosons.

The work in \cite{HKPS00} is essentially the continuation of \cite{ChPT} 
using this identification directly with the output of the renormalization 
group. They calculated $1/N_c$ corrections to the hadronic matrix elements 
of all the operators in the chiral limit and the unfactorized 
contributions for $Q_6$, but their results depend on the choice of the 
euclidean cut-off $\lambda_c$ that can not be fixed unambiguously. In 
this work the authors already found large corrections coming from the 
unfactorized contribution.

Large unfactorized corrections were also found in the approach 
of \cite{epsprime}. They calculated the matrix elements to NLO in 
the $1/N_c$ expansion using the X-boson method. 
This technique allowed a solution to the scale identification and to 
the scheme dependence that appears at two-loops. Another ingredient  
used in this reference is the inclusion of the ENJL model 
-see Chapter \ref{matrixelements}- for the couplings of the X-bosons 
to improve on the high energy behaviour. This method reproduce the 
$\Delta I=1/2$ rule within errors, has no free input parameters and 
have a correct scheme and scale identification at all stages. The 
result is calculated in the chiral limit and eventually corrected by 
estimating the $SU(3)$ breaking effects. 
The general method has been outlined in 
Section \ref{secq6} and a more detailed calculation of the $\Delta I =3/2$ 
contribution is given in Chapter \ref{chq7q8}. The update of this 
calculation made in \cite{BGP03} is reported in Chapter \ref{secepsilonprime}. 

In reference \cite{BEF01}, the authors also estimated the unfactorized 
contribution using a semi-phenomenological approach based on 
the constituent chiral quark model of reference \cite{PR91}. The model 
dependent parameters necessary in this calculation are fixed by fitting 
the $\Delta I=1/2$ rule. The main drawback of this determination is 
that there is not a clear scale and scheme matching. The scales in 
the matrix elements are no precisely identified and the short-distance 
running is neither precisely done.

The work in \cite{BJL93}, recently reanalyzed in \cite{BJ03,Buras03}, 
uses a semi-phenomenological approach to calculate the hadronic matrix 
elements by fitting the data for CP-conserving $K\to\pi\pi$ amplitudes. 
Within this approach it is possible to determine all $(V-A)\times(V-A)$ 
operators in any renormalization scheme, but not the dominant ones 
$\langle Q_6\rangle$ and $\langle Q_8\rangle$. The  gluonic and electroweak 
penguins are then taken around their leading $1/N_c$ values. In 
\cite{BJ03,Buras03}, the value of these two matrix elements constrained by  
the experimental result for $\varepsilon_K'$ is discussed. The results 
obtained within this method are strongly dependent on the value of 
the strange quark mass. Furthermore, the scale dependence of 
the matrix elements is fully governed by the scale dependence of 
$m_{s,d}(\mu)$. 
%Similar approach is followed in \cite{Martinelli} but 
%using matrix elements from lattice calculations instead of the 
%$1/N_c$ results.

One ingredient that turns out to be very important in the evaluation 
of $\varepsilon_K'$ within the SM is the role of 
higher order CHPT corrections and, in particular, of  
FSI as emphasized in \cite{PPS01,PalPich00}. The authors of \cite{PalPich00}  
calculated the FSI corrections to the leading $1/N_d$ result using 
dispersion relation techniques which resulted in an Omn\'es type exponential. 
They found that the strong rescattering for the two final pions can generate 
a large enhancement of $\varepsilon_K'$, through obtaining a 1.3 
enhancement factor in the $\Delta I=1/2$ contribution. In \cite{PPS01} 
a complete reanalysis of $\varepsilon_K'/\varepsilon_K$ taking into 
account the FSI corrections to the amplitudes calculated in 
\cite{PalPich00} is made. They calculate the dominant hadronic matrix 
elements at LO in $1/N_c$ by performing a matching between 
the effective short-distance description of the hamiltonian 
(\ref{hamiltonians1}) and the low energy CHPT prediction coming 
from the LO and NLO lagrangians collected in Chapter \ref{chCHPT} 
-equations (\ref{lagdS1}), (\ref{8deltaS1}), (\ref{27deltaS1}) and 
(\ref{EMdeltaS1}). 
They found an exact scale matching between the matrix elements 
and the corresponding Wilson coefficients at this order. 
A general analysis of isospin 
breaking and electromagnetic corrections to $K\to\pi\pi$ amplitudes 
is given in \cite{CG00,EIMNP00,CGisosbreak}.

The most recent estimate of direct CP violation in $K$ decays within the 
large $N_c$ framework is the one in \cite{HPR03}, in which the results 
in \cite{KPR01} are also used. They calculated 
$\order(N_c^2\frac{n_f}{N_c})$ unfactorized contributions to the 
dominant hadronic matrix element and find that they are large, even 
larger than the factorized contribution in the case of $Q_6$. The 
four-point functions necessary to evaluate such kind of contributions 
are described using a minimal hadronic approximation for the large $N_c$ 
spectrum, that in this case consists in a vector resonance 
and a scalar resonance. This is the minimal ansatz that fulfills the 
short- and long-distance constraints coming from CHPT and the OPE 
expansion of the corresponding Green's functions. These constraints 
are used to fix the free parameters of the model. At these order 
there exist an explicit cancellation of the renormalization scale 
dependence between the Wilson coefficients and the matrix elements.

There is also a lot of work calculating the hadronic matrix elements 
relevant for $\varepsilon_K'$ using lattice QCD techniques. Several 
results existing for the $\Delta I=3/2$ contribution to $\varepsilon_K'$ 
are discussed in Chapter \ref{chq7q8}. They are quite precise although 
the systematic uncertainties are not yet under control. The $\Delta I=1/2$ 
contribution is more problematic. There are many difficulties at present 
to find reliable results for these matrix elements. The results quoted in 
Table \ref{theovareps} for the results using lattice techniques, 
correspond to values of $B_6$ between 0.3 and 0.4.

\begin{table}[htb]
\begin{center}
\begin{tabular}{|c|c|}
\hline
Reference & $\varepsilon_K'/\varepsilon_K \times 10^{3}$\\
\hline
Bijnens, G\'amiz and Prades \cite{BGP03}   & 4.5$\pm$3.0 \\
Hambye, Peris and de Rafael \cite{HPR03,Peris03} &   5$\pm$3        \\
Pallante, Pich and Scimemi \cite{PPS01}  
& 1.7$\pm$0.2$_{-0.5}^{+0.8}$$\pm$0.5\\
Bertolini, Eeg and Fabbrichesi \cite{BEF01} & (0.9,4.8)  \\
Hambye \emph{et al} \cite{HKPS00}     & (0.15,3.16)     \\
Buras \cite{Buras98}   & 0.6$\pm$0.5\\
%\cite{Martinelli} &                                      \\ 
\hline
CP-PACS Coll., \cite{CPPACS01} lattice(chiral) & (-0.7,-0.2)\\
RBC Coll., \cite{RBC01} lattice(chiral)        & (-0.8,-0.4)\\
\hline
\end{tabular}
\end{center}
\caption{\label{theovareps} Recent theoretical determinations of 
$\varepsilon_K'/\varepsilon_K$}
\end{table}

\section{Direct CP Violation in Charged Kaons $K\to 3 \pi$ Decays} 
\label{cpK3pi}

  The decay of a Kaon into three pions has a long history.
The first calculations were done 
 using current algebra methods or tree level Lagrangians,
see \cite{treelevel}  and references therein. 
Then using Chiral Perturbation Theory
(CHPT) \cite{ChPT1,ChPT2} at tree level in \cite{CHPTtree}. 
The basic ingredients of CHPT as well as the lagrangians and definitions 
related to this theory at next-to-leading order (NLO) are given in Chapter 
\ref{chCHPT}
theory an be found there. References on this topic can be found there.

The one-loop calculation was done in \cite{KMW91,KMW90} and used
in \cite{KDHMW92}, unfortunately
the analytical full results were not available. Recently,  there has 
appeared the first full published result in \cite{BDP03}.

CP-violating observables in $K \to 3 \pi$ decays 
have also attracted a lot of work since long time ago 
\cite{HOL69,LW80,AVI81,GRW86,DHV87,BBEL89,DIP91,IMP92,MP95,SHA03,SHA93,DIPP94} and references therein.

At next-to-leading (NLO) there were no exact results available in CHPT 
so that the results presented in \cite{BBEL89,DIP91,IMP92,MP95}
 about the NLO were based  in assumptions about  the behaviour 
of those corrections and/or 
 using model depending results  in \cite{BBEL89}. 
In \cite{SHA03,SHA93} there are partial
 results at NLO within the linear  $\sigma$-model.

%Assumptions were done in  the
%numerical  analysis to relate $K^+ \to 3 \pi$ CP-violating  asymmetries
%to the measured value of $\varepsilon'_K$ in 
%\cite{BBEL89,DIP91,IMP92,MP95}.
%For instance, they neglected the electroweak  penguins contribution
%to $\varepsilon'_K$
%which turn out to be a quite important ingredient.
% The Final State Interactions (FSI) 
%contribution to $\varepsilon'_K$ -which was used as input 
%to fix the gluonic penguin contribution-- was
%not treated correctly in \cite{BBEL89,DIP91,IMP92,MP95,SHA03,SHA93}. 
%The $\Delta I=1/2$ rule  in $K\to \pi \pi$ decays 
%was also assumed to be  exact in some of these references.
%A more careful analysis is therefore needed even at leading 
%order (LO).

The most promising observables in $K\to 3\pi$ are the CP-odd charge 
asymmetries in $K^{\pm}$ decays. As explained in the last section, in the 
Standard Model direct CP violation parameter $\varepsilon_K'$ tends to be 
quite small due to the fact that the dominant gluon penguin contribution and 
the one arising from the electroweak penguin diagrams partially cancel each 
other. The asymmetries in $K^{\pm}\to 3\pi$ have the same two classes of 
contributions but without cancellation between them, what can be used as a 
consistency check between  the theoretical predictions of 
$K^{\pm}\to 3\pi$ and  $\varepsilon_K'$. Furthermore, 
while $\varepsilon_K'$ is essentially suppressed by  
since it is proportional to the small ratio of the $\Delta I=3/2$ to the 
$\Delta I=1/2$ amplitudes --the $\Delta I=1/2$ rule--, 
in $K\to 3\pi$ there are two independent 
$\Delta I=1/2$ amplitudes whose CP-violating interference can avoid this 
suppression. So in principle one could expect an enhanced effect.

In the charged kaon decays into three pions we can study two kinds of 
parameters, namely, asymmetries in the total rate and in the linear slope of 
the Dalitz plot. The later is done by performing an expansion of the 
amplitudes in powers of the Dalitz variables $x$ and $y$
\be \label{gdefinition}
\frac{\left|A_{K^{\pm} \to 3 \pi}(s_1,s_2,s_3)\right|^2}
{\left|A_{K^{\pm} \to 3 \pi}(s_0,s_0,s_0) \right|^2}=
1+ g \, y + h \, y^2 + k \, x^2 + {\cal O}( y  x^2,y^3) \, ,
\ee
where $x$ and $y$ are given by
\ba \label{Dalitzvar}
x\equiv \frac{s_1-s_2}{m_{\pi^+}^2}& \hspace*{0.5cm} {\rm and}
\hspace*{0.5cm} &
y \equiv \frac{s_3-s_0}{m_{\pi^+}^2} \, 
\ea
with 
$s_i\equiv(k-p_i)^2$, $3s_0\equiv m_K^2 + m_{\pi^{(1)}}^2+ 
m_{\pi^{(2)}}^2+m_{\pi^{(3)}}^2$.

The CP-violating asymmetries in the slope $g$ 
are defined as 
\ba
\label{defDeltag}
\Delta g_C \equiv
\frac{g[K^+ \to \pi^+ \pi^+ \pi^-]-g[K^-\to\pi^-\pi^-\pi^+]}
{g[K^+ \to \pi^+ \pi^+ \pi^-]+g[K^-\to\pi^-\pi^-\pi^+]}
\nonumber \\ {\rm and} \hspace*{0.5cm} 
 \Delta g_N \equiv 
\frac{g[K^+ \to \pi^0 \pi^0 \pi^+]-g[K^-\to\pi^0\pi^0\pi^-]}
{g[K^+ \to \pi^0 \pi^0 \pi^+]+g[K^-\to\pi^0\pi^0\pi^-]} \, .
\ea
 A first update at LO of these asymmetries 
was already presented in \cite{GPS03M}. 

The CP-violating asymmetries in the decay rates are defined as
\ba 
\label{defDeltaGam}
\Delta \Gamma_C \equiv
\frac{\Gamma[K^+ \to \pi^+ \pi^+ \pi^-]-\Gamma[K^-\to\pi^-\pi^-\pi^+]}
{\Gamma[K^+ \to \pi^+ \pi^+ \pi^-]+\Gamma[K^-\to\pi^-\pi^-\pi^+]}
 \nonumber \\  
 {\rm and} \hspace*{0.5cm} \Delta \Gamma_N \equiv 
\frac{\Gamma[K^+ \to \pi^0 \pi^0 \pi^+]-\Gamma[K^-\to \pi^0\pi^0\pi^-]}
{\Gamma[K^+ \to \pi^0 \pi^0 \pi^+]+\Gamma[K^-\to \pi^0\pi^0\pi^-]} \, .
\ea

 Recently, two experiments, namely, NA48 at CERN and KLOE at Frascati,
have announced the possibility of measuring
the asymmetry $\Delta g_C$  and 
$\Delta g_N$ with a sensitivity of the order of $10^{-4}$,
i.e., two orders of magnitude better than at present \cite{ISTRA+}, 
 see for instance \cite{NA48} and \cite{KLOE}. 
It is therefore mandatory to have 
these predictions  at  NLO in CHPT. In this thesis we include 
such predictions. The LO analytical expressions for the asymmetries that 
appeared in \cite{GPS03M,GPS03} are collected in Section \ref{CHPTresultsLO} 
and the NLO results in \cite{GPS03} are reported in Chapter \ref{chK3pi}.

%????? poner algo de las asimetrias de desintegraciones de kaones neutros.

\chapter{The Effective Field Theory of the Standard 
Model at Low Energies}
\label{chCHPT}

At low energies there exist a systematic method to analyze 
the structure of 
the Standard Model by performing a Taylor expansion in powers 
of external momenta 
and quark masses over the chiral symmetry breaking scale 
($\Lambda_\chi\sim 1GeV$) \cite{ChPT1,ChPT2}. 
In particular, it allows one to know the 
low energy behaviour of Green's functions built from quark currents 
and/or densities. This kind 
of expansions are carried out in an effective field theory where the quark 
and gluon 
fields are replaced by a set of pseudoscalar fields which describe the 
degrees of freedom of the relevant particles at low energies that are the 
Goldstone bosons $\pi$, $K$ and $\eta$. This formalism is based on two main 
ingredients: the chiral symmetry properties of the Standard Model and the 
concept of Effective Field Theory as the quantum theory described by 
the most general 
Lagrangian built with the operators involving the relevant degrees of freedom 
at low energies and 
compatible with all the symmetries of the original theory. The information on 
the heavier degrees of freedom is encoded in the couplings 
that modulate the operators. The effective field theory which describes 
the strong interactions between the lightest pseudoscalar mesons, 
namely, $\pi$, $K$, $\eta$ and  external vector 
($v^\mu$), axial-vector ($a^\mu$), scalar ($s$) and 
pseudoscalar ($p$) sources   
is called Chiral Perturbation Theory (CHPT). For instance, CHPT can be 
used to describe processes with vector sources as in 
$\pi^0 \to \gamma \gamma$ \cite{WZ71,Witten83} or 
with a scalar source as in $\eta \to \pi^0 h^0$ \cite{PP90}.

Some introductory lectures on CHPT can be 
found in \cite{schools} and  recent reviews in \cite{rafaelTASI,reviews}.
 In this chapter we limit ourselves to collect the 
Lagrangians and definitions corresponding to the chiral effective realization 
of strong,  
electroweak and $\Delta S=1,2$ weak interactions that we need in 
other chapters.

\section{Lowest Order Chiral Perturbation Theory}

To lowest order in CHPT, i.e., order $e^0p^2$ and $e^2p^0$, strong and 
electroweak interactions between $\pi$, $K$ and $\eta$ and vector, 
axial--vector, pseudoscalar and scalar external 
sources are described by
\ba \label{lag2strong}
{\cal L}^{(2)} &=& \frac{F_0^2}{4}\tr\left(u_\mu u^\mu + \chi_+\right)+
e^2\tilde C_2\tr\left(QUQU^\dagger\right)
\ea
with
\ba
u_\mu \equiv i u^\dagger (D_\mu U) u^\dagger = u_\mu^\dagger \; , 
\nonumber \\
\chi_{+(-)}= u^\dagger \chi u^\dagger +(-)u \chi^\dagger u  
\ea 
$\chi= \mbox{diag}(m_u,m_d,m_s)$ 
a 3 $\times$ 3 matrix that explicitly break the chiral symmetry 
through the light quark masses and 
$U\equiv u^2=\exp{(i\sqrt 2 \Phi /F_0)}$ is the exponential
representation incorporating the octet of light pseudo-scalar mesons
in the SU(3) matrix $\Phi$; 
\ba
\Phi\equiv \left(  
\begin{array}{ccc}
\frac{\dis \pi^0}{\dis \sqrt{2}} + 
\frac{\dis \eta_8}{\dis \sqrt{6}} & \pi^+ & K^+ 
\nonumber \\ 
\pi^- & -\frac{\dis \pi^0}{\dis \sqrt{2}} 
+ \frac{\dis \eta_8}{\dis \sqrt{6}} & K^0 
\nonumber \\ 
K^- & \bar{K^0} &- 2 \frac{\dis \eta_8}{\dis \sqrt{6}}  
\end{array}
\right) \, .
\ea
The matrix $Q=\mbox{diag}(2/3,-1/3,-1/3)$ collects the electric charge of the 
three light quark flavours and $F_0$ is the pion decay coupling constant 
in the chiral limit. To this order $f_{\pi}=F_0$=87 MeV.

The covariant derivatives
\be
D_\mu U = \partial_\mu U -ir_\mu U+iUl_\mu \,,\quad \quad
D_\mu U^\dagger = \partial_\mu U^\dagger +i U^\dagger r_\mu 
-il_\mu U^\dagger\,
\ee
and the strength tensors 
\be
F^{\mu\nu}_L =
\partial^\mu \ell^\nu - \partial^\nu \ell^\mu 
- i [ \ell^\mu , \ell^\nu ] \, ,
\quad\;
F^{\mu\nu}_R =
\partial^\mu r^\nu - \partial^\nu r^\mu - i [ r^\mu , r^\nu ]\, ,
\ee
that appear at the next order in the chiral expansion, 
are the only structures involving the gauge fields $v_\mu$ and $a_\mu$ 
that respect the local invariance. Through them we can introduce 
external fields which will allow us to compute the effective 
realization of general Green's functions.

To $\order(p^2)$ and $\order(e^2p^0)$ (the lowest order)  
the chiral Lagrangians describing $|\Delta S|=2$ 
and $|\Delta S|=1$ transitions are
\ba
\label{lagdS2}
{\cal L}^{(2)}_{|\Delta S|=2}&=&
\hat C F_0^4 G_{\Delta S=2}
\, \tr\left (\Delta_{32} u^\mu \right)
\tr \left( \Delta_{32} u_\mu\right) +\mbox{h.c.}
\ea
and 
\ba
\label{lagdS1} 
{\cal L}^{(2)}_{|\Delta S|=1}&=&
C F_0^6 e^2 G_E \,  \tr\left(\Delta_{32}\tilde{Q}\right)
\\&&\hskip-2.0cm+
 C F_0^4 \Bigg[G_8\tr\left(\Delta_{32}u_\mu u^\mu\right)
+G_8^\prime\tr\left(\Delta_{32}\chi_+\right) 
+G_{27}t^{ij,kl}\tr\left(\Delta_{ij}u_\mu\right)
  \tr\left(\Delta_{kl}u^\mu\right)\Bigg] +\mbox{h.c.} \ ; 
\nonumber
\ea
respectively. With $\Delta_{ij} = u\lambda_{ij}u^\dagger$,
$(\lambda_{ij})_{ab} = \delta_{ia}\delta_{jb}$,
$\tilde{Q}=u^\dagger Q u$; and
\ba \label{Cdef}
C&=& -\frac{3}{5} \frac{G_F}{\sqrt 2} \, V_{ud} V_{us}^* \approx
-1.065 \times 10^{-6} \, {\rm GeV}^{-2} \, . 
\ea

The non-zero components of the 
SU(3) $\times$ SU(3) tensor $t^{ij,kl}$  are
\ba
t^{21,13}=t^{13,21}=\frac{1}{3} \, ; \, &  t^{22,23}=t^{23,22}=
-\frac{\dis 1}{\dis 6} \, ;
\nonumber \\
t^{23,33}=t^{33,23}=-\frac{1}{6} \, ; \, 
&  t^{23,11}=t^{11,23}=\frac{\dis 1}{\dis 3} \,.
\ea

The weak couplings $G_8$ and $G_{27}$ and the couplings $c_2$ 
and $c_3$ of \cite{KMW91,KMW90} are related as follows  
\ba
c_2=C F_0^4 \, G_8; \nonumber \\
c_3= -\frac{\dis 1}{\dis 6} C F_0^4 \, G_{27} \, .
\ea
The constant $\hat C$ in (\ref{lagdS2}) is a known function
of  the  $W$-boson, top and charm quark masses and
of Cabibbo-Kobayashi-Maskawa (CKM) matrix elements. 
%--see last reference in \cite{}??? primera referencia de bpp98

The Lagrangians in (\ref{lagdS2}) and (\ref{lagdS1}) 
have the same $SU(3)_L\times SU(3)_R$ transformation properties 
as the corresponding short-distance hamiltonians in 
(\ref{hamiltonianS2}) and (\ref{hamiltonians1}).

In the presence of CP-violation, the couplings $G_8$, $G_{27}$, and
$G_E$ get an imaginary part. In the Standard Model,
$\im  G_{27}$ vanishes and
$\im  G_8$ and $\im  G_E$ are proportional
to $\im  \tau$ with
 $\tau \equiv -\lambda_t /\lambda_u$ and $\lambda_i\equiv
V_{id} V_{is}^*$ and where $V_{ij}$ are CKM matrix elements. See Section 
\ref{LOcounterterms} for a discussion on the value of these couplings.

\section{Next-to-Leading Order Chiral Lagrangians} 
\label{NLOcounterterms}

At NLO in momenta it is necessary to consider two different 
contributions in the calculation of any process
\begin{itemize}
\item one-loop amplitudes generated by the lowest order 
Lagrangian ${\cal L}^{(2)}$ (which will be the one in (\ref{lag2strong})
, (\ref{lagdS2}) or (\ref{lagdS1}) depending on the process)
\item tree level amplitudes obtained with the Lagrangians of order $p^4$ 
and $e^2p^2$. 
\end{itemize}
Another ingredient involved at this order is the Wess--Zumino--Witten 
(WZW) \cite{WZ71,Witten83} functional to account 
for the QCD chiral anomaly. Using it we can compute 
all the contributions generated by the chiral anomaly 
to electromagnetic and semileptonic decays of pseudoscalar mesons. 
Chiral power counting insures that the coefficients of the WZW functional, 
that are completely fixed by the anomaly,  
are not renormalized by next-order contributions. An explicit expression and 
more comments about the anomaly functional can be found in \cite{schools}.

CHPT tree amplitudes are finite and scale independent since the 
couplings in (\ref{lag2strong}), (\ref{lagdS2}) and (\ref{lagdS1}) 
are. However, one-loop graphs with vertices 
generated by these LO Lagrangians and Goldstone bosons propagators in 
the internal lines are in general divergent. The divergences they present 
come from the integration over the moment 
in the loop with logarithms and threshold factors, as required by 
unitarity, and need to be renormalized.   
By symmetry arguments, if we use a regularization that preserves the 
symmetries of the Lagrangian (for example, dimensional regularization), these 
divergences have exactly the same structure as the NLO local terms of order 
$p^4$ and $e^2p^2$ and can be absorbed in a renormalization of the 
counterterms constants occurring in these NLO Lagrangians.  The 
theory is renormalizable order by order in the chiral expansion.

The divergences 
appearing at one loop using the strong part of the Lagrangian in 
(\ref{lag2strong}) are order $p^4$ and 
therefore they are renormalized by the low-energy couplings in  
the SU(3) $\times$ SU(3) strong chiral Lagrangian of order $p^4$ 
\cite{ChPT2}
\ba \label{strongl}
{\cal L}^{(4)} &=& 
L_1 \tr \left( u_\mu u^\mu \right)^2 
+L_2 \tr \left( u^\mu u^\nu \right) \tr \left( u_\mu u^\nu \right)
+L_3 \tr \left( u^\mu u_\mu u^\nu u_\nu\right)
+L_4 \tr \left( u^\mu u_\mu \right) \tr \left( \chi_+ \right)
\nonumber \\
&+& L_5 \tr \left( u^\mu u_\mu \chi_+ \right)
+L_6 \tr \left( \chi_+ \right) \tr \left (\chi_+ \right)
+L_7 \tr \left( \chi_-\right) \tr \left( \chi_-\right)
+\frac{1}{2} L_8 \tr \left( \chi_+ \chi_+ + \chi_- \chi_- \right) 
\nonumber\\
&-&iL_9 \tr \left(F_R^{\mu\nu} D_\mu U D_\nu U^\dagger 
+ F_L^{\mu\nu} D_\mu^\dagger U D_\nu U \right)
+L_{10}\tr \left(U^\dagger F_R^{\mu\nu}UF_{L\mu\nu}\right) 
\nonumber\\
&+&H_1\tr \left(F_{R\mu\nu}F_R^{\mu\nu}+F_{L\mu\nu}F_L^{\mu\nu}\right)
+H_2\tr\left(\chi^\dagger \chi\right)
\ea
Since we will only use the $\order(p^4)$ Lagrangian 
at tree level, the $\order(p^2)$ equations of motion obeyed by $U$ 
have been used to reduce the number of independent terms \cite{ChPT2}. 

The renormalized strong counterterms $L_i^r(\mu)$ one obtains once 
the divergences from the one-loop contributions have been absorbed,  
are given in the dimensional regularization scheme by 
\be \label{Lir}
L_i = L_i^r(\mu) + \Gamma_i {\mu^{2\epsilon}\over 32 \pi^2} \left\{
{1\over \hat{\epsilon}} - 1 \right\} , \quad
H_i = H_i^r(\mu) + \widetilde\Gamma_i {\mu^{2\epsilon}\over 32 \pi^2} 
\left\{ {1\over \hat{\epsilon}} - 1 \right\}\,.
\ee
with $\hat{\epsilon}\equiv {1\over \epsilon} + \gamma_E -\log{(4\pi)}$, 
$D=4 + 2\epsilon$ and D the dimension in dimensional regularization. 
They depend on the scale of dimensional regularization $\mu$, but 
this dependence is canceled by that of the loop amplitude in any 
measurable quantity.

The value of the constants $L_i$ are not fixed only by symmetry 
requirements. They parametrize our ignorance about the details of 
the underlying QCD dynamics and must be determined by experimental data. 
The values obtained for the renormalized constants $L_i^r$ defined in 
(\ref{Lir}) at the scale $\mu=M_\rho\simeq 0.77 GeV$, together with the 
processes used to fix them and the scale factors $\Gamma_i$ that relate 
the bare and the renormalized constants \cite{ChPT2}, 
are reported in Table \ref{tablaLis}. The scale factor 
$\widetilde \Gamma_i$ for the counterterms $H_i^r$ are also listed 
in the same table. We don't give any value for the $H_i^r$ since 
they are not physical quantities that depend on the renormalization 
scheme used to define them. 
At any other renormalization scale, the couplings can 
be obtained through the running implied in (\ref{Lir})
\be
L_i^r(\mu_2) \, = \, L_i^r(\mu_1) \, + \, {\Gamma_i\over 16\pi^2}
\,\log{\left({\mu_1\over\mu_2}\right)}
% , \quad
%H_i^r(\mu_2) \,=\, H_i^r(\mu_2)\,  + \, {\widetilde \Gamma_i\over 16\pi^2}
%\,\log{\left({\mu_1\over\mu_2}\right)\,
.
\ee

%\begin{table}[htb]
%\begin{center}
%\begin{tabular}{|c|c|c|c|}
%\hline
%$i$ & $\Gamma_i$ & $L_i^r(M_\rho) \times 10^3$ & Reference \\
%\hline
%1 & 3/32  & $\hphantom{-}0.4\pm0.3$ & $K_{e4}$, $\pi\pi\to\pi\pi$
%\\
%2 & 3/16  & $\hphantom{-}1.4\pm0.3$ & $K_{e4}$, $\pi\pi\to\pi\pi$
%\\
%3 & 0 & $-3.5\pm1.1$ & $K_{e4}$, $\pi\pi\to\pi\pi$
%\\
%4 & 1/8 & $-0.3\pm0.5$ &  Zweig rule
%\\
%5 & 3/8 & $\hphantom{-}1.4\pm0.5$ & $F_K : F_\pi$
%\\
%6 & 11/144 & $-0.2\pm0.3$ & Zweig rule
%\\
%7 & 0 & $-0.4\pm0.2$ & Gell-Mann--Okubo, $L_5$, $L_8$
%\\
%8 & 5/48 & $\hphantom{-}0.9\pm0.3$ & $M_{K^0} - M_{K^+}$, $L_5$,
%$(m_s - \hat{m}) : (m_d-m_u)$
%\\
%9 & 1/4 & $\hphantom{-}6.9\pm0.7$ & $\langle r^2\rangle^\pi_V$
%\\
%10 & -1/4 & $-5.5\pm0.7$ & $\pi\to e\nu\gamma$
%\\ \hline\hline
%$H_i$ & $\widetilde \Gamma_i$ & &Source \\
%\hline
% $H_1$ & -1/8 & - & Scheme dependent \\
% $H_2$ & 5/24 &$2L_8+H_2=(2.9\pm1.0)\times10^{-3}$& Scheme dependent. \\
%       &      &&$\langle\bar u u +\bar d d\rangle$, Sum Rules 
%\cite{BPR}\\
%\hline
%\end{tabular}
%\end{center}
%\caption{\label{tablaLis}
%Values of the
%renormalized couplings $L_i^r(M_\rho)$  
%and values of $\Gamma_i$ and $\widetilde\Gamma_i$.}
%\end{table}

\begin{table}[htb]
\begin{center}
\begin{tabular}{|c|c|c|c|}
\hline
$i$ & $\Gamma_i$ & $L_i^r(M_\rho) \times 10^3$ & Reference \\
\hline
1 & 3/32  & $0.46$[$0.53\pm0.25$] 
& $\order(p^4)$[$\order(p^6)$] \cite{ABT3}
\\
2 & 3/16  & $1.49$[$0.71\pm0.27$] & 
$\order(p^4)$[$\order(p^6)$] \cite{ABT3}
\\
3 & 0 & $-3.18$[$-2.72\pm1.12$] & $\order(p^4)$[$\order(p^6)$] \cite{ABT3}
\\
4 & 1/8 & $-0.3\pm0.5$ &  Zweig rule
\\
5 & 3/8 & $1.46$[$0.91\pm0.15$] & 
 $\order(p^4)$[$\order(p^6)$] \cite{ABT3}
\\
6 & 11/144 & $-0.2\pm0.3$ & Zweig rule
\\
7 & 0 & $-0.49$[$-0.32\pm0.15$] & 
 $\order(p^4)$[$\order(p^6)$] \cite{ABT3}\\
8 & 5/48 & $1.00$[$0.62\pm0.20$] & 
 $\order(p^4)$[$\order(p^6)$] \cite{ABT3}
\\
9 & 1/4 & $\hphantom{-}5.93\pm0.43$ & \cite{BT1}
\\
10 & -1/4 & $-4.4\pm0.7$ & \cite{BT1,BT2}
\\ \hline\hline
$H_i$ & $\widetilde \Gamma_i$ & &Source \\
\hline
 $H_1$ & -1/8 & - & Scheme dependent \\
 $H_2$ & 5/24 &$2L_8+H_2=(2.9\pm1.0)\times10^{-3}$& Scheme dependent \\
       &      &&$\langle\bar u u +\bar d d\rangle$, Sum Rules 
\cite{BPR}\\
\hline
\end{tabular}
\end{center}
\caption{\label{tablaLis}
Values of the
renormalized couplings $L_i^r(M_\rho)$  
and values of $\Gamma_i$ and $\widetilde\Gamma_i$.}
\end{table}

Analogously to the strong case, the divergences that appear 
in the one-loop diagrams using the LO Lagrangian in (\ref{lagdS1}) 
can be reabsorbed in the couplings counterterms of the NLO order, i.e., 
$\order(p^4)$ and $\order(e^2p^2)$ SU(3) $\times$ SU(3) chiral 
Lagrangian describing $|\Delta S|=1$ transitions. The part of this 
Lagrangian that is relevant for $K\to 3\pi$ decays is 
\ba
\label{8deltaS1}
{\cal L}^{(4)}_{|\Delta S|=1}&=&
 C F_0^2  \re{G_8} \, 
\left\{ N_1 {\cal O}_1^8 + N_2 {\cal O}_2^8 + N_3 {\cal O}_3^8 +
 N_4 {\cal O}_4^8 + N_5 {\cal O}_5^8 + N_6 {\cal O}_6^8 + 
N_7 {\cal O}_7^8 \right. \nonumber \\ 
&+& \left. N_8 {\cal O}_8^8 +  N_9 {\cal O}_9^8 +
N_{10}{\cal O}_{10}^8 +  N_{11} {\cal O}_{11}^8 
+ N_{12} {\cal O}_{12}^8 + N_{13} {\cal O}_{13}^8 +\dots\right\}  + {\rm h.c.}
\ea
for the octet part \cite{KMW90,EF91,EKW93},
\ba
\label{27deltaS1}
{\cal L}^{(4)}_{|\Delta S|=1}&=&
 C F_0^2 G_{27} \, 
\left\{ D_1 {\cal O}_1^{27} + D_2 {\cal O}_2^{27} + 
D_4 {\cal O}_4^{27} + D_5 {\cal O}_5^{27}+ D_6 {\cal O}_6^{27}  +
D_7 {\cal O}_7^{27}  \right. \nonumber \\ 
&+&\left. D_{26} {\cal O}_{26}^{27} +  D_{27} {\cal O}_{27}^{27}  
+ D_{28}{\cal O}_{28}^{27} + D_{29} {\cal O}_{29}^{27}  
D_{30} {\cal O}_{30}^{27} + D_{31} {\cal O}_{31}^{27}+ \dots \right\} 
 + {\rm h.c.}\nonumber\\
\ea
 for the 27-plet part \cite{KMW90,EF91} and 
\ba
\label{EMdeltaS1}
{\cal L}^{(4)}_{|\Delta S|=1}&=&
 C e^2 F_0^4 \re{G_{8}} \, 
\left\{ Z_1 {\cal O}_1^{EW} + Z_2 {\cal O}_2^{EW} + 
Z_3 {\cal O}_3^{EW} + Z_4 {\cal O}_4^{EW} + Z_5 {\cal O}_5^{EW}  
Z_6 {\cal O}_6^{EW} \right.  \nonumber \\ 
&+& Z_{7} {\cal O}_{7}^{EW} + Z_{8} {\cal O}_{8}^{EW}  
+ Z_{9}{\cal O}_{9}^{EW} + Z_{10} {\cal O}_{10}^{EW}  \nonumber \\
&+& \left. Z_{11} {\cal O}_{11}^{EW} + Z_{12} {\cal O}_{12}^{EW} 
+ Z_{13} {\cal O}_{13}^{EW}  + Z_{14} {\cal O}_{14}^{EW} + \dots\right\} 
 + {\rm h.c.}
\ea
 for the electroweak part with the 
dominant octet structure \cite{EIMNP00}. 

The dots in (\ref{8deltaS1}), (\ref{27deltaS1}) and 
(\ref{EMdeltaS1}) stand for operators that, 
although in principle also appear 
in the Lagrangians at this order, are not written here since they don't 
contribute to the processes in which we are interested in this Thesis. 
However, they must be considered where studying different problems as, in 
example, kaon radiative decays.

The renormalized weak counterterms with which we must replace 
those in (\ref{8deltaS1}), (\ref{27deltaS1}) and 
(\ref{EMdeltaS1}) for having finite amplitudes are given in 
the dimensional regularization scheme by
\ba \label{nidizirenorm}
N_i &=& N_i^r(\mu) + {\mu^{2\epsilon}\over 32 \pi^2} \left\{
{1\over \hat{\epsilon}} - 1 \right\} \left[ n_i+\frac{G_8'}{G_8}
n_i'\right]\, ,\nonumber\\
D_i &=& D_i^r(\mu) + {\mu^{2\epsilon}\over 32 \pi^2} \left\{
{1\over \hat{\epsilon}} - 1 \right\} d_i\, ,\nonumber\\
Z_i &=& Z_i^r(\mu) + {\mu^{2\epsilon}\over 32 \pi^2} \left\{
{1\over \hat{\epsilon}} - 1 \right\} \left[z_i+\frac{\tilde C_2}{F_0^4}
z_i'+\frac{G_E}{\re G_8}z_i''\right]\, .\nonumber\\
\ea
The infinites needed in the octet and 27-plet weak sector were 
calculated first in \cite{KMW90} and confirmed in \cite{EF91}. Those relatives 
to the electroweak Lagrangian were obtained in \cite{EIMNP00}. 
The values of these coefficients are collected in Table \ref{tabnidizidiv}.

\begin{table}
\begin{center}
\begin{tabular}{|ccc|cc|cccc|}
\hline
$N_i$   &  $n_i$ & $n_i'$  &   $D_i$    &  $d_i$ & $Z_i$ &  $z_i$ &$z_i'$
& $z_i''$  \\
\hline
 1  &  2      &     0           &1 &-1/6 &1&  -17/12  &-3&3/2\\
 2  &  -1/2 	 &     0           &2 &0   &2& 1  &16/3&1\\
 3  &  0  	 &     0           &4 &3   &3& 3/4 &7&0\\
 4  &  1	 &     0           &5 &1   &4& -3/4  &-7&0 \\
 5  &  3/2    &    3/4         &6 &-3/2  &5&-2  &0 &0\\
 6  &  -1/4 	 &    0            &7 &1   &6&7/2  &5&3/2\\
 7  &  -9/8    &    1/2        &26&-1    &7& 3/2 &5&0\\
 8  &  -1/2  	 &    0           &27&-1/2 &8& -1/2 &0&0\\
 9  &  3/4    &    -3/4         &28&-5/3 &9&-11/6  &4/3&2\\
 10 &  2/3    &    5/12       &29&19/3   &10&-3/2  &-1&0\\
 11 &  -13/18   &    11/18      &30&10/3 &11&-3/2  &-2&0\\
 12 &  -5/12    &    5/12       &31&0   &12& 3/2 &0&0\\
 13 &  0       &    0           &  &       &13&-35/12  &-3&1\\
    &            &                  &  &       &14&3  &15&0\\
\hline
\end{tabular}
\end{center}
\caption{\label{tabnidizidiv} 
Coefficients of the subtraction of the infinite
parts defined in equation (\ref{nidizirenorm}).}
\end{table}

The weak NLO counterterms as much less known than the 
strong NLO counterterms and there doesn't exist a phenomenological 
determination of all of them. Only some combinations can be fixed 
from experiment -see Section \ref{Kcount}. The best that can be done 
is to get the order of magnitude of the counterterms using several 
approaches. Among these approaches  are
factorization plus meson dominance \cite{EGPR89,Eckeretal}.
 If one uses factorization, 
one needs couplings of order $p^6$ from the strong chiral Lagrangian
for some of the $\widetilde K_i$ counterterms, see also \cite{PPS01}.
Not very much is known about these $\order(p^6)$ couplings though. 
 One can use  Meson Dominance 
to  saturate them  but it is not clear that this procedure
 will be in general a good 
estimate.  See for instance \cite{KN01}  for some detailed analysis of 
some order $p^6$  strong counterterms 
 obtained  at large $N_c$ using also short-distance QCD constraints 
and comparison with meson  exchange saturation. 
See also \cite{CENP03}  for a very recent estimate of some relevant
order $p^6$ counterterms in the strong sector using 
Meson Dominance and factorization.

Another more ambitious procedure to predict 
the necessary NLO weak counterterms is to combine short-distance QCD, 
large $N_c$ constraints plus other chiral constraints
and some phenomenological inputs
to construct the relevant $\Delta S=1$ Green's functions, see
\cite{BGLP03,KN01,PPR}. This last program has not yet been
used systematically to get all the $\Delta S=1$
counterterms at NLO.

Finally, we list the operators that appear in (\ref{8deltaS1}), 
(\ref{27deltaS1}) and (\ref{EMdeltaS1}). The octet operators are
\ba
{\cal O}_1^8 = \tr \left( \Delta_{32} u_\mu u^\mu u_\nu u^\nu \right), 
\; &
{\cal O}_2^8  = \tr \left( \Delta_{32} u_\mu u_\nu u^\nu u^\mu \right),
\; \nonumber \\
{\cal O}_3^8 = \tr \left( \Delta_{32} u_\mu u_\nu \right)
\tr \left( u^\mu u^\nu \right), \; &
   {\cal O}_4^8 = \tr \left( \Delta_{32} u_\mu \right) 
\tr \left( u_\nu u^\mu u^\nu \right),  \;  \nonumber \\
{\cal O}_5^8 = \tr \left( \Delta_{32} \left( \chi_+ u_\mu u^\mu
 + u_\mu u^\mu \chi_+\right) \right) , \;  &
  {\cal O}_6^8 = \tr \left( \Delta_{32} u_\mu \right) 
\tr \left( u^\mu \chi_+ \right), \; \nonumber \\ 
{\cal O}_7^8 = \tr \left( \Delta_{32} \chi_+\right) 
\tr \left( u_\mu u^\mu \right), \; & 
  {\cal O}_8^8 = \tr \left( \Delta_{32} u_\mu u^\mu \right) 
\tr \left( \chi_+ \right), \; \nonumber \\
{\cal O}_9^8 = \tr \left( \Delta_{32} \left( \chi_- u_\mu u^\mu
 - u_\mu u^\mu \chi_-\right) \right) , \; & 
  {\cal O}_{10}^8 = \tr \left( \Delta_{32} \chi_+ \chi_+\right), 
 \; \nonumber \\ 
{\cal O}_{11}^8 = \tr \left( \Delta_{32}  \chi_+\right) 
\tr \left( \chi_+ \right), \; & 
  {\cal O}_{12}^8 = \tr \left( \Delta_{32} \chi_- \chi_-\right), 
 \; \nonumber \\ 
{\cal O}_{13}^8 = \tr \left( \Delta_{32} \chi_-\right)
\tr \left(\chi_-\right)\, .  & 
\ea

 The 27-plet operators are
\ba
{\cal O}_1^{27}= t^{ij,kl} \tr \left( \Delta_{ij} \chi_+\right) 
\tr\left( \Delta_{kl} \chi_+\right), \; \nonumber \\
{\cal O}_2^{27}= t^{ij,kl} \tr \left( \Delta_{ij} \chi_-\right) 
\tr\left( \Delta_{kl} \chi_-\right), \; \nonumber \\
{\cal O}_4^{27}= t^{ij,kl} \tr \left( \Delta_{ij} u_\mu\right) 
\tr\left( \Delta_{kl}
\left(  u^\mu \chi_+ + \chi_+ u^\mu \right) \right), \; \nonumber \\
{\cal O}_5^{27}= t^{ij,kl} \tr \left( \Delta_{ij} u_\mu\right) 
\tr\left( \Delta_{kl}
\left(u^\mu \chi_- - \chi_- u^\mu \right) \right), 
\; \nonumber \\
{\cal O}_6^{27}= t^{ij,kl} \tr \left( \Delta_{ij} \chi_+\right) 
\tr\left( \Delta_{kl} u^\mu u_\mu \right), \; \nonumber \\
{\cal O}_7^{27}= t^{ij,kl} \tr \left( \Delta_{ij} u_\mu\right) 
\tr\left( \Delta_{kl} u^\mu \right) \tr \left( \chi_+ \right), \; 
\nonumber \\
{\cal O}_{26}^{27}= t^{ij,kl} \tr \left( \Delta_{ij} u^\mu u_\mu\right) 
\tr\left( \Delta_{kl} u^\nu u_\nu \right) , \; \nonumber \\
{\cal O}_{27}^{27}= t^{ij,kl} \tr \left( \Delta_{ij} \left(u_\mu u_\nu
+ u_\nu u_\mu \right) \right) \tr \left( \Delta_{kl} \left(u^\mu u^\nu
+ u^\nu u^\mu \right) \right), \; \nonumber \\
{\cal O}_{28}^{27}= t^{ij,kl} \tr \left( \Delta_{ij} \left(u_\mu u_\nu
- u_\nu u_\mu \right) \right) \tr \left( \Delta_{kl} \left(u^\mu u^\nu
- u^\nu u^\mu \right) \right), \; \nonumber \\
{\cal O}_{29}^{27}= t^{ij,kl} \tr \left( \Delta_{ij} u_\mu \right) 
\tr \left( \Delta_{kl} u_\nu u^\mu u^\nu\right), \; \nonumber \\
{\cal O}_{30}^{27}= t^{ij,kl} \tr \left( \Delta_{ij} u_\mu \right) 
\tr \left( \Delta_{kl} \left(u^\mu u_\nu u^\nu + 
u_\nu u^\nu u^\mu \right)\right), \; \nonumber \\
{\cal O}_{29}^{27}= t^{ij,kl} \tr \left( \Delta_{ij} u_\mu \right) 
\tr \left( \Delta_{kl} u^\mu \right) \tr \left( u_\nu u^\nu \right).
\ea

 The dominant octet electroweak operators are
\ba
{\cal O}_1^{EW}= \tr \left( \Delta_{32} 
\left\{ u^\dagger Q u ,\chi_+ \right\}\right) , \; 
&  {\cal O}_2^{EW}= \tr \left( \Delta_{32} 
u^\dagger Q u \right) \tr \left( \chi_+\right), \nonumber \\
{\cal O}_3^{EW}= \tr \left( \Delta_{32} 
u^\dagger Q u \right) \tr \left( \chi_+ u^\dagger Q u \right), 
\; &  {\cal O}_4^{EW}= \tr \left( \Delta_{32} \chi_+\right) 
\tr \left(  Q U^\dagger Q U \right), \nonumber \\ 
{\cal O}_5^{EW}= \tr \left( \Delta_{32} u^\mu u_\mu \right) , \; 
&  {\cal O}_6^{EW}=  \tr \left( \Delta_{32} 
\left\{ u^\dagger Q u , u^\mu u_\mu \right\}\right) , \; \nonumber \\
{\cal O}_7^{EW}= \tr \left( \Delta_{32} u^\mu u_\mu \right) 
\tr \left( Q U^\dagger Q U \right) , \; 
&  {\cal O}_8^{EW}=  \tr \left( \Delta_{32}u^\mu \right) 
\tr \left(Q u^\dagger  u_\mu u \right), \; \nonumber \\
{\cal O}_{9}^{EW}=  \tr \left( \Delta_{32}u^\mu \right) 
\tr \left(Q u  u_\mu u^\dagger \right), \; &
{\cal O}_{10}^{EW}=  \tr \left( \Delta_{32}u^\mu \right) 
\tr \left(\left\{u Q u^\dagger,   u^\dagger  Q u \right\} 
u_\mu \right), \; \nonumber \\
{\cal O}_{11}^{EW}=  
\tr \left(\Delta_{32} \left\{u^\dagger  Q u , u^\mu \right\} \right)
\tr \left( u Q u^\dagger u_\mu \right) , \; & 
{\cal O}_{12}^{EW}=  
\tr \left(\Delta_{32} \left\{u^\dagger  Q u , u^\mu \right\} \right)
\tr \left( u^\dagger  Q u  u_\mu \right) , \; \nonumber \\
{\cal O}_{13}^{EW}=  
\tr \left(\Delta_{32} u^\dagger  Q u  \right)
\tr \left( u^\mu u_\mu \right) , \; & 
{\cal O}_{14}^{EW}=  
\tr \left(\Delta_{32} u^\dagger  Q u  \right)
\tr \left( u^\dagger  Q u  u_\mu u^\mu\right) . \nonumber \\
\ea

\section{Leading Order Chiral Perturbation Theory Predictions}
\label{CHPTresultsLO}

The chiral Lagrangian in (\ref{lagdS2}) that describes $|\Delta S|=2$ 
transitions at order $p^2$ can be used to make a prediction for the parameter 
$\hat B_K$ defined in (\ref{bkdef}) in the chiral limit,
\be
\hat B_K^{\chi} = \frac{3}{4} \, G_{\Delta S=2} \, ,
\ee
where $G_{\Delta S=2}$ is the coupling appearing in (\ref{lagdS2}).
%The diagram contributing to this parameter at LO is the one in 
%Figure \ref{FigBKLO}.

The amplitudes $K\to 2\pi$ are fixed at LO in CHPT using the Lagrangian
(\ref{lagdS1}). One gets
\ba
a_0&\equiv& C \left[ G_8 + \frac{1}{9} G_{27} \right]
\, \sqrt{6} \, F_0\, (m_K^2-m_\pi^2) \,  , \nonumber \\
a_2&=&C \, G_{27} 
\, \frac{10 \sqrt 3}{9} \, F_0\, (m_K^2-m_\pi^2) \,  , 
\ea
and 
\ba
\delta_0 = \delta_2 = 0 \, , 
\ea
with the constant $C$ defined in (\ref{Cdef}). 
We have disregarded some tiny electroweak corrections
proportional to $e^2 G_E$. 
The ratios needed to calculate the direct CP-violation parameter 
$|\varepsilon_K'|\simeq
\re a_2/(\sqrt{2}\,\re a_0)\left[\im a_2/\re a_2
-\im a_0/\re a_0\right]$ 
-see equation (\ref{epsilonprimedef})- are then 
\be
\left(\frac{\re a_0}{\re a_2}\right)^{LO} \simeq {\sqrt 2}  \, 
\frac{ 9 \re G_8 + G_{27}}{10 G_{27}} \, ,
\ee
\be
\left(\frac{\im a_0}{\re a_0}\right)^{LO} 
\simeq \frac{\im G_8}{\re G_8 + G_{27}/9} \, , 
\ee
and
\be 
\left(\frac{\im a_2}{\re a_2}\right)^{LO}
 \simeq -\frac{3}{5} \frac{F_0^2}{m_K^2-m_\pi^2}
\frac{\im (e^2 G_{E})}{G_{27}} \, .
\ee

By using as inputs parameters in these expressions the values of the 
couplings discussed in Section \ref{LOcounterterms} and the results in 
(\ref{imgeinput}) and (\ref{img8input}), 
the numerical values of these ratios normalized 
in such a way that we can use them directly to make a prediction for 
$\varepsilon_K'/\varepsilon_K$ is
\ba \label{LOratios}
-\frac{1}{|\varepsilon_K| \sqrt 2} \left(\frac{\re a_2}{\re a_0}
\frac{\im a_0}{\re a_0}\right)^{LO} &\hskip-3ex=\hskip-1ex&
\hskip-1ex -(10.8 \pm 5.4) \, \im \tau \,
\, , 
\nonumber\\
\frac{1}{|\varepsilon_K| \sqrt 2} \left(\frac{\re a_2}{\re a_0}
\frac{\im a_2}{\re a_2}\right)^{LO} &\hskip-3ex=\hskip-1ex& 
(2.7 \pm 0.8 ) \, \im \tau \, 
\nonumber\\&&
\ea
where $\im \tau$ is the combination of CKM matrix elements given in 
(\ref{tau}).

%\subsection{$K\to 3 \pi$ Amplitudes at LO}

We can also made predictions for the $K\to 3 \pi$ amplitudes. 
The numerators of the asymmetries in (\ref{defDeltag}) and 
(\ref{defDeltaGam}) are proportional to strong phases times the
real part of the squared amplitudes. At LO in CHPT, i.e., using the 
Lagrangian in (\ref{lagdS1}), the strong phases 
start at one-loop and  are order $p^4/p^2$ while the real parts
are order $(p^2)^2$. The denominators are proportional 
to the real part of the amplitudes 
which are order $(p^2)^2$,  so the asymmetries for
the slope $g$ and decay rates $\Gamma$ are order $p^2$ in CHPT.

At this order,  
the CP violating asymmetries $\Delta g_{C(N)}$ defined in 
(\ref{defDeltag}) can be written as
\be \label{DeltagLOnum}
\Delta g_{C(N)}^{\rm LO}
 \simeq \frac{m_K^2}{F_0^2} \, B_{C(N)} \, \im G_8 +
D_{C(N)} \, \im (e^2 G_E) \, , 
\ee
where the functions $B_{C(N)}$ and $D_{C(N)}$ only depend
on $\re G_8$, $G_{27}$, $m_K$ and $m_{\pi}$. These functions were 
found in \cite{GPS03M} to be
\ba \label{ACBC}
B_C &=& -\frac{15}{64}\frac{1}{\pi}\,G_{27}
\sqrt{\frac{\mkd-9\mpd}
{\mkd+3\mpd}}\,\nonumber\\
&&\times \frac{14\mkd\mpd-18m_{\pi}^4+5m_K^4}{\mkd(\mkd-\mpd)
(3\re G_8+2G_{27})(13G_{27}-3\re G_8 )}\, ,\nonumber\\
D_C &=& \frac{3}{64}\frac{1}{\pi}\,\sqrt{\frac{\mkd-9\mpd}
{\mkd+3\mpd}}\frac{1}{\mkd(\mkd-\mpd)(3\re G_8 +2G_{27})
(13G_{27}-3\re G_8 )}
\nonumber\\ &&
\times  \left\lbrack3\re G_8 (16\mkd\mpd-18m_{\pi}^4+3m_K^4)
-G_{27}(178\mkd\mpd-234m_{\pi}^4+69m_K^4)\right\rbrack\,,\nonumber\\
\ea
and, in the neutral case,
\ba \label{ANBN}
B_N &=&  -\frac{15}{32}\frac{1}{\pi}\,G_{27}
\sqrt{\frac{\mkd-9\mpd}{\mkd+3\mpd}}\frac{7\mpd+4\mkd}{E}
\,,\nonumber\\
D_N &=& \frac{9}{32}\frac{1}{\pi}\,\re G_{8} 
\sqrt{\frac{\mkd-9\mpd}{\mkd+3\mpd}}\frac{\mpd(18\mpd-7\mkd)}{\mkd E}
\nonumber\\
&&+\frac{3}{32}\frac{1}{\pi}\,G_{27}
\sqrt{\frac{\mkd-9\mpd}{\mkd+3\mpd}}\frac{36m_{\pi}^4-119\mpd\mkd
-60m_K^4}{\mkd E}\,,
\ea
with 
\be 
E  \equiv (3\re G_8 +2G_{27})\left((19\mkd-4\mpd)G_{27}+6(\mkd-\mpd)
\re G_8 \right)\,.
\ee

In order to have simple expressions for $\Delta g_{C(N)}$ we used the 
next relations
\begin{itemize}
  \item $F_0^2\, \re \left(e^2G_E\right)<<m_\pi^2\re G_8$ 
  \item $\im G_8<<\re G_8$
  \item $\im (e^2 G_E)<<\re G_8$
\end{itemize}
Corrections to the terms regarded with the application of these relations 
have been found to be negligible \cite{GPS03}.

In Chapter \ref{chK3pi} we present numerical results for these asymmetries 
and the decay rate asymmetries at LO as well as at NLO. For the numerics given 
there we don't use any simplification as those applied in the analytical 
results.

\section{Couplings of the Leading Order Lagrangian} 
\label{LOcounterterms}

The couplings $G_{\Delta S=2}$, $G_8$, $G_{27}$ and $G_E$ that modulate the 
action of the different operators in the chiral Lagrangians (\ref{lagdS2}) and 
 (\ref{lagdS1}) are not fixed only by symmetry requirements. They are, in 
general, complex unknown 
functions and must be obtained by the calculation of hadronic matrix elements 
-following the different methods pointed out in Chapter \ref{matrixelements}- 
or fits to experimental data. Once the CHPT couplings have been extracted, 
one can make the predictions of the physical quantities at lowest order. One 
can also go further and obtain information on the NLO CHPT weak countertems 
-written in Section \ref{NLOcounterterms}- needed for instance in the isospin 
breaking 
corrections or in the rest of the NLO corrections. We discuss here the value 
of the most recent determinations of these couplings.

In \cite{BDP03},  a fit to all available $K \to \pi \pi$ 
amplitudes at NLO in CHPT \cite{BPP98}  and
$K \to 3 \pi$ amplitudes  and slopes in the $K\to 3 \pi$ 
amplitudes at NLO in CHPT was done.
The result found there for the ratio of the isospin definite [0 and 2]
$K\to \pi \pi$ amplitudes defined in (\ref{A0A2def}) to all orders in CHPT was
\be
\frac{A_0[K\to \pi\pi]}{A_2[K\to\pi\pi]}= 21.8 \, ; 
\ee
giving the infamous $\Delta I=1/2$ rule for Kaons and 
\be
\left[\frac{A_0[K\to \pi\pi]}{A_2[K\to\pi\pi]}\right]^{(2)}= 17.8 \, , 
\ee
to lowest CHPT order $p^2$. I.e., Final State Interactions  and the 
rest of higher order corrections are responsible for 22 \% of the 
$\Delta I= 1/2$ rule. Yet most of this enhancement appears at lowest 
CHPT order! The last result is equivalent to
\ba \label{G8exp}
\re G_8 = \left(7.0 \pm 0.6\right) \, 
\left(\frac{87 {\rm MeV}}{F_0}\right)^4 \, 
{\rm and} \, G_{27}= \left(0.50\pm0.06 \right) \, 
\left(\frac{87 {\rm MeV}}{F_0}\right)^4  \, .
\ea
No information can be obtained for $ \re (e^2 G_E)$ due to its tiny 
contribution to CP-conserving amplitudes. In this normalization, 
$G_{\Delta S=2}=G_8=G_{27}=1$ and $G_E=0$ at large $N_c$.
%, so 
%large $1/N_c$ corrections are needed \cite{Pich89,PichRafael996,JP94} 
%to understand the enhancement of $\re(G_8)$ in (\ref{G8exp}).

CP-conserving observables are fixed by 
physical meson masses, the pion decay coupling $F_0$
and the real part of the counterterms.
 To predict CP-violating asymmetries 
one  also need the values of the imaginary part of 
these couplings. Let us see what we know about them. 
At large $N_c$, all  the contributions 
to $\im G_8$ and  $\im (e^2 G_E)$  are 
factorizable and the  scheme dependence is not  under control.
The unfactorizable topologies are not included at this order and they
bring in unrelated dynamics, so that we cannot give an uncertainty
to the large $N_c$ result. We get
\ba \label{img8Nc}
\im G_8\Big|_{N_c} \!\!\!&=&\!\!\! 1.9 
 \, \im \tau \, , \nonumber \\
\im(e^2 G_E)\Big|_{N_c} \!\!\!&=&\!\!\!  -2.9 \, \im \tau \, ,
\ea
using $L_5(M_\rho)=(1.4\pm0.3)\times 10^{-3}$ and, from \cite{BPR},
\be
\langle 0 | \overline q q | 0 \rangle_{\overline{\rm MS}}(2 {\rm GeV})
=-(0.018 \pm 0.004) \, {\rm GeV}^3 
\ee
which agrees with the most recent sum rule determinations
of this condensate and of light quark masses 
--see \cite{JOP02,JOP02b} for instance-- and the lattice light
quark masses world average \cite{WIT02}. The Wilson coefficients 
necessary to get the results in (\ref{img8Nc}), i.e., $C_4$, $C_6$, 
$C_7$ and $C_8$ are known to two loops \cite{NLOWilscoef,CFMR94} 
as said in Chapter \ref{CPviolation}.
Finally, in the Standard Model
\be \label{tau}
\im \tau \equiv -\im \left( \frac{V_{td}V^*_{ts}}{V_{ud}V^*_{us}} \right) 
\simeq -(6.05 \pm 0.50)\times 10^{-4} \, .  
\ee

There has been recently advances on going beyond the leading order 
in $1/N_c$ in both couplings, $\im G_8$ and $\im (e^2 G_E)$.

In \cite{CDGM02,NAR01,BGP01}, there are recent model independent
calculations of $\im (e^2 G_E)$. The results there are valid to all
orders in $1/N_c$ and NLO in $\alpha_S$. They are obtained using 
the hadronic tau data collected by ALEPH \cite{ALEPH} and OPAL 
\cite{OPAL} at LEP. The agreement is quite good between them and their 
results can be summarized in
\be \label{imgeinput}
\im (e^2 G_E) = - (4.0 \pm 0.9) \left(\frac{87 {\rm MeV}}{F_0}\right)^6
\, \im \tau \, , 
\ee
where the central value is an average and the error
is  the smallest one. In Chapter \ref{chq7q8} we describe in more detail the 
calculation of this coupling performed in \cite{BGP01} and the compatibility 
with other determinations. 
In \cite{KPR01} it was used a Minimal
Hadronic Approximation to large $N_c$ to calculate $\im (e^2 G_E)$, 
they got
\be
\im (e^2 G_E) = -(6.7 \pm 2.0) \left(\frac{87 {\rm MeV}}{F_0}\right)^6
\, \im \tau \, ,
\ee
which is also in agreement though somewhat larger. There are also lattice
results for $\im (e^2 G_E)$ 
using domain-wall fermions \cite{CPPACS01,RBC01,domainwall} and
 using Wilson fermions \cite{wilson}. All of them made the chiral limit 
extrapolations, their results are in agreement between themselves (see
comparison in the tables the results of Section \ref{comparison}) and
their average gives
\be
\im (e^2 G_E) = - (3.2 \pm 0.3) \left(\frac{87 {\rm MeV}}{F_0}\right)^6
\, \im \tau \, .
\ee

There are  also results on  $\im G_8$ at NLO in $1/N_c$.
 In \cite{epsprime}, the authors made a calculation
using a hadronic model which reproduced the $\Delta I=1/2$ rule for Kaons
within 40\% through a very large $Q_2$ penguin-like contribution
-see \cite{BP99} for details. 
The results obtained were
\ba
\re G_8 = \left(6.0 \pm 1.7\right)\left(\frac{87 {\rm MeV}}{F_0}\right)^4
 , \,\,\,   & {\rm and}& G_{27} = \left(0.35 \pm 0.15 \right)
\left(\frac{87 {\rm MeV}}{F_0}\right)^4\, ,
\ea
in very good agreement with the experimental results in (\ref{experiment}).
The result found there was
\be \label{img8input}
\im G_8 = (4.4 \pm 2.2) \left(\frac{87 {\rm MeV}}{F_0}\right)^6\, \im \tau \, . 
\ee
at NLO in $1/N_c$. The uncertainty is dominated by the quark condensate error. 
The hadronic model used there had however some drawbacks
\cite{PPR} which will be eliminated and the work eventually updated 
following the lines in \cite{BGLP03}.

 In \cite{epsprime} there was also  a determination
of $\re (e^2 G_E)$ though very uncertain.

Very recently, using a Minimal Hadronic Approximation to large $N_c$, 
the authors of \cite{HPR03} found qualitatively similar results
to those in \cite{epsprime}.
I.e. enhancement toward the explanation of the $\Delta I=1/2$ rule
through $Q_2$ penguin-like diagrams and a matrix element of the gluonic
penguin $Q_6$ around three times the factorizable contribution. Indications 
of large values of $\im G_8$ were also found in \cite{HKPS00}.

\chapter{Charged Kaons $K \rightarrow 3\pi$ CP Violating 
Asymmetries at NLO in CHPT}
\label{chK3pi}

We report in this chapter the work in \cite{GPS03},  
in which the first full next-to-leading order analytical 
results in Chiral Perturbation Theory for the charged $K \to 3 \pi$
slope $g$ and decay rates CP-violating asymmetries defined in 
(\ref{defDeltag}) and (\ref{defDeltaGam}) respectively 
were found. 
We included the dominant Final State Interactions
at NLO analytically and discussed 
the importance of the unknown countertems. 
The large sensitivity of these asymmetries to the unknown counterterms can
be used to get  valuable information 
on those parameters and on the $\im G_8$ coupling --very important 
for the CP-violating parameter $\varepsilon'_K$ (see (\ref{delta12}))--
from their eventual measurement.

We calculate  the amplitudes 
\ba
\label{defdecays}
K_2(k)&\to&\pi^0(p_1)\pi^0(p_2)\pi^0(p_3)\,,\quad [A^2_{000}]\,,\nonumber\\
K_2(k)&\to&\pi^+(p_1)\pi^-(p_2)\pi^0(p_3)\,,\quad [A^2_{+-0}]\,,\nonumber\\
K_1(k)&\to&\pi^+(p_1)\pi^-(p_2)\pi^0(p_3)\,,\quad [A^1_{+-0}]\,,\nonumber\\
K^+(k)&\to&\pi^0(p_1)\pi^0(p_2)\pi^+(p_3)\,,\quad [A_{00+}]\,,\nonumber\\
K^+(k)&\to&\pi^+(p_1)\pi^+(p_2)\pi^-(p_3)\,,\quad [A_{++-}]\,,
\ea
as well as their CP-conjugated decays
at NLO in  the  chiral expansion (i.e. order $p^4$ in this case)
and in the isospin symmetry limit
$m_u=m_d$. We have also calculated the  contribution of the
$\order(e^2 p^2)$ electroweak octet counterterms. 
In (\ref{defdecays}) we 
have indicated the four momentum carried by each particle
and the symbol we will use for the amplitude.
The states $K_1$ and $K_2$ are defined as
\be
K_{1(2)} \,=\,\frac{K^0-(+)\overline{K^0}}{\sqrt{2}} \, .
\ee

The chiral Lagrangians defined in Chapter \ref{chCHPT} 
are the tools utilized to get these amplitudes. 
Our results for the  octet and 27-plet terms fully agree with
the results found in \cite{BDP03} so that we don't write them again, we 
only give in Appendix \ref{Amplitudes} the relations between the 
functions defined there and those we used to describe the charged kaon 
decays.
 The electroweak (EW)
 contributions to $K \to 3 \pi$ decays  of order $e^2 p^0$ and
$e^2 p^2$ can be found 
in  Appendix B.1 of reference \cite{GPS03}.

Definitions of the asymmetries are in Section
 \ref{cpK3pi} of Chapter \ref{CPviolation}.
 In Section \ref{inputs} we collect the inputs we use
for the weak counterterms in the leading and next-to-leading order
weak chiral Lagrangians.
In Section \ref{CPconserving}
 we give the CHPT predictions  at leading- and next-to-leading 
order for the decay rates and the slopes $g$, $h$ and $k$.  
We discuss the results for the CP-violating
asymmetries at leading order 
  first in Section \ref{LOresults} 
and we discuss them at NLO in Section \ref{NLOresults}.
Finally,  we make comparison 
with earlier work in Section \ref{conclu}.
In  Appendix \ref{Amplitudes} we give the notation
we use for  the $K \to 3 \pi$ amplitudes and  new
results at order $e^2 p^2$. In Appendix \ref{Adeltag}
 we give the analytic formulas needed for the slope $g$ and the asymmetries
$\Delta g$ at LO and NLO 
and in Appendix \ref{ANLO}  the relevant quantities  to calculate
the decay rates $\Gamma$ and  
the CP-violating asymmetries in the decay rates  $\Delta \Gamma$
 also  at LO and NLO. 
In Appendix \ref{FSI6} we give  the  analytical
results for the dominant --two-bubble--  FSI contribution
to the decays of charged Kaons and to the CP-violating 
asymmetries at NLO order, i.e. order $p^6$.

\section{Numerical Inputs for the Weak Counterterms}
\label{inputs}

Here we collect the values of the weak  counterterms we use
in this chapter. For a discussion on the values of these parameters  
see Section \ref{LOcounterterms} in Chapter \ref{chCHPT}. 

At LO we need the next values 
\ba
\label{experiment}
\re G_8 = \left(6.8 \pm 0.6\right) \,  
{\rm and} \, G_{27}= \left(0.48\pm0.06 \right) \,  ,
\ea
for the real part and 
\be
\label{gluonpenguin}
\im G_8 = (4.4 \pm 2.2) \, \im \tau \, , 
\ee
\be
\label{EMpenguin}
\im (e^2 G_E) = - (4.0 \pm 0.9) \, \im \tau \, , 
\ee
for the imaginary part of the couplings. 
For the results in the large $N_C$ limit we use
\ba \label{largeNccouplings}
\im G_8\Big|_{N_c} \!\!\!&=&\!\!\! 1.9 
 \, \im \tau \, , \nonumber \\
\im(e^2 G_E)\Big|_{N_c} \!\!\!&=&\!\!\!  -2.9 \, \im \tau \, .
\ea

In \cite{epsprime} there was a determination
of $\re (e^2 G_E)$ though very uncertain.
However, since the contribution of $\re (e^2 G_E)$ is very small
in all the quantities we calculate in this chapter, we take the value from 
\cite{epsprime} with 100\% uncertainty and
add its contribution  to the error of those quantities. 

\subsection{Counterterms of the NLO Weak Chiral Lagrangian}
\label{Kcount}

To describe $K\to 3 \pi$ at NLO, 
in addition to $\re G_8$, $G_{27}$, $\re (e^2 G_E)$, $\im G_8$ and
$\im (e^2 G_E)$, we also need several other ingredients.
Namely, for the real part we need the chiral logs and the counterterms. 
The relevant counterterm combinations 
were called $\widetilde K_i$ in \cite{BDP03}. 
The chiral logs are fully 
analytically known \cite{BDP03} --we have confirmed them 
in the present work. The real part of the counterterms,
 $\re \widetilde K_i$, can be obtained from 
the fit  of the $K \to 3 \pi$ 
CP--conserving decays  to data done in \cite{BDP03}. The relation of the 
$\widetilde K_i$ counterterms and those defined in Section 
\ref{NLOcounterterms} of Chapter \ref{chCHPT}, 
and the values used for them are in Table \ref{tabKdef} and Table 
\ref{tabKvalues} respectively.
\begin{table}
\begin{center}
\begin{tabular}{|c|c|}\hline
$\widetilde K_1$ & $ \re (G_8) (N_5^r-2N_7^r+2N_8^r+N_9^r)+G_{27}
\left(-\frac{1}{2}D_6^r\right)$\\\hline
$\widetilde K_2$ & $ \re (G_8) (N_1^r+N_2^r)+G_{27}
\left(\frac{1}{3}D_{26}^r-\frac{4}{3}D_{28}^r\right)$\\\hline
$\widetilde K_3$ & $\re (G_8) (N_3^r)+G_{27}
\left(\frac{2}{3}D_{27}^r+\frac{2}{3}D_{28}^r\right)$\\\hline
$\widetilde K_4$ & $G_{27}\left(D_4^r-D_5^r+4D_7^r\right)$\\\hline
$\widetilde K_5$ & $G_{27}\left(D_{30}^r+D_{31}^r+2D_{28}^r\right)$\\\hline
$\widetilde K_6$ & $G_{27}\left(8D_{28}^r-D_{29}^r+D_{30}^r\right)$\\\hline
$\widetilde K_7$ & $G_{27}\left(-4D_{28}^r+D_{29}^r\right)$\\\hline
$\widetilde K_8$ & $ \re(G_8)
(2N_5^r+4N_7^r+N_8^r-2N_{10}^r-4N_{11}^r-2N_{12}^r)
+G_{27}\left(-\frac{2}{3}D_1^r+\frac{2}{3}D_6^r\right)$\\\hline
$\widetilde K_9$ & $\re(G_8)
(N_5^r+N_8^r+N_9^r)+G_{27}
\left(-\frac{1}{6}D_6^r\right)$\\\hline
$\widetilde K_{10}$ & $G_{27}\left(2D_2^r-2D_4^r-D_7^r\right)$\\\hline
$\widetilde K_{11}$ & $G_{27} D_7^r $\\\hline
\end{tabular}
\end{center}
\caption{\label{tabKdef} Relevant combinations of the octet $N_i^r$ 
and 27-plet $D_i^r$ weak counterterms for $K\to 3\pi$ 
decays.}     
\end{table}
\begin{table}
\begin{center}
\begin{tabular}{||c|c|c||}\hline
 & $\re \widetilde K_i(M_\rho)$ from \cite{BDP03}& 
$\im \widetilde K_i(M_\rho)$  from (\ref{assum1})\\  
\hline\hline
$\widetilde K_2(M_\rho)$ & $0.35\pm0.02$
& $\lbrack 0.31 \pm 0.11 \rbrack\, \im \tau$ \\ 
\hline
$\widetilde K_3(M_\rho)$ & $0.03\pm0.01$ 
& $\lbrack 0.023\pm 0.011\rbrack \,\im \tau$ \\
\hline
$\widetilde K_5(M_\rho)$ & $-(0.02\pm0.01)$& 
$0$\\ \hline
$\widetilde K_6(M_\rho)$ & $-(0.08\pm0.05)$ & $0$ \\\hline
$\widetilde K_7(M_\rho)$ & $0.06\pm0.02$ & $0$ \\\hline
\end{tabular}
\end{center}
\caption{\label{tabKvalues} Numerical inputs used for 
the weak counterterms of order $p^4$. 
The values of $\re \widetilde K_i$ and $\im \widetilde K_i$ 
which do not appear are zero. For explanations, see the text.} 
\end{table}

For the imaginary parts
at NLO, we need $\im G_8'$ in addition to $\im G_8$ and $\im (e^2 G_E)$.
To the best of our knowledge, there is just one calculation  at NLO 
in $1/N_c$ at present \cite{epsprime}. The results found there,
using the same hadronic model discussed above, are
\ba
\label{G8pri}
\re G_8' = 0.9 \pm 0.1  \,\, \,  
  & {\rm and} & \im G_8' = (1.0 \pm 0.4) \,
\im \tau \, .
\ea
The imaginary part of the order $p^4$ counterterms, $\im \widetilde K_i$,
is much more problematic.  They cannot be obtained from data and 
there is no available NLO in $1/N_c$ calculation for them.

One can use several approaches to get the order of magnitude
and/or the signs of $\im \widetilde K_i$, such as factorization plus 
meson dominance or the construction of the relevant $\Delta S=1$ Green's 
functions using a determined model; as explained in Section 
\ref{NLOcounterterms} of Chapter \ref{chCHPT}.

We will follow here more naive approaches
that will be enough for our purpose of estimating the effect of the 
unknown counterterms.
 We can assume   that the ratio of the real to the imaginary parts
is  dominated by the same strong dynamics at LO and NLO 
in CHPT, therefore
\be
\label{assum1}
\frac{\im \widetilde K_i}{\re \widetilde K_i}
\simeq \frac{\im G_8}{\re G_8} 
\simeq \frac{\im G_8'}{ \re G_8'} \simeq (0.9 \pm 0.3) \, \im \tau  \, ,
\ee
if we use (\ref{gluonpenguin}) and (\ref{G8pri}).
The results obtained under these  assumptions for the imaginary 
part of the $\widetilde K_i$ counterterms are written in 
Table \ref{tabKvalues}. 
In particular, we set to zero those $\im \widetilde K_i$
whose corresponding $\re \widetilde K_i$ are 
set also to zero in the fit to CP-conserving amplitudes
done in \cite{BDP03}.
Of course, the relation above can only be applied to those 
$\widetilde K_i$ couplings
with non-vanishing imaginary part.
Octet dominance to order $p^4$ is a further
assumption implicit in (\ref{assum1}).
The second equality  in (\ref{assum1})
 is well satisfied by the model calculation
in (\ref{G8pri}). 

The values of $\im \widetilde K_i$
obtained using (\ref{assum1}) 
 will allow us to check the counterterm dependence
of the CP-violating asymmetries. They will also provide us  a good 
estimate of the counterterm contribution to the CP-violating asymmetries
that we are studying. 

We can get a second  piece of information   from the 
variation of the amplitudes when $\im \widetilde K_i$ are put to zero 
and the remaining scale dependence is varied between 
$M_\rho$ and 1.5 GeV.
We use in this case the known scale dependence of 
$ \re \widetilde K_i$ together with their 
absolute value  at the scale $\nu=M_\rho$ from \cite{BDP03}.

\section{CP-Conserving Observables} 
\label{CPconserving}

Here we give the results for the CP-conserving 
slopes $g_C$, $h_C$, and $k_C$
  and the decay rate $\Gamma_C$ of $K^+ \to \pi^+ \pi^+ \pi^-$
and slopes $g_N$, $h_N$, and $k_N$
  and decay rate $\Gamma_N$ of $K^+ \to \pi^+ \pi^0 \pi^0$
within CHPT at LO and NLO.
These results are not new --see \cite{BDP03} and references therein--
but we want to give them again, first as  a check of our analytical 
results and second, to recall the  kind of corrections that one 
expects in the CP-conserving quantities from LO to NLO
for the different observables.

We will use the values of $\re G_8$ and $G_{27}$ in (\ref{experiment}),
and disregard the EW corrections since we
are in the isospin limit  and  they are much smaller than the
octet and  27-plet contributions. For the real part of
the NLO counterterms, we will use the results from a fit to data 
in \cite{BDP03}. So, really these are just checks.

The  values of the NLO counterterms given in \cite{BDP03} were fitted 
without including CP-violating 
contributions in the amplitudes, i.e., taking 
the coupling $G_8$ and the counterterms themselves as real quantities.
The inclusion of an imaginary part for these couplings
does not affect significantly the CP  conserving observables.

To be consistent with the fitted values 
of the counterterms  of the $\order(p^4)$
Lagrangian   we do not consider any 
$\order (p^6)$ contribution to the amplitudes in this section.  
Indeed,  these  counterterms,
 fixed with the use of experimental data and order $p^4$ formulas, 
do contain the  effects of higher order contributions.
We also use the same conventions used in \cite{BDP03}
 for the pion masses, i.e., we use  the average final state
pion mass which for $K^+\to \pi^+\pi^+\pi^-$ is
$m_\pi=$ 139 MeV and for $K^+\to \pi^0\pi^0\pi^+$ is
$m_\pi=$ 137 MeV.
In the following subsections we provide analytic formulas
 at LO  and in  Tables \ref{tabCPcons}  and \ref{tabslopes}
we give the numerical results.

\subsection{Slope  $g$} \label{slopeg}

The slope $g$ is defined in equation (\ref{gdefinition}). 
We give here  the results for
\ba
 \, g_C &\equiv&\frac{1}{2} 
\Big\{g[K^+ \to \pi^+ \pi^+ \pi^-] + g[K^- \to \pi^- \pi^- \pi^+]
\Big\}
\nonumber\\
 \, {\rm and} \, \, \, g_N &\equiv& \frac{1}{2}
\Big\{g[K^+ \to \pi^0 \pi^0 \pi^+]+g[K^-\to \pi^0\pi^0\pi^-] \Big\} .
\ea
Without including  the tiny CP-violating effects  
$g[K^+]_{\rm LO}=g[K^-]_{\rm LO}$, 
\ba \label{gLOCN}
g_{C}^{\rm LO} &=&  \, \, -3\mpd
\frac{3\re G_8-13G_{27}}{\mkd\left(3\re G_8 + 2G_{27}\right)+9F_0^2
\re \left(e^2G_E\right)}\, 
,\nonumber\\
g_{N}^{\rm LO} \, \, &=& \, \, 3\frac{\mpd}{(\mkd-\mpd)}
\frac{(19\mkd-4\mpd)G_{27}+6(\mkd-\mpd) \re G_8+9F_0^2
\re \left(e^2G_E\right)}
{\mkd\left(3\re G_8 + 2G_{27}\right)+9F_0^2\re \left(e^2G_E\right)} .
\nonumber\\
\ea
The value for $\re (e^2 G_E)$ is not very well known. 
However 
its contribution turns out to be negligible and for numerical purposes
we take the result for 
$\re (e^2 G_E)$  from \cite{epsprime} with 100\% uncertainty.
We do not consider its contribution for the central values 
in Table \ref{tabCPcons} and  we add its effect to the quoted 
error.  In addition, the quoted uncertainty for $g^{\rm LO}_{C}$
 and $g^{\rm LO}_{N}$ 
contains  the uncertainties from
$\re G_8$ and $G_{27}$ in (\ref{experiment}).

The analytical NLO formulas are in (\ref{AgNLO}).
It is interesting to observe the impact of the counterterms 
so that we calculate also the slopes at NLO with $\widetilde K_i=0$,
see Table \ref{tabCPcons}. The contribution of the counterterms
at $\mu=M_\rho$ is relatively small for $g_C$ and $g_N$, see Table 
\ref{tabCPcons}.
\begin{table}
\begin{center}
\begin{tabular}{||c|c|c|c|c||}\hline
 &$ g_C$ & $\Gamma_C \; (10^{-18}\ {\rm GeV})$&
$ g_N$ & $\Gamma_N \; (10^{-18}\ {\rm GeV})$\\  \hline\hline
LO & $-0.16\pm0.02$ &$1.2\pm0.2$ &$0.55\pm 0.04$  
& $0.37\pm 0.07$ \\ \hline
NLO,$\widetilde K_i(M_\rho)$ &&&&\\
 from Table \ref{tabKvalues}&
$-0.22 \pm 0.02$ & $3.1 \pm 0.6$ & $0.61 \pm 0.05$ & $0.95 \pm 0.20$ 
\\\hline
NLO,&&&& \\
 $\widetilde K_i(M_\rho)=0$
 & $-0.28 \pm 0.03$ &$1.3 \pm 0.4$ & $0.80 \pm 0.05$ &  $0.41 \pm 0.12$ 
\\\hline
PDG02 &
$-0.2154\pm 0.0035$ & $2.97\pm 0.02$ & $0.652\pm 0.031$ 
& $0.92\pm0.02$\\\hline
 ISTRA+ & -- & -- & $0.627\pm 0.011$ 
& --\\\hline
 KLOE & -- & -- & $0.585\pm0.016$
& $0.95\pm0.01$\\\hline
\end{tabular}
\end{center}
\caption{\label{tabCPcons} CP conserving predictions for the 
slope $g$ and the decay rates. The theoretical errors come from the 
variation in the inputs parameters discussed in Section \ref{inputs}.
In the last three lines, we give the experimental 2002 world
average from PDG \cite{PDG02}, and the recent results
from ISTRA+ \cite{ISTRA+} and the preliminary ones from KLOE \cite{KLOE} 
which are not included in \cite{PDG02}.}
\end{table}

\subsection{Slopes $h$ and $k$}

We can also predict  the slopes $h_{C(N)}$ and $k_{C(N)}$ 
defined  in (\ref{gdefinition}).
At LO,  the slope $k_{C}$  for $K^+ \to \pi^+ \pi^+ \pi^-$
and the slope $k_N$ for $K^+ \to \pi^0 \pi^0 \pi^+$  are identically zero
and the corresponding slopes $h_{C(N)}$  are equal to $g_{C(N)}^2/4$. 
The NLO results are written in Table \ref{tabslopes} 
together with the 
slopes obtained when the counterterms $\widetilde K_i$ are switched off
at $\mu=M_\rho$. We can see 
that  the slopes $h_{C(N)}$ and $k_{C(N)}$
 are dominated by the counterterm 
contribution contrary to what happened with $g_{C(N)}$ which get
the main contributions  at LO. 
\begin{table}
\begin{center}
\begin{tabular}{||c|c|c|c|c||}\hline
 &$ h_C$ & $k_C $&
$ h_N$ & $k_N$\\  \hline\hline
LO & $0.006 \pm 0.001$ &$0$ &$0.075 \pm 0.003$  
& $0$ \\ \hline
NLO,$\widetilde K_i(M_\rho)$ &&&&\\
from Table \ref{tabKvalues}&
$0.012 \pm 0.005$ & $-0.0054 \pm 0.0015$ & $0.069 \pm 0.018$ 
& $0.004 \pm  0.002$ \\\hline
NLO, &&&&\\
$\widetilde K_i(M_\rho)=0$
 & $0.04 \pm 0.01$ & $0.0004 \pm 0.0025$ & $0.15 \pm 0.05$
&$0.008 \pm 0.002$\\\hline
PDG02 &$0.012\pm0.008$ & $-0.0101\pm0.0034$ & $0.057\pm 0.018$ 
& $0.0197\pm0.0054$\\\hline
 ISTRA+ & -- & -- & $0.046\pm 0.013$ 
& $0.001\pm 0.002$ \\\hline
 KLOE & -- & -- & $0.030\pm0.016$ 
& $0.0064\pm0.0032$ \\\hline
\end{tabular}
\end{center}
\caption{\label{tabslopes} CP conserving predictions for the slopes $h$ and $k$. 
The theoretical errors come from the
variation in the inputs parameters discussed in Section \ref{inputs}.
In the last three lines, we give the experimental 2002 world
average from PDG \cite{PDG02}, and the recent results
from ISTRA+ \cite{ISTRA+}  and the preliminary ones from
KLOE \cite{KLOE}
which are not included in \cite{PDG02}.}
\end{table}

\subsection{Decay Rates}

The decay rates $K \to 3 \pi$ with two identical
pions can be written as 
\be \label{Gammadef}
\Gamma_{ijl} \,\equiv \frac{1}{512 \pi^3 m_K^3}\,
\int_{s_{3min}}^{s_{3max}} ds_3\int_{s_{1min}}^{s_{1max}} ds_1\,
|A(K\rightarrow \pi^i\pi^j\pi^l)|^2 ,
\ee
with 
\ba
\label{eq:extrs} 
s_{1max}&=&(E_j^*+E_l^*)^2-\left(\sqrt{E_j^{*2}-m_j^2}
-\sqrt{E_l^{*2}-m_l^2}\right)^2\, ,\nonumber\\
s_{1min}&=&(E_j^*+E_l^*)^2-\left(\sqrt{E_j^{*2}-m_j^2}
+\sqrt{E_l^{*2}-m_l^2}\right)^2 \, ,\nonumber\\
s_{3max}&=&(m_K-m_l)^2\quad{\rm and}\quad s_{3min}=(m_i+m_j)^2\,.
\ea
The energies $E_j^*=(s_3-m_i^2+m_j^2)/(2\sqrt{s_3})$ and 
$E_l^*=(m_K^2-s_3-m_l^2)/(2\sqrt{s_3})$ are those of the pions $\pi^j$ and 
$\pi^l$ in the $s_3$ rest frame.
It is useful  to define 
\ba
\label{defACN}
|A_C|^2&=&\frac{1}{2}\Big\{
|A\left(K^+\rightarrow \pi^+\pi^+\pi^-\right)|^2
+ |A\left(K^-\rightarrow \pi^-\pi^-\pi^+\right)|^2\Big\}\,,\nonumber\\
|A_N|^2&=&
\frac{1}{2} \Big\{|A\left(K^+\rightarrow \pi^0\pi^0\pi^+\right)|^2
+ |A\left(K^-\rightarrow \pi^0\pi^0\pi^-\right)|^2 \Big\}\,.
\ea

At LO and again disregarding the tiny CP-violating effects we get
\ba
\label{GammaLO}
&&\,|A_C^{\rm LO}|^2\,\equiv \, 
|A_{++-}^{\rm LO}|^2\,=\,|A_{--+}^{\rm LO}|^2\,=\, \nonumber \\ 
&& |C|^2 \times \left| \re G_8 \left(s_3-\mkd-\mpd\right) \,+\,
\frac{G_{27}}{3}\left(13\mpd+3\mkd-13s_3\right)\,
+\,\re\left(e^2G_E\right)(-2F_0^2)\right|^2\, , \nonumber\\
&&\,|A_{N}^{\rm LO}|^2\,\equiv\,
|A_{00+}^{\rm LO}|^2\,=\,|A_{00-}^{\rm LO}|^2\,= \, |C|^2 \times
\Big| \re G_8 \left(\mpd-s_3\right) \nonumber\\
&&\hspace{2 cm}+
\frac{G_{27}}{6(\mkd-\mpd)}\left( 5m_K^4+19\mpd\mkd
-4m_{\pi}^4+s_3(4\mpd-19\mkd)\right)\nonumber\\ 
&&  \hspace{2 cm}+\re\left(e^2G_E\right)
\frac{F_0^2}{2(\mkd-\mpd)}\left(5\mpd-\mkd-3s_3\right)\Big|^2\,.
\nonumber \\
\ea
where the constant C is given in (\ref{Cdef}). 
 The amplitudes $|A_{C(N)}|^2$ needed for the NLO prediction are in  
(\ref{eqANLO}) in Appendix \ref{ANLO}.

The  results for $\Gamma_C$ and $\Gamma_N$ at LO and NLO
 are in Table \ref{tabCPcons}.
 The contribution of $\re (e^2 G_E)$ is  very small 
(around 1\%) and we  include  it  in the final uncertainty  as in 
Section \ref{slopeg} together with the rest of input uncertainties.
We have also  included  in Table \ref{tabCPcons}
the results  with  the counterterms $\widetilde K_i=0$ at $\mu=M_\rho$.
 We can conclude from them that the decay widths are 
strongly dependent  on  the NLO counterterms contribution.

\section{CP-Violating Observables at Leading Order} 
\label{LOresults}

In this Section we analyze analytically as well as numerically
 the LO results for the asymmetries 
in the slope $g$ and decay rates $\Gamma$ that are 
defined in (\ref{defDeltag}) and (\ref{defDeltaGam}). As explained 
in Section \ref{CHPTresultsLO} they are order $p^2$ in CHPT.

We have checked that
the effect of   $\re (e^2 G_E)$  is very small
also for the $\Delta g$ and $\Delta \Gamma$ asymmetries. 
 For the numerics, we have used
$\re (e^2 G_E)=0$  and used the value in \cite{epsprime} with 
100\%  variation to estimate its contribution
which we have added to the 
quoted final uncertainty of the asymmetries.
For the  $\re G_8$ and $G_{27}$ we have used always 
the values in (\ref{experiment}). 
For $\im G_8$ and $\im (e^2 G_E)$, we have used
two sets of inputs; namely,  the large $N_c$ limit
predictions in (\ref{largeNccouplings})
and the values  in (\ref{gluonpenguin}) and
(\ref{EMpenguin}). 
For the pion masses we have used the same convention
used in \cite{BDP03} and given here in Section \ref{CPconserving}.
The results are reported  in Table \ref{tabLO}.

\subsection{CP-Violating Asymmetries in the Slope $g$} \label{LODeltag}

At LO, the CP-violating asymmetries in the slope $\Delta g_{C(N)}$
can be written as \cite{GPS03M} 
\be 
\Delta g_{C(N)}^{\rm LO}
 \simeq \frac{m_K^2}{F_0^2} \, B_{C(N)} \, \im G_8 +
D_{C(N)} \, \im (e^2 G_E) \, , 
\ee
where the functions $B_{C(N)}$ and $D_{C(N)}$ only depend
on $\re G_8$, $G_{27}$, $m_K$ and $m_{\pi}$  and can be found in 
(\ref{ACBC}) and (\ref{ANBN}) in Section \ref{CHPTresultsLO}. 
Numerically, 
\ba
\label{eq:gLOeff}
\Delta g_{C}^{\rm LO}
 \simeq \left \lbrack 1.16 \, \im G_8 -  0.12\, \im (e^2 G_E) 
\right \rbrack \times 10^{-2}\, ,
\nonumber \\ 
\Delta g_{N}^{\rm LO}
 \simeq - \left \lbrack 0.52 \, \im G_8 +  0.063\, \im (e^2 G_E) 
\right \rbrack \times 10^{-2}\, .
\ea
\begin{table}
\begin{center}
\begin{tabular}{||c|c|c|c|c||}\hline
&$ \Delta g_C^{\rm LO}(10^{-5})$ &$ 
\Delta \Gamma_C^{\rm LO}(10^{-6})$
&$ \Delta g_N^{\rm LO}(10^{-5})$ &$ \Delta
 \Gamma_N^{\rm LO}(10^{-6})$\\\hline\hline
(\ref{largeNccouplings}) 
& $-1.5   $ &$ -0.2   $ & $0.5   $ & $0.8   $\\
(\ref{gluonpenguin}) and (\ref{EMpenguin})& $-3.4\pm 2.1$&$-0.6\pm 0.4$&
$1.2\pm 0.8$ &$2.0\pm 1.3$\\
\hline
\end{tabular}
\end{center}
\caption{\label{tabLO}CP-violating predictions  at LO in the chiral
  expansion.  The details of the calculation are  
in Section \ref{LOresults}. The inputs used  for $\im G_8$ and
$\im (e^2 G_E)$ are in the first column.
The difference between $\Delta g_C^{\rm LO}$ here and the one 
reported in \cite{GPS03M} comes from updating the values
of $\re G_8$ and $G_{27}$ from \cite{BDP03}.
The error in the first line is not reported for the reasons 
explained in Section \ref{LOcounterterms}.}
\end{table}
 \ From (\ref{eq:gLOeff}) and  the inputs 
discussed in Section  \ref{LOcounterterms}
we conclude that the asymmetries 
$\Delta g_{C (N)}$ are poorly sensitive to $\im (e^2 G_E)$.
This fact makes  an accurate enough measurement of these asymmetries 
 very interesting to check if  
$\im G_8$ can be as large as predicted in \cite{epsprime,HPR03,HKPS00}.
It also makes these CP-violating asymmetries
complementary to the direct CP-violating parameter $\varepsilon'_K$
where there is a  cancellation between the $\im G_8$ 
and $\im (e^2 G_E)$ contributions. 

\subsection{CP-Violating Asymmetries in the Decay Rates}
\label{5.2}

The observables we study here were   defined in
(\ref{defDeltaGam}). We can write them again as follows
\ba
\label{eq:dGLO}
\Delta\Gamma_{C(N)}&=&
\frac{\int_{s_{3min}}^{s_{3max}}ds_3\int_{s_{1min}}^{s_{1max}}ds_1
   \, \Delta |A_{C(N)}|^2}{
\int_{s_{3min}}^{s_{3max}}ds_3\int_{s_{1min}}^{s_{1max}}ds_1
 \, |A_{C(N)}|^2 }
\ea
 where the extremes of integration are 
in (\ref{eq:extrs}), the quantities $|A_{C(N)}|^2$ were defined
in (\ref{defACN}) and $\Delta |A_{C(N)}|^2$ are defined by
\ba \label{Deltapm}
\Delta |A_C|^2&=& \frac{1}{2}
\Big\{ |A\left(K^+\rightarrow \pi^+\pi^+\pi^-\right)|^2
- |A\left(K^-\rightarrow \pi^-\pi^-\pi^+\right)|^2\Big\} \,,\nonumber\\
\Delta |A_N|^2&=&\frac{1}{2} 
\Big\{|A\left(K^+\rightarrow \pi^0\pi^0\pi^+\right)|^2
-|A\left(K^-\rightarrow \pi^0\pi^0\pi^-\right)|^2\Big\}\,.
\ea

At LO we get,  
\ba \label{DGammaLO}
\Delta |A_{C(N)}^{\rm LO}|^2 &=& 2 \,
\Bigg \lbrack \im G_8 \left\{ \,G_{27}\left(B_{8}^{(2)}C_{27}^{(4)}
-B_{27}^{(2)}C_8^{(4)}\right)\,\right.
\nonumber \\ && \left.
\hspace{1 cm}+\,\re \left(e^2G_{E}\right)\left(B_{8}^{(2)}C_{E}^{(4)}
-B_{E}^{(2)}C_8^{(4)}\right)\right\}\nonumber\\
&&+\im \left(e^2G_E\right) \,\left\lbrace \re G_8 \left(B_E^{(2)}C_8^{(4)}
-B_8^{(2)}C_E^{(4)}
\right)\right.
\nonumber \\ && \left. 
\hspace{1 cm}+ G_{27}\left(B_E^{(2)}C_{27}^{(4)}-B_{27}^{(2)}C_E^{(4)}\right)
\right\rbrace\hspace{1.5 cm}\Bigg \rbrack\, ,
\ea
where we  do not show explicitly the  $s_j$ 
dependence of the functions $B_i^{(2)}$ and $C_i^{(4)}$
nor the subscript $C$ or $N$  in $B_i^{(2)}$ and  $B_i^{(4)}$
for the sake of simplicity.
The analytical expressions for  the functions $B_i^{(2)}$ and $C_i^{(4)}$
are reported in Appendix \ref{ANLO}.

In (\ref{DGammaLO}),  we have
used  consistently the  LO  result for  the denominator of 
(\ref{eq:dGLO}) though its value is very different
from the experimental number, see Table \ref{tabCPcons}.
 
The numerics  for the asymmetries in the decay rates are
\ba
\Delta \Gamma_{C}^{\rm LO}
 \simeq \left\lbrack 0.24  \, \im G_8 +  0.03\, 
\im (e^2 G_E) \right\rbrack \times 10^{-3} \, ,
\nonumber \\ 
\Delta \Gamma_{N}^{\rm LO} \simeq -\left\lbrack 0.88  \, \im G_8 + 
0.13\, \im(e^2 G_E)  \right\rbrack \times 10^{-3}\, .
\ea
The results using the two sets of inputs discussed in Section
\ref{LOcounterterms} for
$\im G_8$ and $\im (e^2 G_E)$ are reported in Table \ref{tabLO}.
The asymmetries  in the width are also poorly sensitive to 
$\im (e^2 G_{E})$
thus also their accurate measurement
 will provide important information on  $\im G_8$.

In \cite{AVI81},  it was noticed that the asymmetry 
$\Delta \Gamma_{C}$ increases if a cut on the energy of the 
pion with charge opposite that of the decaying Kaon is made. Afterward, 
 the authors in  \cite{GRW86}  claimed that if this cut is made 
at $s_3=1.1\times 4\mpd$, the asymmetry 
 is enhanced by one  order of magnitude.
We checked that the decay rate asymmetry $\Delta \Gamma_C$
at LO changes from its value in Table \ref{tabLO} 
 to $\Delta \Gamma_{C}=-5.6\times10^{-6}$, i.e.
one order of magnitude enhancement when we perform such a cut in the 
integration, 
in agreement with \cite{GRW86}. It remains to see if this
enhancement persists  at NLO and how feasible is 
to perform this cut at the experimental side.
We will come back to this issue in the conclusions 
in Section \ref{conclu}. This enhancement
does not occur for $\Delta \Gamma_N$.

\section{CP-Violating Observables at Next-to-Leading Order} 
\label{NLOresults}

At NLO one needs the real parts at order $p^4$, 
i.e. at one-loop, for which we have
the exact expression, see Appendix \ref{Amplitudes}. 
To make the full discussion about CP-violating asymmetries at NLO
in CHPT we also need the FSI at order $p^6$
that would imply to calculate  $K \to 3 \pi$ amplitudes
at two-loops. However, one can use the optical  theorem 
and the one-loop and tree-level  $\pi \pi$ scattering and $K \to
3 \pi$ results 
 to get the imaginary part  of the dominant two-bubble
contributions. The results for these dominant two-bubble
FSI are presented in the next subsection.

\subsection{Final State Interactions at NLO}
\label{6.1}

Though the complete analytical FSI at NLO are unknown at present, 
 one can do a very good job using the known results at order 
$p^2$ and order $p^4$ for $\pi \pi$ scattering and for $K \to 3 \pi$ 
together with the optical theorem
to get analytically the  order $p^6$ imaginary parts that come
from two-bubbles. 
These contributions are expected to be dominant to a very good accuracy.
 We are disregarding  three-body re-scattering since they cannot be 
written as a bubble resummation. One can expect them to be rather 
small being suppressed by the available phase space \cite{DIPP94}.

Making use of the Dalitz variables defined in (\ref{Dalitzvar}) the 
amplitudes in (\ref{defdecays}) 
[without isospin breaking terms] 
can be written as expansions in powers of  $x$ and $y$, 
\ba \label{amp1}
A_{++-} &=& (-2\alpha_1+\alpha_3)\,-\,(\beta_1-\frac{1}{2}\beta_3+\sqrt 3
\gamma_3)\,y \,+\,\order(y^2,x)\, ,\nonumber\\
A_{00+} &=& \frac{1}{2}(-2\alpha_1+\alpha_3)\,-\,(-\beta_1
+\frac{1}{2}\beta_3+\sqrt 3\gamma_3)\,y  \,+\,\order(y^2,x)\,,\nonumber\\
A^2_{+-0} &=& (\alpha_1+\alpha_3)^R\,- \,(\beta_1+\beta_3)^R\,y
 \,+\,\order(y^2,x)\, ,\nonumber\\
A^1_{+-0} &=&  (\alpha_1+\alpha_3)^I\,- \,(\beta_1+\beta_3)^I\,y
 \,+\,\order(y^2,x)\, ,\nonumber\\
A^2_{000} &=& 3\,(\alpha_1+\alpha_3)^R \,+\,\order(y^2,x) \, ,\nonumber\\
A^1_{000} &=& 3\,(\alpha_1+\alpha_3)^I \,+\,\order(y^2,x) \, ,
\ea
where the parameters $\alpha_i$, $\beta_i$ and $\gamma_i$ are  
functions of the 
pion and Kaon masses, $F_0$, the lowest order $\Delta S=1$ Lagrangian
couplings $G_8$, 
$G_8'$, $G_{27}$, $G_E$ and the counterterms  appearing 
at order $p^4$, i.e., $L_i's$, $\widetilde K_i's$. 
We do not add here
EW corrections since we expect them to be small and of the same size of 
isospin breaking effects in quark masses which we have not considered.
 The $\order ( e^2 p^0)$ and $\order (e^2 p^2)$
contributions can be found in Appendix B.1 of reference \cite{GPS03}.

 In (\ref{amp1}), superindices $R$  and $I$ mean that either  the 
real part of the counterterms or their imaginary part 
appear, respectively. In the remainder,  the 
superscript ${(+-0)}$ will refer to the 
amplitude $A(K^0\to\pi^+\pi^-\pi^0)= 
(A^2_{+-0}+A^1_{+-0})/\sqrt{2}$, that is proportional to 
the full couplings and not 
only to the real or the imaginary part of such couplings.

If we do not consider FSI, the complex parameters $\alpha_{i}^{{\rm NR}}$, 
$\beta_{i}^{{\rm NR}}$ and $\gamma_{i}^{{\rm NR}}$ 
--with the superscript ${\rm NR}$ meaning that 
re-scattering effects have not been included-- can be written at NLO 
in terms of the order $p^2 $ and $p^4$ 
counterterms and the constants  $B_{i,0(1)}=
B_{i,0(1)}^{(2)}+B_{i,0(1)}^{(4)}$ 
and $H^{(4)}_{i,0(1)}$  defined in (\ref{F4F6def}),  
(\ref{functionsNLO}) and (\ref{yexpansion}). They can be obtained
from Appendix \ref{ANLO} by expanding the corresponding functions
$B_i$, $C_i$ and $H_i$ as in (\ref{yexpansion}). We get
\ba \label{alphas}
\alpha_{1}^{\rm NR} &=&  \ -G_8\,\frac{1}{2}\,B_{8,0}^{(++-)}
\,+\,\frac{1}{3}\,\sum _{i=27,E}G_{i}\,\left(B_{i,0}^{(+-0)}-
B_{i,0}^{(++-)}\right)\nonumber\\ 
&&\hspace{2 cm}+  \frac{1}{3}\,\sum_{i=1,11}\left(H_{i,0}^{(4)(+-0)}-
H_{i,0}^{(4)(++-)}\right)\widetilde K_i 
  \, ,
\nonumber\\
\alpha_{3}^{\rm NR} &=& \sum _{i=27,E} G_{i}\,\frac{1}{3}
\,\left(B_{i,0}^{(++-)}+2B_{i,0}^{(+-0)}\right) 
+  \frac{1}{3}\,\sum_{i=1,11}\left(H_{i,0}^{(4)(++-)}
+2H_{i,0}^{(4)(+-0)}\right)\widetilde K_i  \, , \nonumber \\
\beta_{1}^{\rm NR} &=&   -G_8\,B_{8,1}^{(++-)}
\,-\,\frac{1}{3}\, \sum _{i=27,E} G_{i}\left(B_{i,1}^{(+-0)}+
B_{i,1}^{(++-)}-B_{i,1}^{(00+)}\right)\nonumber\\
&-& \frac{1}{3}\,\sum_{i=1,11}\left(H_{i,1}^{(4)(+-0)}
+H_{i,1}^{(4)(++-)}-H_{i,1}^{(4)(00+)}\right)\widetilde K_i \, ,\nonumber\\
\beta_{3}^{\rm NR} &=& \frac{1}{3}\,   \sum _{i=27,E} G_{i}
\left(B_{i,1}^{(++-)}-B_{i,1}^{(00+)}-2B_{i,1}^{(+-0)}\right)\nonumber\\  
&&+  \frac{1}{3}\,\sum_{i=1,11}\left(H_{i,1}^{(4)(++-)}
-H_{i,1}^{(4)(00+)}-2H_{i,1}^{(4)(+-0)}\right)\widetilde K_i\,,\nonumber\\
\sqrt 3 \gamma_{3}^{\rm NR} &=&  
- \frac{1}{2}\,\sum _{i=27,E} G_{i}\left( B_{i,1}^{(++-)}
+ B_{i,1}^{(00+)}\right)
- \frac{1}{2}\,\sum_{i=1,11}\left(H_{i,1}^{(4)(++-)}
+H_{i,1}^{(4)(00+)}\right)\widetilde K_i\,.\nonumber\\
\ea

At LO the expressions above give
\ba 
\alpha_{1}^{\rm LO} &=& iC\left \lbrack G_8 \frac{\mkd}{3}\,+\, 
G_{27}\frac{\mkd}{27}\,+ \,e^2G_E\frac{2}{3} F_0^2\right\rbrack\,,
\nonumber\\
\alpha_{3}^{\rm LO} &=& iC\left \lbrack 
-G_{27}\frac{10\mkd}{27}\,-\,e^2G_E\frac{2}{3} F_0^2\right\rbrack\,,
\nonumber\\
\beta_{1}^{\rm LO} &=& iC\left \lbrack -G_8 \mpd\,-\, 
G_{27}\frac{\mpd}{9}\right\rbrack\,,
\nonumber
\ea
\ba
\beta_{3}^{\rm LO} &=& iC\left \lbrack -G_{27} \frac{5\mpd}{18(\mkd-\mpd)}
\left(5\mkd-14\mpd\right)\,
+ \,e^2G_E F_0^2\frac{3\mpd}{2(\mkd-\mpd)}\right\rbrack\,,\nonumber\\
\sqrt{3}\gamma_{3}^{\rm LO} 
&=& iC\left \lbrack G_{27} \frac{5\mpd}{4(\mkd-\mpd)}
\left(3\mkd-2\mpd\right)\,
+ \,e^2G_E F_0^2\frac{3\mpd}{4(\mkd-\mpd)}\right\rbrack\, ,
\ea
with the constant $C$ defined in (\ref{Cdef}).

The strong FSI  mix the two final states with isospin $I=1$
and leaves unmixed  the isospin 
$I=2$ state. The mixing in the isospin $I=1$ 
decay amplitudes is taken into account by introducing the
strong re-scattering 2 $\times$ 2 matrix $\mathbb{R}$ \cite{DIP91}. 
The amplitudes in (\ref{defdecays}) including  the FSI effects 
can be written as follows at all orders,  
\ba
\label{Rescat}
T_c\left(\begin{array}{c}A_{++-}^{(1)}
\\A_{00+}^{(1)}\end{array}\right)_{\rm R} &=& 
\Big(\, {\mathbb{I}} 
+i\,\mathbb{R}\,\Big) T_c\left(\begin{array}{c}A_{++-}^{(1)}
\\A_{00+}^{(1)} \end{array}\right)_{\rm NR} \, ,\nonumber\\
T_n\left(\begin{array}{c}A_{+-0}^{(2)}
\\ A_{000}^{(2)} \end{array}\right)_{\rm R} &=& 
\Big(\,{\mathbb{I}}+i\,\mathbb{R}
\,\Big)T_n\left(\begin{array}{c}A_{+-0}^{(2)}
\\ A_{000}^{(2)}\end{array}\right)_{\rm NR} \, ,\nonumber\\
A_{++-}^{(2)}|_{\rm R} &=&  \left(\,1\,+\,i\,\delta_2\,\right)A_{++-}^{(2)}
|_{\rm NR}\, ,\nonumber\\
\ea
with the matrices
\ba
T_c\,=\,\frac{1}{3}\,\left(\begin{array}{cc}1&1\\1&-2 \end{array}\right)
\,,\hspace{2 cm} 
T_n\,=\,\frac{1}{3}\,\left(\begin{array}{cc}0&1\\-3&1 \end{array}\right)
\ea
projecting the final state with $I=1$ into the symmetric--non-symmetric 
basis \cite{DIP91}. The subscript R (NR) means that the re-scattering
effects  have (not) been included. 
In these definitions the matrix $\mathbb{R}$, $\delta_2$ 
and the amplitudes $A^{(i)}$ depend on $s_1$, $s_2$ and $s_3$. 

Up to linear terms in $y$, equation (\ref{Rescat})  is equivalent to
\ba \label{Rdelta2def}
\left(\begin{array}{c}-\alpha_1+\frac{1}{2}\alpha_3
\\-\beta_1+\frac{1}{2}\beta_3 \end{array}\right)_{\rm R} &=& 
\left(\,{\mathbb{I}}+
i\,\mathbb{R}\,\right)\left(\begin{array}{c}-\alpha_1+\frac{1}{2}\alpha_3
\\-\beta_1+\frac{1}{2}\beta_3 \end{array}\right)_{\rm NR}\,, \nonumber\\
\left(\begin{array}{c}\alpha_1+\alpha_3
\\ \beta_1+\beta_3 \end{array}\right)_{\rm R} &=& 
\left(\,{\mathbb{I}}+i\,\mathbb{R}\,\right)
\left(\begin{array}{c}\alpha_1+\alpha_3
\\ \beta_1+\beta_3 \end{array}\right)_{\rm NR} \,,\nonumber\\
\gamma_{3,{\rm R}} &=&  
\left(\,1\,+\,i\,\delta_2\,\right)\gamma_{3,{\rm NR}}\,.
\nonumber\\
\ea
Here, the matrix $\mathbb{R}$ and
 $\delta_2$ are functions of the meson masses and the pion decay coupling. 
At lowest order in the chiral counting they are  given by
\ba
\mathbb{R}^{\rm LO}
 &=&\frac{1}{32 \pi F_0^2}\sqrt{\frac{\mkd-9\mpd}{\mkd+3\mpd}}
\left(\begin{array}{cc}\frac{1}{3}(9\mpd+2\mkd)&0\\
\frac{1}{3}\mkd&-5\frac{\mpd(m_K^4-27m_{\pi}^4)}{(\mkd+3\mpd)(\mkd-9\mpd)}
\end{array}\right)
\ea
and
\be
\delta_2^{\rm LO} \,=\,-\frac{1}{96\pi F_0^2}\,\mkd\, 
\sqrt{\frac{\mkd-9\mpd}{\mkd+3\mpd}}
\ee
in agreement with \cite{DIPP94}.

If we substitute the values of the masses and the coupling constant 
$F_0$, we get
\be
\mathbb{R}^{\rm LO}=\left(\begin{array}{cc}0.136&0
\\0.050&-0.143\end{array}\right)
\,,\hspace{1.5cm}\delta_2^{\rm LO}\,=\,-0.050\,.
\ee

We have also obtained the phase $\delta_2^{\rm NLO}$ 
and two combinations of the  $\mathbb{R}^{\rm NLO}$ matrix elements 
at NLO  when including the dominant FSI from 
two-bubbles obtained as explained before. 
The determination  of all the elements of $\mathbb{R}^{\rm NLO}$ 
would require the  calculation of the FSI at NLO for all the 
amplitudes in (\ref{amp1}) --we only have done the charged Kaon decays.
 The analytical expressions for these NLO 
quantities are given in Appendix \ref{phasesNLO}. 
Numerically, we  get
\ba \label{RmatrixNLO}
\frac{\left(-\alpha_1+\frac{1}{2}\alpha_3\right)_{\rm R}}
{\left(-\alpha_1+\frac{1}{2}\alpha_3\right)_{\rm NR}}
\Bigg|^{\rm NLO}=1+i\,0.156
\,,\hspace{2 cm}\nonumber\\
\frac{\left(-\beta_1+\frac{1}{2}\beta_3\right)_{\rm R}}
{\left(-\beta_1+\frac{1}{2}\beta_3\right)_{\rm NR}}\Bigg|^
{\rm NLO}=1+i\,0.569
\,, \hspace{1.5cm} {\rm and}
\hspace{1.5cm}\delta_2^{\rm NLO}\,=\,-0.104\,.
\ea

\subsection{Results on the Asymmetries in the Slope $g$}

As we have seen 
in Section \ref{LOresults}, the electroweak contribution
to $\Delta g$ at LO proportional to $\im(e^2 G_E)$ is at most around 
10\% of the leading contribution proportional to $ G_8$
while $\re(e^2 G_E)$ generates a negligible contribution.  
We include in our results the NLO absorptive part of the
electroweak  amplitude which is proportional to
$\im (e^2 G_E)$. The rest of the electroweak amplitude is just used
in the estimate of the errors.\footnote{The expressions for the
order $e^2p^0$ and $e^2p^2$ contributions to all the decay $K \to 3 \pi$
amplitudes are in Appendix B.1 of reference \cite{GPS03}.}. 

In  order to study the NLO
effects in $g_{C(N)}$  and $\Delta g_{C(N)}$, 
it is convenient  to introduce 
\ba
\label{amp2}
\vert A(K^+ \rightarrow 3\,\pi) \vert ^2 &=& A^+_0 \,+\,y\,A^+_y
\,+\,\order(x,y^2)\,,\nonumber\\
\vert A(K^- \rightarrow 3\,\pi) \vert ^2 &=& A^-_0 \,+\,y\,A^-_y
\,+\,\order(x,y^2)\,,
\ea
so that
\ba \label{AmpDeltag}
&&\hspace{0.7 cm}g[K^{+(-)}\to 3\pi] \,=\, \frac{A^{+(-)}_y}{A^{+(-)}_0} 
\,,\nonumber\\
&&\Delta g \,=\, \frac{A^+_yA^-_0 - A^+_0 A^-_y}
{A^+_yA^-_0 + A^+_0A^-_y}\,.
\ea
Notice that the numerator and denominator in 
(\ref{AmpDeltag}) are not the same as the difference 
$g[K^+\to 3\pi]-g[K^-\to 3\pi]$ and the sum 
$g[K^+\to 3\pi]+g[K^-\to 3\pi]$ respectively.
At NLO, the sum 
$A^+_yA^-_0 + A^+_0A^-_y$ does not contain
the FSI at NLO since they are part of the
order $p^6$ contributions, i.e. of the next-to-next-to-leading
order effects for the real parts.
However,   the difference   $A^+_yA^-_0 - A^+_0 A^-_y$
is proportional to the imaginary part of the amplitudes, 
therefore to have it  at NLO we must take into 
account the FSI phases, i.e. we need to include the FSI at NLO only 
in the imaginary part. 

The analytical expressions of the 
functions $A^{+(-)}_0$ and $A^{+(-)}_y$ at 
NLO are collected for the charged and the neutral Kaon cases in Appendix 
\ref{Adeltag}. From these expressions, we get the following
 numerical results 
\ba
\label{eq:gNLOeff}
\Delta g_{C}^{\rm NLO}\simeq \left \lbrack 0.66 \, 
\im G_8 +4.33\, \im \widetilde K_2 -18.11\, \im \widetilde K_3 
 -  0.07\,\im (e^2 G_E) \right \rbrack \times 10^{-2}\, ,
\nonumber \\ 
\Delta g_{N}^{\rm NLO}\simeq  - \left \lbrack 0.04 \, \im G_8 
+ 3.69\, \im \widetilde K_2  
+ 26.29\, \im \widetilde K_3 + 0.05 \, \im (e^2 G_E) 
\right \rbrack \times 10^{-2}\, . \nonumber \\
\ea
Where we have used the values for $\re \widetilde K_i$
from the fit to CP-conserving $K\to 3 \pi$ amplitudes \cite{BDP03}.
The NLO counterterms $\im G_8$, $\im (e^2 G_E)$ and
 $\im \widetilde K_3$ are scale independent.
In (\ref{eq:gNLOeff}), we have fixed the remaining scale dependence from 
$\im \widetilde K_2$ at  $\mu=M_\rho$.  
For the only two unknown counterterms 
$\im \widetilde K_2$ and $\im \widetilde K_3$, 
we have made two estimates of their effects.
First, using (\ref{assum1}) 
as explained in Section \ref{Kcount}.
 The other  estimate  of the effects of $\im \widetilde K_2$
and $\im \widetilde K_3$ is to put them to zero and to vary their
known scale dependence between 
$\mu= M_\rho$ and $ \mu = 1.5$ GeV. We include the induced variation
as a further uncertainty in our predictions.

Our final results for the slope $g$ asymmetries at NLO are
in Table \ref{tabNLO}.
\begin{table}
\begin{center}
\begin{tabular}{||c|c|c|c|c||}\hline
&$ \Delta g_C^{ \rm NLO}(10^{-5})$ & 
$ \Delta \Gamma_C^{ \rm NLO}(10^{-6})$  &
$ \Delta g_N^{\rm NLO}(10^{-5})$
&$ \Delta \Gamma_N^{\rm NLO}(10^{-6})$  
\\\hline\hline
$\widetilde K_i(M_\rho)$ from Table \ref{tabKvalues}
& $-2.4\pm 1.2$& $\lbrack -11,9
\rbrack$&$1.1 \pm 0.7 $ & $\lbrack -9,11\rbrack$ \\
$\widetilde K_i(M_\rho)=0$& $-2.4\pm 1.3$& $1.0 \pm 0.7 $&
$0.9  \pm 0.5 $  & $4.0\pm  3.2 $ \\
\hline
\end{tabular}
\end{center}
\caption{\label{tabNLO}CP-violating predictions for the slope $g$ and the
decay  rates $\Gamma$ at NLO in CHPT.
 The details of the calculation are  
in Section \ref{NLOresults}.  The inputs used for
$\im G_8$ and $\im (e^2 G_E)$ are in   
(\ref{gluonpenguin}) and (\ref{EMpenguin}), respectively.}
\end{table}
The central values  are obtained with the input values in Table 
\ref{tabKvalues} and the uncertainty includes the uncertainties 
of $\im G_8$, $\im (e^2 G_E)$,  the uncertainties
of the counterterms  quoted in Table \ref{tabKvalues},
 the variation due to the scale explained above and the error due to
the electroweak corrections.

The contribution  of the order $p^4$ 
counterterms $\im \widetilde K_i$
to $\Delta g_C$ is around 25\% using the values in Table
\ref{tabKvalues} and  the dominant contribution 
 is  the term  proportional to  $\im G_8$.  
 For $\Delta g_N$  we find a much larger dependence
on the values of the $\im \widetilde K_i$. 
Of course, since $\im \widetilde K_i$ are unknown
these results should be taken just as order of magnitude results,
a factor of two  or three could not be unreasonable
for $\Delta g_C$ and $\Delta g_N$.
The contribution
of $\im (e^2 G_E)$ is smaller than a 10\% of the dominant one
for both $\Delta g_C$ and $\Delta g_N$.

\subsection{Results on the Asymmetries in the Decay Rates}

We also only include NLO absorptive 
electroweak effects proportional to $\im (e^2 G_E)$
 for the same reasons explained in the previous subsection.
The analytical functions $|A_{C(N)}^{NLO}|^2$ and  
$\Delta |A_{C(N)}^{NLO}|^2$ 
needed to obtain the asymmetries in (\ref{eq:dGLO}) 
at NLO are given in (\ref{eqANLO}). 
Also  as explained in the previous subsection,
one should consistently not include 
FSI at NLO, which are  order $p^6$,
in the squared amplitudes $|A_{C(N)}^{NLO}|^2$  
since  they are part of the next-to-next-to-leading order
 corrections.
On the contrary, one has to include FSI at NLO
 in the differences $\Delta |A_{C(N)}^{NLO}|^2$ since 
these differences  are proportional to the FSI phases. 

The results obtained numerically from (\ref{eqANLO}) in terms of the 
imaginary  part of the counterterms are
\ba
\Delta \Gamma_{C}^{\rm NLO}\simeq \left \lbrack -2.8 \, 
\im G_8 + 49.2\, \im \widetilde K_2 + 103.6\, \im \widetilde K_3 
 +  0.2\,\im (e^2 G_E) \right \rbrack \times 10^{-3}\, ,
\nonumber \\ 
\Delta \Gamma_{N}^{\rm NLO}\simeq   \left \lbrack -3.1 \, \im G_8 
+45.7 \, \im \widetilde K_2  
+56.3 \, \im \widetilde K_3+0.12 \, \im (e^2 G_E) 
\right \rbrack \times 10^{-3}\, . \nonumber \\
\ea
In both cases the final value of the asymmetry 
is strongly dependent on the 
exact value of the $\im \widetilde K_i$ 
due  to large cancellations in the contribution
proportional to $\im G_8$. If we use the 
uncertainties quoted in Table \ref{tabNLO} 
for $\im \widetilde K_i$, the  induced errors in $\Delta \Gamma_{C}$
and $\Delta \Gamma_{N}$ are over 100\%.
In Table \ref{tabNLO}, we just quote therefore ranges
for the two decay rates CP-violating asymmetries.

\section{Comparison with Earlier Work}
\label{conclu}

The asymmetries $\Delta g_C$ and $\Delta g_N$ 
have been discussed in the literature before finding
conflicting results.
The rather large result 
\be
\left|\Delta g_C\right|\simeq \left| \Delta g_N \right| 
\simeq 140.0 \times 10^{-5}\ ,
\ee
was found in \cite{BBEL89}.

The upper bounds
\be
\left| \Delta g_C \right| \leq 0.7 \times 10^{-5} \, , 
\ee
at LO  and 
\be
\left| \Delta g_C \right| \leq 4.5  \times 10^{-5} \, , 
\ee
at NLO were found in \cite{DIP91}. The NLO bound was obtained
making plausible assumptions  since  no full NLO  result in CHPT was used.
 
In  \cite{IMP92}, 
\be
\Delta g_C  \simeq - 0.16 \times 10^{-5} \ ,
\ee 
was found at LO and 
\ba
\Delta g_C  \simeq - (0.23 \pm 0.06) \times 10^{-5} \, 
\, {\rm and} \, \, \Delta g_N  \simeq  (0.13 \pm 0.04) \times 10^{-5} 
\ea  
in \cite{MP95} also at LO.
 The authors of \cite{IMP92,MP95}  also made some estimate 
of the NLO corrections and arrived to the conclusion that they
  could increase their LO result up to one order of magnitude.
But again no full NLO calculation in CHPT was used.

The asymmetries  $\Delta \Gamma_C$ and $\Delta \Gamma_N$ have also
been discussed before and the results found were also in conflict
among them: 
\ba
\left|\Delta \Gamma_C\right|\simeq 31.0 \times 10^{-6} 
&\hspace*{0.5cm} {\rm and} \hspace*{0.5cm} 
& \left| \Delta \Gamma_N \right| 
\simeq 100.0 \times 10^{-6}\ ,
\ea
in \cite{BBEL89} ,
\be
\label{resIMP}
\Delta \Gamma_C  \simeq - 0.04 \times 10^{-6} \ ,
\ee 
in \cite{IMP92}, 
\ba
\label{resMP}
\Delta \Gamma_C  \simeq - (0.06 \pm 0.02) \times 10^{-6} 
\, \, {\rm and} \, \, 
\Delta \Gamma_N  \simeq (0.24 \pm 0.08) \times 10^{-6} 
\ea 
in \cite{MP95}, and  
\be
\label{resSHA}
\Delta \Gamma_C  \simeq - 1.0 \times 10^{-6} \ ,
\ee
in \cite{SHA93} --where we have used $\sin (\delta_{SM}) \simeq 0.85$
\cite{PDG02}.
The result in \cite{BBEL89}
 was claimed to be at one--loop, however
they did not use CHPT fully at one--loop. 
 We find, in general, that the results in \cite{BBEL89}
are overestimated as already pointed out in
 \cite{DIP91,IMP92,MP95,SHA03,SHA93}. See \cite{DIP91}
where some explanations for this large discrepancy are 
discussed.

The results in \cite{DIP91,IMP92,MP95} were reviewed
in \cite{DI96}.
They used factorizable values for $\im G_8$ and $\im G_E$,
i.e. the couplings in (\ref{largeNccouplings}), 
therefore their results have to be compared with the first row
in Table \ref{tabLO}.  
The reason of the difference between their results and ours 
is due mainly to the fact that these authors obtain the value 
of $\re G_8$ using the experimental value
for the isospin I=0 $K\to \pi \pi$
amplitude $\re a_0$. This amplitude $\re a_0$
contains  large higher order in CHPT
corrections. Corrections of similar size occur also 
in $\im a_0$ when considered at all orders.
However the authors used analytic LO formulas
for $\im a_0$ as well as for the imaginary parts of $K \to 3\pi$
amplitudes. This asymmetric procedure of considering the
real parts of the amplitudes experimentally and
the imaginary parts  analytically just at LO 
leads to a value for $\re G_8$  which is overestimated.
Therefore the CP violating asymmetries at LO
are underestimated in \cite{DIP91,IMP92,MP95}. 
The same comments apply to the predictions
of $\varepsilon_K'$ in those references as emphasized in \cite{PPS01}.
Our result in Tables \ref{tabLO} and \ref{tabNLO}
fulfill numerically the upper bound found in \cite{DIP91}
for $\Delta g_C$ at NLO but not the upper bound found
there at LO because of the same reason explained above.

The results in \cite{SHA93} were obtained at NLO using the
linear $\sigma$-model.
Recently, there was an update of those results in \cite{SHA03}:
\be
\label{7.9}
\Delta g_C  \simeq - (3.4 \pm 0.6)\times 10^{-5} \ ,
\ee 
at LO and
\be
\label{7.10}
\Delta g_C  \simeq - (4.2 \pm 0.8)\times 10^{-5} \ ,
\ee 
at NLO in the linear $\sigma$-model.  It is , however,  unclear from 
the text, the values used for 
the gluonic and the  electroweak penguins matrix elements
to get those results.
Though the LO result in (\ref{7.9}) agrees numerically
 with our result in Table \ref{tabLO}, we   do not agree analytically
with the results in \cite{SHA03} when the author
says that the electroweak penguins contribution at LO is as much as
34\% of the gluonic penguins contribution.
 We find that the electroweak
penguin contribution is one order of magnitude suppressed
with respect to the gluonic one.

\chapter{QCD Short-Distance Constraints and Hadronic Approximations}
\label{matrixelements}

According to what we have discussed in Chapter \ref{CPviolation}, present
main uncertainties in the study of CP violation stems from  
the evaluation of hadronic matrix elements. So that their precise calculation 
is urgently needed to test the validity of the SM and unveil the 
possibility of new physics. 
There are two important things 
to take into account in such kind of evaluations. 
One is the need to take care exactly of the scale and scheme dependences 
of these quantities 
(the value of $\{\gamma_\mu \gamma_5\}$, the evanescent operators $\dots$ )
 in such a way 
that these dependences are fully eliminated with the same dependences of the 
Wilson coefficients. Another problem is the treatment of the strong interactions 
at low and intermediate energies. At short distance we can use perturbative 
QCD that give good results down to energies around (1.5-2)$\,\gev$. 
However, at low and 
intermediate energies when asymptotic freedom appears, we can not longer work 
in the perturbative regime. Formulating a consistent hadronic approximation 
to QCD to deal with the strong interactions in the non-perturbative regime is 
an old an very difficult problem. 
Solving these two problems can be done in several ways. 

A first estimation of the matrix elements that must be considered 
as an order of magnitude result is naive-factorization. 
This approximation assume the 
factorization of the four quark operators in products of currents and densities. 
But this procedure does not exhibit a consistent matching of the scale and scheme 
dependences of the Wilson coefficients. The result potentially has large 
systematic uncertainties \cite{Bertolini00}. An improvement of naive-factorization 
is made by taking a 
matrix element between a particular quark and gluon external state. This 
removes the scale and scheme dependences but introduces a dependence on the 
particular external state chosen \cite{BJW90,NLOWilscoef,CFMR94}. 
More sophisticated methods are
\begin{itemize}
\item \textbf{Lattice calculations}: They stay in QCD but with a discretized 
space.  
There are two methods used in lattice calculations to compute the 
matrix elements $\langle \pi \pi|Q_i|K\rangle (\mu)$
\begin{itemize}
\item Compute first $\langle 0|Q_i|K\rangle(\mu)$ and  
$\langle \pi|Q_i|K\rangle(\mu)$ and then relate them to the wanted one using 
soft pion theorems.
\item Compute directly $\langle \pi \pi |Q_i|K\rangle(\mu)$. 
\end{itemize}

There are many 
difficulties associated with this approach at present. In the first case the 
$K\to\pi\pi$ amplitudes can be evaluated only at the lowest order of the 
chiral expansion, what doesn't take into account Final State Interactions 
(FSI) effects that may have large influence in the results. The direct 
computation has not been attempted for a long time due to the Maiani-Testa 
no-go theorem \cite{MT90}. However, since the important step toward the 
solution of the problem given by Lellouch and L\"uscher \cite{LL01}, there 
have been more work in this direction.
Some reviews where further references can be found are \cite{latticerev}.

\item \textbf{QCD Sum Rules}: They are based on the method of \cite{SVZ79}, 
for reviews 
see \cite{QCDSR}. QCD sum rules relate the hadronic 
regimes --the low energy world of resonances-- with the high energy world 
of QCD. They generally use two-point functions --although three-point like are 
also used-- which are studied at 
low energies through dispersion relations related to data and at high 
energies through their operator product expansion (OPE) and power 
suppressed corrections modulated by condensates. Examples of the calculation 
of matrix elements by using QCD sum rules to determine more inclusive 
quantities that can be related to them are in \cite{JP94,PR91} and 
references therein.

\item \textbf{Large $N_C$}: There are several ways of dealing with this method, 
combining it with something like the $X$-boson method explained above. 
The difference is mainly in the treatment of the low-energy 
hadronic physics. The most important approaches are
\begin{itemize}
\item CHPT (Chiral Perturbation Theory): Originally proposed by 
Bardeen-Buras-G\'erard \cite{ChPT}. 
Although it is the solution for the problem 
of having a consistent hadronic approximation to QCD 
at low energies, it has mainly two problems. Its domain of validity is fairly 
limited and there tend to be a rather number of parameters that needs to be 
dealt with. It cannot be simply extended to the intermediate energy domain. 
For further details, references and Lagrangians of leading and 
next-to-leading order see Chapter \ref{chCHPT}. 

\item ENJL: Extended Nambu-Jona-Lasinio model \cite{ENJL,ENJLreview}. 
They start with Lagrangians that are purely fermionic, containing 
the kinetic terms for the fermion and four-fermion interaction terms, 
and the hadronic fields are generated by the models themselves. 
The features and drawbacks of the ENJL model have been raised several times in 
\cite{epsprime,BPPx}. They have the advantage of needing only an small 
number of parameters and of generating the spontaneous breakdown of chiral 
symmetry by themselves. The most important drawback is that it does 
not contain all the QCD constraints \cite{PPR}.

\item MHA (Minimal hadronic approximation) : At large $N_C$ the spectrum of 
the theory consists in an infinite number of narrow stable meson states. The 
MHA keep only a finite number of resonances whose residues and masses are 
fixed by matching to the first few terms of both the chiral and the OPE 
expansions of the relevant Green's functions. Examples of these kind of 
calculations can be seen in \cite{KPR01,PR00,Rafael} and references therein. 

\item LRA (Ladder Resummation Approximation): Formulated in 
\cite{BGLP03}, the goal of this approach is getting a complete set of Green's  
functions compatible with as many QCD short-distance, large $N_C$ 
constraints and hadronic observables as possible. 
This approximation naturally reproduces the successes of the single 
meson per channel saturation models (e.g. VMD) and, in addition, contains 
the good features of NJL based models, i.e., 
some short-distance QCD constraints, CHPT up to order $p^4$, good 
phenomenology, $\dots$. It is 
treated more extensively in the next sections.
\end{itemize}
\item \textbf{Dispersive methods}: Some matrix elements corresponding to 
two-point functions can be related to experimental spectral functions. 
A good example is the mass difference between the charged and the neutral pion 
in the chiral limit that can be related to a dispersive integral over the 
difference of the vector and axial--vector spectral functions \cite{Das67}, 
for which we have information from $\tau$ decay data. Directly from these 
data the matrix element of $Q_7$ can also be extracted. The matrix element of 
$Q_8$ need a deeper analysis. An explanation of the calculation of the 
hadronic matrix elements of these two operators in the chiral limit is given 
in Chapter \ref{chq7q8}.
\end{itemize}

In all of these methods the so-called $X$-boson \cite{bosonX} can be 
used. It is based on the replacement  of the 
four--quark vertices coming from the hamiltonian in (\ref{hamiltonians1}) by 
the exchange of a series of colourless X--boson between currents or/and 
densities. Examples of calculations carried out with the $X$-boson 
method technique applied to one of the methods enumerated above are 
in \cite{BGP01,epsprime,BP95bk,BP99,scheme}. 
In Section \ref{secq6} we outlined the general 
structure carried out in these references and in Chapter \ref{chq7q8} 
we report carefully the calculation in \cite{BGP01}.

\section{Basics of the Model and Two-Point Functions}
\label{twopoint}

In the next sections we describe an approach based on a few simple 
assumptions, which we called LRA in the methods listed above. 
This fits naturally in  the limit
of large number of colours ($N_c$).
In this limit
and assuming confinement, QCD is known
to reduce to a theory of stable hadrons interacting only at tree level
\cite{largeN}. So the only singularities in amplitudes are produced
by the various tree-level poles occurring. This has long been a problem
for various variants of models incorporating some notion of constituent quarks
like the Nambu--Jona-Lasinio (NJL) models \cite{ENJLreview,NJL,NJLreviews}
or the chiral quark model \cite{CQM}.

The main idea in this section is to take the underlying
principle of ladder resummation
approaches to hadronic physics and make two successive approximations
in this. First we treat the rungs of the ladder as a type of general
contact interaction and second the remaining loop-integrations that occur,
which are always products of one-loop integrations, we treat
as general everywhere analytic functions. The only singularities that
occur then are those generated by the resummations and we naturally end up
with a hadronic large $N_c$ model.

This is also very close to the treatment
of the ENJL models
as given in \cite{ENJL,BRZ,BP1} where $n$-point Green's 
functions\footnote{In the remainder these are
often referred to as $n$-point functions.}
are seen as chains of one-loop bubbles connected by a one-loop
with three or more vertices.
The one-loop bubbles can be seen as one-loop Green's functions as well.
The full Green's functions there are thus composed of one-loop Green's functions
glued together by the (ENJL) couplings $g_V$ and $g_S$. 
One way to incorporate confinement in these ENJL models is by
introducing an infinite number of counterterms to remove all the
unwanted singularities \cite{PPR}. 
In \cite{PPR} it was then argued that
the ENJL approach was basically identical
to a one resonance saturation approach. They then proposed a minimal hadronic
ansatz where one resonance saturation is the underlying principle
and all couplings should be
determined from QCD short-distance and chiral constraints
with the relevant
short-distance constraints those that result from order parameters.
Order parameters are
quantities which would be fully zero if only perturbative QCD without
quark masses and condensates is considered.
This approach has been further discussed for two-point Green's functions
in \cite{GP} and applied to some three-point functions in \cite{KN01},
see also the discussions in \cite{Moussallam} for earlier 
similar uses of order parameters.
Problems appear for $n$-point Green's functions in that not necessarily
all freedom in the parameters can be fixed by the long-distance chiral 
constraints and/or short-distance
constraints or involve too many unknown constants in the chiral constraints.

In this approach we follow a different scheme.
We {\em assume} that the Green's functions are produced by a
ladder-resummation
like ansatz. They consist of bubble-diagrams put together from one-loop
Green's functions. We do {\em not} use the (constituent) quark-loop
expressions for these one-loop Green's functions but instead consider
them as constants or low-order polynomials in the kinematic variables.
This set of assumptions turns out to be rather constraining in the type
of model that can be constructed. In particular the gap equation for
spontaneous symmetry breaking follows from the requirements of resummation
and the full Ward identities as shown in this section.
The link with constituent quark models is the
fact that given the full Ward identities one can define a constituent
quark mass, obeying a gap equation,
and the one-loop Green's functions satisfy the Ward identities
with {\em constituent} quark-masses. In the two-point function
sector this naturally reduces to the approach of \cite{PPR} but it
allows to go beyond two-point functions in a more systematic manner.

In the next subsections we discuss the buildup of the model
and the two-point functions. We first work in the chiral limit and then
add corrections due to current quark masses. Chiral Perturbation
Theory, or low-energy, constraints are naturally satisfied in our approach
which is chiral invariant from the start. Also large $N_c$ constraints are 
satisfied naturally. We show how the short-distance constraints can be 
included.
Section \ref{threepoint} treats several three-point functions and includes
here short-distance constraints coming from form factors and from
the more suppressed combinations of short-distances.

Numerical results are presented in Section \ref{numerics}.
We find a reasonable
agreement for the predictions.

Going beyond the one-resonance saturation
in this approach is difficult as explained in Subsection \ref{trouble}.
Another point raised is that hadronic models will in general have problems
with QCD short-distance constraints, even if the short-distance behaviour is
an order parameter, we discuss in detail how the
pseudo-scalar--scalar--pseudo-scalar three-point function is a typical
example of this problem in Section \ref{SDProblem}.

We consider this class of models still useful even with the problems
inherent in it. They provide a consistent framework to address the
problems of nonleptonic matrix-elements where in general very many
Green's functions with a large number of insertions is needed. The present
approach offers a method to \textsl{analytically} calculate these
Green's functions and
thus study the effects of the various ingredients on the final results.
One motivation for this work was to understand many of the rather surprising
features found in the calculations using the ENJL model of the $B_K$ parameter,
the $\Delta I=1/2$ rule, gluonic and electroweak Penguins,
electroweak effects and the muon anomalous magnetic
moment \cite{epsprime,BP99,scheme,BPP98,BPPx,BPx} and improve on 
those calculations. In Section \ref{scbk} we describe the status of the 
study of the $\hat B_K$ parameter in which we are working at this moment. 
Finally, in the last section referring to OPE computation of 
Green's functions we give an example of the calculation of one of the 
functions we have analyzed at long (CHPT) and short (OPE) distance, 
to fit with our approach predictions.

\subsection{General}
\label{twopointgeneral}

The Lagrangian for the large $N_c$ ENJL model is
\ba
{\cal L}_{ENJL}
&=&
\sum_{i,j,\alpha}
\overline q_\alpha^i\left\{ \gamma^\mu\left(i\partial_\mu\delta^{ij}+v_\mu^{ij}
+a_\mu^{ij}\gamma_5\right)
-{\cal M}^{ij}-s^{ij}+ip^{ij}\gamma_5\right\}q^j_\alpha
\nonumber\\&&
+2 g_S\sum_{i,j,\alpha,\beta}\left(\overline q^i_{R\alpha} q^j_{L\alpha}\right)
\left(\overline q^j_{L\beta} q^i_{R\beta}\right)
\nonumber\\&&
- g_V\sum_{i,j,\alpha,\beta}
\left(\overline q^i_{L\alpha}\gamma_\mu q^j_{L\alpha}\right)
\left(\overline q^j_{L\beta}\gamma^\mu q^i_{L\beta}\right)
- g_V\sum_{i,j,\alpha,\beta}
\left(\overline q^i_{R\alpha}\gamma_\mu q^j_{R\alpha}\right)
\left(\overline q^j_{R\beta}\gamma^\mu q^i_{R\beta}\right)
\ea
with $i,j$ flavour indices, $\alpha,\beta$ colour indices
and $q_{R(L)} = (1/2)(1+(-)\gamma_5)q$.
The flavour matrices $v,a,s,p$ are external fields and can be used to
generate all the Green's functions we will discuss.
The four-quark interactions can be seen as an approximation for the
rungs of a ladder-resummation scheme.

The Green's functions generated by functional differentiation w.r.t.
$v^{ij}(x)$, $a^{ij}(x)$, $s^{ij}(x)$, $p^{ij}(x)$ correspond 
to Green's functions of the currents
\ba
\label{currents}
V_\mu^{ij}(x) &=& \overline q^i_\alpha(x) \gamma_\mu q^j_\alpha(x)\,,
\nonumber\\
A_\mu^{ij}(x) &=& \overline q^i_\alpha(x) \gamma_\mu\gamma_5 q^j_\alpha(x)\,,
\nonumber\\
S^{ij}(x) &=& -\overline q^i_\alpha(x)  q^j_\alpha(x)\,,
\nonumber\\
P^{ij}(x) &=& \overline q^i_\alpha(x) i\gamma_5 q^j_\alpha(x)\,.
\ea
An underlying assumption is that these currents can be
identified with the QCD ones.

In the remainder of this section we will discuss the two-point functions
\ba
\Pi^V_{\mu\nu}(q)^{ijkl}
&=&
i\int d^dx\,e^{i\, q\cdot x}
\langle 0|T\left(V^{ij}_\mu(x)V^{kl}_\nu(0)\right)|0\rangle\,,
\nonumber\\
\Pi^A_{\mu\nu}(q)^{ijkl}
&=&
i\int d^dx\,e^{i\, q\cdot x}
\langle 0|T\left(A^{ij}_\mu(x)A^{kl}_\nu(0)\right)|0\rangle\,,
\nonumber\\
\Pi^S_{\mu}(q)^{ijkl}
&=&
i\int d^dx\,e^{i\, q\cdot x}
\langle 0|T\left(V^{ij}_\mu(x)S^{kl}(0)\right)|0\rangle\,,
\nonumber\\
\Pi^P_{\mu}(q)^{ijkl}
&=&
i\int d^dx\,e^{i\, q\cdot x}
\langle 0|T\left(A^{ij}_\mu(x)P^{kl}(0)\right)|0\rangle\,,
\nonumber\\
\Pi^S(q)^{ijkl}
&=&
i\int d^dx\,e^{i\, q\cdot x}
\langle 0|T\left(S^{ij}(x)S^{kl}(0)\right)|0\rangle\,,
\nonumber\\
\Pi^P(q)^{ijkl}
&=&
i\int d^dx\,e^{i\, q\cdot x}
\langle 0|T\left(P^{ij}(x)P^{kl}(0)\right)|0\rangle\,.
\ea
The other possibilities vanish because of parity. The large $N_c$ limit
requires these to be proportional to $\delta^{il}\delta^{jk}$
and Lorentz and translational invariance allow them to be written in
terms of functions that only depend on $q^2$ and the flavour index
$i,j$.
\ba \label{twopointdef}
\Pi_{\mu\nu}^V(q)_{ijkl} &=&
\left\{\left(q_\mu q_\nu-g_{\mu\nu} q^2\right) \Pi_{Vij}^{(1)}(q^2)
+ q_\mu q_\nu \Pi_{Vij}^{(0)}(q^2)\right\}
\delta^{il}\delta^{jk}\,,
\nonumber\\
\Pi_{\mu\nu}^A(q)_{ijkl} &=&
\left\{\left(q_\mu q_\nu-g_{\mu\nu} q^2\right) \Pi_{Aij}^{(1)}(q^2)
+ q_\mu q_\nu \Pi_{Aij}^{(0)}(q^2)\right\}
\delta^{il}\delta^{jk}\,,
\nonumber\\
\Pi_{\mu}^S(q)_{ijkl} &=&
 q_\mu  \Pi_{Sij}^{M}(q^2)
\delta^{il}\delta^{jk}\,,
\nonumber\\
\Pi_{\mu}^P(q)_{ijkl} &=&
 i q_\mu  \Pi_{Pij}^{M}(q^2)
\delta^{il}\delta^{jk}\,,
\nonumber\\
\Pi^S(q)_{ijkl} &=&
 \Pi_{Sij}(q^2)
\delta^{il}\delta^{jk}\,,
\nonumber\\
\Pi^P(q)_{ijkl} &=&
 \Pi_{Pij}(q^2)
\delta^{il}\delta^{jk}\,.
\ea
These functions satisfy Ward-identities following from chiral symmetry
and the QCD equations of motion
\ba
\label{WI}
q^2 \Pi_{Vij}^{(0)}(q^2) &=& \left(m_i-m_j\right) \Pi_{Sij}^{M}(q^2)\,,
\nonumber\\
q^2 \Pi_{Sij}^{M}(q^2) &=& \left(m_i-m_j\right) \Pi_{Sij}(q^2)
+\cond_i -\cond_j\,,
\nonumber\\
q^2 \Pi_{Aij}^{(0)}(q^2) &=& \left(m_i+m_j\right) \Pi_{Pij}^{M}(q^2)\,,
\nonumber\\
q^2 \Pi_{Pij}^{M}(q^2) &=& \left(m_i+m_j\right) \Pi_{Pij}(q^2)
+\cond_i+\cond_j\,.
\ea
Here we use $\cond_i =
 \sum_\alpha\langle 0|\overline q^i_\alpha q^i_\alpha | 0 \rangle$.

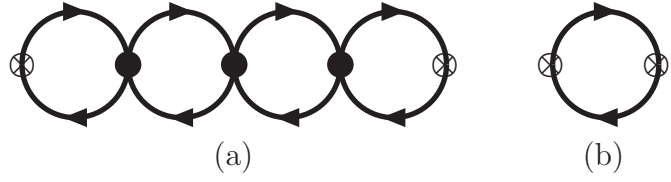
\begin{figure}
\begin{center}
\setlength{\unitlength}{1pt}
\begin{picture}(250,60)(0,-15)
\SetScale{1.}
\SetWidth{2.}
\ArrowArcn(25,25)(20.,0.,180.)
\ArrowArcn(25,25)(20.,180.,0.)
\ArrowArcn(65,25)(20.,0.,180.)
\ArrowArcn(65,25)(20.,180.,0.)
\ArrowArcn(105,25)(20.,0.,180.)
\ArrowArcn(105,25)(20.,180.,0.)
\ArrowArcn(145,25)(20.,0.,180.)
\ArrowArcn(145,25)(20.,180.,0.)
\Text(5,25)[]{{\boldmath$\otimes$}}
\Text(165,25)[]{{\boldmath$\otimes$}}
\Vertex(45,25){5}
\Vertex(85,25){5}
\Vertex(125,25){5}
\Text(85,-10)[]{(a)}
\ArrowArcn(225,25)(20.,0.,180.)
\ArrowArcn(225,25)(20.,180.,0.)
\Text(205,25)[]{{\boldmath$\otimes$}}
\Text(245,25)[]{{\boldmath$\otimes$}}
\Text(225,-10)[]{(b)}
\end{picture}
\end{center}
\caption{\label{figtwopoint}The type of diagrams in large $N_c$ that contribute
to the two-point function. {\boldmath$\otimes$} indicates and insertion
of an external current and $\bullet$ indicates the ENJL four-quark vertex.
(a) the full two-point function. (b) The one-loop two-point function.}
\end{figure}
The type of diagrams that contribute in large $N_c$ to the two-point functions
is depicted in Fig.~\ref{figtwopoint}(a). The contribution from only the
one-loop diagram is depicted in Fig.~\ref{figtwopoint}(b) and we will
generally denote these as $\overline \Pi$.

Under interchange of $i$ and $j$, $\smnf(q^2)$ is anti-symmetric, all others
are symmetric. The one-loop equivalents have the same symmetry properties.

In Refs.~\cite{BRZ,BP1} it was shown that the full two-point functions
can be obtained from the one-loop ones via a resummation procedure
\ba
\label{vssector}
\vtf (q^2)&=& \frac{\vt (q^2)}{1-q^2 g_V \vt(q^2)}\nonumber \\
\vlf (q^2)&=& \frac{1}{\Delta_S(q^2)}[(1-g_S\s (q^2))\vl 
(q^2)+g_S(\SMN(q^2))^2]\nonumber \\
\smnf (q^2)&=& \frac{1}{\Delta_S(q^2)}\SMN(q^2) \nonumber\\
\ssf (q^2) &=& \frac{1}{\Delta_S(q^2)}[(1+q^2 g_V \vl (q^2))\s 
(q^2)-q^2 g_V (\SMN(q^2))^2]\nonumber\\
\Delta_S(q^2)&=&(1+q^2 g_V \vl(q^2) )(1-g_S \s (q^2))+q^2 g_S g_V (\SMN(q^2))^2
\ea
\ba
\label{apsector}
\atf (q^2)&=& \frac{\at (q^2)}{1-q^2 g_V \at(q^2)}\nonumber \\
\alf (q^2)&=& \frac{1}{\Delta_P(q^2)}[(1-g_S\p (q^2))\al 
(q^2)+g_S(\PMN(q^2))^2]\nonumber \\
\pmnf (q^2)&=& \frac{1}{\Delta_P(q^2)}\PMN(q^2) \nonumber\\
\ppf (q^2) &=& \frac{1}{\Delta_P(q^2)}[(1+q^2 g_V \al (q^2))\p 
(q^2)-q^2 g_V (\PMN(q^2))^2]\nonumber\\
\Delta_P(q^2)&=&(1+q^2 g_V \al(q^2) )(1-g_S \p (q^2))+q^2 g_S g_V (\PMN(q^2))^2
\ea
This resummation is only consistent with the Ward Identities, Eq. (\ref{WI}), 
if
the one-loop two-point functions obey the Ward Identities
of Eq.~(\ref{WI}) with the current quark masses $m_i$ replaced
by the constituent quark masses $M_i$ given by
\be
\label{gap}
M_i = m_i -g_S\cond_i\,,
\ee
known as the gap equation. The {\em assumption} of resummation thus leads
to a constituent quark mass picture and one-loop Ward identities
with constituent quark masses.

Using the gap equation and the one-loop Ward identities
the resummation formulas can be simplified
using
\ba
\Delta_S(q^2)&=&1-g_S \s(q^2)+g_V(m_i-m_j)\SMN(q^2)\,,
\nonumber\\
\Delta_P(q^2)&=&1-g_S \p(q^2)+g_V(m_i+m_j)\PMN(q^2)\,.
\ea

Our model assumption is to choose the one-loop functions as basic
parameters rather than have them predicted via the constituent
quark loops. This allows for a theory that has confinement built in
a simple way and at the same time keeps most of the successes of the
ENJL model in low-energy hadronic physics.

We now choose the two-point functions as far as possible as constants
and have thus as parameters in the two-point sector
\be
\label{par2}
\cond_i, g_S, g_V, \PMN,\alt, \SMN,\vlt
\ee
and the remaining one-loop two-point functions can be obtained
from the one-loop Ward identities. As discussed below,
more input will be needed for the three-point functions.
We do not expand higher in momenta in the one-loop two-point functions.
The reason for this is that assuming that $g_V$ and $g_S$ are constants,
expanding the one-loop two-point functions higher in momenta causes
a gap in the large $q^2$ expansion between the leading and the non-leading
terms. Such a gap in powers is not present as we know from perturbative QCD.

\subsection{Chiral Limit}
\label{twopointc}

In the chiral limit, the Ward identity for $\s(q^2)$ becomes singular
and it is better to choose instead as parameters
\be
\label{par2c}
\condc,\Delta,g_S,g_V,\PMNc,\altc,\Gamma,\vltc
\ee
with the parameters $\Delta$, $\Gamma$ defined via
\ba
\label{defDelta}
\cond_i &=& \condc + m_i \Delta + m_i^2 \epsilon + \order(m_i^3)\,,
\nonumber\\
\s(q^2) &=& q^2\Gamma -\frac{\Delta}{1-g_S\Delta} + \order(m_i,m_j)\,.
\ea 

\subsubsection{Short-Distance}
\label{chiralshort}

We define $\Pi_{LR} = \Pi_V - \Pi_A$
and $\Pi_X^{0+1} = \Pi_X^{(0)}+ \Pi_X^{(1)}$ for $X=LR,V,A$ then
the first and third Weinberg sum rules\cite{Weinberg},
\be
\lim_{q^2 \rightarrow -\infty}(q^2\Plr^{(0+1)QCD}(q^2))=0
\quad\mbox{and}\quad
\lim_{q^2 \rightarrow -\infty}(q^4\Plr^{(0)QCD}(q^2))=0\,,
\ee
are automatically satisfied but the second one\,,
\be
\lim_{q^2 \rightarrow -\infty}(q^4\Plr^{(1)QCD}(q^2))=0\,,
\ee
implies the relation
\be
\label{SDcVA}
\altc = \vltc\,.
\ee

Analogs of the Weinberg sum rules exist in scalar-pseudoscalar sector.
With $\Pi_{SP} = \Pi_S-\Pi_P$ we have \cite{Moussallam,RRY}
\be
\lim_{q^2\to-\infty}
\Pi_{SP\,ij}^{QCD}(q^2) = 0\quad\mbox{and}\quad
\lim_{q^2\to-\infty}
(q^2\Pi_{SP\,ij}^{QCD}(q^2)) = 0\,.
\ee
The first one is the equivalent of the first Weinberg sum rule and
is automatically satisfied. The second one implies
\be
\label{SDcSP}
\Gamma = \frac{-\PMNc}{2g_S\condc\left(1-2g_S g_V\condc\PMNc\right)}\,.
\ee

The short-distance relation found in Eq.~(\ref{SDcSP}) does not satisfy
the heat kernel relation for the one-loop two-point functions derived
in \cite{BRZ} in the chiral limit. Note that that heat kernel relation
was the underlying cause of the relation
$m_S = 2 M_q$ between the scalar mass and the constituent quark mass
in ENJL models \cite{BRZ,BP1}.

\subsubsection{Intermediate-Distance}
\label{chiralintermediat}

The two-point functions in the chiral limit can be written as
\ba \label{twopointchiral}
\Pi_V^{(1)\chi}(q^2) &=& \frac{2 f_V^2 m_V^2}{m_V^2-q^2}\,,
\nonumber\\
\Pi_A^{(1)\chi}(q^2) &=& \frac{-2 F_0^2}{q^2}+
\frac{\dsp 2 f_A^2 m_A^2}{\dsp m_A^2-q^2}\,,
\nonumber\\
\Pi_P^{M\chi}(q^2) &=& \frac{\dsp 2\condc}{q^2}\,,
\nonumber\\
\Pi_S^\chi(q^2) &=& K_S+\frac{2 F_S^2 m_S^2}{m_S^2-q^2}
\nonumber\\
\Pi_P^\chi(q^2) &=& K_P-\frac{2 F_0^2 B_0^2}{q^2}
\ea

{}From the poles in the two-point functions we can find the various masses.
There is a pole at $q^2=0$ corresponding to the massless pion.
The scalar, vector and axial-vector masses are given by
\ba
m_S^2 &=& \frac{1}{g_S\Gamma\left(1-g_S\Delta\right)}\,,
\nonumber\\
m_V^2 &=& \frac{1}{g_V\vltc}\,,
\nonumber\\
m_A^2 &=& \frac{1-2 g_S g_V \condc \PMNc}{g_V\altc} = 
\left(1-2 g_S g_V \condc \PMNc\right) m_V^2\,.
\ea
The residues at the poles lead to
\ba
2 f_V^2 &=& \vltc\,,
\nonumber\\
2 f_A^2 &=&  \frac{\vltc}{\left(1-2 g_S g_V \condc \PMNc\right)^2}\,,
\nonumber\\
2 F_0^2 &=& \frac{-2 g_S\condc\PMNc}{1-2 g_S g_V \condc \PMNc}\,,
\nonumber\\
K_S &=&K_P = -\frac{1}{g_S}\,,
\nonumber\\
2 F_S^2 &=& \frac{1-g_S\Delta}{g_S}\,,
\nonumber\\
B_0^2 F_0^4 &=& \condc^2
\ea
The short distance constraints lead as expected to
\ba
f_V^2 m_V^2 &=& f_A^2 m_A^2+F_0^2\,,
\nonumber\\
f_V^2 m_V^4 &=& f_A^2 m_A^4\,,
\nonumber\\
K_S &=& K_P\,,
\nonumber\\
F_S^2 m_S^2 &=& F_0^2 B_0^2\,.
\ea

\subsubsection{Long-Distance}
\label{chirallong}

The two-point functions in the chiral limit can be determined from Chiral
Perturbation Theory. This lead to the identification of $B_0$, $F_0$
with the quantities appearing there and in addition
\ba
\label{L10}
L_{10} &=& -\frac{1}{4}\left(f_V^2-f_A^2\right)\,,   \quad
H_1 = -\frac{1}{8}\left(f_V^2+f_A^2\right)\,,
\nonumber\\
32 B_0^2 L_8 &=& 2 F_S^2\,,\quad\quad\quad\quad\quad
 16 B_0^2 H_2 = 2 K_S+ 2 F_S^2
\ea

\subsubsection{Parameters}

Notice that from the six input parameters we can only determine five
from the two-point function inputs.
A possible choice of input parameters is $m_V$, $m_A$, $F_0$, $m_S$ and $F_S$.
The last can be traded for $B_0$ or $\condc$.
The remaining parameter could in principle be fixed from $K_S$ but that
is an unmeasurable quantity.

\subsection{Beyond the Chiral Limit}
\label{twopointbeyond}

The resummation formulas of Sect.~\ref{twopointgeneral} remain valid.
What changes now is that we have values for the current quark masses $m_i$
and corresponding changes in the one-loop functions.
An underlying expectation is that the vertices $g_S$ and $g_V$
are produced by purely gluonic effects and have no light quark-mass dependence.
The first order the quark-mass dependence of $g_V$ and $g_S$
must be zero from short-distance constraints as shown below.

The input parameters are now given by Eq. (\ref{par2}) and we will below
expand them as functions in $m_q$.

\subsubsection{Intermediate-Distance}

The resummation leads to expressions for the two-point functions
which can again be written as one resonance exchange.
\ba 
\label{mesondominance}
\vtf (q^2)&=& -\frac{2\,f_{Sij}^2}{q^2}\,
+\,\frac{2\,f_{Vij}^2 m_{Vij}^2}{m_{Vij}^2-q^2} \,,
\nonumber\\
\vlf(q^2)&=& 2f_{Sij}^2\left(\frac{1}{m_{Sij}^2-q^2}+\frac{1}{q^2}\right) \,,
\nonumber\\
\atf (q^2)&=& -\frac{2\,f_{ij}^2}{q^2}\,
+\,\frac{2\,f_{Aij}^2 m_{Aij}^2}{m_{Aij}^2-q^2}  \,,
\nonumber\\
\alf(q^2)&=& 2f_{ij}^2\left(\frac{1}{m_{ij}^2-q^2}+\frac{1}{q^2}\right) \,,
\nonumber\\
\smnf(q^2)&=&\frac{2F_{Sij}m_{Sij}f_{Sij}}{m_{Sij}^2-q^2} \,,
\nonumber\\
\pmnf(q^2)&=&\frac{2B_{ij}f_{ij}^2}{m_{ij}^2-q^2} \,,
\nonumber\\
\ssf (q^2) &=& K_{Sij}\,+\frac{2\,F_{Sij}^2m_{Sij}^2}{m_{Sij}^2-q^2} \,,
\nonumber\\
\ppf (q^2) &=& K_{Pij}\,+\frac{2\,f_{ij}^2 B_{ij}^2}{m_{ij}^2-q^2} \,.
\ea
These satisfy the Ward Identities (\ref{WI}).
The values of the couplings and masses are given by
\ba \label{LMDparameters}
m_{Vij}^2 &=& \frac{1+g_V\SMN (M_i-M_j)}{g_V \vlt}  \,,
\nonumber\\
m_{Aij}^2 &=& \frac{1+g_V\PMN (M_i+M_j)}{g_V \alt}  \,,
\nonumber\\
m_{Sij}^2 &=& \frac{m_i-m_j}{\SMN}\,\frac{ (1+g_V\SMN (M_i-M_j))}{g_S} \,,
\nonumber\\
m_{ij}^2&=&(m_i+m_j)\frac{1+g_V(M_i+M_j)\PMN}{g_S\PMN} \,,
\nonumber\\
2\,f_{Vij}^2 &=& \frac {\vlt} {(1+g_V\SMN (M_i-M_j))^2} \,,
\nonumber\\
2\,f_{Aij}^2 &=&  \frac {\alt} {(1+g_V\PMN (M_i+M_j))^2} \,,
\nonumber\\
2\,F_{Sij}^2 &=& \frac{M_i-M_j}{g_S\,(m_i-m_j)} \,,
\nonumber\\
2\,f_{Sij}^2 &=& \frac{(M_i-M_j)\SMN}{1+g_V\SMN (M_i-M_j)} \,,
\nonumber\\
2f_{ij}^2&=&\frac{(M_i+M_j)\PMN}{1+g_V\PMN (M_i+M_j)} \,,
\nonumber\\
K_{Sij} &=& K_{Pij} = -\frac{1}{g_S} \,,
\nonumber\\
B_{ij} &=&\frac{1+g_V(M_i+M_j)\PMN}{g_S\PMN} \,.
\ea
\vspace{3 cm}

\subsubsection{Short-Distance}

In order to proceed we have to expand the input parameters of
Eq. (\ref{par2}) in the quark masses $m_q$.
\ba
\vlt &=& \vltc+(m_i+m_j)\vlti+\order(m_q^2)\,,
\nonumber\\
\alt &=& \altc+(m_i+m_j)\alti+\order(m_q^2)\,,
\nonumber\\
\PMN &=& \PMNc+(m_i+m_j)\PMNi+\order(m_q^2)\,,
\nonumber\\
\s(q^2) &=& q^2\left(\Gamma+(m_i+m_j)\Gamma^I\right)-\frac{\Delta}{1-g_S\Delta}
-\frac{\epsilon}{\left(1-g_S\Delta\right)^2}(m_i+m_j)+\order(m_q^2)
\, ,\nonumber\\
 \ea
The parameters $\epsilon$ and $\Delta$ are defined in the first line of
Eq.~(\ref{defDelta}). The other one-loop two-point functions are derivable
from the one-loop Ward identities.

The chiral limit short-distance constraints Eqs. (\ref{SDcVA})
 and (\ref{SDcSP})
remain valid but there are new constraints on the coefficients of the
quark mass expansions. The derivatives w.r.t. the quark masses of the
two-point functions allow to construct more order parameters
than  $\Pi_{LR}$ and $\Pi_{SP}$. In particular we have\footnote{We have derived
these expressions but they can also be found in \cite{RRY}.}
\ba
\label{SDi2point}
\lim_{q^2\to-\infty}\lim_{m_q\to 0}
\left(q^4\frac{\partial}{\partial m_i}\vtf(q^2)\right) &=& \condc\,,
\nonumber\\
\lim_{q^2\to-\infty}\lim_{m_q\to 0}
\left(q^4\frac{\partial}{\partial m_i}\vlf(q^2)\right) &=& 0\,,
\nonumber\\
\lim_{q^2\to-\infty}\lim_{m_q\to 0}
\left(q^4\frac{\partial}{\partial m_i}\atf(q^2)\right) &=& -\condc\,,
\nonumber\\
\lim_{q^2\to-\infty}\lim_{m_q\to 0}
\left(q^4\frac{\partial}{\partial m_i}\alf(q^2)\right) &=& 2\condc\,,
\nonumber\\
\lim_{q^2\to-\infty}\lim_{m_q\to 0}
\left(q^2\frac{\partial}{\partial m_i}\ssf(q^2)\right) &=&-\frac{3}{2}\condc\,,
\nonumber\\
\lim_{q^2\to-\infty}\lim_{m_q\to 0}
\left(q^2\frac{\partial}{\partial m_i}\ppf(q^2)\right) &=&\frac{1}{2}\condc\,.
\ea
The ones with lower powers of $q^2$ must vanish.
The second and fourth are automatically satisfied
as a consequence from the
Ward identities. $\vlf(q^2)$ only starts at $\order(m_q^2)$
and the $m_i+m_j$ term in $\alf(q^2)$ follows from the Ward identity
and the chiral limit form of $\pmnf$.

The vanishing of those with lower powers of $q^2$ requires that
\be
\lim_{m_q\to0} \frac{\partial g_V}{\partial m_i} =
\lim_{m_q\to0} \frac{\partial g_S}{\partial m_i} = 0\,.
\ee
The first, third, fifth and sixth identities give
\ba
\label{2pointSD}
\vlti &=& g_V^2\left(\vltc\right)^2\condc\,,
\nonumber\\
\alti &=& -g_V^2\left(\vltc\right)^2\condc\,,
\nonumber\\
\Gamma^I &=& -\frac{3}{2} g_S^2\Gamma^2\condc\,,
\nonumber\\
\PMNi &=&  -\frac{1}{4}g_S\left(\PMNc\right)^2
-\frac{1-g_S\Delta}{2 g_S\condc}\, \PMNc
\left(1-4 g_V g_S\condc\PMNc\right)\,.
\ea
This implies that the only new parameter that appears
to include quark masses to first order is $\epsilon$.
The last constraint turns out to be incompatible with short-distance
constraints from three-point functions as discussed below.

\subsubsection{Long-Distance}

The long-distance expansion of our results to $\order(p^4)$
in Chiral Perturbation Theory allows in addition to those already obtained
in the chiral limit also
\be
\label{result1}
L_5 = \frac{1}{16}F_0^6\Big \lbrack \frac{\PMNc(g_S\Delta-1)+2g_S\condc
\PMNi} {(\PMNc)^2g_S^2\condc^3}\Big \rbrack \,.
\ee

\subsubsection{Intermediate-Distance}

The short-distance constraints lead to several relations between resonance
parameters also beyond the chiral limit to first order in current quark masses.
In the vector sector we obtain
\ba
\label{Vrelations}
f_{Vij}^2 m_{Vij}^2 &=& f_{Vkl}^2 m_{Vkl}^2\,,
\nonumber\\
f_{Vij}^2 m_{Vij}^4 - f_{Vkl}^2 m_{Vkl}^4 &=&
-\frac{1}{2}\condc(m_i+m_j-m_k-m_l)\,.
\ea
$V_{ij}$ stands here for the vector degree of freedom built
of quarks with current mass $m_i$ and $m_j$.

The corresponding axial relations are
\ba
\label{Arelations}
f_{Aij}^2 m_{Aij}^2+f_{ij}^2 &=& f_{Akl}^2 m_{Akl}^2+f_{kl}^2\,,
\nonumber\\
f_{Aij}^2 m_{Aij}^4 - f_{Akl}^2 m_{Akl}^4 &=&
\frac{1}{2}\condc(m_i+m_j-m_k-m_l)\,.
\ea

\section{Three-Point Functions}
\label{threepoint}

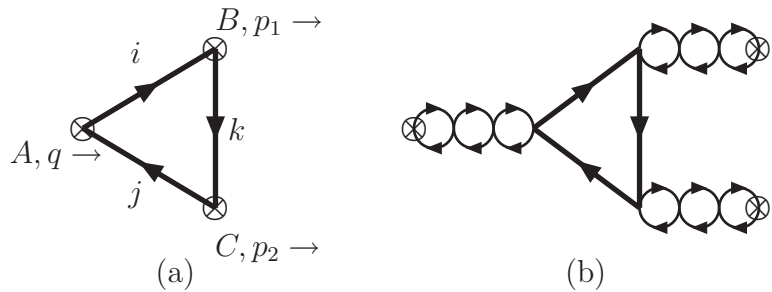
\begin{figure}
\begin{center}
\setlength{\unitlength}{1pt}
\begin{picture}(100,100)(-10,-10)
\SetScale{1.0}
\SetWidth{2.}
\ArrowLine(5,50)(55,80)
\ArrowLine(55,80)(55,20)
\ArrowLine(55,20)(5,50)
\Text(5,50)[]{{\boldmath$\otimes$}}
\Text(55,80)[]{{\boldmath$\otimes$}}
\Text(55,20)[]{{\boldmath$\otimes$}}
\Text(-5,40)[]{$A, q\rightarrow$}
\Text(55,85)[lb]{$B, p_1\rightarrow$}
\Text(55,10)[lt]{$C, p_2\rightarrow$}
\Text(25,75)[b]{$i$}
\Text(25,30)[t]{$j$}
\Text(60,50)[l]{$k$}
\Text(40,-5)[]{(a)}
\end{picture}
~~~~~~
\setlength{\unitlength}{1pt}
\begin{picture}(100,100)(0,-10)
\SetScale{1.0}
\SetWidth{1.}
\ArrowArcn(12.5,50)(7.5,0.,180.)
\ArrowArcn(12.5,50)(7.5,180.,0.)
\ArrowArcn(27.5,50)(7.5,0.,180.)
\ArrowArcn(27.5,50)(7.5,180.,0.)
\ArrowArcn(42.5,50)(7.5,0.,180.)
\ArrowArcn(42.5,50)(7.5,180.,0.)
\SetWidth{2.}
\ArrowLine(50,50)(90,80)
\ArrowLine(90,80)(90,20)
\ArrowLine(90,20)(50,50)
\SetWidth{1.}
\ArrowArcn(97.5,80)(7.5,0.,180.)
\ArrowArcn(97.5,80)(7.5,180.,0.)
\ArrowArcn(112.5,80)(7.5,0.,180.)
\ArrowArcn(112.5,80)(7.5,180.,0.)
\ArrowArcn(127.5,80)(7.5,0.,180.)
\ArrowArcn(127.5,80)(7.5,180.,0.)
\ArrowArcn(97.5,20)(7.5,0.,180.)
\ArrowArcn(97.5,20)(7.5,180.,0.)
\ArrowArcn(112.5,20)(7.5,0.,180.)
\ArrowArcn(112.5,20)(7.5,180.,0.)
\ArrowArcn(127.5,20)(7.5,0.,180.)
\ArrowArcn(127.5,20)(7.5,180.,0.)
\Text(5,50)[]{{\boldmath$\otimes$}}
\Text(135,80)[]{{\boldmath$\otimes$}}
\Text(135,20)[]{{\boldmath$\otimes$}}
\Text(70,-5)[]{(b)}
\end{picture}
\end{center}
\caption{\label{figthreepoint} The $\Pi^+$ contribution to a generic
three-point function. (a) The flavour and momentum flow indicated on a one-loop
diagram. (b) A generic large $N_c$ diagram with the resummation in terms
of bubbles. Note that the resummation leads to full two-point functions.}
\end{figure}

A generic three-point function of currents $A,B,C$ chosen from
the currents in Eq. (\ref{currents}) is defined as
\be \label{gener3point}
\Pi^{ABC}(p_1,p_2)^{ijklmn}
= i^2\int d^dx d^dy \,e^{ip_1\cdot x}e^{i p_2\cdot y}
\langle 0 | T\left(A^{ij}(0)B^{kl}(x)C^{mn}(y)\right)|0\rangle\,.
\ee
In the large $N_c$ limit these can only have two types of
flavour flow
\be
\label{defpm}
\Pi^{ABC}(p_1,p_2)^{ijklmn}
=
\Pi^{ABC+}(p_1,p_2)^{ijk}\delta^{il}\delta^{jm}\delta^{kn}
+\Pi^{ABC-}(p_1,p_2)^{ijl}\delta^{in}\delta^{jk}\delta^{lm}
\ee
and they satisfy
\be
\Pi^{ABC-}(p_1,p_2)^{ijl}
=
\Pi^{ACB+}(p_2,p_1)^{ijl}\,.
\ee
The flavour and momentum flow of $\Pi^{ABC+}(p_1,p_2)^{ijk}$
is indicated in Fig.~\ref{figthreepoint}{(a)}. In the remainder we
will always talk about the $\Pi^+$ part only but drop the superscript $+$.
We also use $q = p_1+p_2$.
A generic contribution to the three-point function is shown in
Fig.~\ref{figthreepoint}(b). The internal vertices are given
by $g_V$ and $g_S$. In Ref.~\cite{BP1} it was shown on two examples
how this resummation can be performed for some three-point functions.
Many other cases were worked out for the work on non-leptonic
matrix-elements in Refs.\cite{epsprime,BP99,scheme,BPP98,BPPx,BPx}.

Here we will make the assumption of resummation for the
three-point functions just as we did for the two-point
functions in Sect.~\ref{twopoint}. It can again be shown that the Ward
identities for the full three-point functions and the resummation
together require that the one-loop
three-point functions satisfy the one-loop Ward identities with the
constituent masses given by the gap equation (\ref{gap}).

We will once more assume that the three-point functions are
constants or low-order polynomials of the kinematical variables, in 
agreement with the large $N_c$ limit Green's functions structure. 
It turns out that the combination of one-loop Ward identities
and short distance constraints is very powerful in restricting the
number of new free parameters appearing in the three-point functions.
This could already be seen in Sect.~\ref{twopointbeyond}, since the
derivative w.r.t. a quark mass of a two-point function is a three-point
function with one of the momenta equal to zero.

A full analysis of three-point functions is in progress. Here we give
a few representative examples.

\subsection{The Pseudoscalar-Scalar-Pseudoscalar Three-Point Function
and the Scalar Form Factor}
\label{sectPSP}

The Pseudoscalar-Scalar-Pseudoscalar three-point function can be calculated
from the class of diagrams depicted in Fig.~\ref{figthreepoint}(b)
using the methods of \cite{BP1} and reads for the case of $m_i=m_k$
\ba
\label{PSPp}
\Pi^{PSP}(p_1,p_2)^{ijk}
& \equiv &\Big \lbrace 1+g_S\Pi_S(p_1^2)_{ki} \Big \rbrace
\nonumber\\
&&\times \Big \lbrace \overline \Pi^{PSP}(p_1,p_2)^{ijk}\,
  (1+g_S\Pi_P(q^2)_{ji})\,
  (1+g_S\Pi_P(p_2^2)_{jk})\nonumber\\
&&+\overline\Pi^{ASP}_{\mu}(p_1,p_2)^{ijk}\,(-g_V i q^{\mu}\Pi^M _P(q^2)_{ji})
(1+g_S\Pi_P(p_2^2)_{jk})\nonumber\\
&&+\overline\Pi^{PSA}_{\nu}(p_1,p_2)^{ijk}\, (1+g_S\Pi_P(q^2)_{ji})
(g_V i p_2^{\nu}\Pi^M _P(p_2^2)_{jk})\nonumber\\
&&+\overline\Pi^{ASA}_{\mu \nu}(p_1,p_2)^{ijk}\,
(-g_V i q^{\mu}\Pi^M _P(q^2)_{ji})
(g_V i p_2^{\nu}\Pi^M _P(p_2^2)_{jk})\Big \rbrace \, .
\ea
The general case has also terms involving one-loop three-point functions
with a vector ($V$) instead of a scalar ($S$).
The one-loop Ward identities can be used to rewrite $\overline\Pi^{ASP}$,
$\overline\Pi^{PSA}$ and $\overline\Pi^{ASA}$ in terms of
$\overline\Pi^{PSP}$ and one-loop two-point functions.

The one-loop three-point function  $\overline\Pi^{PSP}$ is in turn fully fixed
by the one-loop Ward Identities. Let us illustrate the derivation,
one Ward Identity is
\be
i p_2^\mu \overline\Pi^{PSA}_\mu(p_1,p_2)^{ijk} =
-(M_j+M_k) \overline\Pi^{PSP}(p_1,p_2)^{ijk}
+\overline\Pi_{Sik}(p_1^2)-\overline\Pi_{Pij}(q^2)\,.
\ee
Putting $p_1^2=p_2^2=q^2=0$ this determines
\be
\overline\Pi^{PSP}(0,0)^{ijk}
= \frac{1}{M_j+M_k}\left\{
\frac{\cond_k-\cond_i}{M_i-M_k}+\frac{\cond_i+\cond_j}{M_i+M_j}
\right\}
\,.
\ee
The same result follows from the identities for
$q^\mu\overline\Pi^{ASP}_\mu(p_1,p_2)^{ijk}$ and 
$p_1^\mu\overline\Pi^{PVP}_\mu(p_1,p_2)^{ijk}$.

The next term, linear in $q^2, p_1^2, p_2^2$, can be derived as well,
since the relevant combinations of the three-point functions with
one vector or axial-vector
can be determined from Ward identities involving three-point
functions with two vector or axial-vector currents.

We only quote here the chiral limit result
\ba
\label{PSPc}
\overline\Pi^{PSP}(p_1,p_2)^{\chi}
&=&\frac{1}{2g_S^2\left(1-g_S\Delta\right)\condc}
-\frac{p_1^2}{8g_S\condc}\left(4\Gamma- \frac{2\PMNi}{\left(1-g_S\Delta\right)}
+\frac{\PMNc}{g_S\condc}
\right)
\nonumber\\&&
-\frac{q^2+p_2^2}{8 g_S\condc}
\left( \frac{2\PMNi}{\left(1-g_S\Delta\right)}+\frac{\PMNc}{g_S\condc}\right)
\,.
\ea

{}From the $q^2$ dependence of the full Green's function at low energies
we can also derive $L_5$, the result agrees with Eq.~(\ref{result1}) as
it should.

We can look at two different types of short-distance constraints.
First, using the methods of exclusive processes
in perturbative QCD \cite{BrodskyLepage}, it can be shown that the scalar
form factor in the chiral limit should decrease as $1/p_1^2$. 
Phenomenologically, this short-distance behaviour has been also imposed in 
\cite{JOP01,Pich02} to calculate the scalar form factor. It was checked that 
this behaviour agrees with data. 
Using the LSZ reduction formulas the scalar form factor
of the pion in the chiral limit is
\be
F^\chi_S(p_1^2) = \lim_{q^2,p_2^2\to 0}
\frac{q^2 p_2^2}{-2 F_0^2 B_0^2} \Pi^{PSP}(p_1,p_2)^{\chi}
\ee
and it can be written in a simpler form\footnote{Notice that in order
to have the usual
scalar form factor we need to add the
$\Pi^+$ and $\Pi^-$ of Eq.~(\ref{defpm}). The formulas here refer only
to $\Pi^+$.}
\be
F^\chi_S(p_1^2) = B_0 \frac{m_S^2}{m_S^2-p_1^2}
\left(1+p_1^2 \left(\frac{4 L_5}{F_0^2}-\frac{1}{m_S^2}\right)\right)\,.
\ee
The short-distance requirement on $F_S^\chi(p_1^2)$
thus requires $L_5$ to have its
resonance dominated value
\be
L_5 = \frac{F_0^2}{4 m_S^2}\,.
\ee
This gives a new relation between the input parameters, after using
Eq.~(\ref{SDcSP}),
\be
\label{3pointSD}
\PMNi = \frac{(1-g_S \Delta)\PMNc}{2 g_S\condc}
\left(-1+4g_V g_S \condc\PMNc\right)\,.
\ee
This constraint is not compatible with Eq.~(\ref{2pointSD}).

The three-point function $\Pi^{PSP}(p_1,p_2)^{ijk}$ is an order parameter
in the sense described above. Its short-distance properties can thus be used
to constrain the theory. The short-distance behaviour is
\be
\label{SD_PSP}
\lim_{\lambda\to \infty} \Pi^{PSP}(\lambda p_1,\lambda p_2)^{\chi}
= 0\,.
\ee
This is automatically satisfied by our expression (\ref{PSPcf}).

The entire $\Pi^{PSP\chi}$ can be written in a simple fashion
\be
\label{PSPcf}
\Pi^{PSP}( p_1, p_2)^{\chi}
= -2 F_0^2 B_0^3\frac{m_S^2}{q^2 p_2^2 (m_S^2-p_1^2)}
\left(1 + b (q^2 + p_2^2-p_1^2)\right)
\ee
with
\be
 b = 0 (\mbox{~Eq.~(\ref{3pointSD})})
\quad\mbox{or}\quad
 b = \frac{F_0^4}{8\condc^2} = \frac{1}{8 B_0^2}
  (\mbox{~Eq.~(\ref{2pointSD})})\,.
\ee
The short distance relation $\lim_{\lambda\to\infty}F_S^\chi(\lambda p_1^2)
=0$ has no $\alpha_S$ corrections. We therefore consider the constraint
Eq.~(\ref{3pointSD}) to be more reliable than the one from Eq.~(\ref{2pointSD}).

\subsection{The Vector-Pseudoscalar-Pseudoscalar Three-Point Function
and the Vector Form Factor}
\label{sectVPP}

We can repeat the analysis of Sect.~\ref{sectPSP} now for the $VPP$
three-point function. The results will be very similar to there and apply
to the vector (electromagnetic) form factor.
We keep here to the simpler case of $m_i=m_j$.
The resummation leads to\cite{BP1}

\ba 
\label{VPPgeneral}
\Pi^{VPP}_{\mu } (p_1,p_2)^{ijk} & = &
\Big \lbrace g^{\mu \nu}
-g_V\Pi^{V}_{\mu \nu}(q)^{ij}\Big \rbrace \nonumber\\
&& \times \Big \lbrace \overline \Pi^{VPP}_{\nu}(p_1,p_2)^{ijk}
\Big(1+g_S \Pi_{Pik}(p_1^2)\Big)\Big(1+g_S \Pi_{Pkj}(p_2^2)\Big)\nonumber\\
&& +\overline\Pi_{\nu\beta}^{VPA}(p_1,p_2)^{ijk}
   \Big(1+g_S \Pi_{Pik}(p_1^2)\Big)
  \Big(g_V\,i\,p_2^{\beta} \,\Pi_{Pkj}^M(p_2^2)\Big)\nonumber\\
&&+\overline\Pi_{\nu\alpha}^{VAP}(p_1,p_2)^{ijk}
\Big(g_V\,i\,p_1^{\alpha} \,\Pi_{Pik}^{M} (p_1^2)\Big)\,
\Big(1+g_S \Pi_{Pkj}(p_2^2)\Big)\nonumber\\
&&+\overline \Pi_{\nu\alpha\beta}^{VAA}(p_1,p_2)^{ijk}\,
\Big(g_V\,i\,p_1^{\alpha} \,\Pi_{Pik}^M (p_1^2)\Big)\,
\Big(g_V\,i\,p_2^{\beta} \,\Pi_{Pkj}^M (p_2^2)\Big)\Big\rbrace\,.
\ea
We can again use the Ward Identities to rewrite this in terms of
two-point functions and $\overline\Pi^{VPP}_{\mu}(p_1,p_2)^{ijk}$
only. 

We now expand in $p_1^2, p_2^2$
and $(p_1+p_2)^2=q^2$.
\be
\label{VPPexpansion}
\overline\Pi^{VPP}_{\mu}(p_1,p_2)^{ijk}
= p_{1\mu} \overline\Pi^{VPPijk}_1
+ p_{2\mu}\overline\Pi^{VPPijk}_2
+ C^{VPP}_{ijk} \left(q\cdot p_2\, p_{1\mu}-q\cdot p_1\, p_{2\mu}\right)\,.
\ee
The one-loop WI imply
\ba
\overline\Pi^{VPPijk}_1 &=&
  \frac{-\SMN+\overline\Pi_{Pik}^M}{M_j+M_k}
\nonumber\\
\overline\Pi^{VPPijk}_2 &=&
\frac{-\SMN-\overline\Pi_{Pjk}^M}{M_i+M_k}\,.
\ea
The next term in the expansion depends only on one constant. This follows
from the assumption (in the previous subsection) that $\overline\Pi^{SPP}$
contains no terms more than linear in $p_1^2,p_2^2,q^2$. The form
given in Eq.~(\ref{VPPexpansion}) includes this assumption already.
This extra constant can be determined from the fact that the pion vector
factor should decrease as $1/q^2$ for large $q^2$. Extracting the chiral
limit\footnote{This argument is also valid outside the chiral limit.}
vector form factor
via
\be
F^\chi_V(q^2) = \lim_{p_1^2,p_2^2\to0}\frac{p_1^2 p_2^2}{2F_0^2B_0^2}
\Pi^{VPP}_1(p_1,p_2)^\chi\,.
\ee
The subscript one means the coefficient of $p_{1\mu}$ in the
expansion
\be
\Pi_\mu^{VPP}(p_1,p_2)= p_{1\mu}\Pi^{VPP}_1(p_1,p_2)
+p_{2\mu}\Pi^{VPP}_2(p_1,p_2).
\ee
The short-distance requirement then determines
\be
\label{valueCVPP}
C_{\chi}^{VPP} = \left(\PMNc\right)^2 g_V^2\vltc.
\ee
The ChPT expression for the pion vector form factor yields then
\be
\label{L9}
L_9 = \frac{F_0^2}{2} g_V \vltc = \frac{1}{2}\frac{F_0^2}{m_V^2}\,.
\ee

The full chiral limit three-point function can be written in a simple fashion
\be
\label{VPPcf}
\Pi_{\mu}^{VPP}(p_1,p_2)^\chi = \frac{-2 F_0^2 B_0^2}{p_1^2 p_2^2}
\frac{m_V^2}{m_V^2-q^2} 
\left(p_{1\mu}-p_{2\mu}+ A (p_2^2 - p_1^2 )(p_{1\mu}+p_{2\mu})\right)\,,
\ee
with
\be
A = g_V \vltc = \frac{1}{m_V^2}\,.
\ee

\subsection{The Scalar-Vector-Vector Three-Point function}
\label{SVV}

The Scalar-Vector-Vector three-point function has been used
to discuss the properties of the scalars in Ref.~\cite{Moussallam}.
The relation between the full and the one-loop functions in the case of
all masses equal is
\ba
\Pi^{SVV}_{\mu \nu}(p_1,p_2)^{iii}& \equiv &
 \lbrace g_{\mu \alpha}-g_V\Pi^V_{\mu\alpha}(p_1)^{iiii}\rbrace \times 
\lbrace g_{\nu \beta}-g_V\Pi^{V}_{\nu\beta}(p_2)^{iiii}\rbrace
\nonumber\\
&&\times \Big \lbrace 1+g_S\Pi_{Sii}(q^2) \Big \rbrace
\overline\Pi^{SVV}_{\alpha \beta}(p_1,p_2)^{iii}\,.
\ea
In the equal mass case both the full and the one-loop three-point function are
fully transverse.

The one-loop two-point functions expanded to second order in the momenta
is fully determined from the Ward Identities via
\ba
\overline\Pi^{SVV}_{\mu\nu}(p_1,p_2)^{ijk} &=&
\overline\Pi^{SVVijk}_1 g_{\mu\nu}
+\overline\Pi^{SVVijk}_2 \left(p_{2\mu} p_{1\nu}-p_1\cdot p_2\,g_{\mu\nu}
\right)\,,
\nonumber\\
\overline\Pi^{SVVijk}_1 &=&
\frac{1}{M_j-M_i}\left\{\left(M_i-M_k\right)\overline\Pi^M_{Sik}
-\left(M_j-M_k\right)\overline\Pi^M_{Sjk}\right\}\,,
\nonumber\\
\overline\Pi^{SVVijk}_2 &=&
\frac{\overline\Pi^{(0+1)}_{Vik}-\overline\Pi^{(0+1)}_{Vjk}}{M_j-M_i}\,.
\ea
In the chiral limit these expressions reduce to
\ba
\nonumber\\
\overline\Pi^{SVV\chi}_1 &=&
0\,,
\nonumber\\
\overline\Pi^{SVV\chi}_2 &=&
-\frac{\vlti}{1-g_S\Delta}\,.
\ea
The expression for the chiral limit full three-point functions is very simple
\be
\label{SVVcf}
\Pi^{SVV}_{\mu\nu}(p_1,p_2)^\chi = A \frac{m_S^2}{m_S^2-q^2}
\,\frac{m_V^2}{m_V^2-p_1^2}
\,\frac{m_V^2}{m_V^2-p_2^2}\left(p_{2\mu} p_{1\nu}-p_1\cdot p_2\, g_{\mu\nu}
\right)\,.
\ee
with
\be
A = -\vlti = -\frac{\condc}{m_V^4}\,.
\ee
This also satisfies the QCD short-distance requirement
\be
\lim_{\lambda\to\infty}\Pi^{SVV}_{\mu\nu}(\lambda p_1, \lambda p_2)^\chi
= 0\,.
\ee

\subsection{The Pseudoscalar--Vector--Axial-vector Three-Point Function}
\label{PVA}

This three-point functions has been studied in a related way
in Ref.~\cite{KN01}. The expression for the full
Pseudoscalar--Vector--Axial-vector three-point function in terms of the
one-loop one and two-point functions is in the case of $m_i = m_k$:
\ba
\label{PVAf}
\Pi^{PVA}_{\mu \nu} (p_1,p_ 2)^{ijk} & = &
\lbrace g^{\mu \beta}-g_V\Pi_{V\,ik}^{\mu \beta}(p_1)\rbrace
 \nonumber\\
&& \times \Bigg \lbrace 
(1+g_S \Pi_{Pij}(q^2))
(g^{\nu \gamma}-g_V\Pi^{A\nu \gamma}(p_2)^{kj})
\overline\Pi^{PVA}_{\beta \gamma}(p_1,p_2)^{ijk}
\nonumber\\
&& + g_V\,\Pi_{Pij}^M(q^2)(g^{\nu \gamma}
-g_V\Pi^{A\nu \gamma}(p_2)^{kj})
\nonumber\\
&&\times \Big(-(M_i+M_j)\,
   \overline\Pi^{PVA}_{\beta \gamma}(p_1,p_2)^{ijk}
  -i\,\overline\Pi^{V}_{\beta\gamma}(p_1)^{ik}
  +i\,\overline\Pi^{A}_{\beta\gamma}(p_2)^{jk}\Big)
\nonumber\\
&&+(1+g_S\Pi_{Pij}(q^2))g_S\,i\,p_2^{\nu}\Pi_{Pkj}^M(p_2^2)
 \overline\Pi^{PVP}_{\beta}(p_1,p_2)^{ijk}
\nonumber\\
&&+g_S\,g_V\, i \,p_2^{\nu}\Pi_{Pij}^M(q^2) \Pi_{Pkj}^M(p_2^2)
\nonumber\\
&&\times \Big ( -(M_i+M_j)\,\overline\Pi^{PVP}_{\beta}(p_1,p_2)^{ijk}
+\overline\Pi^S_{\beta}(p_1)^{ik}-i\,\overline\Pi^P_{\beta}(p_2)_{jk}\Big)
\Bigg \rbrace\,,
\ea
where we have used the Ward identities
\ba
-i\,q^{\alpha}\,\overline\Pi^{AVA}_{\alpha \beta \gamma}(p_1,p_2)^{ijk}
 & = & -(M_i+M_j)\,\overline\Pi^{PVA}_{\beta \gamma}(p_1,p_2)^{ijk}
-i\,\overline\Pi^{V}_{\beta\gamma}(p_1)^{ik}
+i\,\overline\Pi^{A}_{\beta\gamma}(p_2)^{jk}\,,
\nonumber\\
-i\,q^{\alpha}\,\overline\Pi^{AVP}_{\alpha \beta}(p_1,p_2)^{ijk} & = & 
  -(M_i+M_j)\,\overline\Pi^{PVP}_{\beta}(p_1,p_2)^{ijk}
+\overline\Pi^S_{\beta}(p_1)^{ik}
-i\,\overline\Pi^P_{\beta}(p_2)^{jk}\,.
\ea

The one-loop three-point function up to second order in the momenta is
determined fully from the one-loop Ward Identities.
\ba
\overline\Pi^{PVA}_{\mu\nu}(p_1,p_2)^{ijk} &=&
\overline\Pi^{PVAijk}_1 \, g_{\mu\nu}
+ \overline\Pi^{PVAijk}_2\left(p_{2\mu}p_{1\nu}-p_1\cdot p_2\,g_{\mu\nu}\right)
\nonumber\\&&
+ C^{PVA}_{ijk}
\left(q\cdot p_1\,g_{\mu\nu}-p_{1\mu}p_{1\nu}-p_{2\mu} p_{1\nu}
\right)
\ea
with
\ba
\overline\Pi^{PVAijk}_1 &=&\frac{i}{M_i+M_j}
\left\lbrace\left(M_j+M_k\right)\,\overline\Pi^M_{Pjk}
-\left(M_i-M_k\right)\,\overline\Pi^M_{Sik}\right\rbrace\,,
\nonumber\\
\overline\Pi^{PVAijk}_2 &=&\frac{i}{M_i+M_j}
\left(\overline\Pi^{(0+1)}_{Vik}-\overline\Pi^{(0+1)}_{Ajk}\right)\,,
\nonumber\\
C^{PVA}_{ijk} &=& i (M_j+M_k) C^{VPP}_{kij}\,.
\ea

This expression can be worked out in the chiral limit using the values
obtained earlier and compared with the chiral limit ChPT 
expression for this amplitude
(see e.g. Ref.~\cite{KN01}).
\ba
\Pi_{PVA}^{\mu\nu\,ChPT}&=& 2\,i\condc \Big\lbrace\Big\lbrack
\frac{(p_1+2p_2)^{\mu}p_2^{\nu}}{p_2^2q^2}-\frac{g^{\mu\nu}}{q^2}\Big\rbrack
\nonumber\\&&
+(p_2^{\mu}p_1^{\nu}-\frac{1}{2}(q^2-p_1^2-p_2^2)g^{\mu\nu}) 
\frac {4}{F_0^2q^2}(L_9+L_{10})\nonumber\\&&
+(p_1^2p_2^{\mu}p_2^{\nu}+p_2^2p_1^{\mu}p_1^{\nu}-\frac{1}{2}(q^2-p_1^2-p_2^2)
p_1^{\mu}p_2^{\nu}-p_1^2p_2^2g^{\mu\nu})\frac{4}{F_0^2p_2^2q^2}L_9
\ea
and leads to values of $L_9$ compatible with those obtained
in Eq.~(\ref{L9}) and
\be
L_{10} = -\frac{1}{2} F_0^4g_V \vltc \frac{(g_V g_S\condc\PMNc-1)} 
{g_S \condc \PMNc}\,,
\ee
which is the same  as Eq.~(\ref{L10}).

The three-point function in the chiral limit has a simple
expression of the form
\ba
\label{PVAc}
\Pi^{PVA}_{\mu\nu}(p_1,p_2)^\chi
&=& 
-\frac{2i\condc}{\left(p_1^2-m_V^2\right)q^2}
\left\lbrace\frac{P_{\mu\nu}(p_1,p_2)\,(m_A^2-m_V^2)
+Q_{\mu\nu}(p_1,p_2)}{p_2^2-m_A^2}
-\frac{2 Q_{\mu\nu}(p_1,p_2)}{p_2^2}
\right\rbrace
\nonumber\\&&+\frac{-2 i\condc}{p_2^2 q^2}
\left(p_{1\mu}p_{2\nu}+2p_{2\mu}p_{2\nu}-p_2^2 g_{\mu\nu}\right)\,.
\ea
The tensors $P_{\mu\nu}$ and $Q_{\mu\nu}$ are transverse and defined by 
\ba
P_{\mu\nu} (p_1,p_2) &=&
 p_{2\mu}p_{1\nu}-p_1 \cdot p_2\,g_{\mu\nu}
\nonumber\\
Q_{\mu\nu} (p_1,p_2) &=&
p_1^2\,p_{2\mu}p_{2\nu}+p_2^2\,p_{1\mu}p_{1\nu}-
p_1\cdot p_2\, p_{1\mu} p_{2\nu}-p_1^2 p_2^2\,g_{\mu\nu}\,.
\ea

By construction, this function satisfies the chiral Ward identities 
(see e.g. \cite{KN01})
\ba
p_{1 \mu}\,\Pi_{PVA}^{\mu \nu }(p_1,p_2) &=& -2\,i \, \condc 
\Big \lbrack \frac {p_2^{\nu}}{p_2^2}- \frac{q^{\nu}}{q^2}\Big \rbrack 
\nonumber\\
p_{2\nu}\,\Pi_{PVA}^{\mu \nu }(p_1,p_2) &=& -2\,i \, \condc 
\frac {q^{\mu}}{q^2}
\ea
that are the same as those involving the one-loop function 
$\bar \Pi_{PVA}^{\mu \nu}$ but replacing the constituent masses by current 
quark masses.
The QCD short-distance relation
\be
\lim_{\lambda\to\infty}\Pi^{PVA}_{\mu\nu}(\lambda p_1,\lambda p_2)^\chi = 0\,.
\ee
is also obeyed.

\subsection{The Pseudoscalar--Axial-Vector--Scalar Three-Point Function}
\label{PAS}

Another order parameter is the sum of the Pseudoscalar--Axial-vector--Scalar
and Scalar--Axial-vector--Pseudoscalar three-point functions. 
These functions can be written in terms of the corresponding 
one-loop functions and the two-point functions following the same method 
as in the other sections

For the simpler case $m_j=m_k$
\ba
\label{PASf}
\Pi^{PAS}_{\mu} (p_1,p_2)^{ijk} & = & 
\lbrace 1 + g_S \Pi_{Sjk}(p_2^2)\rbrace \nonumber\\
&& \times \Big \lbrace 
\overline\Pi^{PAS\gamma}(p_1,p_2)^{ijk}
\Big(1+g_S \Pi_{Pij}(q^2)\Big) 
\Big(g_{\mu \gamma}-g_V\Pi^{A}_{\mu \gamma}(p_1)^{ki}\Big)
\nonumber\\
&& +\overline\Pi^{AAS\alpha\gamma}(p_1,p_2)^{ijk}
\Big(-g_V\,i\,q_{\alpha} \,\Pi_P^{Mij} (q^2)\Big)
\Big(g_{\mu \gamma}-g_V\Pi^{A}_{\mu \gamma}(p_1)^{ki}\Big)
\nonumber\\
&& +\overline\Pi^{PPS}(p_1,p_2)^{ijk}
\Big(1+g_S \Pi_{Pij}(q^2)\Big) 
\Big(g_S \,i\,p_{1\mu}\,\Pi_{Pki}^M (p_1^2)\Big)
\nonumber\\
&& +\overline\Pi^{APS}_{\alpha}(p_1,p_2)^{ijk}
\Big(-g_V\,i\,q^{\alpha} \,\Pi_P^{Mij} (q^2)\Big)
\Big(g_S \,i\,p_{1\mu}\,\Pi_{Pki}^M (p_1^2)\Big) \Big\rbrace
\ea
and for the case $m_i=m_j$
\ba
\label{SAPf}
\Pi^{SAP}_{\mu} (p_1,p_2)^{ijk} & = & 
\lbrace 1 + g_S \Pi_{Sij}(q^2)\rbrace \nonumber\\
&& \times \Big \lbrace 
\overline\Pi^{SAP\gamma}(p_1,p_2)^{ijk}
\Big(1+g_S \Pi_{Pjk}(p_2^2)\Big) 
\Big(g_{\mu \gamma}-g_V\Pi^{A}_{\mu \gamma}(p_1)^{ki}\Big)
\nonumber\\
&& +\overline\Pi^{SAA{\alpha\gamma}}(p_1,p_2)^{ijk}
\Big(g_V\,i\,p_{2\alpha} \,\Pi_P^{Mij} (p_2^2)\Big)
\Big(g_{\mu \gamma}-g_V\Pi^{A}_{\mu \gamma}(p_1)^{ki}\Big)
\nonumber\\
&& +\overline\Pi^{SPP}(p_1,p_2)^{ijk}
\Big(1+g_S \Pi_{Pjk}(p_2^2)\Big) 
\Big(g_S \,i\,p_{1\mu}\,\Pi_{Pki}^M (p_1^2)\Big)
\nonumber\\
&& +\overline\Pi^{SPA\alpha}(p_1,p_2)^{ijk}
\Big(g_V\,i\,p_{2\alpha} \,\Pi_P^{Mjk} (p_2^2)\Big)
\Big(g_S \,i\,p_{1\mu}\,\Pi_{Pki}^M (p_1^2)\Big)
\ea

The most general expressions for the one-loop three-point functions 
$\overline\Pi^{SAP}_{\gamma}(p_1,p_2)^{ijk}$ and 
$\overline\Pi^{SAP}_{\gamma}(p_1,p_2)^{ijk}$ up to order $O(p^3)$ and 
compatible with all the symmetries
\be
\label{PASexpansion}
\overline\Pi^{PAS}_{\mu}(p_1,p_2)^{ijk}
= p_{1\mu} \overline\Pi^{PASijk}_1
+ p_{2\mu}\overline\Pi^{PASijk}_2
+ C^{PAS}_{ijk}\,\left(p_1^2\,p_{2\mu}
-p_1\cdot p_2\, p_{1\mu}\right)\,
\ee
\be
\label{SAPexpansion}
\overline\Pi^{SAP}_{\mu}(p_1,p_2)^{ijk}
= p_{1\mu} \overline\Pi^{SAPijk}_1
+ p_{2\mu}\overline\Pi^{SAPijk}_2
- C^{PAS}_{kji} \left(p_1^2\,p_{2\mu}
-p_1\cdot p_2\, p_{1\mu}\right)\,
\ee

There is only one constant at order $O(p^3)$ that remains unknown when we 
apply all the symmetry criteria. The functions in the term of order 
$O(p)$ are fully determined by the use of the one-loop Ward identities
 
\ba
\overline\Pi^{PASijk}_1 &=& 
i\,\frac{\overline\Pi_{Pij}^M -\overline\Pi_{Pik}^M }{M_j-M_k}
\nonumber\\
\overline\Pi^{PASijk}_2 &=&
i\,\frac{\overline\Pi_{Sjk}^M +\overline\Pi_{Pik}^M }{M_i+M_j}
+i\,\frac{\overline\Pi_{Pij}^M -\overline\Pi_{Pik}^M }{M_j-M_k}
\nonumber\\
\overline\Pi^{SAPijk}_1 &=& 
i\,\frac{\overline\Pi_{Sij}^M -\overline\Pi_{Pik}^M }{M_j+M_k}
\nonumber\\
\overline\Pi^{SAPijk}_2 &=& 
i\,\frac{\overline\Pi_{Sij}^M -\overline\Pi_{Pik}^M }{M_j+M_k}
+i\,\frac{\overline\Pi_{Pjk}^M -\overline\Pi_{Pik}^M }{M_i-M_j}
\ea

Using the values of the coupling constants $L_5$ and $L_8$ we obtained 
from two-point functions, the functions $\Pi^{PAS}_{\mu} (p_1,p_ 2)^{ijk}$ 
and $\Pi^{SAP}_{\mu} (p_1,p_ 2)^{ijk}$ have the correct behaviour at 
long distance as described by Chiral Perturbation Theory. In this limit 
the unknown constant $C^{PAS}_{ijk}$ is not involved. 

The sum of the two three-point functions in the chiral limit
can be written in a fairly simple
fashion
\ba
\Pi^{PAS+SAP}_{\mu} (p_1,p_ 2)^{\chi} &=& i B_0^2 F_0^2
\frac{m_S^2}{(m_S^2-q^2)(m_S^2-p_2^2)(m_A^2-p_1^2)p_2^2q^2p_1^2}\nonumber\\
&&\times\Bigg\{ p_{2\mu}\,4\,
(m_A^2 +  D^{PAS}_{\chi}p_1^2)p_1^2(q^2-p_2^2)\nonumber\\
&&+p_{1\mu}\Big\lbrack-2m_S^2(q^2+p_2^2)(m_A^2-p_1^2)-2m_A^2(p_1^2(p_2^2-q^2)
-2q^2p_2^2)\nonumber\\
&&-2 p_1^2 (p_2^4+q^4)
-2 D^{PAS}_\chi p_1^2 (q^2-p_1^2-p_2^2)(q^2-p_2^2)
\Big\rbrack
\Bigg\}
\ea
with
\be
D^{PAS}_{\chi} = i C^{PAS}_\chi \frac{g_S\condc}{g_V\PMNc\vltc}\,.
\ee

\section{Comparison with Experiment}
\label{numerics}

The input we use for $\condc$ is the value derived from sum rules
in Ref.~\cite{BPR}, 
which is in agreement with most recent sum rules determinations of this 
condensate and of light quark masses -see \cite{JOP02} for instance- 
and the lattice light quark masses world average in \cite{Kaneko02}. 
The value of $F_0$ is from Ref.~\cite{ABT3} and the remaining masses are 
those from the PDG.
\ba
F_0 = (0.087\pm0.006)~\mbox{GeV}\,, && m_V= 0.770~\mbox{GeV}\,,
\nonumber\\
m_A = 1.230~\mbox{GeV} \,,&& m_S = 0.985 ~\mbox{GeV}\,,
\nonumber\\
\frac{\langle \bar u u + \bar d d\rangle^{\overline{MS}}(m_V)}{2}&=& \condc
^{\overline{MS}} (m_V)
= -(0.0091\pm0.0020)~\mbox{GeV}^3\,.
\ea
Putting in the various relations, we immediately obtain
\ba
\label{numresults}
f_V &=& 0.15 ~\lbrack0.20\rbrack ~\cite{ENJL,ENJLreview} \,,
\nonumber\\
f_A &=& 0.057 ~\lbrack0.097\pm0.022\rbrack ~\cite{ENJL,ENJLreview}\,,
\nonumber\\
L_5 (m_V)&=& 1.95\cdot10^{-3} ~\lbrack(1.0\pm0.2)\cdot10^{-3} \rbrack
~\mbox{\cite{ABT3}} \,,\nonumber\\
L_8  (m_V) &=& 0.5\cdot10^{-3}  ~\lbrack(0.6\pm0.2)\cdot10^{-3}\rbrack 
~\mbox{\cite{ABT3}}\,,\nonumber\\
L_9  (m_V)&=& 6.8\cdot10^{-3}  ~\lbrack(5.93\pm0.43\rbrack\cdot10^{-3}
\rbrack ~\mbox{\cite{BT1}} \,,\nonumber\\
L_{10}  (m_V)&=& -4.4\cdot10^{-3} ~\lbrack(-4.4\pm0.7)\cdot 10^{-3} \rbrack~
\mbox{\cite{BT1,BT2}}\,.
\ea
These numbers\footnote{The value for $L_{10}$ used the
values of $L_9$ from \cite{BT1}, the $2l_5-l_6$ value from \cite{BT2}
and the $p^4$ relation $2l_5-l_6 = 2 L_9+2 L_{10}$.} 
are in reasonable agreement with the experimental values
given in brackets
with the possible exception of $L_5$ which is rather high. We expect to 
have an uncertainty between $30~\%$ and $40~\%$ in our hadronic predictions.
The values in Eq.~(\ref{numresults})
do not depend on the value of the quark condensate.

We cannot determine $\Delta$ at this level. The three-point functions
$PSP$, $VPP$, $SVV$ and $PVA$ can be rewritten
in terms of the above inputs. There is more freedom in those functions
by expanding the underlying $\overline\Pi$ functions to higher order.
These extra terms can usually be determined from the short-distance
constraints up to the problem discussed in Sect.~\ref{SDProblem}.

\section{Difficulties in Going Beyond the One-Resonance Approximation}
\label{trouble}

An obvious question to ask is whether we can easily go beyond the one
resonance per channel approximation used above using the general
resummation based scheme. At first sight one would have said that
this can be done simply by including higher powers in the expansion
of the one-loop two-point functions and/or giving $g_S,g_V$ a momentum
dependence. Since we want to keep the nice analytic behaviour expected
in the large $N_c$ limit with only poles {\em and} have simple
expressions for the one-loop functions and $g_S,g_V$, it turns out to be
very difficult to accomplish. We have tried many variations but essentially
the same type of problems always showed up, related to the fact that
the coefficients of poles of two-point functions obey positivity
constraints. Let us concentrate on the scalar two-point function in the
chiral limit to illustrate the general problem.

In this limit the full two-point function can be written in terms of the
one-loop function as
\be
\Pi_S(q^2) = \frac{\overline\Pi_S(q^2)}{1-g_S\overline\Pi_S(q^2)}\,.
\ee
If we want to give $g_S$ a polynomial dependence on $q^2$ this two-point
function generally becomes far too convergent in the large $q^2$ limit.
The other way to introduce more poles is to expand $\overline\Pi(q^2)$
beyond what we have done before to quartic or higher order.
For the case of two-poles this means we want
\be
1-g_S\overline\Pi_S(q^2) = a (q^2-m_1^2)(q^2-m_2^2)\,.
\ee
However that means we can rewrite
\be
\label{signs}
\Pi_S(q^2) = -\frac{1}{g_S}+\frac{1}{g_S a (m_1^2-m_2^2)}
\left(\frac{1}{q^2-m_1^2}-\frac{1}{q^2-m_2^2}\right)\,.
\ee
{}From Eq.~(\ref{signs}) it is obvious that the residues of the two poles
will have opposite signs, thus preventing this simple approach for
including more resonances. We have illustrated the problem here for the
simplest extensions but it persists as long as both $g_S,g_V$
and the one-loop two-point functions are fairly smooth functions.

\section{A General Problem in Short-Distance Constraints in Higher Green
Functions}
\label{SDProblem}

At this level we have expanded our one-loop two-point functions
to at most second nontrivial order in the momenta and we found that
it was relatively easy to satisfy the short-distance constraints
involving exact zeros. However, if we check the short-distance relations
for the three-point functions that are order parameters given in
Eqs. (\ref{PSPSD}), (\ref{SVVSD}) and (\ref{PVASD}) and compare
with short-distance properties of our model three-point functions of
(\ref{PSPcf}), (\ref{SVVcf}) and (\ref{PVAc}), we find that they
are typically too convergent. In this subsection we will discuss
how this cannot be remedied
in general without spoiling the parts we have already
matched. In fact, we will show how in general this cannot be
done using a single or any finite number of resonances
per channel type of approximations.
An earlier example where single resonance
does not allow to reproduce all short-distance constraints was
found in Ref.~\cite{KN01}.

First look at the function $\Pi^{PSP}$ and see whether by adding terms
in the expansion in $q^2, p_1^2, p_2^2$ to $\overline\Pi^{PSP}(p_1,p_2)^\chi$
beyond those considered in Eq.~(\ref{PSPc}) we can satisfy the short-distance
requirement of Eq.~(\ref{PSPSD}).
It can be easily seen that setting
\ba
\overline\Pi^{PSP}(p_1,p_2)^\chi
&=&
\left.\overline\Pi^{PSP}(p_1,p_2)^\chi\right|_{\mbox{Eq.~(\ref{PSPc})}}
+\overline\Pi_5^{PSP\chi}\left(q^4+p_2^4-2 q^2 p_2^2 -p_1^4\right)\,,
\nonumber\\
\overline\Pi_5^{PSP\chi} &=& 
\frac{\left(\PMNc\right)^3}{16\condc^2\left(1-2g_S g_V\condc\PMNc\right)}
\ea
makes the short-distance constraint Eq.~(\ref{PSPSD}) satisfied.
However, a problem is that now we obtain a very bad short-distance behaviour
for the pion scalar form factor $F_S^\chi(p_1^2)$ which diverges as
$p_1^2$ rather than going to zero.
Inspection of the mechanism behind this shows that this is a general problem
going beyond the single three-point function and model discussed here.

The problem is more generally a problem between the short-distance requirements
on form factors and cross-sections, many of which can be qualitatively
derived from the quark-counting rules or more quantitatively using the methods
of Ref.~\cite{BrodskyLepage}, with the short-distance properties of general
Green's functions.

The quark-counting rules typically require a form factor,
here $F_S^\chi(p_1^2)$, to vanish as $1/p_1^2$ for large $p_1^2$.
The presence of the short-distance part proportional to $p_1^2/(q^2 p_2^2)$
in the short distance expansion of $\Pi^{PSP}(p_1,p_2)^\chi$ then
requires a coupling of the hadron in the $P$ channel to the
$S$ current proportional to $p_1^2$ (or via a coupling to a hadron in the
$S$ channel which in turn couples to the $S$ current, this complication
does not invalidate the argument below).
In the general class of models
with hadrons coupling with point-like couplings the negative powers
in Green's functions can only be produced by a hadron propagator.
The positive power present in the short-distance expression must thus
be present in the couplings of the hadrons. This in turn implies that this
power is present in the form factor of at least some hadrons. The latter
is forbidden by the quark-counting rule.

It is clear that for at most a single resonance in each channel there is no
solution to this set of constraints. In fact, as will show below,
there is no solution to this problem for any finite number of resonances
in any channel. This shows that even for order parameters the
approach of saturation by resonances might have to be supplemented
by a type of continuum. 
We will illustrate the problem for the PSP three-point function.
The general expression, labeling resonances in the first $P$-channel by
$i$, in the $S$-channel by $j$ and in the last $P$-channel by $k$
is 
\ba
\Pi^{PSP}(p_1,p_2)^\chi &=& 
f_0(q^2,p_1^2,p_2^2)
+\sum_i \frac{f_{1i}(p_1^2,p_2^2)}{(q^2-m_i^2)}
+\sum_j \frac{f_{2j}(q^2,p_2^2)}{(p_1^2-m_j^2)}
+\sum_k \frac{f_{3k}(q^2,p_2^2)}{(p_2^2-m_k^2)}
\nonumber\\&&
+\sum_{ij} \frac{f_{4ij}(p_2^2)}{(q^2-m_i^2)(p_1^2-m_j^2)}
+\sum_{ik} \frac{f_{5ik}(p_1^2)}{(q^2-m_i^2)(p_2^2-m_k^2)}
\nonumber\\&&
+\sum_{jk} \frac{f_{6jk}(q^2)}{(p_1^2-m_j^2)(p_2^2-m_k^2)}
+\sum_{ijk} \frac{f_{ijk}}{(q^2-m_i^2)(p_1^2-m_j^2)(p_2^2-m_k^2)}\nonumber\\
\ea
The couplings $f_i$ are polynomials in their respective arguments.
The short-distance constraint now requires
$f_0(q^2,p_1^2,p_2^2) = 0$ and various cancellations between coefficients
of the other functions. The presence of the term $p_1^2/(q^2 p_2^2)$
now requires the presence of at least a nonzero term of order $p_1^2$
in one of the $f_{5ik}(p_1^2)$. However the Green's function can then be
used to extract the scalar (transition) form factor between hadron $i$ and $k$
which necessarily increases as $p_1^2$ which is forbidden by the
quark-counting rules for this (transition) scalar form factor.
The terms with $p_2^2/(q^2 p_1^2)$ and $q^2/(p_1^2 p_2^2)$ obviously
leads to similar problems but in other (transition) form factors.

We have discussed the problem here for one specific three-point function
but it is clear that this is a more general problem for three-point
functions. 
For Green's functions with more than three insertions similar 
conflicts with the
quark counting rules will probably arise also from hadron-hadron scattering
amplitudes.

\section{Applications: Calculation of $\hat B_K$}
\label{scbk}

As a first application, we intend to use the hadronic approach 
introduced in this chapter to calculate $\hat B_K$ -defined in 
Section \ref{scepsilonk}- outside the chiral limit \cite{BGLP03b}. 
In particular, we plan to investigate the origin of the large  
chiral corrections to this quantity found in \cite{BP95bk}.

The method followed is similar to the one used in \cite{BP95bk} 
but replacing the ENJL by our hadronic approach in the calculation 
of the relevant Green's functions. The object we study is the 
$\Delta S=2$ two-point function
\be
G_F\,\Pi_{\Delta S=2}(q^2) \, \equiv \,
i^2 \int d^4 x \, e^{iq\cdot x}
\langle 0 | T \left( P^{ds}(0)P^{ds}(x) \Gamma_{\Delta S =2}
\right)| 0 \rangle
\ee
in the presence of strong interactions. The action $\Gamma_{\Delta S=2}$, 
that is given by
\be
\Gamma_{\Delta S=2} \equiv - G_F\,
\int d^4 y \, Q_{\Delta S = 2} (y) \, ,
\ee
with the operator $Q_{\Delta S = 2}$ defined in (\ref{operatorDS2}),  
can be rewritten as 
\ba
\Gamma_{\Delta S=2} \, = \,
 - 4\,G_F \, \int \frac{d^4 r}{(2\pi)^4}
\int d^4 x_1 \int d^4 x_2 \,
e^{-i r \cdot(x_2 - x_1)} \left[\bar s_{\alpha}\gamma_{\mu}
d_{\alpha}\right]_L(x_1)
\left[\bar s_{\beta}\gamma^{\mu}d_{\beta}\right]_L(x_2)\,,
\ea
with $\left[\bar s_{\beta}\gamma^{\mu}d_{\beta}\right]_L$ defined as in 
(\ref{operatorDS2}). This allows us to describe the operator 
by the exchange of a heavy colourless $X$ boson 
with $\Delta S=2$ and moment $r$. Using the inhomogeneous renormalization 
group equation involving $\Gamma_{\Delta S=2}$, one can do de upper part of 
the integral in the moment $r$ with the result \cite{BP95bk}
\be \label{2.7}
\Gamma_{\Delta S=2} = - 4\,G_F \, C(\mu) \int_0^\mu \frac{d^4 r}{(2\pi)^4}
\int d^4 x_1 \int d^4 x_2 \, e^{-ir\cdot(x_2 - x_1)}
\left[\bar s_{\alpha}\gamma_{\mu}
d_{\alpha}\right]_L(x_1)
\left[\bar s_{\beta}\gamma^{\mu}d_{\beta}\right]_L(x_2)\, ,
\ee
where $C(\mu)$ is the corresponding Wilson coefficient.

\begin{figure}
\begin{center}
\epsfig{file=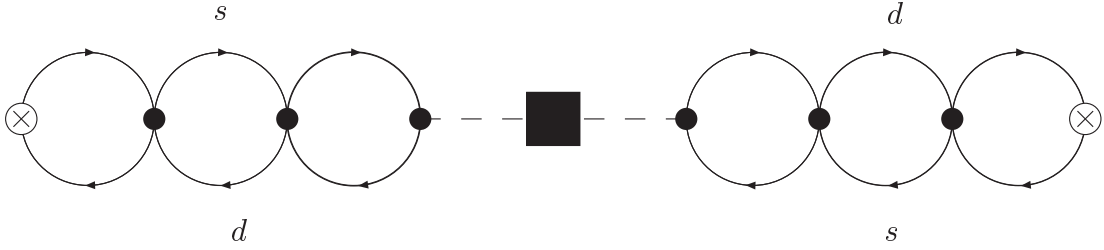,width=15cm}
\caption{\label{LObk} The leading $1/N_c$ contribution to 
$\Pi_{\Delta S=2}(q^2)$ in our ladder resummation approximation. The crosses 
are insertions of the pseudoscalar current, $P^{ds}$, 
and the box is a $\Gamma_{\Delta S=2}$ insertion. The resummation in terms 
of bubbles leads to full two-point functions.}
\end{center}
\end{figure}

In our model, the only kind of diagram that can contribute to 
$\Pi_{\Delta S=2}(q^2)$ in the large $N_c$ limit is the 
one depicted in Figure 
\ref{LObk}. It corresponds to the product of two two-point functions, 
namely, two mixed pseudoscalar--axial-vector $\Pi_P^\mu(q)_{sd}$, that 
are connected by the exchange of a $X$ boson between the two left currents.
 
When the $X$ boson propagator is cut, as in (\ref{2.7}), 
this diagram factorizes and gives the LO in $N_c$ result 
\be \label{6.2}
\Pi_{\Delta S=2}(q^2) = -2\,\Pi^\mu_P (-q)_{sd} \Pi_\mu^P(q)_{sd} \, .
\ee
For the two-point functions in (\ref{6.2}) we can use the expressions in 
(\ref{twopointdef}), (\ref{twopointchiral}) in the chiral limit and 
(\ref{mesondominance}) out of the chiral limit. After performing the 
appropriate reduction of $\Pi_{\Delta S=2}(q^2)$ in (\ref{6.2}) to the matrix 
element in (\ref{bkdef}), these functions lead to the known 
$1/N_c$ result $\hat B_K=3/4$.

The NLO in the $1/N_c$ expansion contribution to $\Pi_{\Delta S=2}(q^2)$ 
in our approximation can be related to the general four-point function
\be
\Pi_{PLPL}^{\mu \nu} (p_1,p_2,p_3)= i^3\int d^4 x \int d^4y 
\int d^4z \,e^{ip_1\cdot x}e^{ip_2\cdot y}e^{ip_3\cdot z}  
\langle 0|P^{ds}(0)L^\mu_{sd}(x)P^{ds}(y)L^{\nu}_{sd}(z)|0\rangle \, ,
\ee
with certain constraints in the moments and Lorentz indices. 
The left-handed currents are defined as $2L_{\mu}^{ij}=
\bar q_i \gamma_\mu(1-\gamma_5)q_j$. 

In the case of four-point functions there are 
2 classes of contributions within the method we are using: 
3-point-functions like contributions and 4-point-functions 
like contributions. The former consists of two full 3-point 
functions with two sources each. They 
are obtained by gluing to the one-loop 3-point functions full 2-point function 
legs to obtain the full structure. Then the third leg of both full 3-point 
functions is removed and the two 3-point functions are pasted together with 
a propagator. This propagator can have any Dirac structure compatible with the 
strong interaction symmetries. The 4-point like functions contribution 
to the generalized four-point function 
consist in full four point functions with the same flavour and Dirac structure 
as the generalized four-point function we are calculating. Each of these 
full-four functions is constructed by gluing to the one-loop four-point 
function four sources with the full two-point functions 
permitted by the symmetries of 
the strong interactions that gives the required structure.

This implies the calculation of many one-loop four-point functions, 
as well as 
three-point functions with contributions from the chiral anomaly, such as 
$\Pi_{PVV}^{\mu \nu}(p_1,p_2)$ or $\Pi_{PAA}^{\mu \nu}(p_1,p_2)$. This 
method will allow us to calculate the $\Pi_{PLPL}^{\mu \nu}$ function 
analytically and investigate the cancellations between the different 
contributions coming from scalar, pseudoscalar, vector and axial-vector 
resonances. 
Work in progress is in this direction \cite{BGLP03b}. The first step, i.e., 
the calculation of $\hat B_K$ in the chiral limit to be compared with the 
results in \cite{BP95bk} is expected to be finished soon.

%\section{An Example of ChPT Green's Function Calculation}

\section{An Example of OPE Green's Function Calculation}
\label{OPEexample}

The operator product expansion (OPE) of local operators was first formulated 
by Wilson \cite{WilsonOPE}, who proposed an expansion of the following form
\be \label{OPEx}
\langle a|A(x)B(0)|b \rangle \stackrel{{x ^{\mu}\to 0}}{\sim}
\langle a|\sum_n C_n(x) O_n(0)|b\rangle
\ee
where $A$, $B$ and $O_n$ are local operators. $C_n(x)$ are c-number 
functions which 
can have singularities of the form $(x^2-i\varepsilon x^0)^{-p}$, with 
p being any real number and also involve logarithms of $x^2$. The functions 
$C_n(x-y)$ are determined, except for some constants, by scale invariance 
(if it is the case). They must be homogeneous functions of order $n-d(A)-d(B)$ 
in $(x-y)$ where $n$, $d(A)$ and $d(B)$ are the canonical dimensions of 
the operators $O_n$, $A$ and $B$ respectively. 

In general, there is an infinite number of non-singular operators $O_n$ 
entering in an OPE; however, to a finite order in $x$ they reduce to a finite 
number. This kind of expansions are valid also for time ordered products, 
commutators or any others combination of elementary or composite local fields 
of a free theory as well as of a renormalized interacting field theory to all 
orders in perturbation theory.

In this Thesis we are interested in the OPE of Green's functions constructed 
with two-quark densities and currents, and within the Standard Model. The 
simplest case is a two point function of the type
\ba
\Pi(q^2)&\equiv&
i \, \int {\rm d}^4 y \, e^{i y \cdot q} \, 
\langle 0 | T \left[ A(y) B(0) \right] | 0 \rangle \, ,
\ea
where A and B are densities. An example of this kind of functions is 
$\Pi_{SS+PP}^{(a)}(p^2)$, that appears in the calculation of the $\Delta I=3/2$ 
contribution to $\varepsilon_K'$ (see Chapter \ref{chq7q8}). For euclidean 
values Q of the moment ($q^2=-Q^2$), for which $Q^\mu\to \infty$ 
implies $Q^2\to-\infty$, we can apply the expansion in (\ref{OPEx}) and 
pass strictly from the limit $x^\mu\to\infty$ to the limit 
$Q^\mu\to\infty$ in the next way
\ba \label{OPEq}
\lim_{Q\to\infty}\,\Pi(Q^2) &=& 
%i\sum_n\langle 0 | O_n(0) | 0 \rangle
%\int  {\rm d}^4 y \, e^{-y \cdot Q} \,C_n(x) \nonumber\\&&
\sum_{n=0}^\infty \sum_{i=1}\,
\langle 0 | O_{2n}^{(i)}(0) | 0 \rangle(\nu)
C_{2n}^{(i)}(\nu,Q^2) \, ,
%\frac{C_{2n}^{(i)}(\nu,Q^2)}{Q^{2n-2}} \, ,
\ea
with $O_{2n}$ gauge invariant operator of dimension 2n constructed from 
quark and gluon fields.

In the standard perturbative theory only the unit operator survive in the sum 
in (\ref{OPEq}). Non-vanishing values of the vacuum expectation values of 
operators of higher dimension, the so called condensates, are pure 
non-perturbative effects. Therefore, they correct the perturbative 
calculation by introducing terms of the type $(1/Q^2)^n$ with $n\ge 1$. 
Once they are renormalized, the vacuum condensates depend on the 
renormalization scale $\nu$ in such a way that they cancel the scale 
dependence of the coefficients $C_{2n}^{(i)}(\nu,Q^2)$ if the complete OPE 
is considered.

As an example of the calculation of the Wilson coefficients 
$C_{2n}^{(i)}(\nu,Q^2)$, we briefly outline here the method followed to 
obtain the results in Appendix \ref{apendixshortdist}. For detailed examples 
of this kind of calculations see \cite{PT}. 
We consider the combination of three-point functions 
\be
\Pi^{PAS+SAP}_\mu(p_1,p_2)\equiv \Pi^{PAS}_\mu(p_1,p_2)+
\Pi^{SAP}_\mu(p_1,p_2)\,,
\ee
that is known to be an order parameter. The two functions $\Pi^{PAS}$ and  
$\Pi^{SAP}$ are defined as in (\ref{gener3point}). The first non-vanishing 
contribution to $\Pi^{PAS+SAP}_\mu(p_1,p_2)$ in the chiral limit, 
 $\Pi^{(PAS+SAP)\,\chi}_\mu(p_1,p_2)$ is proportional 
to the quark condensate squared. 

\begin{figure}
\begin{center}
\epsfig{file=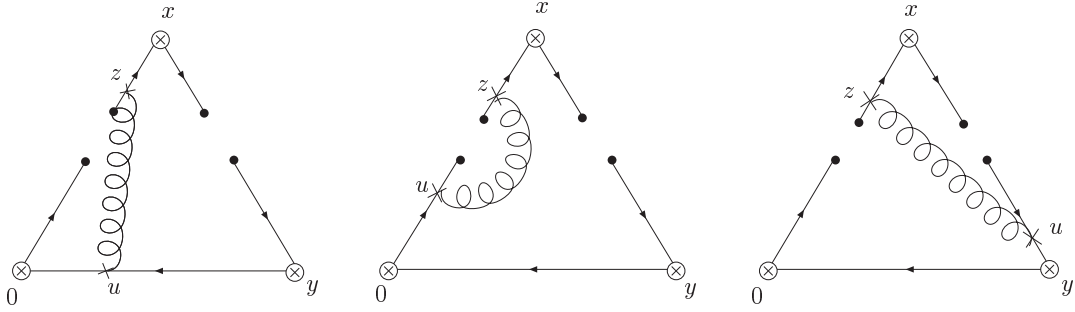,width=15cm}
\caption{\label{figOPE} The three types of diagrams that contribute to 
the dominant term in the OPE of the function $\Pi^{PAS+SAP}_\mu(p_1,p_2)$. For 
explanations, see the text.}
\end{center}
\end{figure}

The diagrams contributing to this order in the OPE  
belong to one of the three groups showed in Figure \ref{figOPE}. In total, 
we must calculate 18 diagrams for each of the three-point functions involved 
in the combination $\Pi^{(PAS+SAP)\,\chi}_\mu(p_1,p_2)$. Since it is quite 
a cumbersome task, we give here, only as an example, the calculation 
of the first diagram in Figure \ref{figOPE} for the function  
$\Pi^{PAS}_\mu(p_1,p_2)^\chi$, which we will denote by 
$\Pi^{PAS}_\mu(p_1,p_2)^\chi _{(1)}$ .  
This diagram comes from the Green function 
with two insertion of the interacting lagrangian
\ba \label{OPE2cond}
i^2\frac{i^2}{2}\int 
d^4x\,d^4y\,d^4z\,d^4u\,e^{ip_1\cdot x}e^{ip_2\cdot y}
%\nonumber\\&&
\langle 0| T\left[P(0)A_\mu(x)S(y)L_I(z)L_I(u)\right]|0\rangle\,,
\ea
where
\ba
P(0)&=&:\bar q^A_{\alpha i}(0) \left(i\gamma_5\right)_{ij}q^A_{\alpha j}(0):
\,,\nonumber\\
A_\mu(x)&=& :\bar q^B_{\beta k}(x) \left(\gamma_\mu\gamma_5\right)_{kl}
q^B_{\beta l}(x):
\,,\nonumber\\
S(y)&=&:\bar q^C_{\gamma }(y)(-1)q^C_{\gamma}(y):
\,,\nonumber\\
L_I(z)&=&\frac{1}{2}g:\bar q_{\delta m}^D (z) \lambda_{\delta\sigma}^a
\left(\gamma_\nu\right)_{m n} q_{\sigma n}^D(z) B_a^\nu(z):
\,,\nonumber\\
L_I(u)&=&\frac{1}{2}g:\bar q_{\rho o}^E (u) \lambda_{\rho\eta}^b
\left(\gamma_\xi\right)_{op} q_{\eta p}^E(u) B_b^\xi(u):\,.
\ea
In these definitions, the capital letters are flavour indices,  
%(??), 
the Greek letters are colour indices, $B_b^\mu(x)$ is a gluon field and 
$\lambda$ are the Gell-Mann matrices in colour space with 
\be \label{GMrelation}
\lambda^a_{\delta\sigma}\lambda^b_{\rho\eta}=2\left[\delta_{\delta\eta} 
  \delta_{\sigma\rho}-1/N_c\delta_{\delta\sigma} \delta_{\eta\rho}\right]
\delta^{a b}\,.
\ee
Now we do in (\ref{OPE2cond}) the contractions of the fields corresponding 
to the first diagram, applying the relation in (\ref{GMrelation}) and 
the definitions of the fermion and gluon propagators
\be
q \underbrack{^A_{\alpha i}(x) \bar q^B_{\beta j}}(y)\equiv 
i\delta_{AB}\delta_{ij}S_{ij}^A(x-y)\,,\quad\quad
B\underbrack{_a^\nu(z) B_b^\xi}(u)\equiv \delta_{ab}D_{\nu\xi}(z-u)
\,,
\ee
that are given by
\ba
S^A_{ij}(x-y) &=& \int\frac{d^4k}{(2\pi)^4}S_{ij}^A(k)e^{-ik\cdot(x-y)}
\,,\quad\quad S^A(k)\equiv\frac{1}{\ksls-m_A+i\eta}\nonumber\\
D_{\mu\nu}(x-y) &=& \int\frac{d^4k}{(2\pi)^4}D_{\mu\nu}(k)e^{-ik\cdot(x-y)}
\,,\quad D_{\mu\nu}(k)\equiv\frac{1}{k^2+i\eta}
\left[-g_{\mu\nu}+(1-a)\frac{k_\mu k_\nu}{k^2+i\eta}\right]\,.\nonumber\\
\ea
The result we obtain from the first diagram is then 
\ba
\Pi^{PAS}_\mu(p_1,p_2)^\chi _{(1)}&=& \frac{i^9}{4}g^2\sum_{ABC}\int 
d^4x\,d^4y\,d^4z\,d^4u\,\int \frac{d^4k}{(2\pi)^4}\frac{d^4l}{(2\pi)^4}
\frac{d^4v}{(2\pi)^4}\frac{d^4w}{(2\pi)^4}\frac{1}{k^2}(-g_{\nu\xi})\nonumber\\
&&\times e^{ip_1\cdot x}e^{ip_2\cdot y}e^{-ik\cdot(z-u)}e^{il\cdot u}
e^{-iv\cdot (u-y)}e^{-iw\cdot (x-z)} 
\left[\gamma_5S(l)^A\gamma_\xi S(v)^C\gamma_\mu \gamma_5S(w)^B\gamma_\nu\right]
\nonumber\\
&&\left \lbrace \langle 0|:\bar q_{\sigma i}^A(0)\bar q_{\gamma k}^B(x)
q_{\gamma r}^A(y) q_{\sigma n}^B(z):|0\rangle -\frac{1}{N_c}
\langle 0|:\bar q_{\gamma i}^A(0)\bar q_{\sigma k}^B(x)
q_{\gamma r}^A(y) q_{\sigma n}^B(z):|0\rangle \right\rbrace.\nonumber\\
\ea

After doing the trace of the Dirac matrices, this expression in the chiral limit 
simplifies to
\ba 
 \Pi^{PAS}_\mu(p_1,p_2)^\chi _{(1)}&=&\frac{i^9}{4}g^2\sum_{ABC}\frac{1}{p_1^2}
\frac{-8(p_1^2p_2^\mu+p_2^2p_1^\mu)}{p_1^2p_2^2(p_1+p_2)^2}\,
\frac{1}{144}\frac{1-N_c}{N_c}\condc^2\nonumber\\
&=&i\alpha_s\condc^2\frac{18}{27}\frac{p_1^2p_2^\mu+p_2^2p_1^\mu}
{p_1^2p_2^2(p_1+p_2)^2}\,,
\ea
that is the final result for the contribution from the first diagram in 
Figure \ref{figOPE} for the PAS part of the dominant term in the OPE of 
$\Pi^{PAS+SAP}_\mu(p_1,p_2)^\chi$. 
The next step to get the final value in Appendix \ref{apendixshortdist} 
will be to calculate the SAP part from this first diagram. Then, we will have  
to repeat the calculation for the other 17 diagrams.

%\chapter{$B_K$} \label{bk}

%\section{The $\hat B_K$ Parameter}

\chapter{Hadronic Matrix Elements of the Electroweak Operators $Q_7$ 
and $Q_8$} 
\label{chq7q8}

In this chapter we report recent advances on the computation
of the matrix elements of the electroweak penguins $Q_7$ and $Q_8$
which are relevant for the $\Delta I=3/2$ contribution to
 $\epp_K$ in the chiral limit, as said in Section \ref{secq7q8}. 
These matrix elements can be calculated without using any model, 
but relating them with integrals over spectral functions for which 
there exist experimental data. This calculation is also an example 
of the use of the X-boson method at all orders in the $1/N_C$ 
expansion.

The method followed in the derivation of these matrix elements starts 
from the effective action (\ref{effective}) whose derivation 
from the Standard 
Model using short-distance renormalization group methods 
has been described in \ref{imai}. In a general way 
\be
\label{step1}
\Gamma_{SD} = \sum_{i=7,8} C_i(\mu_R) Q_i(\mu_R)\,.
\ee 
This effective action can be used directly in lattice calculations
but is less easy to use in other methods.
What we know how to identify are currents and densities. We therefore
go over to an equivalent scheme using only densities and currents
whereby we generate (\ref{step1}) by the exchange of colourless $X$-bosons
(Eq. (\ref{Xeffective}))
\be
\label{step2}
\Gamma_X = \sum_i g_i(\mu_C) X_i^I (\bar q' \gamma_I q)
\ee
where the coupling constants $g_i$ can be determined using short-distance
calculations only. The result is Eqs. (\ref{g7}) and (\ref{g8}).
At this step the scheme-dependence in the calculation of
the Wilson coefficients $C_i$
is removed but we have now a dependence on $M_X$ and the scheme used
to calculate with $\Gamma_X$.

We then need to evaluate the matrix-elements of (\ref{step2}).
For the case at hand this simplifies considerably. In the chiral limit,
the relevant matrix element can be related to  vacuum matrix elements (VEVs).
The disconnected contributions are just two-quark condensates.
The connected ones 
can be expressed as integrals over two-point functions (or correlators) as
given in Eq. (\ref{GE}), which we evaluate in Euclidean space.
The two relevant integrals are Eqs. (\ref{Q7}) and (\ref{Q8}).

Both of the integrals are now dealt with in a similar way.
We split them into two pieces at a scale $\mu$ via Eq. (\ref{split}).
The two-point function to be integrated over is replaced by its
spectral representation, which we assume known.

The long-distance part of the $Q^2$ integral can be evaluated
and integrals of the type (\ref{Q7LD})
and (\ref{Q8LD}) remain.

The short-distance part we evaluate in a somewhat more elaborate way
which allows us to show that the residual dependence on the 
$X$-boson mass disappears and that the correct behaviour given by
the renormalization  group is also incorporated.
To do this, we split the short-distance
integral in the part with the lowest dimensional operator,
which is of dimension six for both $Q_7$ and $Q_8$, 
and the remainder,  the latter is
referred to as the contribution from higher-order operators \cite{CDG00}.

The dimension six part can be evaluated using the known QCD 
short-distance 
behaviour of the two-point functions at this order. It is
vacuum expectation values of  dimension six operators over $Q^6$
for $Q_7$  and over $Q^4$ for $Q_8$
times a known function of $\alpha_S$. The vacuum expectation values
can be rewritten again as integrals over two-point functions
and the resulting integrals are precisely those needed to cancel
the remaining $M_X$-dependence.
For the contribution from {\em all} higher order operators we again
perform simply the relevant $Q^2$ integrals over the 
{\em same} two-point functions as for dimension six and they are 
the ones needed to match long- and short-distances exactly.

This way we see how our procedure precisely cancels  
all the scheme- and scale-dependence and fully relates the results to
known spectral functions.

The chapter consists out of two parts. In the first part, Sections 
\ref{Q7Q8}--\ref{bag},  
we discuss how the $X$-boson approach takes care of the scheme-dependence
in the chiral limit independent of the large $N_c$ expansion we used
in our previous work. We also show precisely how the 
needed matrix-elements in the chiral limit
are related to integrals over spectral functions. This clarifies 
and extends
the previous work on this relation \cite{DG00,NAR01,KPR01,KPR99}.
Equation (\ref{imge}) is our main result, but we also present
the expression in terms of the usual bag parameters
in Section \ref{bag}.

In the second part, Sections \ref{PILR}--\ref{comparison}, 
we present numerical results and compare our results
with those obtained by others and our previous work.
Sections \ref{PILR} and \ref{SSPP} describe
the experimental and theoretical information
on both $\im \Pi_{LR}^T(Q^2)$  and the 
scalar--pseudo-scalar 
$\im \Pi_{SS+PP}^{(0-3)}(Q^2)$ spectral functions and give the values
of the various quantities needed. The comparison with earlier results
is Section \ref{comparison}.

In addition, in the Appendix \ref{AppA} we derive the NLO in $\alpha_S$
coefficient of the leading order term in the OPE of the needed
correlators in the same scheme as used for the short-distance weak
Hamiltonian. This coefficient was previously only known in a different
scheme \cite{LSC86}.

\section{The $Q_7$ and $Q_8$ Operators}
\label{Q7Q8}
      
The imaginary part of $G_E$, the coupling defined in (\ref{lagdS1}) that 
modulates the chiral Lagrangian of order $e^2p^0$, is dominated by
the short-distance electroweak effects and can thus be reliably
estimated  from the purely strong matrix-elements of
the $|\Delta S|=1$ effective hamiltonian in (\ref{hamiltonians1}) with only 
the $Q_7$ and $Q_8$ operators present
\ba
\label{effective}
\Gamma_{{\rm eff}} &=& -\frac{G_F}{\sqrt 2} \, 
V_{ud} V_{us}^* \, {\dis \sum_{i=7,8}}  \, \im C_i \,\,
 \int {\rm d}^4 x \,  Q_i(x)
\ea
with $ \im  C_i =  y_i \, \im  \tau$ the imaginary part
of the Wilson coefficients in (\ref{hamiltonians1}). Up to $O(\alpha_S^2)$, 
the $Q_7$ and $Q_8$  
operators only mix  between themselves below the charm quark mass
via the strong interaction.

 The QCD anomalous dimension matrix $\gamma(\nu)$ 
in regularizations like Naive Dimensional Regularization (NDR)
or 't Hooft-Veltman (HV) which do not mix
operators of different dimension, is defined as\footnote{In a cut-off
regularization one has, on the right hand side, an infinite
series of higher dimensional operators suppressed by powers of the cut-off.
Explicit expressions for the matrices $\gamma^{ji}(\nu)$
are in App. \ref{AppA}.}
($i=$ 7,8)
\ba
\label{gamma}
\nu \, \frac{{\rm d}}{{\rm d} \nu } Q_i(\nu) &=&
- {\dis \sum_{j=7,8}} \gamma^{ji}(\nu) \, Q_j(\nu) \, ; \quad\quad
\gamma(\nu)=\sum_{n=1} \gamma^{(n)}a^n(\nu) \, 
\ea
where $a(\nu) \equiv \alpha_S(\nu) /\pi$.

At low energies, 
it is convenient to describe the $\Delta S=1$ transitions
with an effective action $\Gamma_{LD}$ which
uses hadrons, constituent quarks, or other objects
to describe the relevant degrees of freedom.
A four-dimensional regularization scheme like an Euclidean cut-off,
separating long-distance physics from integrated out short-distance
physics, is also more practical. In addition, the 
color singlet Fierzed 
operator basis becomes  useful for identifying QCD currents 
and densities.
The whole procedure has been explicitly done in \cite{scheme,epsprime}
and reviewed in \cite{benasque}.

At low energies, the effective action (\ref{effective}) is therefore 
replaced by the equivalent
\ba
\label{Xeffective}
\Gamma_X&=& \ g_7(\mu_C,\cdots)
  X_7^\mu \, \left( (\overline s \gamma_\mu d )_L
 + \frac{3}{2} e_q 
{\dis \sum_{q=u,d,s}} (\overline q \gamma_\mu q )_R \right) 
\nonumber  \\  &+& 
g_8(\mu_C,\cdots) {\dis \sum_{q=u,d,s}} X_{q,8} \left( 
(\overline q d)_L 
+ (-2) \frac{3}{2} e_q (\overline s q)_R 
\right) \, .
\ea
Here all colour sums are performed implicitly inside the brackets.
There is also a kinetic term for the X-bosons which we take
to be all of the same mass for simplicity. 

The couplings $g_i$ are determined as functions of the 
Wilson coefficients $C_i$ by taking matrix elements of both sides
between quark and gluon external states as explained in 
\cite{scheme,epsprime,benasque}. We obtain
\ba
\label{g7}
\frac{|g_7(\mu_C)|^2}{M_X^2}&=&
\im C_7(\mu_R) \, \left[ 1 + a(\mu_C) \, \left( 
\gamma^{(1)}_{77} \ln \frac{M_X}{\mu_R} + \Delta r_{77}  \right)
 \right]
 \nonumber \\
&+&   \im C_8(\mu_R) \,  \left[ a(\mu_C) \,  \Delta r_{78}  \right]
+ O \big( a(\mu_R)-a(\mu_C) \big)  
\ea
and
\ba
\label{g8}
\frac{|g_8(\mu_C)|^2}{M_X^2}&=&
\im C_8(\mu_R) \, \left[ 1 + a(\mu_C) \, \left( 
\gamma^{(1)}_{88} \ln \frac{M_X}{\mu_R}  
+ \tilde \gamma^{(1)}_{88} \ln \frac{\mu_C}{M_X} + 
\Delta r_{88}  \right) \right] \nonumber \\
&+& \im C_7(\mu_R) \,  \left[ a(\mu_C) \, \left( 
\gamma^{(1)}_{87} \ln \frac{M_X}{\mu_R} + \Delta r_{87}  \right)
 \right]
+ O \big( a(\mu_R)-a(\mu_C) \big) \, . \nonumber \\
\ea
 $\tilde \gamma_{ij}^{(1)}$ is due to the anomalous
 dimensions of the two-quark color-singlet  
densities or currents.  It vanishes for conserved currents.
In our case $\tilde \gamma_{88}^{(1)}= -2 \gamma_m^{(1)}$,
where $\gamma_m^{(1)}$ is the QCD anomalous dimension of the quark mass
in the regularization used in (\ref{Xeffective}).
The values of $\Delta r_{ij} \equiv (r-\tilde r)_{ij}$  
have been calculated in \cite{epsprime}. 

The effective action to be used at low-energies is now specified 
completely.
Notice that singlet color currents and densities are connected
by the exchange of a colourless X-boson and therefore are well
identified also in the low energy effective theories, and
the finite terms which appear
due to the correct identification of currents and densities.

The coupling $G_E$ is defined in the chiral limit so that
we can use soft pion theorems to calculate the relevant matrix-elements,
and relate them to a vacuum-matrix-element\footnote{In the real 
$K\to\pi\pi$
case we would need to evaluate integrals over strong-interaction five
point functions, three meson legs and two $X$-boson legs. For
vacuum matrix-elements this reduces to integrals over 
two-point functions, the two $X$-boson legs.
The same is not possible for $G_8$  and $G_{27}$
since the corresponding terms are order $p^2$
and have zero vacuum matrix elements.}.
For the contribution of $Q_7$ and $Q_8$, we obtain 
\ba
\label{GE}
-\frac{3}{5} \, 
e^2 \, F_0^6 \, \im G_E &=&  -|g_7(\mu_C, \cdots)|^2 \, 3 \, 
\, i \, \int \frac{{\rm d}^4 p_X}{(2\pi)^4} \, \frac{1}{p_X^2-M_X^2}\, 
g_{\mu \nu} \, \Pi_{LR}^{\mu\nu}(p_X^2) \nonumber \\
&+&    |g_8(\mu_C, \cdots)|^2  \, 
i \, \int \frac{{\rm d}^4 p_X}{(2\pi)^4} \, \frac{1}{p_X^2-M_X^2}\, 
\left( \Pi_{SS+PP}^{(0)}(p_X^2)- \Pi_{SS+PP}^{(3)}(p_X^2) \right)
 \, . \nonumber \\
\ea
Where $\Pi_{LR}^{\mu\nu}(p^2)$ is 
the following two-point function in the chiral limit 
\cite{KPR99,DG00}:
\ba
\Pi_{LR}^{\mu\nu}(p)&\equiv&
\frac{1}{2} \, i \, \int {\rm d}^4 y \, e^{i y \cdot p } \, 
\langle 0| T \left[ L^\mu(y) R^{\nu\dagger}(0) \right] | 0 \rangle
 \equiv \left[ p^\mu p^\nu -g^{\mu\nu} p^2\right] \Pi_{LR}^T(p^2)
\nonumber \\ 
&+& p^\mu p^\nu  \Pi_{LR}^L(p^2)\,.
\ea 
In Eq. (\ref{GE}) we used the chiral limit
so  SU(3) chiral symmetry is exact.
$L(R)^\mu= (\overline u \gamma^\mu  d)_{L(R)}$
or  $L(R)^\mu= (\overline d \gamma^\mu  s)_{L(R)}$,
$\Pi^L_{LR}(p^2)$ vanishes
and $\Pi_{SS+PP}^{(a)}(p^2)$ 
is the two-point function
\ba
\Pi_{SS+PP}^{(a)}(p^2)&\equiv&
i \, \int {\rm d}^4 y \, e^{i y \cdot p} \, 
\langle 0 | T \left[ (S+iP)^{(a)}(y) (S-iP)^{(a)}(0) \right] | 0 \rangle
\ea
with  
\ba
S^{(a)}(x)&=& -\overline q(x) \frac{\lambda^{(a)}}{\sqrt 2} q (x), \, \, 
P^{(b)}(x)= \overline q(x) i \gamma_5 
\frac{\lambda^{(a)}}{\sqrt 2} q (x)\,.
\ea
The 3 $\times$ 3 matrix 
$\dis\lambda^{(0)} = \sqrt{2}\,{I}/{\sqrt 3}$ 
and the rest are the Gell-Mann matrices normalized to
$
\tr \left( \lambda^{(a)}  \lambda^{(b)} \right) = 2 \delta^{ab}  
$.
An alternative form for the last term in (\ref{GE}) is,
\be
\label{alternative}
   |g_8(\mu_C, \cdots)|^2  \, 
3 \, i \, \int \frac{{\rm d}^4 p_X}{(2\pi)^4} \, \frac{1}{p_X^2-M_X^2}\, 
 \Pi_{SS+PP}^{(ds)}(p_X^2)  
\ee
with 
\be
 \Pi_{SS+PP}^{(ds)}(p^2) = 
i \, \int {\rm d}^4 y \, e^{i y \cdot p} \, 
\langle 0 | T \left[ (\overline d d)_L (y)
          (\overline s s)_R(0) \right] | 0 \rangle
\ee
and $ \left( \overline q  q \right)_{L(R)}
=\overline q (1-(+)\gamma_5) q$.

\section{Exact Long--Short-Distance Matching at NLO in $\alpha_S$}
\label{exactmatching}

\subsection{The $\mathbf{Q_7}$ Contribution}
In Euclidean space, the term multiplying $|g_7|^2$ 
in the rhs of (\ref{GE}) can be written as
\ba
\label{Q7}
-\frac{9}{16\pi^2} \int^\infty_0 {\rm d} Q^2 \frac{Q^4}{Q^2+M_X^2}
\, \Pi_{LR}^T(Q^2)
\ea
with $Q^2=-q^2$.
We split the integration into a short-distance
and a long-distance part by
\ba
\label{split}
\int^{\infty}_0 \, {\rm d} Q^2  &=&
 \int^{\mu^2}_0 \, {\rm d} Q^2  +
 \int^{\infty}_{\mu^2} \, {\rm d} Q^2 
\ea
with $M_X^2 >>\mu^2$. 
In QCD, $\Pi_{LR}^T(Q^2)$ obeys an unsubtracted dispersion relation
\ba
\label{KL}
\Pi_{LR}^T(Q^2)&=& \int_0^\infty    
 {\rm d} t \, \frac{1}{\pi} 
\frac{\im \Pi_{LR}^T(t)}{t+Q^2} \, .
\ea  

\subsubsection{$\mathbf{Q_7}$ Long-distance}

Putting (\ref{KL}) in (\ref{Q7}) and performing the integral
up to $\mu^2$ gives
\be
\label{Q7LD}
-\frac{9}{16\pi^2} \int_0^\infty dt
\frac{t^2}{M_X^2} \ln\left(1+\frac{\mu^2}{t}\right)
 \frac{1}{\pi}{\im \Pi_{LR}^T(t)}+{\cal O}
\left(\frac{\mu^2}{M_X^4}\right)\,,
\ee
with the use of the Weinberg Sum Rules
\cite{Weinberg}, Eqs. (\ref{WSRS}).

\subsubsection{$\mathbf{Q_7}$ Short-distance}

At large $Q^2$ in the chiral limit, $\Pi_{LR}^T(Q^2)$ 
behaves in QCD as \cite{SVZ79} 
\ba
\label{SVZLR}
\Pi_{LR}^T(Q^2)&\to& 
\sum_{n=0}^\infty \,
\sum_{i=1} \,
 \frac{ C^{(i)}_{2(n+3)}(\nu, Q^2)}{Q^{2(n+3)}} \, 
\langle 0|  O^{(i)}_{2(n+3)}(0) | 0 \rangle (\nu)
\ea
where $O^{(i)}_{2(n+3)}(0)$ are dimension $2(n+3)$ 
gauge invariant operators.
\ba
\label{D6_1}
 O^{(1)}_6(0) &=&    \frac{1}{4} \, L^\mu(0) \, R_\mu(0) =
 \frac{1}{4} \left( \overline s \gamma^\mu  d \right)_L (0)
\left( \overline d \gamma_\mu  s \right)_R (0) \, ; 
\nonumber\\
 O^{(2)}_6(0) &=&  
(S+iP)^{(0)}(0) \, (S-iP)^{(0)}(0) 
- (S+iP)^{(3)}(0) \, (S-iP)^{(3)}(0)  \nonumber \\
&=& 3 \, (\overline d d)_L (0)
(\overline s s)_R (0)\,.
\ea
The coefficients $C_6^{(i)}(\nu,Q^2)$ are related to
the anomalous dimension matrix defined in (\ref{gamma}).
This can be used to obtain the NLO in $\alpha_S$ part of the coefficient
with the same choice
of evanescent operators as in \cite{NLOWilscoef,BBL96},
calculations of the $\alpha_S^2$ term in other schemes and choices of 
evanescent operators are in\cite{LSC86}.
Our calculation and results are in App. \ref{AppA}.
At the order we work
we only need the
lowest order \cite{SVZ79}
\ba
\label{D6_2}
C^{(1)}_{6}(\nu,Q^2)&=& 
-\frac{16 \pi^2 a(\nu)}{3}
%\left\{ 
\gamma_{77}^{(1)}
%+ a(\nu)
%\left[ \gamma_{77}^{(1)}+
%\gamma_{77}^{(2)} + \frac{\gamma_{77}^{(1)}}{2} 
%\left( \beta_1 - \gamma_{77}^{(1)}\right) 
%\left(\ln \left( \frac{Q^2}{\nu^2} \right) -\frac{1}{3} 
%\right) \right]  \right\} \, ; 
\nonumber \\
C^{(2)}_{6}(\nu,Q^2)&=&  
\frac{8 \pi^2 a(\nu)}{9}
%\left\{
\gamma_{87}^{(1)} 
%+ a(\nu)  
%\left[ \gamma_{87}^{(1)}+\gamma_{87}^{(2)}  +  \frac{\gamma_{87}^{(1)}}{2}
% \left( \beta_1 - \gamma_{88}^{(1)} - \gamma_{77}^{(1)} \right)
%\left(\ln \left( \frac{Q^2}{\nu^2} \right) -\frac{1}{3}
%\right)\right] \right\}  .
\ea

The values of the coefficients of the power corrections 
are physical quantities and can be determined
with global duality
FESR\footnote{The specific form (\ref{FESR}) is only true to lowest
order in $\alpha_S$ due to the $\ln(Q^2)$ dependence at higher orders.},
 \cite{ALEPH,OPAL,FESRtalks},
\ba
\label{FESR}
{\dis \sum_{m=0}^\infty \, 
\sum_{i=1}} (-1)^m \, 
\langle 0| O^{(i)}_{2(m+3)}(0) | 0 \rangle(s_0) \, 
\frac{1}{2 \pi i} \, \oint_{C_{s_0}} \,{\rm d} s \, 
\frac{C^{(i)}_{2(m+3)}(s_0,-s)}{s^{1+m-n}} 
\nonumber \\ = M_{n+2}\equiv  \int_0^{s_0} {\rm d} t \, t^{n+2} \, 
\frac{1}{\pi}  \, \im \Pi_{LR}^T(t) \, , 
\ea
with $n \geq 0$. $s_0$ is the threshold for local
duality\footnote{A discussion of the value of the local duality onset
is in Section  \ref{PILR}.}.
At leading order in $\alpha_S$ only $n=m$ survive
and we can rewrite the short-distance 
contribution to (\ref{Q7}) as 
\ba
\label{shortQ7}
\lefteqn{
-\frac{9}{16\pi^2} \int^\infty_{\mu^2} {\rm d} Q^2 \frac{Q^4}{Q^2+M_X^2}
\, \Pi_{LR}^T(Q^2) =}&&\nonumber\\
&=&a(\mu) \, \ln \frac{\mu}{M_X}  
 i \int \frac{{\rm d}^4 \tilde q}{(2\pi)^4}  \, 
\left( \frac{1}{M_X^2} \right) \, \,  
\left[ \gamma^{(1)}_{77} \, 3 g_{\mu\nu} \, \Pi_{LR}^{\mu\nu}(\tilde q) -
 \gamma^{(1)}_{87} \,
\left( \Pi_{SS+PP}^{(0)}(\tilde q^2)- 
\Pi_{SS+PP}^{(3)}(\tilde q^2) \right) 
\right] \nonumber \\ 
&&+  \left(\frac{-1}{M_X^2}\right)
\frac{9}{16 \pi^2} \, \sum_{n=1}^{\infty} \frac{1}{n} \,  
\sum_{i=1} \frac{ C^{(i)}_{2(n+3)}}{\mu^{2n}} 
\left[ 1 + O\left( \frac{\mu^2}{M_X^2} \right) \right]
\langle 0| O^{(i)}_{2(n+3)}(0) | 0 \rangle + O(a^2) + \cdots
\nonumber \\
&=&  a(\mu)\, \ln \frac{\mu}{M_X}  
 i \int \frac{{\rm d}^4 \tilde q}{(2\pi)^4}  \, 
\left( \frac{1}{M_X^2} \right) \, 
\, \left[ \gamma^{(1)}_{77} 
\, 3g_{\mu\nu} 
\, \Pi_{LR}^{\mu\nu}(\tilde q) - \gamma^{(1)}_{87} \,
\left( \Pi_{SS+PP}^{(0)}(\tilde q^2)- 
\Pi_{SS+PP}^{(3)}(\tilde q^2) \right) 
\right]  \nonumber \\ 
&&+  \frac{9}{16 \pi^2} \, 
\int_0^{s_0} {\rm d} t \, 
\frac{t^2}{M_X^2} \, \ln \left( 1+ \frac{t}{\mu^2} \right) 
\, \frac{1}{\pi}  \, \im \Pi^T_{LR}(t) 
 + {\cal O}\left( \frac{\mu^2}{M_X^4} \right)
+ {\cal O}(a^2)
\ea
where we have used
\ba
\label{Tordered}
 \int \frac{{\rm d}^4 \tilde q }{(2\pi)^4}  \, 
\int {\rm d}^4 x  \, e^{i x\cdot\tilde q}
\, \langle 0|  T \left[ J(x) \, \tilde J(0) \right] | 0 \rangle
&\equiv& \langle 0|  J(0) \, \tilde J(0) | 0 \rangle \, .  
\ea

\subsection{The $\mathbf{Q_8}$ Contribution}

In Euclidean space,  
the term multiplying $|g_8|^2$  in the rhs of (\ref{GE}), is
\ba
\label{Q8}
\frac{1}{16 \pi^2} \int^\infty_{0} 
{\rm d} Q^2 \frac{Q^2}{Q^2+M_X^2} \, \Pi_{SS+PP}^{(0-3)}(Q^2) \, 
 \ea
with
\ba
 \Pi_{SS+PP}^{(0)}(Q^2)- \Pi_{SS+PP}^{(3)}(Q^2) 
&\equiv&  \Pi_{SS+PP}^{(0-3)}(Q^2) \, . 
\ea
This two-point function has a disconnected contribution,
corresponding to what is usually called the factorizable
contribution\footnote{For the other operators this
correspondence does not hold and even for $Q_8$
it is only valid in certain schemes, including ours.}.
We split off that part explicitly:
\ba
\label{Q8conn}
 \frac{1}{16 \pi^2}
\int^\infty_{0} {\rm d} Q^2 \frac{Q^2}{Q^2+M_X^2}
\, \Pi_{SS+PP}^{(0-3)}(Q^2) \, 
= \frac{1}{M_X^2} \,  \big| \langle 0 | S^{(0)}(0) | 0 \rangle \big|^2 \,  
\nonumber \\ +  \frac{1}{16 \pi^2}
\int^\infty_{0} {\rm d} Q^2 \frac{Q^2}{Q^2+M_X^2}
\, \Pi_{SS+PP}^{(0-3)~conn}(Q^2) \,. 
\ea 

\subsubsection{The Disconnected Contribution}
\label{disconnected}

We have included in (\ref{g8})
all the $O(\alpha_S)$ logs and finite terms 
that take into account passing the four-quark
matrix element from the cut-off $\mu_C$ 
regulated $X$-boson effective theory to the $\overline{MS}$ one.
Therefore to the order needed
\be 
\langle 0 | S^{(0)}(0) | 0 \rangle^2 = 3 \, 
\langle 0 | \overline q q | 0 \rangle \big|^2_{\overline{MS}}(\mu_C)
\ee
and from now on the quark condensate is understood to be in
the $\overline{MS}$ scheme.
As shown in \cite{BG86,deR89}, $\gamma_{88}^{(1)} = - 2 \gamma_m^{(1)}$
where $\gamma_m^{(1)}$ is the one-loop quark mass anomalous
dimension\footnote{ See App. \ref{AppA} for the explicit expressions.
It can be seen there that no such relation holds
for $\gamma_{88}^{(2)}$.}. 
This cancels exactly the scale $\mu_C$ dependence in (\ref{g8})
to order $\alpha_S$\cite{BG86,deR89}.

The disconnected contribution to $\im G_E$ is thus
\be
\label{Q8disc}
-\frac{3}{5} \, e^2 F_0^6 \, \im G_E^{Fact} = 
 3 \, \langle 0 | \bar q q | 0 \rangle ^2 (\mu_C)
 \frac{|g_8(\mu_C)|^2}{M_X^2}
\ee
but now with
\ba
\frac{|g_8(\mu_C)|^2}{M_X^2}
&=&  \im C_8(\mu_R) 
\left[ 1 + a(\mu_C)  \left( \gamma^{(1)}_{88}
\ln \frac{\mu_C}{\mu_R}  + \Delta r_{88}  
\right) \right] 
\nonumber \\
&+&   \im C_7(\mu_R) a(\mu_R)  
\left( \gamma^{(1)}_{87}
\ln \frac{M_X}{\mu_R}  + \Delta r_{87} 
\right)
+  {\cal O}(a^2)\,.
\ea
Here one can see that the factorizable contribution
is not well defined. It is due to the mixing of $Q_7$
and $Q_8$ and  is reflected here in the $\ln(M_X/\mu_R)$.
This $M_X$ dependence  cancels with
the {\em non-factorizable} contribution of $Q_7$ 
in (\ref{shortQ7}). Notice
that the contribution of both terms, $\im C_8$  and $\im C_7$, 
to $ \im G_E$ are of the same order in $1/N_c$. 
 It is then necessary to add the non-factorizable
term  to have $\im G_E$  well defined.
Since $\im G_E$ is a physical quantity,
factorization is \underline{not} well defined for $Q_8$.
This was also shown to be the case for $Q_6$ in  \cite{BP99}. 
Of course, the leading term of the $1/N_c$
expansion is well defined but that approximation
would  miss a completely new topology, namely
the non-factorizable contributions.

\subsubsection{The Connected Contribution}
\label{connected}

{}From the leading high energy behaviour,
the scalar--pseudo-scalar spectral functions satisfy
in the chiral limit \cite{ChPT2,BDLW75}
\be
\label{weinSP1}
\int_0^{\infty} dt\, \frac{1}{\pi}  \, 
\left[ \im \Pi_{SS}^{(0)}(t) - \im \Pi_{PP}^{(3)}(t) \right]
= 0 =
\int_0^{\infty} dt\, \frac{1}{\pi}  \, 
\left[ \im \Pi_{SS}^{(3)}(t) - \im \Pi_{PP}^{(0)}(t) \right]  
\ee
which are analogous to  Weinberg Sum Rules.
Therefore the connected part of  
$\Pi^{(0-3)}_{SS+PP}(Q^2)$
satisfies an unsubtracted dispersion relation in the chiral limit,
\be
\label{dispersionSS}
\Pi^{(0-3) \,\rm conn}_{SS+PP}(Q^2)= 
{\dis \int^\infty_0} {\rm d}t  \, \frac{1}{\pi} \, 
\frac{\im \Pi^{(0-3)}_{SS+PP}(t)}{t+Q^2} \, .
\ee

Also in the chiral limit, the scalar and pseudo-scalar $(0-3)$ 
combinations satisfy other Weinberg-like Sum Rules 
as shown in \cite{MOU00} for the scalar\footnote{In \cite{MOU00} it
was the alternative form of Eq. (\ref{alternative}) which was 
used.}   and in \cite{LEU90} for the pseudo-scalar, 
\be
\label{weinSP2}
\int_0^{\infty} dt\, \frac{1}{\pi}  \,  \im \Pi_{SS}^{(0-3)}(t)= 0 =
\int_0^{\infty} dt\, \frac{1}{\pi}  \,  \im \Pi_{PP}^{(0-3)}(t)\,.
\ee

We also know that the spectral functions $\im \Pi_{SS(PP)}(Q^2)$
depend on scale due to the non-conservation of the quark densities. 
\ba
\label{scaleSS}
\mu_C \frac{{\rm d} }{{\rm d} \mu_C } \,  
\im \Pi_{SS(PP)}^{(a)}(t)
 &=&  2 \gamma_m(\mu_C) \, \im \Pi_{SS(PP)}^{(a)}(t)  
\, .
\ea
This scale dependence is analogous to the one of the disconnected
part (\ref{Q8disc}) 
and  cancels the $\mu_C$ dependence in $|g_8(\mu_C)|^2$ 
also for the connected part.

We now proceed as for  $Q_7$ and split the
integral in (\ref{Q8conn}) at $\mu^2$.

\subsubsection*{\ref{connected}.a $\mathbf{Q_8}^{conn}$ Long-Distance}

We perform simply the integral and obtain
\be
\label{Q8LD}
-\frac{1}{16\pi^2} \int_0^\infty dt
\frac{t}{M_X^2} \ln\left(1+\frac{\mu^2}{t}\right)
 \frac{1}{\pi}{\im \Pi_{SS+PP}^{(0-3)}(t)}+{\cal O}
\left(\frac{\mu^2}{M_X^4}\right)\,.
\ee

\subsubsection*{\ref{connected}.b $\mathbf{Q_8}^{conn}$
Short-distance}

Using the unsubtracted dispersion relation in (\ref{dispersionSS}),
$\Pi_{SS+PP}^{(0-3) \, \rm conn}(Q^2)$ in the chiral limit
behaves at  large $Q^2$  in QCD as
\ba
\label{SVZSP}
\Pi_{SS+PP}^{(0-3)\, \rm conn}(Q^2) &\to& 
\sum_{n=0}^\infty \, \sum_{i=1} \,
 \frac{ \tilde C^{(i)}_{2(n+3)}(\nu, Q^2)}{Q^{2(n+2)}} \, 
\langle 0|  \tilde O^{(i) \rm}_{2(n+3)}(0) | 0 \rangle
 (\nu) 
\ea
where $\tilde O^{(i)}_{2(n+3)}(0)$ are dimension $2(n+3)$ 
gauge invariant operators.
\be
 \tilde O^{(1)}_6(0) =    O^{(1)}_6(0)\, ; \quad\quad
 \tilde O^{(2)}_6(0) =  O^{(2)}_6(0)  \,.
\ee

Using the information on the  mixing of $Q_7$ and $Q_8$ 
in (\ref{gamma}), the scale dependence  (\ref{scaleSS}),
it is easy to obtain the leading power behavior in (\ref{SVZSP})
(see Appendix \ref{AppB})
\ba
\label{M1sumrule}
\tilde C^{(1)}_{6}(\nu,Q^2)&=& 
\frac{45\pi^2}{2}  a(\nu)^2  + O(a^3)  
\, ; \nonumber \\
\tilde C^{(2)}_{6}(\nu,Q^2)&=&  
 \frac{211\pi^2}{4}  a(\nu)^2  + O(a^3)  \, .
\ea

Again the values of the coefficients of the power corrections
in (\ref{SVZSP}) can be calculated using global duality FESR, 
\ba
{\dis \sum_{m=0}^\infty \, \sum_{i=1}} (-1)^{m+1}
\langle 0| \tilde O^{(i)}_{2(m+3)}(0) | 0 \rangle(\tilde s_0) \, 
\frac{1}{2 \pi i} \, \oint_{C_{\tilde s_0}} {\rm d} s \,  
\frac{\tilde C^{(i)}_{2(m+3)}(\tilde s_0,-s)}{s^{1+m-n}}
 \nonumber \\ =
\tilde M_{n+1} \equiv 
\int_0^{\tilde s_0} {\rm d} t \, t^{n+1} \, \frac{1}{\pi}  \, 
 \im \Pi^{(0-3)}_{SS+PP}(t)  \, ,
\ea
with $n\geq 0$, and $\tilde s_0$ the  threshold for local duality
for this two-point function.
Again only terms with $n=m$ survive at $O(\alpha_S)$
and one gets
\ba
\label{shortQ8}
\lefteqn{\frac{1}{16 \pi^2}
\int^\infty_{\mu^2} {\rm d} Q^2 \frac{Q^2}{Q^2+M_X^2}
\, \Pi_{SS+PP}^{(0-3)~conn}(Q^2)=}&&
\nonumber \\ 
&&\frac{1}{16 \pi^2} \,  
\int_0^{\tilde s_0} {\rm d} t \, 
\frac{t}{M_X^2} \, 
\ln \left( 1+ \frac{t}{\mu^2} \right)  \, 
\frac{1}{\pi}  \, \im \Pi^{(0-3)}_{SS+PP}(t)
 + {\cal O}\left( \frac{\mu^2}{M_X^4} \right) 
+ {\cal O}(a^3).
 \ea

\subsection{Sum}

We now  add all the contributions of Eqs. (\ref{Q7LD}),
(\ref{shortQ7}), (\ref{Q8disc}), (\ref{Q8LD})
and (\ref{shortQ8}) to obtain the full result.
Notice in particular that all contributions contain
the correct logarithms of $M_X$ to cancel that dependence
in Eqs. (\ref{g7}) and (\ref{g8}).

The integrals over the spectral functions in the respective long
and short-distance regime can in both cases be combined to
give a simple $\ln(t/\mu^2)$.

Therefore, when summing everything to ${\cal O}(\alpha_S)$ 
and {\em all} orders in $1/N_c$, we obtain
\ba
\label{imge}
-\frac{3}{5} \, e^2 F_0^6 \im G_E &=& 
   \Bigg\{ \im C_7(\mu_R) 
\left[1+ a (\mu_C)
\left( \gamma^{(1)}_{77} \ln \frac{\mu}{\mu_R} + \Delta r_{77}  
\right)\right] 
\nonumber \\
 &&+   \im C_8 (\mu_R) a(\mu_C)
 \Delta r_{78}  \Bigg\}   \frac{9}{16\pi^2}   {\cal A}_{LR}(\mu) \nonumber \\ 
&+&  \left\{\im  C_8(\mu_R) \left[1+ a(\mu_C)
\left( \gamma^{(1)}_{88} \ln \frac{\mu_C}{\mu_R}
 + \Delta r_{88}  \right) \right] \right. 
\nonumber \\
&&+ \left. \im C_7(\mu_R) a(\mu_C) 
\left( \gamma^{(1)}_{87} \ln \frac{\mu}{\mu_R} + 
\Delta r_{87}\right) \right\} \times
\nonumber \\
&& \quad\times
\left( 3 \,  \langle 0 | \bar q q | 0 \rangle^2(\mu_C) 
+ \frac{1}{16\pi^2} {\cal A}_{SP}(\mu,\mu_C)) \right)\, ;
\ea
where
\ba
\label{ALRSP}
{\cal A}_{LR}(\mu) &\equiv&
 \int_0^{s_0} {\rm d} t \, t^2 \, \ln \left(\frac{t}{\mu^2} \right) 
\, \frac{1}{\pi}  \, \im \Pi^T_{LR}(t) \, ; \nonumber \\
{\cal A}_{SP}(\mu,\mu_C) &\equiv&
\int_0^{\tilde s_0}{\rm d}t \, t
 \ln\left(\frac{t}{\mu^2}\right)
\frac{1}{\pi}\im \Pi_{SS+PP}^{(0-3)}(t) \, .
\ea

To obtain this result we have used the local duality relations
\ba
\int^{\infty}_{s_0} {\rm d} t \, t^2 \, 
\ln \left(\frac{t}{\mu^2} \right) 
\, \frac{1}{\pi}  \, \im \Pi^T_{LR}(t) &=& O(a^2)  \, , \nonumber  \\
\int^{\infty}_{\tilde s_0} {\rm d} t \, t \, 
\ln \left(\frac{t}{\mu^2} \right) 
\, \frac{1}{\pi}  \, \im \Pi^{(0-3)}_{SS+PP}(t) &=&O(a^2)  \, .
\ea

The expression in (\ref{imge})
 is exact in the chiral limit, at NLO in $\alpha_S$,
all  orders in $1/N_c$
and without electromagnetic corrections. 
It doesn't depend on any scale nor scheme
at that order analytically.  
The dependence on $M_X$ also nicely cancels out.
The $\mu_C$ dependence cancels against the $\mu_C$
dependence of the densities.
Notice that in this final result
we have taken into account the contribution
of all  higher order operators.

%????? comprobar si lo siguiente es cierto

%This result differs from the ones  in  
%\cite{KPR99,DG00,NAR01,KPR01} in the finite terms 
%$\Delta r_{88}$ and $\Delta r_{78}$. They are necessary
%to cancel the scheme dependence. In addition,
%\cite{DG00,NAR01} only take into account the dimension eight
%corrections and \cite{KPR01} uses a hadronic large $N_c$ Ansatz
%to estimate them.

%??????

As noticed in \cite{epsprime},  the connected
scalar--pseudo-scalar two-point function is
exactly zero in U(3)  symmetry, i.e.  is  $1/N_c$ suppressed. 
We used this fact to disregard this contribution there.
We will check later the quality of this approximation from a 
phenomenological analysis of its value.

\section{Bag Parameters}
\label{bag}

We now re-express our main result (\ref{imge}) in terms of the
usual definition of the bag parameters
\ba
-\frac{3}{5} \, e^2 F_0^6 \im G_E &\equiv& 
\langle 0 | \bar q q | 0 \rangle^2(\mu_C)
\left[ \im C_7(\mu_R) B_{7\chi}(\mu_C,\mu_R) + 
3 \im C_8(\mu_R) \, B_{8\chi}(\mu_C,\mu_R) \right]
\nonumber \\
&\equiv&  - 6 \, \im C_7(\mu_R) \,  
\langle 0 | O_6^{(1)} | 0 \rangle_\chi (\mu_R) +
\im C_8(\mu_R) \,  \langle 0 | O_6^{(2)} | 0 \rangle_\chi (\mu_R) \, ,
\ea
where the subscript $\chi$ means in the chiral limit.

This definition coincides with the one in \cite{epsprime}
and gives 
\ba
B_{7\chi}
(\mu_C,\mu_R) &=& \left[ 1 +  \Delta r_{77} \, a(\mu_R) \right]  
\, \frac{9}{16 \pi^2} \, 
\frac{1}{\langle 0 | \bar q q | 0 \rangle^2(\mu_C)}\, 
{\cal A}_{LR}(\mu_R) \nonumber \\ 
&+& 3 \,  a(\mu_C) \, \Delta r_{87} 
\left[ 1 + \frac{1}{48 \pi^2} \, 
\frac{1}{\langle 0 | \bar q q | 0 \rangle ^2(\mu_C)}
\, {\cal A}_{SP}(\mu_R,\mu_C) \right] 
\, ; \nonumber \\
B_{8\chi}
(\mu_C, \mu_R)&=& \left[ 1+ 
\left( \gamma^{(1)}_{88} \ln 
\left( \frac{\mu_C}{\mu_R} \right) + \Delta r_{88} 
\right) a(\mu_C)\right] \times \nonumber\\
&& \times
\left[ 1  + \frac{1}{48 \pi^2} \, 
\frac{1}{\langle 0 | \bar q q | 0 \rangle ^2(\mu_C)}
\, {\cal A}_{SP}(\mu_R,\mu_C) \right] \nonumber \\ 
&+& \frac{a(\mu_C)}{\langle 0 | \bar q q | 0 \rangle^2(\mu_C)}
\, \Delta r_{78} \, \frac{3}{16 \pi^2} 
\, {\cal A}_{LR}(\mu_R) \, .
\ea
The finite terms that appear in the
matching between the X-boson effective theory with a cut-off
and the  Standard Model regularized with the NDR scheme were calculated 
in \cite{epsprime}, 
\be
\label{rNDR}
\Delta r^{\rm NDR}_{77} = \frac{1}{8 N_c} \, ,
\quad
\Delta r^{\rm NDR}_{78} = -\frac{3}{4} \, ,
%\nonumber\\
\Delta r^{\rm NDR}_{87} = - \frac{1}{8} \, ,
\quad %&\quad&
\Delta r^{\rm NDR}_{88}   %= \frac{5}{4} C_F + \frac{3}{4N_c}  
= \frac{5}{8} N_c + \frac{1}{8N_c} \, .
\ee
The finite terms to pass from NDR to HV in the same
 basis and evanescent operators we use can be found in \cite{CFMR94}.
In the HV scheme of \cite{NLOWilscoef,BBL96} \footnote{There is a finite
renormalization from the HV scheme of \cite{CFMR94} to the HV scheme
of \cite{NLOWilscoef,BBL96}. 
If one uses the Wilson coefficients in the HV scheme
including the $C_F$ terms from the renormalization
of the axial current as \cite{CFMR94}, one has to add
$ - C_F$ to the diagonal terms $\Delta r_{ii}$ in (\ref{rHV}) and
$ - \beta_1 \, C_F $ to the diagonal terms in the two-loop
anomalous dimensions in (\ref{HV}).} these finite terms are
\be
\label{rHV}
\Delta r^{\rm HV-NDR}_{77} = - \frac{3}{2 N_c} \, , 
\quad %&\quad&
\Delta r^{\rm HV-NDR}_{78} = 1  \, ,
%\nonumber \\
\Delta r^{\rm HV-NDR}_{87} =  \frac{3}{2} \, , 
\quad
\Delta r^{\rm HV-NDR}_{88}  %=  C_F - \frac{1}{N_c}  
= \frac{N_c}{2}- \frac{3}{2N_c} \, .
\ee
The results for the scheme 
dependent terms  $\Delta r_{77}$ and $\Delta r_{87}$
\cite{epsprime} agree with those in \cite{DG00,KPR01}.

The $B_7$ and $B_8$ bag parameters
are  independent of $\mu$ but depend on 
$\mu_R$ and $\mu_C$, 
and these dependences only cancel in the physical value
of $\im G_E$. The $\mu_C$ dependence is artificial and 
a consequence of the normalization of the bag parameters to
the quark condensate.

At NLO in $1/N_c$ we get
\ba
\label{resbag1}
B^{NDR}_{7\chi}(\mu_C,\mu_R) 
&=& \left( 1 + \frac{a(\mu_R)}{24}  \right) 
\frac{9}{16 \pi^2} \, 
\frac{1}{\langle 0 | \bar q q | 0 \rangle^2(\mu_C)}
{\cal A}_{LR}(\mu_R)
\nonumber \\ 
&-& \frac{3}{8} \, a(\mu_C)  
\left[ 1 + \frac{1}{48 \pi^2} 
\frac{1}{\langle 0 | \bar q q | 0 \rangle ^2(\mu_C)} \, 
{\cal A}_{SP}(\mu_R,\mu_C) \right] \, ;
\nonumber \\ 
B^{NDR}_{8\chi}(\mu_C,\mu_R)
&=& \left[ 1+  \frac{1}{12} \, 
\left( 54 \ln \left( \frac{\mu_C}{\mu_R} \right) + 23  \right) 
a(\mu_C) \right]\times 
\nonumber\\
&&\times\left[ 1 + \frac{1}{48 \pi^2} 
\frac{1}{\langle 0 | \bar q q | 0 \rangle ^2(\mu_C)}
{\cal A}_{SP}(\mu_R,\mu_C) \right] 
\nonumber \\ 
&-& \, \frac{9}{64 \pi^2} \, 
 \, \frac{a(\mu_R)}{\langle 0 | \bar q q | 0 \rangle^2(\mu_C)}
{\cal A}_{LR}(\mu_R)
\ea
and in the HV scheme\cite{NLOWilscoef,BBL96}
\ba
\label{resbag2}
B^{HV}_{7\chi}(\mu_C,\mu_R) 
&=& \left( 1 - \frac{11}{24} a(\mu_R) \right) 
\frac{9}{16 \pi^2} \, 
\frac{1}{\langle 0 | \bar q q | 0 \rangle^2(\mu_C)} \, 
{\cal A}_{LR}(\mu_R)
\nonumber \\ 
&+& \frac{33}{8} \, a(\mu_C)  
\left[ 1 + \frac{1}{48 \pi^2} 
\frac{1}{\langle 0 | \bar q q | 0 \rangle ^2(\mu_C)}
{\cal A}_{SP}(\mu_R,\mu_C) \right] \, ; 
\nonumber \\ 
B^{HV}_{8\chi}(\mu_C,\mu_R)
&=& \left[ 1+  \frac{1}{12} \, 
\left( 54  \ln \left(\frac{\mu_C}{\mu_R} \right)+  35  \right) 
a(\mu_C) \right] \,\times
\nonumber\\&&\times
 \left[ 1 + \frac{1}{48 \pi^2} 
\frac{1}{\langle 0 | \bar q q | 0 \rangle ^2(\mu_C)}
{\cal A}_{SP}(\mu_R,\mu_C) \right]  
\nonumber\\
&+& \frac{3}{64 \pi^2} \, 
\frac{a(\mu_R)}{\langle 0 | \bar q q | 0 \rangle^2(\mu_C)}
{\cal A}_{LR}(\mu_R) \, . 
\ea
We find an exact result for these B-parameters in QCD in the chiral
limit including the effects of higher dimensional
operators to all orders. The scheme dependence is also
fully taken into account. To our knowledge this
is the first time these fully model independent
expressions bag parameters are presented.
%In \cite{DG00,NAR01} only the dimension eight corrections 
%were included and the scheme dependence of the matrix elements 
%of $Q_8$ was   not included neither there nor in \cite{KPR01}.

\section{The $\,\Pi_{LR}^T(Q^2)\,$ 
Two-Point Function and Integrals over It}
\label{PILR}

There are very good data for $\Pi_{LR}^T$ \cite{ALEPH,OPAL} in the 
time-like region below the tau lepton mass.
They have  been extensively used previously
\cite{DG00,NAR01,DHGS98,PPR01}, see the talks 
\cite{FESRtalks} for recent reviews. We consider it a good 
approximation to take this data as the chiral limit data.
Nevertheless, one can  estimate
the effect of the chiral corrections with Cauchy's
integrals around a circle of radius $4 m_\pi^2$ of the type
\ba
\oint_{4 m_\pi^2} {\rm d} s \, s^n\,  \ln^m(s) \, \Pi_{LR}^T(-s) \, 
\ea
with $n>0$, $m=0,1$.
For all the integrals we use, we have checked
that these contributions are negligible using the CHPT expressions
for $\Pi_{LR}^T(Q^2)$ at one-loop\cite{ChPT2}.
The discussion below is focused on the ALEPH data
but we present the OPAL results as well.

We reanalyze here the first and the second Weinberg Sum Rules 
(WSRs) \cite{Weinberg}, which are properties of QCD in the chiral limit
\cite{FSR79},
\ba
\label{WSRS}
\int_0^{\infty} {\rm d} t \,  \frac{1}{\pi}  \, \im \Pi_{LR}^T(t) =
 \int_0^{s_0} {\rm d} t \,  \frac{1}{\pi}  \, \im \Pi_{LR}^T(t) 
 + O(a^2)  = f_\pi^2 \, ; \nonumber \\
\int_0^{\infty} {\rm d} t \, t \, \frac{1}{\pi}  \, \im \Pi_{LR}^T(t)=  
  \int_0^{s_0} {\rm d} t \,  t \, \frac{1}{\pi}  \, \im \Pi_{LR}^T(t) 
 + O(a^2)
% + 2 (m_u+m_d)\langle 0 | \overline q q | 0 \rangle
 = 0 \, 
\ea
where we used the perturbative QCD result for the imaginary part
at energies larger than $s_0$, i.e. we assumed local duality above $s_0$.
 These two WSRs determine the threshold of perturbative QCD $s_0$.
We used the experimental value
for the pion decay constant  $f_\pi=(92.4 \pm 0.4)$ MeV. 

\begin{figure}
\includegraphics[angle=-90,width=0.49\textwidth]{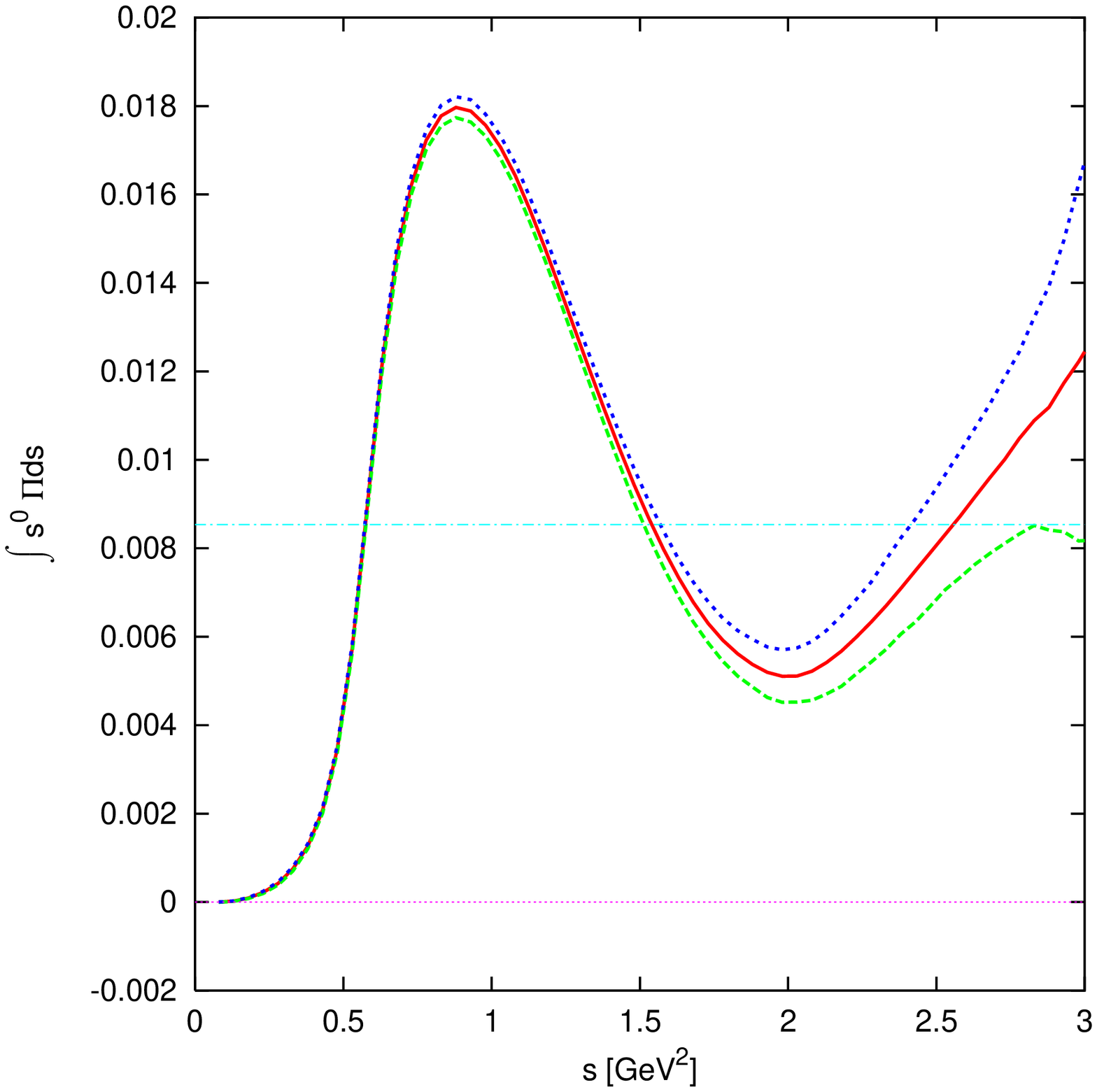}
\includegraphics[angle=-90,width=0.49\textwidth]{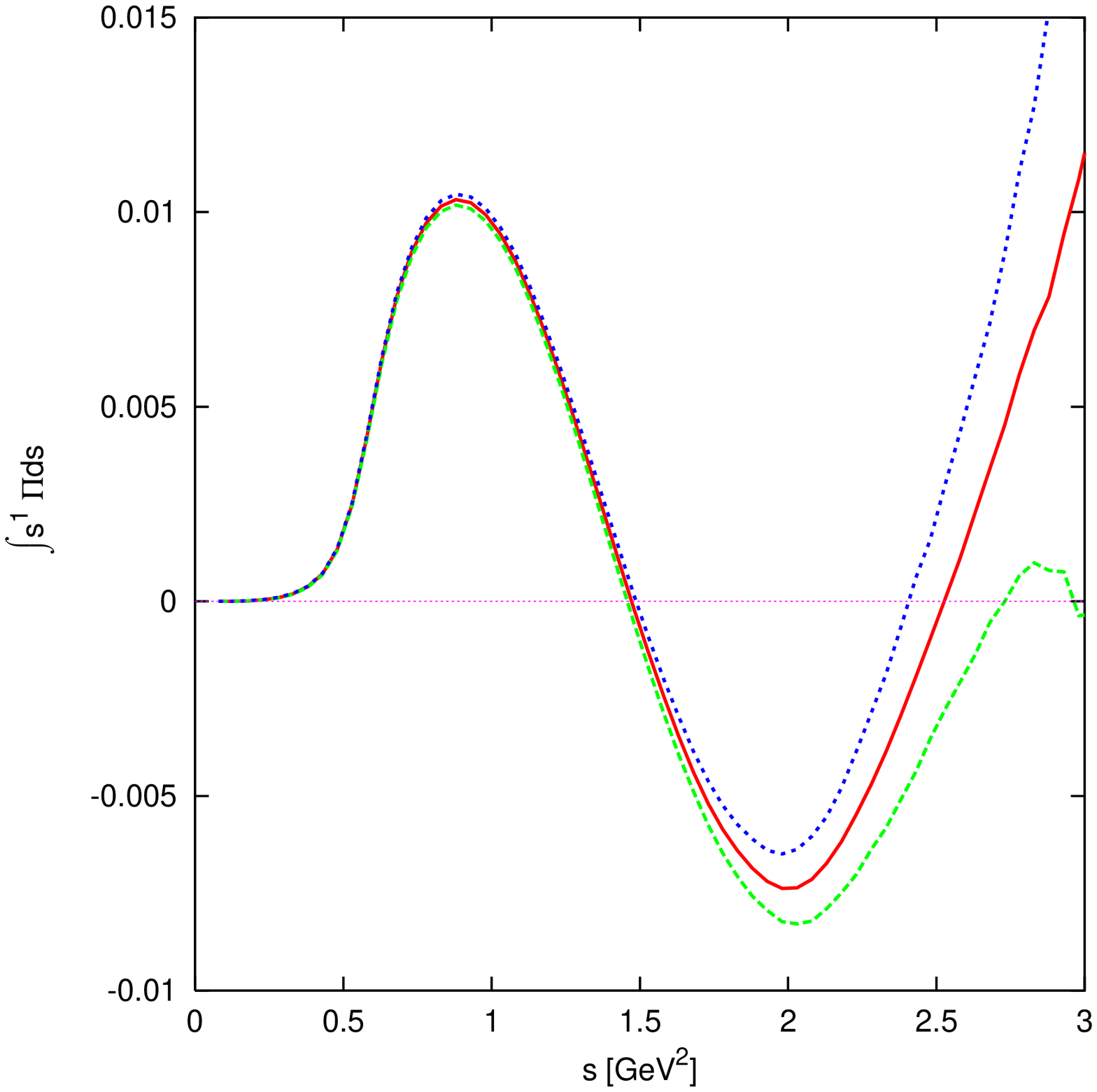}
\caption{\label{Fig1} The first and second Weinberg sum rule
as a function of the upper integration variable $s$. The central
curve corresponds to the central values of \cite{ALEPH} while
the upper and lower curve are the one sigma errors calculated 
as described in the text.}
\end{figure}%FIGURE
These two sum rules are plotted in Fig. \ref{Fig1} for the central data
values and the one sigma errors. These latter are calculated by 
generating a
distribution of spectral functions distributed according to 
the covariance
matrix of \cite{ALEPH}. We then take the one sigma error 
to be the value
where 68\% of the distributions fall within.  All errors in the numbers
of this section and in the plots shown are
calculated in this way.

At this point, we would like to discuss where
local duality sets in: $s_0$.
As we can see from Fig. \ref{Fig1} for $s < M_\tau^2$
there are two points where (\ref{WSRS}) are satisfied,
the first one around 1.5 GeV$^2$ and the second around 2.5 GeV$^2$. 
Of course, this does not mean that  local duality
is already settled at these points as the oscillations show.
One can expect however that the violations of local duality are small
at these points. It is also obvious that local duality
will be better when the value of $s_0$ is larger.
The procedure to determine the value of $s_0$ is
repeated for each of the spectral functions generated before
and we use consistently a spectral function together
with {\em its} value of the onset of local duality.

There are several points worth making.
Though for every distribution the first duality point  
in the 1st WSR is very near the corresponding one of the 2nd WSR,
they differ by more than their error.
Numerically, when used in other sum-rules they produce results
outside the naive error.
The second duality point, $s_0\approx 2.5$ GeV$^2$
yields more stable results.
There is no a priori
reason for the value of $s_0$ to be exactly the same for different
sum rules.

Though the change from the 1st WSR
to the second is small, and even smaller if one looks at 
negative moments, when one uses large positive 
moments (the ones we need here), the deviations  
are quite sizable as we will show. This is because positive 
large moments weigh more the higher energy region
and the negative moments essentially use only information
of the low energy region.

Probably in the second duality point, local duality
has not been reached either
but certainly we should be closer to the asymptotic regime.
We therefore take the highest global duality point available,
the solution of Eq. (\ref{WSRS}), around $2.5~$GeV$^2$.
Fortunately, for the physical matrix elements, 
the additional $\log(t/\mu^2)$ in the integrand 
reduces the contribution of the data points near
the real axis for $t$ around $\mu^2$. 
This makes these sum rules much more reliable
than the single moments used in~\cite{DG00,NAR01}.

The second, and highest value with good data, value
of $s_0$ where the WSRs are satisfied
runs roughly between 2.2 GeV$^2$ and 3.0 GeV$^2$.
But not all of these values are equally probable.
If we look at the distribution of the $s_0$ values, there is a clear
peak situated around the value calculated with the central data points
but there are tails towards higher $s_0$. The widths of the peak are
essentially the same as the errors we quote.
The $s_0$ where the second WSR are mainly in the area
\ba
\label{s0}
s_0 &=& (2.53^{+0.13}_{-0.12}) \,    {\rm GeV}^2~\mbox{(ALEPH)} \,,
\quad\quad
s_0 = (2.49^{+0.17}_{-0.13}) \,    {\rm GeV}^2~\mbox{(OPAL)} \,.
\ea
and where the first WSR is satisfied in
\ba
s_0 &=& (2.56^{+0.15}_{-0.14}) \,    {\rm GeV}^2~\mbox{(ALEPH)} \,,
\quad\quad
s_0 = (2.53^{+0.17}_{-0.12}) \,    {\rm GeV}^2~\mbox{(OPAL)} \,.
\ea
These errors have been  obtained as explained above. 
In the analysis below we use all experimental
distributions with their associated value
of $s_0$ and not only those with $s_0$
in the intervals above.

The OPE of the $\Pi_{LR}^T(Q^2)$ was studied using 
the same data \cite{ALEPH} in \cite{DHGS98}.
They obtained a quite precise determination of the
dimension six and eight higher dimensional
operators from a fit to different moments of the energy distribution.
This procedure has in principle smaller errors since one
can use the tau decay kinematic factors which suppresses the data near
the real axis but has a different local duality error.
They use $M_\tau^2$ as upper limit of the hadronic moments, 
we agree with \cite{PPR01} that one should 
use the $s_0$ where there is global duality with QCD
to eliminate possible effects of the lack of local duality 
at $M_\tau^2$. 
Another comment is that as noticed in \cite{KPR01}
the $\alpha_S^2$ corrections used in \cite{DHGS98} are
in a  different scheme \cite{LSC86}. These corrections
in the scheme used in \cite{BBL96}
are presented in the appendices.

\begin{figure}
\includegraphics[angle=-90,width=0.49\textwidth]{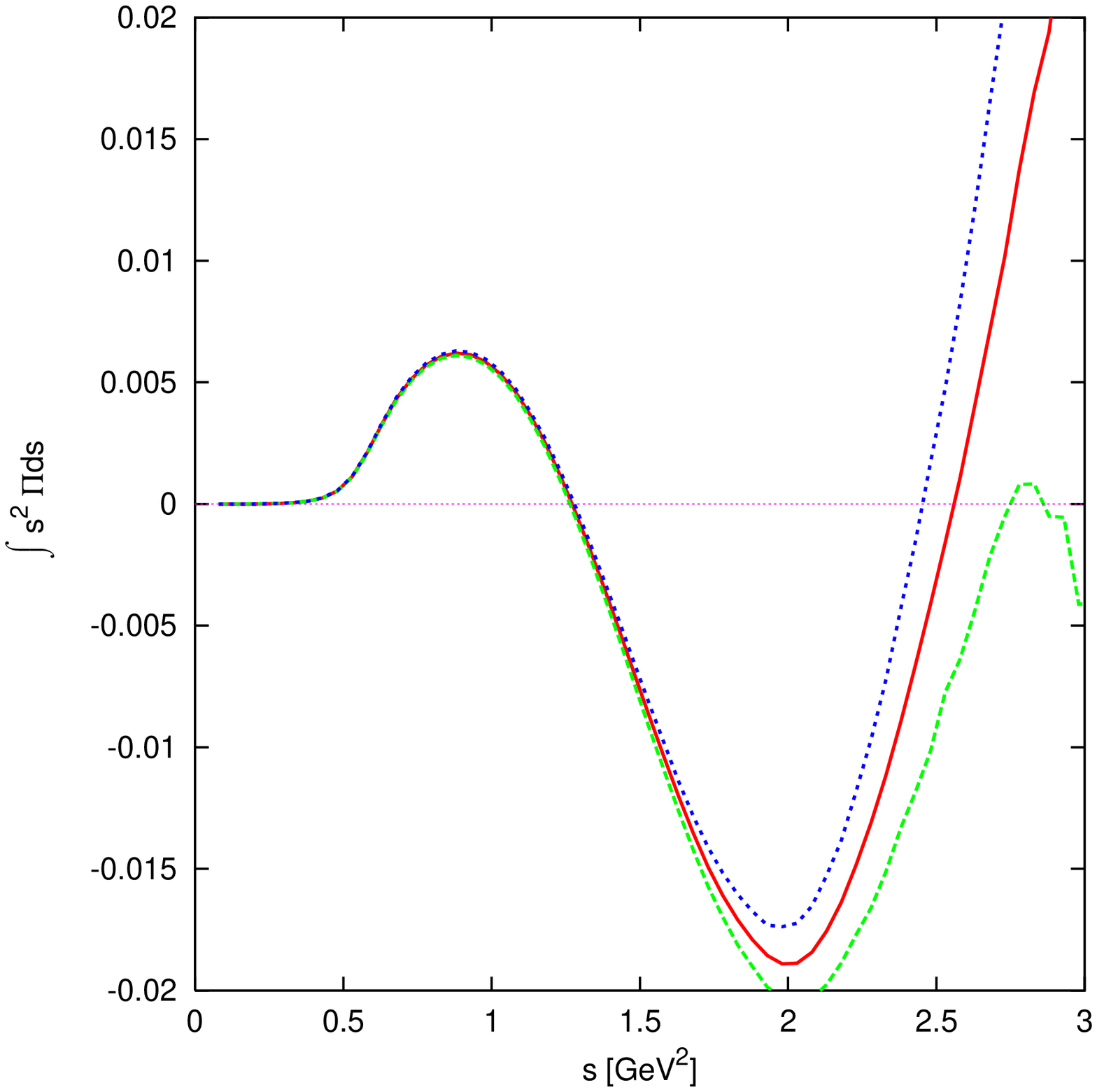}
\includegraphics[angle=-90,width=0.49\textwidth]{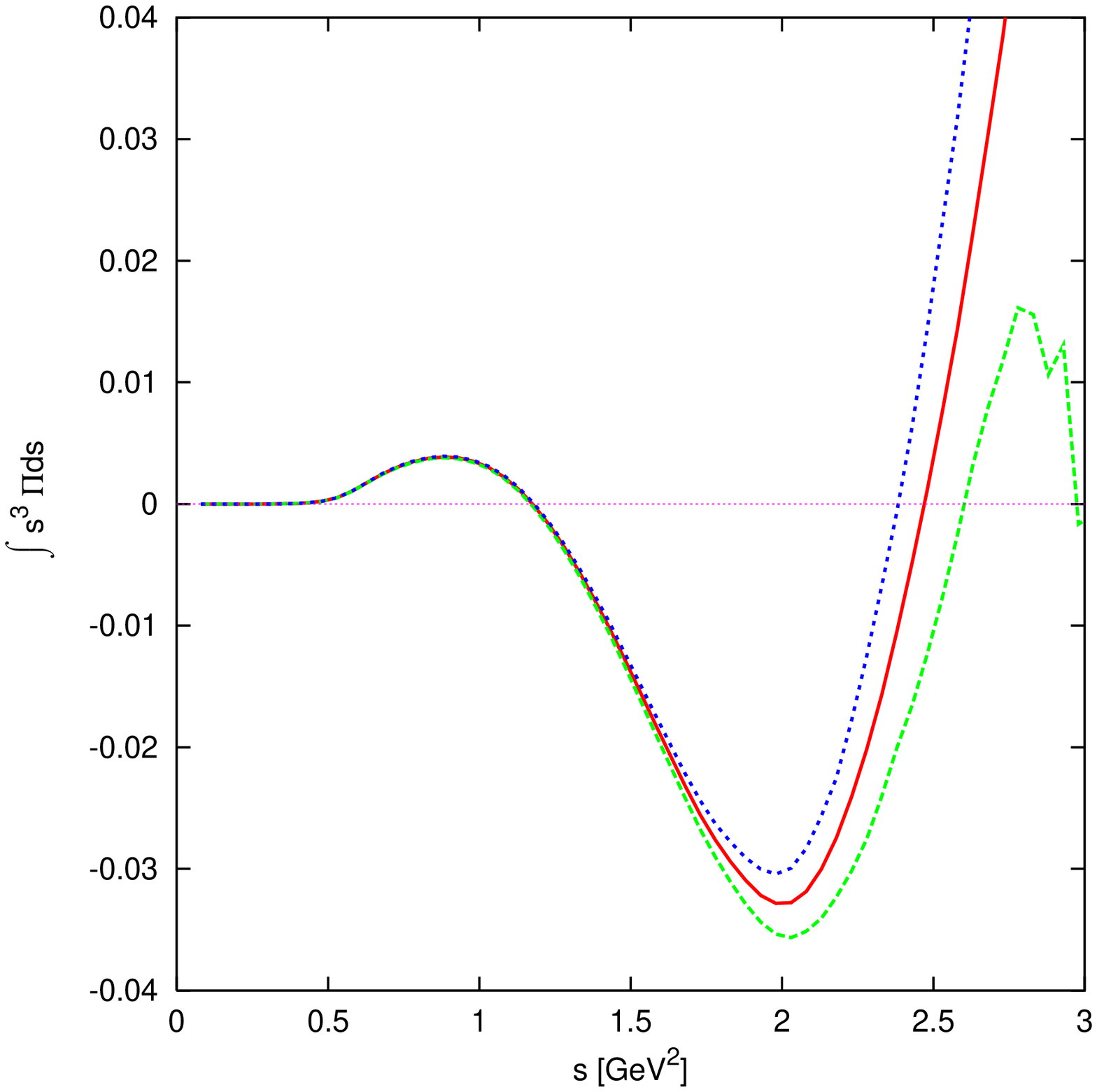}
\caption{\label{fig2} The second and third moment as a function
of the upper limit of integration $s_0$ and the one sigma variation.}
\end{figure}

We can determine the following higher dimensional operator
contributions (\ref{SVZLR}) 
\ba
\label{dimsix}
M_2 &\equiv& \int_0^{s_0} {\rm d} t \, t^2 \, \frac{1}{\pi}  \, 
\im \Pi_{LR}^T(t) \nonumber \\ &=& {\dis \dis \sum_{m=0}^\infty \, 
\sum_{i=1}} \langle 0| O^{(i)}_{2(m+3)}(0) | 0 \rangle(s_0) 
\, (-1)^m \, \frac{1}{2 \pi i} \, \oint_{C_{s_0}} \,{\rm d} s \, 
\frac{C^{(i)}_{2(m+3)}(s_0,-s)}{s^{1+m}} \, \nonumber \\
M_3 &\equiv& \int_0^{s_0} {\rm d} t \, t^3 \, \frac{1}{\pi}  \, 
\im \Pi_{LR}^T(t) \nonumber \\ &=& {\dis \dis \sum_{m=0}^\infty \, 
\sum_{i=1}} \langle 0| O^{(i)}_{2(m+3)}(0) | 0 \rangle(s_0) 
\, (-1)^m \, \frac{1}{2 \pi i} \, \oint_{C_{s_0}} \,{\rm d} s \, 
\frac{C^{(i)}_{2(m+3)}(s_0,-s)}{s^{m}} \, .
\ea
In Fig. \ref{fig2} we have plotted the value of $M_2$ and $M_3$
as a function of $s_0$ used in the integration, together with the one sigma
error band.
It is immediately obvious that the main uncertainty is the choice of
$s_0$ to be used. This uncertainty is increasingly important with the
increase of the moment.
%figure

Using ALEPH data on $V-A$ spectral functions
we get for the dimension six and eight
FESR using the value for $s_0$ where the second WSR is satisfied
\ba
\label{data}
M_2 &=& -(1.7^{+1.2}_{-1.0}) \cdot 10^{-3} \, {\rm GeV}^{6} \nonumber \\
M_3 &=&  (7.2^{+5.2}_{-4.0}) \cdot 10^{-3} \, {\rm GeV}^{8} \, .
\ea
The error bars are obtained by taking 68\% of the generated distributions
within this value, only including those where the WSR can be satisfied.
The error is smaller than one would judge from Fig. \ref{fig2}
since the value of $M_2$ and $M_3$ at the value of $s_0$ where the
spectral function satisfies a WSR is much more stable than the variation
at a fixed value of $s_0$. 

Using the OPAL data we get,
\ba
\label{OPALdata}
M_2 &=& -(2.0^{+1.0}_{-1.0}) \cdot 10^{-3} \, {\rm GeV}^{6} \nonumber \\
M_3 &=& (5.2^{+4.0}_{-3.2})  \cdot 10^{-3} \, {\rm GeV}^{8} \, .
\ea

Eq. (\ref{data}) and (\ref{OPALdata})
should be compared to other model independent determinations of $M_2$ and 
$M_3$ using the ALEPH and OPAL data. 
In \cite{ALEPH,OPAL,DHGS98} the results obtained were 
\ba
\label{ALEPH}
M_2 &=& - (3.2 \pm 0.9) \cdot 10^{-3} \, {\rm GeV}^{6} \nonumber \\
M_3 &=& - (4.4 \pm 1.2) \cdot 10^{-3} \, {\rm GeV}^{8} \, .
\ea
The value for $M_2$ is compatible within errors 
but $M_3$ differs even in sign.
Our error bars take into account the variation of $s_0$ 
but our result for $M_3$ at the second duality point is always positive.
 
This indicates a potential problem in the determination 
of $M_3$ and higher moments (and of smaller importance in $M_2$).
As said before violations of local duality can be sizable
for higher moments like $M_3$  even at $t\simeq M_\tau^2$
used  as upper limit of the moment integrals as done in these references.
Other source of discrepancy here is the fact that these three analysis assumed 
implicitly that contributions with dimension $d>8$ were negligible in all 
cases, while in our results we include the effect of all higher order 
dimension operators. The value of the moments $M_2$ and $M_3$ obtained in 
\cite{ALEPH,OPAL,DHGS98} must contain higher dimension contamination 
that can account for the differences between our result and those in 
(\ref{ALEPH}) \cite{CDGM02}.

In the most recent analysis \cite{CDGM02}, in which several Finite Energy Sum 
Rules were studied, the effect of higher dimension 
operators do was taken into account. The results they got were
\ba
\label{CGMALEPH}
M_2 &=& - (2.27 \pm 0.42 \pm 0.09) \cdot 10^{-3} \, {\rm GeV}^{6} \, ,
\nonumber \\
M_3 &=&  (2.85 \pm 1.86 \pm 0.32) \cdot 10^{-3} \, {\rm GeV}^{8} \, 
\ea
with the ALEPH data and 
\ba
\label{CGMOPAL}
M_2 &=& - (2.53 \pm 0.45 \pm 0.06) \cdot 10^{-3} \, {\rm GeV}^{6} \, ,
\nonumber \\
M_3 &=&  (1.56 \pm 1.91 \pm 0.23) \cdot 10^{-3} \, {\rm GeV}^{8} \,
\ea
with the OPAL data. The values for $M_2$ and $M_3$ here are in agreement with 
ours in (\ref{data}) and (\ref{OPALdata}) within errors, although the central 
value of $M_3$ calculated by Cirigliano \emph{et al.} is much smaller than 
our. Despite the fact of this quantitative difference, in \cite{CDGM02} was 
confirmed the sign of $M_3$, first obtained in \cite{BGP01}.

The integrals  which are needed for Eq. (\ref{imge}) can 
 be evaluated from the ALEPH data in the same way. We need
\ba
\label{numALEPH}
{\cal A}_{LR}(\mu_R) &\equiv& 
{\dis \int^{s_0}_0} \,
{\rm d} t \, t^2 \ln \left(\frac{t}{\mu_R^2} \right)\,  \frac{1}{\pi}
\, \im \, \Pi_{LR}^T(t) = 
(4.7_{-0.4}^{+0.5})\cdot10^{-3}~{\rm GeV}^6 \, ; 
\nonumber\\
 {\cal A}^{\rm Lower}_{LR}(\mu_R) &\equiv& 
- {\dis \int^{s_0}_0} \,
{\rm d} t \, t^2 \ln \left(1+\frac{\mu_R^2}{t} \right)\,  \frac{1}{\pi}
\, \im \, \Pi_{LR}^T(t) = 
(3.7_{-0.4}^{+0.5})\cdot10^{-3}~{\rm GeV}^6 \, ; 
\nonumber\\
 {\cal A}^{\rm Higher}_{LR}(\mu_R) &\equiv& 
{\dis \int^{s_0}_0} \,
{\rm d} t \, t^2 \ln \left(1+\frac{t}{\mu_R^2} \right)\,  \frac{1}{\pi}
\, \im \, \Pi_{LR}^T(t) =
(1.0_{-0.7}^{+0.9}) \cdot10^{-3}~{\rm GeV}^6
\ea
\begin{figure}
\begin{center}
\includegraphics[angle=-90,width=0.49\textwidth]{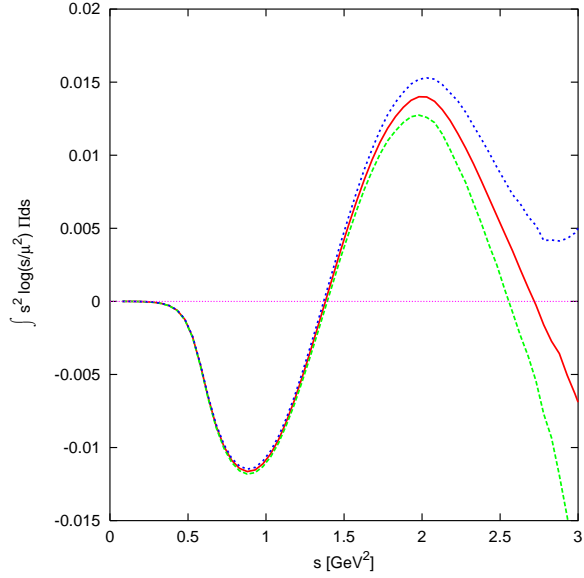}
\end{center}
\caption{\label{figalr} The integral over
the spectral function needed for $\im G_E$.}
\end{figure}%figure
at  $\mu_R$ = 2 GeV and using for each distribution its second duality
point $s_0$. Notice  the much smaller error of 
${\cal A}_{LR}$ and  ${\cal A}^{\rm Lower}_{LR}$ 
when compared with $M_2$ and $M_3$.
These values are all taken at the second duality point $s_0$ where the
second WSR is satisfied. 
We plot ${\cal A}_{LR}$
as a function of $s_0$ in Fig. \ref{figalr}.
The OPAL data give instead
\ba
{\cal A}_{LR}(\mu_R) &=& 
(4.4_{-0.3}^{+0.4})\cdot10^{-3}~{\rm GeV}^6 \, ; 
\nonumber\\
 {\cal A}^{\rm Lower}_{LR}(\mu_R) &=&
(3.8_{-0.5}^{+0.4})\cdot10^{-3}~{\rm GeV}^6 \, ; 
\nonumber\\
 {\cal A}^{\rm Higher}_{LR}(\mu_R) &=& 
(0.6_{-0.6}^{+0.8}) \cdot10^{-3}~{\rm GeV}^6
\nonumber\\
\ea

As a test, we can also calculate the electromagnetic 
pion mass difference in the chiral limit \cite{Das67}, 
\ba
\label{EMpion}
{\cal B}_{LR}&=&{\dis \int^{s_0}_0} \,
{\rm d} t \, t \,  \ln \left(\frac{t}{\mu_R^2} \right)\,  \frac{1}{\pi}
\, \im \, \Pi_{LR}^T(t) 
=\frac{4 \pi F_0^2}{3 \alpha_{QED}} \, 
\left(m_{\pi^0}^2- m_{\pi^+}^2\right) 
\nonumber \\ 
&=&  - (5.2 \pm 0.5) \cdot10^{-3}~{\rm GeV}^4 ~\mbox{(ALEPH)}\,; 
\nonumber\\
&=& - (5.2 \pm 0.6) \cdot10^{-3}~{\rm GeV}^4 ~\mbox{(OPAL)}\,; 
\ea
where we also used the value of $s_0$ given by the 2nd WSR.
Notice that ${\cal B}_{LR}$ does not depend on $\mu_R$
due to the second WSR (\ref{WSRS}).
The experimental number is 
\ba
\frac{4 \pi F_0^2}{3 \alpha_{QED}} \, 
\left( m_{\pi^0}^2- m_{\pi^+}^2 \right)_{\rm E.M.}
&=&
- (5.15 \pm 0.90) \cdot 10^{-3}\, {\rm GeV}^4 \, . 
\ea
where we used $F_0= (87 \pm 6)$ MeV as the chiral limit value
of the pion decay constant and 
removed the QCD contributions  \cite{ABT3}.

For comparison we quote the central values using as $s_0$ the 
second duality  point where the first WSR is satisfied
\ba
\begin{array}{rclrcl}
s_0 &=& (2.56^{+0.15}_{-0.14})~\mbox{GeV}^2\,,
&
{\cal A}_{LR}(2 {\rm GeV}) &=& (4.0_{-0.7}^{+0.6}) \cdot 10^{-3}
\, {\rm GeV}^{6} \,,\\
M_2 &=& -(0.1_{-2.8}^{+3.0}) \cdot 10^{-3}\, {\rm GeV}^{6} , 
&
 {\cal A}^{Lower}_{LR}(2 {\rm GeV}) &=& 
(2.2 \pm 2.1) \cdot 10^{-3}\, {\rm GeV}^{6} ,
\\
M_3 &=& (11_{-7}^{+9}) \cdot 10^{-3}\, {\rm GeV}^{8}\,,
&
{\cal A}^{Higher}_{LR}(2 {\rm GeV}) &=& 
 (1.8_{-1.6}^{+0.5})\cdot 10^{-3}\, {\rm GeV}^{6}.
\end{array}
\ea
for ALEPH and for OPAL
\ba
\begin{array}{rclrcl}
s_0 &=& (2.53^{+0.17}_{-0.12})~\mbox{GeV}^2\,,
&
{\cal A}_{LR}(2 {\rm GeV}) &=& (3.4_{-0.8}^{+0.7}) \cdot 10^{-3}
\, {\rm GeV}^{6} \,,\\
M_2 &=& (0.1_{-2.3}^{+2.8}) \cdot 10^{-3}\, {\rm GeV}^{6} , 
&
 {\cal A}^{Lower}_{LR}(2 {\rm GeV}) &=& 
(1.7^{+1.4}_{-1.3}) \cdot 10^{-3}\, {\rm GeV}^{6} ,
\\
M_3 &=& (10_{-6}^{+9}) \cdot 10^{-3}\, {\rm GeV}^{8}\,,
&
{\cal A}^{Higher}_{LR}(2 {\rm GeV}) &=& 
 (1.7_{-1.3}^{+1.8})\cdot 10^{-3}\, {\rm GeV}^{6}.
\end{array}
\ea

The errors are larger here. The value of $s_0$ where 
the first WSR is satisfied varies more and is somewhat larger
than the $s_0$ where the second WSR
is satisfied, this makes the last results more dependent on the
spectral function at high $t$ which have large errors.

If one tried to see the results using the first duality point, 
where less duality with QCD is expected, we get
that using the one from the 2nd WSR
\ba
\label{data1st}
\begin{array}{rclrcl}
s_0 &=& (1.47\pm 0.02)\,   {\rm GeV}^2 \;, &
{\cal A}_{LR}(2 {\rm GeV}) &=& 
(3.3 \pm 0.1) \cdot 10^{-3}\, {\rm GeV}^{6} ,
\\
M_2 &=& -(6.6\pm0.2) \cdot 10^{-3}\, {\rm GeV}^{6} , 
 &
  {\cal A}^{Lower}_{LR}(2 {\rm GeV}) 
&=&  (5.9 \pm 0.2) \cdot 10^{-3}\, {\rm GeV}^{6} ,\quad
\\ 
M_3 &=& -(12_{-2}^{+1}) \cdot 10^{-3}\, {\rm GeV}^{8} , & 
{\cal A}^{Higher}_{LR}(2 {\rm GeV})  
&=& -(2.6\pm 0.1)\, {\rm GeV}^{6}.
\end{array}
\ea
Notice that $M_2$ is not compatible with (\ref{data}) 
 with the central values
differing by more than twice the error.
The moment $M_3$ changes even sign with respect to
the second duality point showing the
problems of local duality violations for larger moments
more dramatically.
As argued before one should take the largest value of $s_0$ to ensure
better local duality.
However, the physical relevant moment ${\cal A}_{LR}$ 
is much more stable  with $s_0$.

\section{The Scalar--Pseudo-Scalar Two-Point Function $\,$ 
$\,\Pi_{SS+PP}^{(0-3)}(Q^2)\,$}
\label{SSPP}

In this section we discuss some of the knowledge of the spectral
function $\im \Pi_{SS+PP}^{(0-3)}(t)$ which governs
 the connected contribution to the matrix element of $Q_8$.
In the large $N_c$ limit there is no difference between the singlet and
triplet channel so the integral in (\ref{SPintegral}) is $1/N_c$ 
suppressed  and its contribution to $\im G_E$ is NNLO.
But in the scalar-pseudo-scalar sector, violations of the large $N_c$
behaviour can be larger than in the vector-axial-vector channel. 
It is therefore interesting to determine the size of this 
contribution as well.

After adding the short-distance part to the long-distance part,
the relevant integral is (\ref{imge})
\be
\label{SPintegral}
\, \frac{1}{48 \pi^2} 
\frac{1}{\langle 0 | \bar q q | 0 \rangle ^2(\mu_C)} \, 
\int_0^{\tilde s_0} 
{\rm d} t \, t \, \ln \left(\frac{t}{\mu^2} \right) 
\, \frac{1}{\pi}  \,  \im \Pi_{SS+PP}^{(0-3)}(t)  \, , 
\ee
This is the contribution
of the connected part relative to the disconnected one.
$\tilde s_0$ is the scale where in
this channel QCD duality sets in. The scale $\mu$ is the cut-off
scale. The dependence on this scale being NNLO in $1/N_c$ {\em cannot}
match the present NLO order Wilson coefficients. 

\begin{figure}
\includegraphics[angle=-90,width=0.49\textwidth]{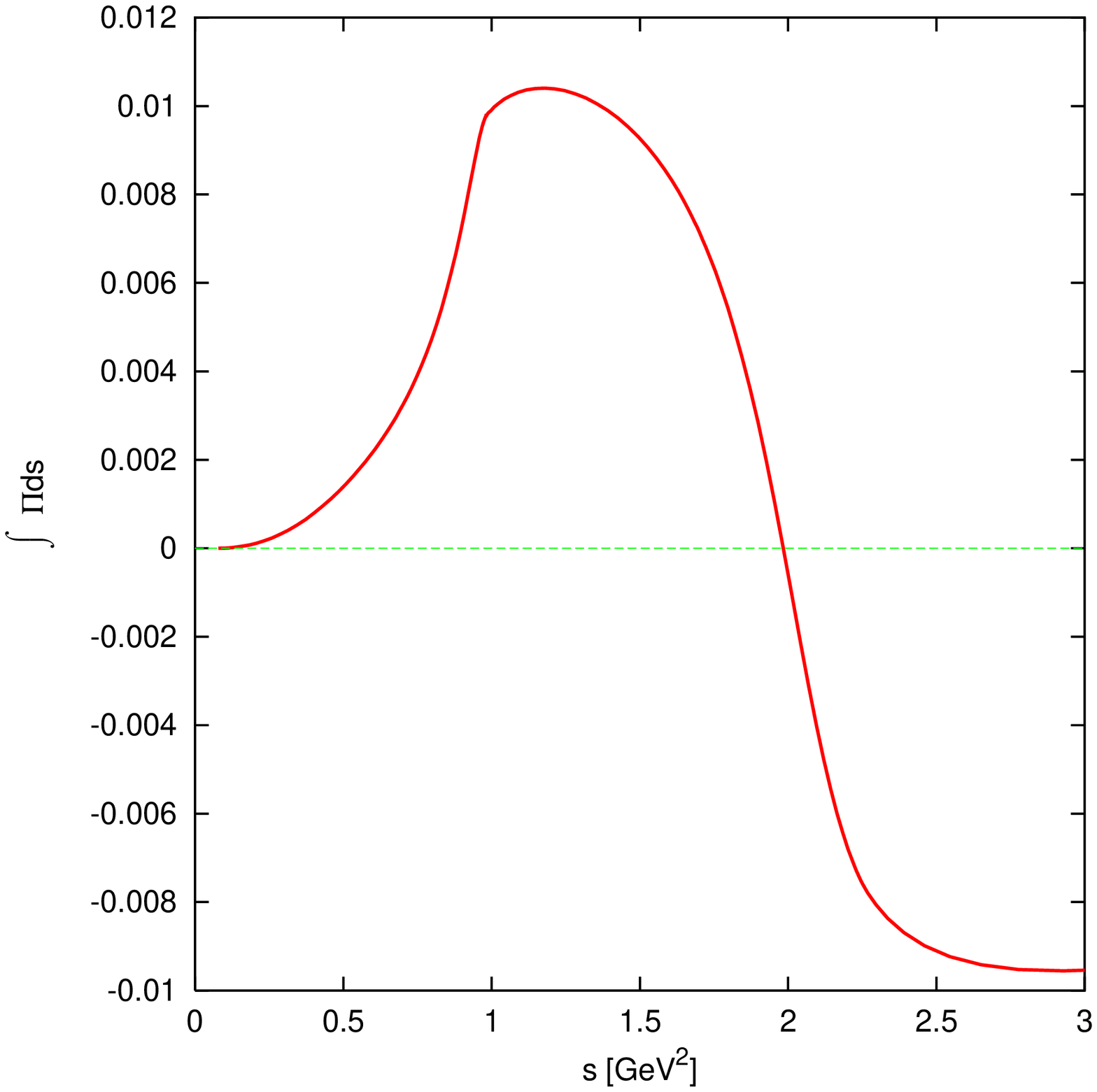}
\includegraphics[angle=-90,width=0.49\textwidth]{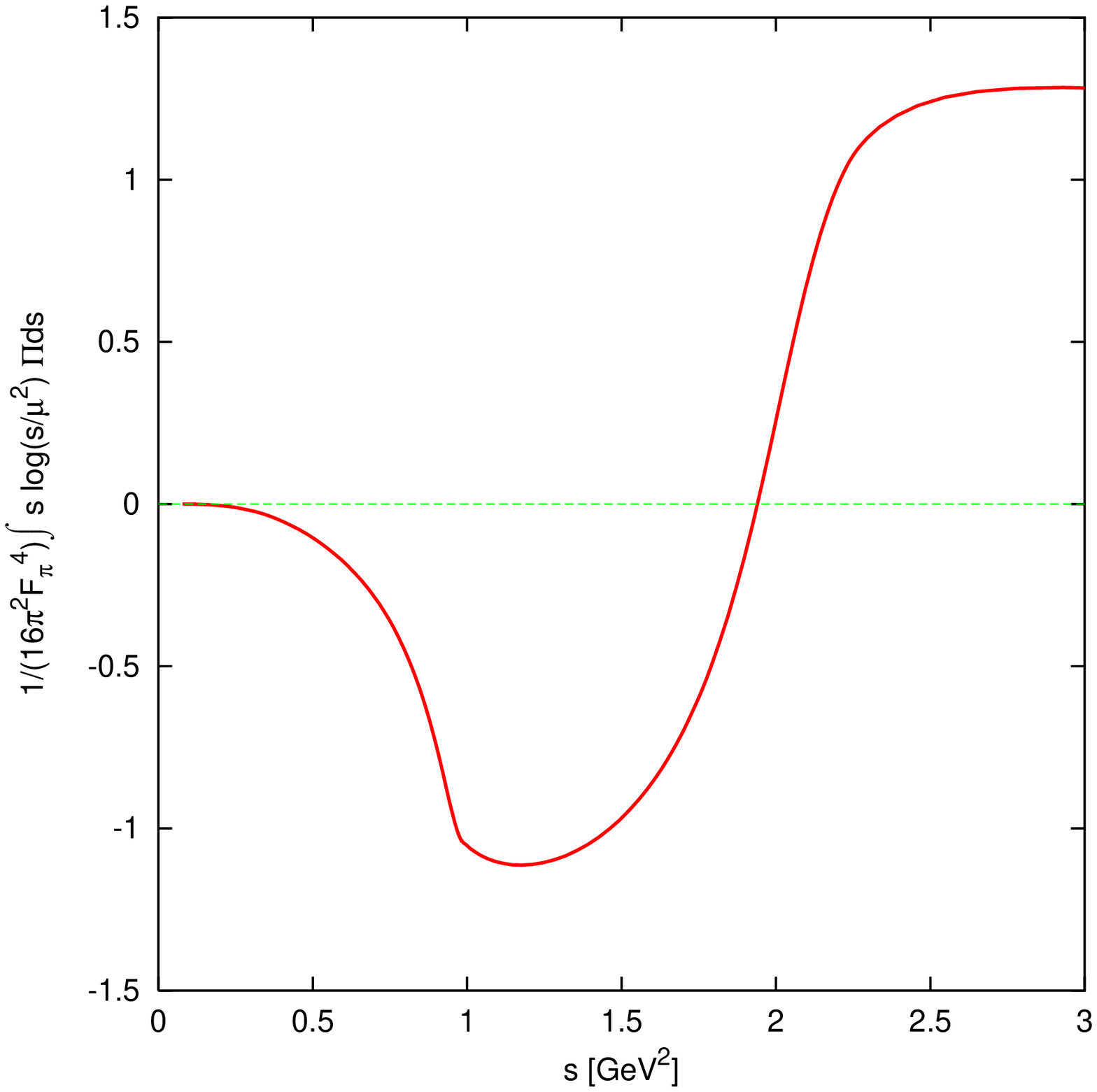}
\caption{\label{fig3} The sum rule (\ref{weinSP2})
as a function of $\tilde s_0$ for the scalar part
and the relative correction to the
disconnected contribution for $Q_8$ using the same parametrization
as a function of $\tilde s_0$ for $\mu_R=\mu=$ 2 GeV.
Notice that the correction is
small in the region where the sum rule is satisfied.}
\end{figure}

We can use models like the ones in  \cite{MOU00} 
to evaluate the scalar part of the
integrals. We only use the model there using the KLM \cite{KLM} analysis,
since only it ratifies
(\ref{weinSP2}) at a reasonable value of $\tilde s_0$ and, in addition, 
produce values for $L_4$, $L_6$ and $L_8$ compatible with phenomenology. 
In Fig. \ref{fig3} we plotted for that parameterization the sum rule
and the relative correction from the scalar part to the disconnected
contribution $3 \langle 0 | \overline q q | 0 \rangle^2(\mu_R)$
for $\mu=\mu_R=2$ GeV.
%figure
The value of the scalar part of Eq. (\ref{SPintegral}) is about 0.18 at
$\tilde s_0 = (1.41$ GeV)$^2$.

In the large $N_c$ limit, a sensible alternative estimate 
is to use meson pole dominance.
In the pseudo-scalar sector, the U(3)$\times$U(3) symmetry
is  broken by  the chiral anomaly splitting the singlet $\eta_1$ mass
away from the zero mass for the Goldstone boson octet.

Three meson intermediate states are not studied enough to be included 
at this level, we include instead the first $\pi'$ resonance.
This means including a massless Goldstone boson 
plus the first $\pi'$ resonance
and the singlet $\eta_1$. The pseudo-scalar sum rule 
in (\ref{weinSP2}) requires the
following relation between the octet and the singlet 
couplings to the pseudo-scalar current for 
$\tilde s_0 \simeq 2.0$ GeV$^2$,
\be
F_0^2 + F_{\pi'}^2  =  F_{\eta_1}^2 \, .
\ee
Phenomenologically $F_{\pi'}^2/F_0^2<<1$ \cite{DR87}
and $F_0 \simeq F_{\eta_1}$.

We can introduce a scalar meson octet $S_8$ and a singlet $S_1$
using the methods of \cite{Eckeretal}.
The coupling constant for the octet can be denoted by 
$c_m$ and has been
estimated in \cite{Eckeretal,ENJL} to be about $(43\pm 14)$ MeV.
In fact, the sum rule (\ref{weinSP1}) is a property of QCD
and relates in this approximation $c_m$ to $F_0$
\be
c_m^2 = \frac{1}{8} \left[ F_0^2 + F_{\pi'}^2 + \cdots \right]
= \frac{F_{\eta_1}^2}{8} \simeq \frac{F_0^2}{8} 
\ee
which numerically agrees quite well with the phenomenological 
estimate.

The scalar sum rule in (\ref{weinSP2}) requires 
the singlet and the octet 
components to have the same coupling leading to
a relative correction from the scalar integral to the disconnected
contribution of
\be
\frac{1}{12\pi^2} \frac{F_{\eta_1}^2}{F_0^4} \, 
\left[M_{S_1}^2 \ln\left(\frac{M_{S_1}}{\mu}\right)
-M_{S_8}^2\ln\left(\frac{M_{S_8}}{\mu}\right)\right].
\ee
using both sum rules and the lowest meson dominance approximation.

The contribution from the pseudo-scalar 
connected two-point function  relative the disconnected
contribution can then be evaluated to
\ba
\frac{1}{12\pi^2} \frac{F_{\eta_1}^2}{F_0^4} \,
\left[ 
 M_{\eta_1}^2 \, \ln\left(\frac{M_{\eta_1}}{\mu}\right)\, 
-\frac{F_{\pi'}^2 }{F_{\eta_1}^2} \, M_{\pi'}^2
\ln\left(\frac{M_{\pi'}}{\mu}\right) 
\right] \, .
\ea
The contribution of the $\pi'$ is negligible. Using a Breit-Wigner shape for 
the $\pi'$ contribution do not change much the result due to the small 
coupling of the $\pi'$.

As said before the scale $\mu$ is free and cannot
be at present matched with OPE QCD since it is a NNLO order in $1/N_c$
effect.
The scale independence is  reached when the sum rule
\ba
\label{MTILDE1}
\tilde M_1 &=& 
M_{S_8}^2-M_{S_1}^2-M_{\eta_1}^2+\frac{F_{\pi'}^2}{F_{\eta_1}^2}
M_{\pi'}^2 =0
\ea
which is $O(N_c^2 \alpha^2)$ (\ref{M1sumrule}) is fulfilled. 
This sum rule is very well satisfied in the linear $\sigma$ model,
see e.g. \cite{MOU00}. 
%Being $O(\alpha^2)$, it is also
%well satisfied for the values of $Q_7$ and $Q_8$ operators
%in the literature \cite{DG00,NAR01,KPR01,CDGM01,DGGM99}.

The masses $M_{\eta_1}\simeq 0.86$ GeV (chiral limit value)
$M_{\pi'}\simeq 1.3$ GeV are known.
The masses of the singlet and octet of scalars are not so well known.
Using  $M_{S_1} = M_{\sigma}\simeq 0.5$ GeV and 
$M_{S_8} = M_{a_0(980)}
\simeq 0.98$ GeV the correction to the disconnected contribution
is almost independent of the scale $\mu$, neglecting the $\pi^\prime$
and is independent of $\mu$ for $F_{\pi^\prime}^2/F_0^2 = 0.017$.

The scalar form factor in \cite{MOU00} is obtained from data and dispersion 
relations up to  1~GeV and Breit-Wigner shapes above. The result of using 
these models agreed with the results of naive narrow widths for the lowest 
scalar resonances. These were constructed to fulfill the short-distance QCD 
constraints and also produces reasonable values for $L_6$ and $L_8$. Here, we 
have also tried Breit-Wigner shapes for the scalar mesons instead of narrow 
widths and again find results in the same ball park. Now, we also have used 
the scalar form factors obtained in \cite{MO01} where the lowest scalar 
triplet and singlet resonances are generated dynamically for energies up to 
1~GeV and Breit-Wigner shapes above and we get similar results.

In all the estimates, we got negative corrections to the disconnected 
part in the region between -10\% to -30\%. Though the connected 
scalar-pseudoscalar contribution in (\ref{SPintegral}) cannot be used at a 
quantitative level at present, the results above indicate that is difficult to 
have corrections larger than $\pm$30\%  to the disconnected part.

Of course, the scale dependence
left in (\ref{SPintegral}) is unsatisfactory in principle 
but small since the sum rule (\ref{MTILDE1}) is quite well satisfied.

\section{Numerical Results for the Matrix-Elements and Bag Parameters}

The vacuum expectation value in the chiral limit of $Q_7$ itself
is related directly to $B_{7 \chi}$.  This allows us 
to obtain\footnote{The
analytical formulas  are  in agreement with \cite{DG00,NAR01,KPR01}
for the  scheme dependent terms in $Q_7$ matrix elements but 
not for the $Q_8$ ones in \cite{DG00,NAR01} and they were 
not included  in \cite{KPR01}.}
\ba
\label{numresult}
\langle 0 | O_6^{(1)}(0) | 0 \rangle^{NDR} (\mu_R) &=&
- \left(1+\frac{1}{24}  a(\mu_R) \right) \, \frac{3}{32\pi^2} 
{\cal A}_{LR}(\mu_R) \nonumber \\
&+& \frac{1}{48}\, a(\mu_R) \, 
\left[ 3  \langle 0 | \overline q q | 0 \rangle^2  (\mu_R)
+ \frac{1}{16 \pi^2} 
{\cal A}_{SP}(\mu_R,\mu_C) \right] \, ;
\ea
\ba
\label{numresultHV}
\langle 0 | O_6^{(1)}(0) | 0 \rangle^{HV} (\mu_R) &=&
-\left(1- \frac{11}{24} a(\mu_R)  \right)
\frac{3}{32\pi^2} \, {\cal A}_{LR}(\mu_R) \nonumber \\
&-& \frac{11}{48}\, a(\mu_R) \, 
\left[ 3 \langle 0 | \overline q q | 0 \rangle^2  (\mu_R) \, 
 + \frac{1}{16 \pi^2} 
{\cal A}_{SP}(\mu_R,\mu_C) \right] \, .
\ea

For the numerics, 
we use the value of the condensate obtained in the $\overline{MS}$
scheme in \cite{BPR}, 
\ba
\label{factor6}
\langle 0 | \overline q q | 0 \rangle (2 {\rm GeV})&=&
- (0.018\pm 0.004) \, {\rm GeV}^3 \,,
\ea
the numerical results of Eq. (\ref{numALEPH}),
\be
a(2~GeV) = 0.102
\ee
and neglect, in first approximation, 
the integral over the scalar--pseudo-scalar two-point function.

The weighted average of the first and second WSR results
for ${\cal A}_{LR}(2 {\rm GeV})$ from ALEPH data is
\be
{\cal A}^{\rm ALEPH}_{LR}(2{\rm GeV})
= (4.5\pm0.5) \cdot10^{-3}~{\rm GeV}^6  
\ee
 and from OPAL data
\be
{\cal A}^{\rm OPAL}_{LR}(2{\rm GeV})
= (4.2\pm0.4) \cdot10^{-3}~{\rm GeV}^6  \, . 
\ee
Though the systematic errors aver very correlated, since the
central values are very similar we take the simple average of
 both results as our result 
\be
{\cal A}_{LR}(2{\rm GeV})
= (4.35\pm0.50) \cdot10^{-3}~{\rm GeV}^6  \,  
\ee
and obtain
\ba
\label{numO1}
\lefteqn{\langle 0 | O_6^{(1)}(0) | 0 \rangle^{NDR} ( 2 {\rm GeV}) =
-(4.0\pm0.5)\cdot10^{-5}~{\rm GeV}^6}&&
\nonumber\\
 &=&
\left(-(4.2\pm0.5)+(0.2\pm0.1)\right)
\cdot10^{-5}~{\rm GeV}^6 
\nonumber\\
&=&
\left((-3.3\pm0.5)+(-0.9\pm0.8)+(0.2\pm0.1)\right)
\cdot10^{-5}~{\rm GeV}^6
\ea
and
\ba
\label{numO2}
\lefteqn{\langle 0 | O_6^{(1)}(0) | 0 \rangle^{HV} ( 2 {\rm GeV}) =
-(6.2\pm1.0)\cdot10^{-5}~{\rm GeV}^6}&&
\nonumber\\
 &=&
\left((-3.9\pm0.5)-(2.3\pm0.9)\right)
\cdot10^{-5}~{\rm GeV}^6 
\nonumber\\
&=&
\left((-3.1\pm0.5)+(-0.8\pm0.7)-(2.3\pm0.9)\right)
\cdot10^{-5}~{\rm GeV}^6
\ea
where we quote, namely, the total result, the integral and the vacuum 
expectation value
separately and in the last case also the long and 
short-distance part of the  integral separately. 

The short-distance part of the integral,
the second term in the above, is the contribution of
all higher dimensional operators. We find that its contribution
is between a few \% up to 35 \% depending on the value of $\mu$.
At $\mu=2$ GeV it is somewhat larger than the error
on the integral cut-off at $\mu$.

Similarly, the matrix-element of $Q_8$ is directly related
to $B_8$ and we obtain\footnote{We disagree in this case with the results
in \cite{DG00,NAR01,KPR01} because of the scheme dependent
terms. These references also disagree with each other.}
\ba
\label{O6operator}
 \langle 0 |  O_6^{(2)}(0) | 0 \rangle^{NDR} ( 2 {\rm GeV} ) 
&=& \left[ 1+ \frac{23}{12} \, a(2 {\rm GeV}) \right] \times
\nonumber\\ && \times 
\left[ 3 \, \langle 0 | \overline q q | 0 \rangle^2  (2 {\rm GeV})
+ \frac{1}{16 \pi^2} {\cal A}_{SP}(2 \rm{GeV},2\rm{GeV}) \right]\, 
\nonumber \\ 
&-&   \frac{27}{64\pi^2} \, a(2 {\rm GeV})  \, 
{\cal A}_{LR}( 2 {\rm GeV}) \, .
\ea
\ba
 \langle 0 |  O_6^{(2)}(0) | 0 \rangle^{HV} ( 2 {\rm GeV} ) 
&=& 
\left[ 1+ \frac{35}{12} \, a(2 {\rm GeV}) \right] \times
\nonumber\\&&\times
\left[ 3 \, \langle 0 | \overline q q | 0 \rangle^2  (2 {\rm GeV})
+ \frac{1}{16 \pi^2} {\cal A}_{SP}(2 \rm{GeV},2\rm{GeV}) \right]\, 
\nonumber \\  &+&   
\frac{9}{64\pi^2} \, a(2 {\rm GeV})  \,  {\cal A}_{LR}( 2 {\rm GeV})
\,  .
\ea

Using the same input as above we obtain
\ba
 \langle 0 |  O_6^{(2)}(0) | 0 \rangle^{NDR} ( 2 {\rm GeV} ) 
&=& (1.2 \pm 0.5 ) \cdot 10^{-3} \, {\rm GeV}^6 \,, \nonumber \\
 \langle 0 |  O_6^{(2)}(0) | 0 \rangle^{HV} ( 2 {\rm GeV} ) 
&=& (1.3 \pm 0.6 ) \cdot 10^{-3} \, {\rm GeV}^6 \,, 
\ea
where the contribution of the integral over $\im \Pi_{LR}^T$
is at the 1\% level and thus totally negligible.

Another combination of these two matrix-elements can also
be obtained from an integral over the ALEPH data\cite{NLOWilscoef,BBL96}
by  putting  (\ref{D6_2}) and (\ref{SVZLR}) in (\ref{dimsix}) 
including also the $\alpha_S$ correction of the appendix
\footnote{We thank Vincenzo Cirigliano, John Donoghue,
Gene Golowich, Marc Knecht, Kim Maltman, Santi Peris, and
Eduardo de Rafael for pointing out an error in the matching 
coefficients in the previous version of our paper.
Our result agrees with the result found in \cite{CDGM01}}:
\ba
\label{sumrule6}
M_2 &=&\int_0^{s_0} {\rm d} t \, t^2 \, \frac{1}{\pi}  \, 
\im \Pi_{LR}^T(t) = {\dis \sum_{i=1}} \, C^{(i)}_{6}(s_0,s_0) \, 
O^{(i)}_6(s_0) \nonumber \\ 
&=& - \frac{4 \pi^2}{3} 
\, a(s_0) \left[ 2 \left(1 +  \frac{13}{8} \, a(s_0) \right) \, 
\langle 0 |  O_6^{(1)}(0) | 0 \rangle^{NDR} (s_0) 
\right. \nonumber \\
&+& \left.  \, \left(1+ \frac{25}{8} a(s_0) \right) 
\, \langle 0 | O_6^{(2)}(0) | 0 \rangle^{NDR} (s_0) \right] \, 
\nonumber \\ 
&=& - \frac{4 \pi^2}{3} 
\, a(s_0) \left[ 2 \left(1 +  \frac{41}{8} \, a(s_0) \right) \, 
\langle 0 |  O_6^{(1)}(0) | 0 \rangle^{HV} (s_0) 
\right. \nonumber \\
&+& \left.  \, \left(1+ \frac{21}{8} a(s_0) \right) 
\, \langle 0 | O_6^{(2)}(0) | 0 \rangle^{HV} (s_0) \right] \, .
\nonumber \\
&=& - 4 \pi^2 a(s_0) \left[ \left(1+\frac{61}{12} a(s_0) \right)
\, \left[ \langle 0 | \overline q q | 0 \rangle^2  (s_0)
+ \frac{1}{48 \pi^2} {\cal A}_{SP}(s_0,s_0) \right]\,\right.
\nonumber \\
&-& \left. \left( 1 +\frac{47}{12} a(s_0) \right) \frac{1}{16 \pi^2}
{\cal A}_{LR}(s_0)\right]
\ea
The right hand-side is physical and we checked that
is independent of the scale $s_0$ and scheme. 
We can therefore evaluate it at $s_0=4$~GeV$^2$.
The contribution from 
$\langle 0 |  O_6^{(1)}(0) | 0 \rangle^{NDR}$ $(s_0)$
is numerically very small and we obtain 
\ba
M_2 = -(2.0\pm 0.9)\cdot10^{-3}~\mbox{GeV}^6\,, 
\ea
perfectly compatible within errors both 
with the result obtained from
the data in Eq. (\ref{data}) and
with the result (\ref{ALEPH}). This confirms our
results on the size of the 
integral over $\im \Pi_{SS+PP}(t)$, which can therefore
be considered negligible within the present accuracy
of the disconnected contribution and $M_2$.

There is another sum rule which combines the two
matrix elements, 
\ba
\tilde M_1 &=& 
\int_0^{\tilde s_0} {\rm d} t \, t \, \frac{1}{\pi}  \, 
\im \Pi_{SS+PP}(t) = - 
{\dis \sum_{i=1}} \, \tilde C^{(i)}_{6}(s_0,s_0) \, 
  O^{(i)}_6(s_0) \nonumber \\ 
&=& - \frac{\pi^2}{4} \, a(s_0)^2 \, \left[ 
211 \, \langle 0 |  O_6^{(2)}(0) | 0 \rangle  (\tilde s_0) 
+  90 \, 
\langle 0 |  O_6^{(1)}(0) | 0 \rangle  (\tilde s_0) \right] 
+ O(a^3) \, . \nonumber \\
\ea
For the calculation of the coefficients see Appendix \ref{AppB}.
This sum rule is much less accurate than $M_2$
since the leading terms are  $\alpha_S^2$
and the value of $\tilde M_1$ is not known directly either. 
Therefore we don't use it.

The numerical estimates of the disconnected part, ${\cal A}_{SP}$,
given above change these numbers somewhat but within the 
errors quoted.

These results can also be expressed in terms of the bag parameters:
\ba
\label{numbag}
B_{7 \chi}^{NDR}(2 {\rm GeV})&=& 0.75 \pm 0.20 \; ;\quad
B_{7 \chi}^{HV}(2 {\rm GeV})= 1.15 \pm 0.30\;  \nonumber \\
B_{8 \chi}^{NDR}(2 {\rm GeV})&=& 1.2 \pm 0.3  \; ; \quad
B_{8 \chi}^{HV}(2 {\rm GeV})= 1.3 \pm 0.4\,.
\ea
We can also express it in terms of $\im G_E$:
\ba
F_0^6~\im G_E &=& \im\tau (-2.1\pm0.9)~10^{-6}~\mbox{GeV}^6\,
\ea
which is quite compatible with the estimate in \cite{epsprime}.
 
\section{Comparison with earlier results}
\label{comparison}

To compare with other results
in the literature we propose to use the VEVs
$\langle 0| O_6^{(1)} | 0 \rangle$ and
$\langle 0| O_6^{(2)} | 0 \rangle$. The reason is that
these quantities are what \cite{DG00,NAR01,KPR01} and we directly 
compute.  The matrix elements of $K\to \pi\pi$ through
$Q_7$ and $Q_8$, in the chiral limit,\footnote{See \cite{epsprime} for
the definition of $M_2[Q_7]$ and $M_2[Q_8]$.} can be expressed as follows
\cite{epsprime}
\ba
M_2[Q_7](\mu_R)&=&
\langle (\pi\pi)_{I=2}| Q_{7} | K^0 \rangle (\mu_R) =
-\sqrt{\frac{2 }{3}} \, 
\frac{\langle 0|\overline q q | 0 \rangle^2(\mu_C)}{F_0^3} \, 
\, B_{7\chi}(\mu_C,\mu_R)  \, \nonumber  \\
&=&  2 \sqrt 6 \,\frac{ \langle 0| O_6^{(1)} | 0 \rangle_\chi (\mu_R)}
{F_0^3}
\, ; \nonumber \\
M_2[Q_8](\mu_R)&=&
\langle (\pi\pi)_{I=2}| Q_{8} | K^0 \rangle (\mu_R) = 
-\sqrt 6 \, \frac{\langle 0|\overline q q | 0 \rangle^2(\mu_C)}{F_0^3} \, 
\, B_{8\chi}(\mu_C,\mu_R)  \, \nonumber  \\
&=& - \frac{\sqrt 6}{3} \, \frac{\langle 0| O_6^{(2)} | 0 \rangle_\chi 
(\mu_R)}{F_0^3} \, .
\ea

The different results obtained in the literature using 
analytical and lattice methods for the matrix elements in the 
NDR scheme and at $\mu=2\gev$ are collected in Tables \ref{q7NDR}  
and \ref{q8NDR}
% and \ref{qsHV}.

%The lattice results \cite{DGGM99} are from computed $K \to \pi$ matrix elements
%and use the physical values of $f_K$ and $f_\pi$
%to convert into $K\to \pi \pi$. 
%Since we and  \cite{DG00,NAR01,KPR01} compute
%in the chiral limit, this amounts to a large 
%spurious factor $f_K f_\pi^2/F_0^3 \simeq 1.6$ 
%of difference when comparing $K \to \pi \pi$ matrix elements
%or  $f_K f_\pi /F_0^2 \simeq 1.4$ when comparing
%$K \to  \pi $ matrix elements.
%Usually this factor is not taken into account.
%Moreover each group uses different conventions,
%either the chiral limit value of $f_{\pi}$ and $f_K$ or their
%physical value. 
%We give the lattice results for $K\to\pi$ rescaling with the factor 
%above.
\begin{table}
\begin{center}
\begin{tabular}{|c|c|}
\hline
Reference&$- 6\times 10^{4} \langle 0| O_6^{(1)}|0 \rangle^{NDR}_\chi
\, {\rm GeV}^{-6}$
 \\ 
\hline
$B_7^{\chi}$(2~GeV)=$B_8^{\chi}$(2~GeV)=1 &$3.2\pm1.3$   \\
This work, \cite{BGP01} (SS+PP=0)&$2.4\pm0.3$  \\  
This work, \cite{BGP01} (Data \& Duality FESR)   &$2.4\pm0.3$  \\   
Cirigliano et al., \cite{CDGM02} (Data \& Fitted FESR)  
&$2.2\pm0.5$  \\  
Knecht et al.,  \cite{KPR01} $N_c\to \infty$,MHA     &$1.1\pm0.3$   \\  
Narison, \cite{NAR01} Data \& Tau-like FESR     &$2.1\pm0.6$     \\  
\hline
CP-PACS Coll., \cite{CPPACS01} lattice(chiral) & $2.4\pm0.3$ (stat.)   \\
RBC Coll., \cite{RBC01} lattice(chiral)        & $2.8\pm0.4$ (stat.)   \\ 
${\rm SPQ_{CD}R}$ Coll., \cite{papinutto} lattice(Wilson) &$1.4\pm0.1$ (stat.)\\
\hline
\end{tabular}
\end{center}
\caption{\label{q7NDR} The values of the VEV 
$\langle 0| O_6^{(1)}|0 \rangle_\chi$ 
 in the NDR scheme at $\mu_R=2$ GeV.}
\end{table}%table
\begin{table}
\begin{center}
\begin{tabular}{|c|c|}
\hline
Reference&
$10^{3}\langle 0| O_6^{(2)}|0 \rangle^{NDR}_\chi
\, {\rm GeV}^{-6}$ 
 \\ 
\hline
$B_7^{\chi}$(2~GeV)=$B_8^{\chi}$(2~GeV)=1& $1.0 \pm 0.4$   \\
This work, \cite{BGP01} (SS+PP=0)& $1.2  \pm 0.5$  \\  
This work, \cite{BGP01} (Data \& Duality FESR)& $1.2  \pm 0.7$  \\   
Cirigliano et al., \cite{CDGM02} (Data \& Fitted FESR)  
& $1.5 \pm 0.3$   \\  
Knecht et al., \cite{KPR01} $N_c\to \infty$,MHA & $2.3 \pm 0.7$   \\ 
Narison, \cite{NAR01}  Data \& Tau-like FESR     & $1.4 \pm 0.4$      \\  
\hline
CP-PACS Coll., \cite{CPPACS01} lattice(chiral) & $1.0\pm0.2$ (stat.)   \\
RBC Coll., \cite{RBC01} lattice(chiral)        & $1.1\pm0.2$ (stat.)   \\  
${\rm SPQ_{CD}R}$ Coll., \cite{papinutto} lattice(Wilson) &$0.8\pm0.1$ (stat.)\\
\hline
\end{tabular}
\end{center}
\caption{\label{q8NDR} The values of the VEV 
$\langle 0| O_6^{(2)}|0 \rangle_\chi$ in 
the NDR scheme at $\mu_R=2$ GeV.}
\end{table}%table
%\begin{table}
%\begin{center}
%\begin{tabular}{|c|c|c|}
%\hline
%Reference&$-6\times 10^{4} \langle 0| O_6^{(1)}|0 \rangle^{HV}_\chi
%\, {\rm GeV}^{-6}$&
%$10^{3}\langle 0| O_6^{(2)}|0 \rangle^{HV}_\chi
%\, {\rm GeV}^{-6}$ 
% \\ 
%\hline
%$B_7=B_8=1$                      &$3.2\pm1.3$ & $1.0 \pm 0.4$    \\
%This work,\cite{BGP01} (SS+PP=0) &$3.7\pm 0.6$& $1.3 \pm 0.6$    \\  
%This work,\cite{BGP01}(Data\&Duality FESR) &$3.7\pm 0.6$  &$1.3\pm0.8$  \\ 
%Cirigilano et al.\cite{CDGM02}   &$8.2\pm0.9$& $2.4\pm0.7$      \\  
%Knecht et al.  \cite{KPR01}      &$11.0\pm2.0$& $3.5 \pm 1.1$      \\   
%\hline
%CP-PACS Coll., \cite{CPPACS01} lattice(chiral) & $4.0\pm0.5$       \\
%RBC Coll., \cite{RBC01} lattice(chiral)        & $4.5\pm0.5$       \\ 
%\hline
%\end{tabular}
%\end{center}
%\caption{\label{qsHV} The values of the VEVs in the 
%HV scheme at $\mu_R=2$ GeV.}
%\end{table}%table  

Within the present accuracy of $\langle 0| \bar q q|0\rangle$, the 
disconnected 
contribution ${\cal A}_{SP}$ in (\ref{ALRSP}) --second line 
in the tables-- is 
perfectly compatible with our full result --third line in the tables--, 
so that 
we cannot conclude a large deviation from the large $N_C$ 
result within the present accuracy. Notice that we include in this result-
third line in Table \ref{q8NDR}- $\order(\alpha_S)$ corrections 
that are indeed leading order in $1/N_C$ (see (\ref{O6operator})) which 
are usually disregarded in the factorization approaches, this makes the 
chiral limit $B_8^{\chi}(2GeV)$ parameter larger than one by around 20\% 
to 30\%.

We agree with the analytical results in \cite{NAR01,CDGM02}. 
We are borderline 
within  errors with the value of 
$\langle 0| O_6^{(2)}|0 \rangle^{NDR}_\chi$ in \cite{KPR01} 
though their central value is twice ours, but our results for 
$\langle 0| O_6^{(1)}|0 \rangle^{NDR}_\chi$ are not compatible.

The agreement between our results and the lattice results 
is quite good 
in the case of the chiral fermion determinations but not 
so good in the 
results coming from calculations using Wilson fermions. 
 The lattice determinations are made in the 
quenched approximation and the quoted errors -which are much 
smaller than ours for $\langle 0| O_6^{(2)}|0 \rangle^{NDR}_\chi$-
 are only statistical ones.
The present lattice results are larger than the earlier 
estimates and the value of $\langle 0| O_6^{(2)}|0 \rangle^{NDR}_\chi$ 
in \cite{CDGM02} is smaller than their previous result in 
\cite{CDGM01}. This makes possible the agreement with our results.

We have not quoted the results from the lattice calculation using 
staggered fermions in \cite{BFKGLS03}, since they give their 
results in terms of the bag parameters. In the chiral limit, the values 
they obtained for the bag parameters in the NDR scheme 
are \cite{LeePC} $B_7^\chi(2\gev)=0.85\pm0.04$ and 
$B_8^\chi(2\gev)=0.89\pm0.05$.

%We have not quoted the results from the CHPT large $N_c$ approach 
%-second reference in \cite{HKPS00}- 
%and the chiral quark model \cite{Trieste} because they are away from the
%chiral limit.

%Our results are very compatible with the work in \cite{epsprime}, where 
%the ENJL model and a very low value of $\mu$ were used 
%to estimate the same matrix-elements.
%The underlying reason for the agreement is that the operators $Q_7$ and $Q_8$
%mix quite strongly and the values of the matrix-element of $Q_8$
%at the low scale $\mu\approx 0.8$ GeV
%dominate the values of the matrix-elements
%of $Q_7$ and $Q_8$ at the higher scale $\mu=2$~GeV.
%The matrix element of $Q_8$ obtained here is the same
%as the one we used in \cite{epsprime}, since the effect
%of ${\cal A}_{SP}$ is estimated to be moderate here.

\subsection{R\^ole of Higher Dimensional Operators}
\label{higherdim}

We clarified here the role of the higher than six dimensional operators,
an issue raised in \cite{CDG00}. In our scheme they remove
the $\mu$-dependence which is not covered by the renormalization group. 
The role of these operators was analyzed in \cite{BGP01} and later in 
\cite{CDGM02}

\begin{figure} [thb]
\begin{center}
\includegraphics[angle=-90,width=0.49\textwidth]{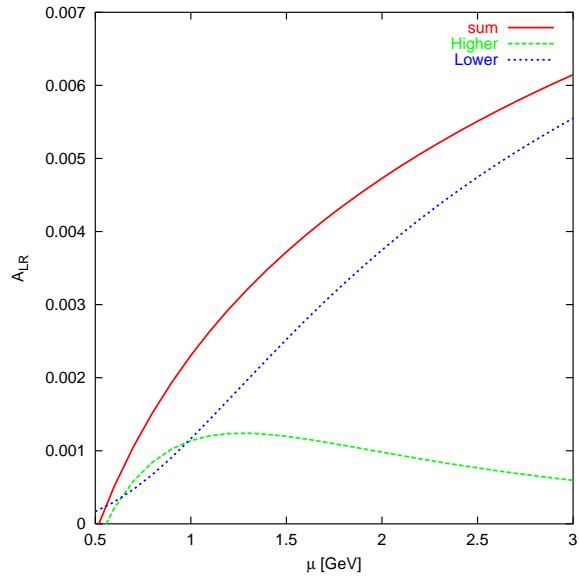}
\end{center}
\caption{\label{fig4} The separate contributions
to ${\cal A}_{LR}$ and the sum. ``Higher'' labels the effect of
the higher than six dimensional operators in the short-distance
contribution and ``Lower'' the long-distance part.}
\end{figure}

The effect of higher dimension operators in our approach is to add 
${\cal A}_{LR}^{\rm Higher}(\mu)$
to the low energy contribution ${\cal A}_{LR}^{\rm Lower} (\mu)$,
these are defined in Eq. (\ref{numALEPH}),
\ba
{\cal A}_{LR}(\mu) &=&
{\cal A}_{LR}^{\rm Higher}+{\cal A}_{LR}^{\rm Lower}
= \int_0^{s_0} {\rm d} t \, t^2 \, \ln \left(\frac{t}{\mu^2} \right) 
\, \frac{1}{\pi}  \, \im \Pi_{LR}^T(t) \, 
\ea
where $\mu$ is an Euclidean cut-off. 
It is clear than the contribution of higher than 
dimension six operators
is less important only for values of $\mu^2$ larger than $s_0$,
where $\im \Pi^T_{LR}$ vanishes because of local duality.
 In Figure \ref{fig4} we plot the two separate contributions
and the sum 
as a function of $\mu$.

%figure 

{}From the figure
we can see for $\mu$ larger than 2 GeV the contribution
of {\em all} higher dimensional operators is less than 25 \%.
We agree with \cite{CDG00} that for the matrix elements that involve
integrals of $\im \Pi^T_{LR}$ one has to go to
such values of $\mu$ to disregard the contribution of 
higher dimensional operators. The contribution we find is somewhat smaller
than in \cite{CDG00} since we include the effect of {\em all}
higher order operators, not just dimension eight.

The high value of $\mu$ is set by the threshold of
perturbative QCD $s_0$ which depends very much on the 
spectral function and on the integrand behaviour.
In fact, from \cite{PR00}
one can  see that  relevant spectral function for the 27-plet coupling
reaches the perturbative QCD behaviour very soon, from 0.7 GeV
to 1 GeV.  The  OPE 
matched impressively well with the hadronic ansatz at such low values
with just dimension six operators.
Therefore  though higher dimensional operators appear
one  can expect smaller contributions
in cases like $G_{27}$ and $ \re G_8$.

The matrix-elements studied in this paper might 
be special in the sense
that they follow from integrals over spectral functions which
have no contributions at short-distances from the unit operator
or the dimension four operators. As the good
matching at low scales in the example in \cite{PR00}
shows, the other quantities which have these contributions might
have much smaller higher dimension effects.

\chapter{Application to $\varepsilon_K'$}
\label{secepsilonprime}

One of the applications of the hadronic model introduced in 
Chapter \ref{matrixelements} is the calculation of $\im G_8$ 
using the $X$-boson method. This can be used to confirm  
the large value found in \cite{epsprime} and to study its origin 
analytically. 

At a first step, we will discuss  
the update presented in \cite{BGP03} of the results of \cite{epsprime}. 
The new things we 
would like to input are the non-FSI corrections which after the work in 
\cite{PPS01,BDP03,BPP98,CG00} are known. We also use the recent complete 
isospin breaking result of \cite{isosbreak}. The result in \cite{epsprime} did 
contain the FSI corrections but not the non-FSI which were unknown at that 
time. The other new input is the $\Delta I=3/2$ contribution calculated 
in the chiral limit and at NLO in \cite{BGP01} and 
discussed in Chapter \ref{chq7q8}.

The chiral corrections to the LO in CHPT predictions for the ratios 
of amplitudes in (\ref{epsilonprimedef})
\be 
\varepsilon_K' \, \simeq\, \frac{i}{\sqrt 2}
\frac{\re  a_2 }{\re  a_0}\left[\frac{\im  a_2}{\re  a_2}
-\frac{\im  a_0}{\re  a_0} \right]e^{i(\delta_2-\delta_0)}
\ee
are introduced in (\ref{ratio12}) and 
(\ref{ratio32}) through the factors $C_0$ and $C_2$
\be 
\frac{\im a_0}{\re a_0}=
\left(\frac{\im a_0}{\re a_0}\right)^{LO} 
{\cal C}_0 \, ,\quad
\frac{\im a_2}{\re a_2}=
\left(\frac{\im a_2}{\re a_2}\right)^{LO} 
{\cal C}_2+\Omega_{{\rm eff}}\frac{
\im a_0}{\re a_0}\, .
\ee
The parameter $\Omega_{eff}$ includes the effects of isospin breaking -see 
Section \ref{secq6}.

In the ratio 
$\re  a_0/\re  a_2$ the chiral corrections can be also introduced by a 
multiplicative factor $C_{\Delta I=1/2}$ as follows
\be  
 \frac{\re a_0}{\re a_2}=
\left(\frac{\re a_0}{\re a_2}\right)^{LO} 
{\cal C}_{\Delta I=1/2} \,.
\ee

We get from the fit to experimental $K \to \pi\pi$ amplitudes
in \cite{BDP03} 
\be 
C_{\Delta I=1/2} = \frac{{\cal S}_0}
{{\cal S}_2} = \frac{1.90 \pm 0.16}{1.56 \pm 0.19} = 1.22 \pm 0.15\, 
\, \, 
\ee
Where ${\cal S}_{\rm I}$ are the chiral corrections to $\re a_I$
to all orders
while we call $\cal{T}_{\rm I}$ to the chiral corrections to $\im a_I$
to all orders.   Therefore, they contain the FSI corrections
which were exhaustively studied in \cite{PPS01} plus the non-FSI
corrections which are a sizable effect and of opposite direction.
All these chiral corrections  contain
the large overall known factor $f_K f_\pi^2 /F_0^3 \simeq 1.47$ from
wave function renormalization. 

The imaginary parts $\im a_I$ get FSI corrections identical to $\re a_I$
owing to Watson's theorem. In addition, 
both due to octet dominance in  $\re a_0$ and $\im a_0$
and to the numerical dominance of the non-analytic terms at NLO 
in $K\to \pi\pi$ amplitudes \cite{PPS01,BDP03,BPP98}
\be
C_0= \,{{\cal T}_0}/{{\cal S}_0}\,\simeq 1.0 \pm 0.2 \, ,
\ee
to a good approximation.

The situation is quite different for $\im a_2$ which is proportional to
$\im (e^2 G_E)$ at lowest order since $\im G_{27}=0$ in the
Standard Model.
{}From the works \cite{PPS01,BDP03,CG00} we also know that
\ba
\label{C2}
C_2&=&\frac{{\cal T}_2}{{\cal S}_2} 
\simeq \frac{0.70 \pm 0.21 - 0.73 L_4/10^{-3}}{1.56 \pm 0.18}
\nonumber \\
&=& 0.45 \pm0.15 - 0.47 \frac{L_4}{10^{-3}}\, . 
\ea

\begin{figure}[thb]
\begin{center}
\includegraphics[height=0.7\textwidth,angle=270]{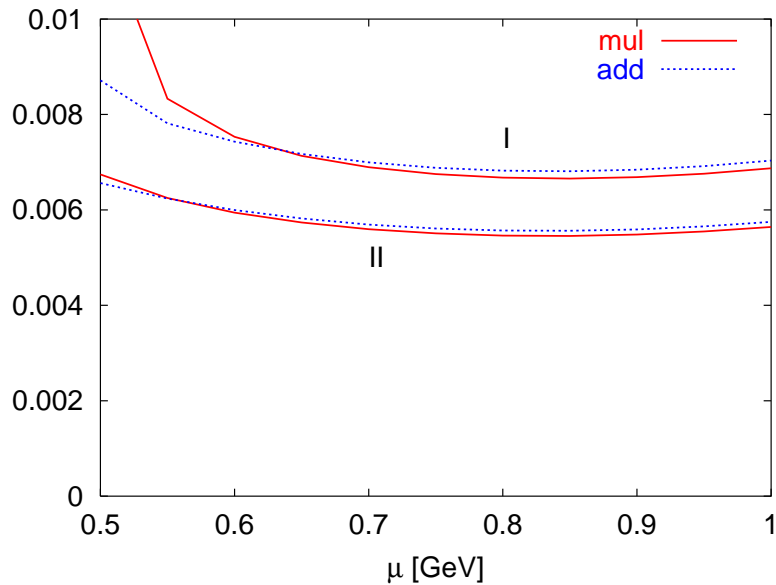}
\caption{\label{eps}  Matching of the short-distance scale
dependence of our LO in CHPT $\varepsilon'_K/\varepsilon_K$
prediction. Labels I and II are for two
different values of $\alpha_S$. The two curves
for two choices of perturbative  matching,
see \cite{epsprime}. Notice the quality of the matching.}
\end{center}
\end{figure}

At LO in CHPT, the result is obtained by substituting directly in  
(\ref{epsilonprimedef}) the value of the ratios given in (\ref{LOratios})
 \cite{BGP03}
\ba
\label{LOeps}
\left(\frac{\varepsilon'_K}{\varepsilon_K}\right)^{LO} &=&
((-10.8 \pm 5.4) + (2.7 \pm 0.8)) \, \im \tau \, 
\nonumber \\ 
&=& -(8.1 \pm 5.5) \,  \im \tau 
= (4.9 \pm 3.3) \times 10^{-3} \, . 
\ea
This result 
is scheme independent and very stable against the
short-distance scale as can be seen in Figure \ref{eps}.
The difference with the result in Figure \ref{eps} \cite{epsprime}
is due to the new values of $\im \tau$ in (\ref{tau})
and $\re G_8$ and $G_{27}$ in (\ref{G8exp}).

Including the known and estimated higher order CHPT corrections, we get 
\cite{BGP03}
\be
-\frac{1}{|\varepsilon_K| \sqrt 2} \, \frac{\re a_2}{\re a_0}
\, \frac{\im a_0}{\re a_0} = -(8.9 \pm 4.8) \, \im \tau \, 
\ee
and
\ba \label{eti1}
\frac{1}{|\varepsilon_K| \sqrt 2} \, \frac{\re a_2}{\re a_0}
\, \frac{\im a_2}{\re a_2} =
 ((1.0 \pm 0.3) + (0.5 \pm 0.7) ) 
\, \im \tau \,= (1.5 \pm 0.8) \, \im \tau \, . \nonumber
\ea
where the second part comes from isospin breaking contribution
with $\Omega_{eff} = 0.06\pm0.08$ \cite{isosbreak},
and we used $L_4=0$ in (\ref{C2}).
And therefore, 
\ba
\label{finaleps}
\frac{\varepsilon'_K}{\varepsilon_K} &=&
((-8.9 \pm 4.8) + (1.5 \pm 0.8) ) \, \im \tau \,  
\nonumber \\
&=& -(7.4 \pm 4.9) \,  \im \tau 
=(4.5 \pm 3.0) \times 10^{-3} 
\ea
to be compared to the world average \cite{NA4802,KTEV03}
\be
\label{expeps}
\re \left( \frac{\varepsilon'_K}{\varepsilon_K}\right)
{\Bigg|}_{\rm exp} = (1.66 \pm 0.16) \times 10^{-3}  \,.
\ee

Though the central value of our Standard Model
prediction in (\ref{finaleps}) is a factor around 3 
too large, within the big uncertainties it is still
compatible with the experimental result.
Two immediate consequences of the analysis above,
namely,  the LO CHPT prediction (\ref{LOeps})
 is actually very close of the
the final result (\ref{finaleps}) and second, the part with $\Delta I=1/2$
dominates  when all higher order CHPT corrections
are included.

The large final uncertainty
we quote in (\ref{finaleps}) is mainly due to the uncertainties of
(i)  the chiral limit
quark condensate, which is not smaller than 20\%,
(ii)  $L_5$, which is around 30\%,  and (iii)
the NLO in $1/N_c$ corrections to  the matrix element
of $Q_6$, which is around 20\%.
All of them together make the present prediction for
the $\Delta I=1/2$  contribution to  $\varepsilon'_K$
to have  an error around  55\%. 
Reduction in the uncertainty of all these inputs, especially of
the quark condensate and $L_5$ is needed to obtain
a reasonable final uncertainty.

We substituted the value  used in \cite{epsprime}
 for $\im (e^2 G_E)$ 
by the one in (\ref{imgeinput}), notice however that numerically 
they coincide within errors.
There are also large uncertainties in the $\Delta I=3/2$
component coming from  isospin breaking \cite{isosbreak},
 and more moderate in
 the non-FSI corrections to $\im a_2$,
fortunately the impact of them in the final result is not as large as the
ones associated to the $\Delta I=1/2$ component.

 Assuming the $\Delta I=3/2$ component of $\varepsilon'_K$
  is  fixed to be the one in 
(\ref{imgeinput}) as indicated by  the analytic methods 
\cite{BGP01,NAR01,KPR01,CDGM02} and
lattice  \cite{CPPACS01,RBC01,domainwall,wilson}, 
one can try to extract the value of $\im G_8$ from the experimental
result in (\ref{expeps}). We get
\be
\im G_8 = (2.0 ^{+0.7}_{-0.4}) \, 
\left(\frac{87 {\rm MeV}}{F_0}\right)^4 \, \im \tau 
\ee
 i.e. the central value coincides with the large $N_c$ result
for $\im G_8$  within 30\% of uncertainty. This value has 
to be compared to the result in \cite{epsprime} 
$\im G_8=(4.4\pm 2.2)\left(87 {\rm MeV}/F_0\right)^4\,\im \tau$.

More work is needed to confirm the large value of $\im G_8$ obtained 
in \cite{epsprime}. We plan to analyze the origin of this value using 
the $X$-boson method as done in \cite{epsprime} but substituting 
the ENJL model by the hadronic model described in Chapter 
\ref{matrixelements}, that allow us to perform all calculations 
analytically. We intend to see the effect of the vector, axial-vector and 
scalar sources as well as the possible cancellations of their contributions. 
This will shed light on the leading large hadronic contributions 
which produce this $\im G_8$.

\section{$\varepsilon_K'$ with a Different Set of Input Parameters}

Alternatively, we can calculate $\varepsilon_K'$ using as input parameters 
the values of $\re G_8$ and $G_{27}$ obtained in the second reference in 
\cite{isosbreak} where the leading isospin breaking effects are included. 
The values of the couplings found there, after normalizing them 
to $F_0=87\mev$ -in \cite{isosbreak} a lower value $F_0=85.7\mev$ is used- 
are
\be \label{G8exp2}
\re G_8 = \left(6.5 \pm 0.26\right) \, 
\left(\frac{87 {\rm MeV}}{F_0}\right)^4 \, 
{\rm and} \, G_{27}= \left(0.54\pm0.02 \right) \, 
\left(\frac{87 {\rm MeV}}{F_0}\right)^4  \, .
\ee
Notice that the value of $G_{27}$ and $\re G_8$ in (\ref{G8exp2}) are compatible 
with those in (\ref{G8exp}), but the central values are somewhat different. 
With these inputs, we get 
\be
-\frac{1}{|\varepsilon_K| \sqrt 2} \, \frac{\re a_2}{\re a_0}
\, \frac{\im a_0}{\re a_0} = -(10.8 \pm 5.4) \, \im \tau \, 
\ee
and
\ba
\frac{1}{|\varepsilon_K| \sqrt 2} \, \frac{\re a_2}{\re a_0}
\, \frac{\im a_2}{\re a_2} =
 ((1.2 \pm 0.3) + (0.6 \pm 0.9) ) 
\, \im \tau \,= (1.8 \pm 0.9) \, \im \tau \, . \nonumber
\ea
where the second part, as in (\ref{eti1}), comes from isospin 
breaking contribution with $\Omega_{eff} = 0.06\pm0.08$ \cite{isosbreak}.

The final value for the ratio $\varepsilon_K'/\varepsilon_K$ from 
this set of input parameters is 
\ba
\frac{\varepsilon'_K}{\varepsilon_K} &=&
((-10.8 \pm 5.4) + (1.8 \pm 0.9) ) \, \im \tau \,  
\nonumber \\
&=& -(8.9 \pm 5.4) \,  \im \tau 
=(5.4 \pm 3.3) \times 10^{-3} 
\ea
that is compatible with our result in (\ref{finaleps}). The difference 
between both results mainly comes from the different values of $\re G_8$ 
obtained by the fits to experimental data in \cite{BDP03} and \cite{isosbreak}.

\chapter{Conclusions}

\label{conclusions0}

{\bf A. Charged Kaon $K\to 3\pi$ CP-Violating Asymmetries}

We have performed the first full analysis
at NLO in CHPT of the CP-violating asymmetries
in the slope $g$ and the decay rate $\Gamma$ for the
disintegration  of charged Kaons into three pions.
 We have done the full order $p^4$ calculation
for $K \to  3 \pi$ and completely agree with the recent results
in \cite{BDP03}.
 To give the CP-asymmetries at NLO, one needs the FSI phases at NLO
also, i.e. at two loops. This is not available at present.
We have calculated the dominant two-bubble contributions
using the optical theorem and the known one-loop and tree level
results in Appendix \ref{FSI6} as explained in
Section \ref{6.1}.  Due to the small phase
 space available for the 
re-scattering effects of the final tree pions one expects the rest 
of the FSI  to be  very suppressed.
We have included this contribution in our final numbers.
As a byproduct, we have  predicted the isospin I=2  FSI phase at NLO
and two combinations of matrix elements of
the isospin I=1 FSI re-scattering matrix $\mathbb{R}$ at NLO.
They can be found numerically  in 
Section \ref{6.1} and analytically in Appendix \ref{phasesNLO}.
We have given  analytical expressions for all the results in the 
Appendices \ref{Amplitudes}, \ref{Adeltag}, \ref{ANLO},
and  \ref{FSI6}.  

 Our final results at LO can be found in Table \ref{tabLO}
and at NLO in Table \ref{tabNLO}. If we use the counterterms
in Table \ref{tabKvalues}, we find  NLO corrections of 
the expected size, i.e. around 20\%,    for $\Delta g_C$
and $\Delta g_N$.
With those values for the NLO counterterms, the CP-violating 
asymmetry $\Delta g_C$ is dominated by  the value of $\im G_8$
while the rest of the CP-violating asymmetries studied here, 
namely,  $\Delta g_N$, $\Delta \Gamma_C$ and $\Delta \Gamma_N$, 
are dominated by the value of $\im \widetilde K_2$
and $\im \widetilde K_3$. 

Of course, our results in Table \ref{tabNLO} 
depend on the size of
$\im \widetilde K_2$ and $\im \widetilde K_3$.
If their values are within a factor two to three
the ones in Table \ref{tabKvalues} then the central
value of $\Delta g_C$
changes within the quoted uncertainties for it,
while the central value for $\Delta g_N$ doubles.
 The asymmetries in the decay rates
$\Delta \Gamma_C$ and $\Delta \Gamma_N$ can change even sign
if we vary $\im \widetilde K_2$ and $\im \widetilde K_3$ within
the uncertainties quoted in Table \ref{tabKvalues}.
Therefore, we have presented for them just ranges.

We partially disagree with
references \cite{DIP91,MP95,DI96} when  the authors claim that one 
could expect one order of magnitude enhancement at NLO
in all the asymmetries studied here. 
We find that for $\Delta g_C$ and $\Delta g_N$ the NLO corrections
are of the order of 20\% to 30\% . Only $\Delta \Gamma_C$
and $\Delta \Gamma_N$  can vary of one order
of magnitude and even change sign depending on the value
of $\im \widetilde K_2$ and $\im \widetilde K_3$.
We also find that $\Delta g_C$ can be as large as $-4 \times
10^{-5}$  both at LO and NLO while in the
conclusions of \cite{DIP91,MP95,DI96}  it was
claimed that any of these asymmetries could not exceed $10^{-5}$
within the Standard Model.
 
In Section \ref{5.2}, we found 
that making the cut proposed in \cite{AVI81,GRW86}
for the energy of the pion with charge opposite to the
decaying Kaon, there is one order of magnitude enhancement
for $\Delta \Gamma_C$
in agreement with the claims in those references. 
This result is however  valid for our LO calculation. 
It remains unclear whether the cut can provide a real
advantage at NLO  since in this case the cancellation among the various 
counterterm contributions can mask the effect. In addition, 
it remains to see how feasible is  to perform this cut experimentally. 
We do not find this enhancement for $\Delta \Gamma_N$. 

The measurement of these CP-violating asymmetries
 by NA48 at CERN, by KLOE at Frascati, by OKA at Protvino 
and/or elsewhere at the level of $10^{-4}$ to 
$10^{-5}$ will  be extremely interesting
for many reasons.
 The combined analysis of all four CP-violating asymmetries
$\Delta g_C$, $\Delta g_N$, $\Delta \Gamma_C$ and 
$\Delta \Gamma_N$ can allow to  obtain more information
on the  values of the presently poorly known $\im G_8$,
and the  unknown $\im \widetilde K_2$ and $\im \widetilde K_3$.
Due to the different dependence on these parameters, 
if the measurement  is  good enough, 
one can  try to fix $\im \widetilde K_2$ and 
$\im \widetilde K_3$ from the measurement of the asymmetries
$\Delta g_N$, $\Delta \Gamma_C$ and $\Delta \Gamma_N$
which are dominated by  the order $p^4$ counterterms and
use them to predict more accurately $\Delta g_C$.

 The large dependence of the  asymmetry $\Delta g_{C}$ 
 of $\im G_8$ at NLO can also be used as 
 consistency check between the theoretical 
predictions for  $\Delta g_C$ and  for the CP-violating
parameter $\varepsilon_K'$.  Any prediction
for $\varepsilon_K'$ has to be also able to predict the CP-violating
asymmetries discussed here.  In particular, 
the measurement of $\Delta g_C$ may  also
 shed light on a possible large  value for $\im G_8$ 
as found in calculations
at NLO in $1/N_c$ --see for instance \cite{epsprime,HPR03,HKPS00}.

Moreover, it seems that some  models beyond the Standard Model can reach
values not much larger than $1\times 10^{-4}$ for the 
CP-violating  asymmetries, see for instance \cite{DIM00}. 
Our results can help to distinguish  new physics effects 
from the Standard Model ones in these  observables
and unveil beyond the Standard Model physics.

\hspace{1 cm}

\noindent{\bf B. $\Delta I=3/2$ Contribution to Direct CP Violation}

In Chapter \ref{chq7q8} we have calculated in a model independent
way the matrix elements of the $\Delta S=1$ operators
$Q_7$ and $Q_8$ in the chiral limit. We have done it
to  all orders in $1/N_c$ and NLO in $\alpha_s$.

The scheme dependence has been taken into account
exactly at NLO using the $X$ boson method
as proposed and used in \cite{epsprime,BP99,scheme}. In fact, these two
operators are a submatrix of the ten by ten done 
in \cite{epsprime}. 

%We would like to mention some issues
%sometimes mixed up in the literature.
%First, the $X$ boson method has nothing to do with using
%or not large $N_c$. It can be used without the large $N_c$
%approximation as well, as shown again in this paper.
%Second, our method of treating the scheme
%dependence is consistent and we never 
%mix up two different schemes, cut-off and $\overline{MS}$ schemes.
%We do an analytic matching between a cut-off 
%regularization and dimensional regularization
%in a well-defined scheme at perturbative scales first.
%The finite parts arising in this matching appear in the methods
%using dimensional regularization to the end as well as explained in the
%appendix.

We obtain exact matching in an Euclidean-cut-off regularization
and  analytical cancellation exact of (all) infrared and UV 
scheme dependences.

For the contribution of higher order operators
discussed in \cite{CDG00} and \cite{NAR01} 
we clarify how to include all higher dimensional operators
and  exact scheme dependence at NLO in $\alpha_S$ of both 
the $Q_7$ and $Q_8$ matrix elements. As a result we find 
smaller corrections due to this effects as discussed 
in Section \ref{higherdim}. In our approach the effect 
of the higher order operators is to remove the
remaining dependence on the Euclidean cutoff $\mu$ beyond 
the RGE evolution. The result of resuming  all
higher dimensional operators in the case of
$Q_7$ makes its prediction much less sensitive
to the choice of $s_0$. The scale cancellation is only possible if both 
contributions have the same hadronic content.

As noticed in \cite{epsprime,KPR01}, ${\cal A}_{SP}$ is zero
in the large $N_c$ limit and therefore is Zweig suppressed.
We find  no sizable violation of the dimension six FESR
using factorization for $Q_8$.

We find that the moment $M_2$ is very sensitive to the 
spectral function around 2~GeV$^2$.

Our main analytical results are the expression for the
matrix-elements (\ref{imge}), the bag parameters (\ref{resbag1}),
(\ref{resbag2}) and the expansion coefficients of the 
spectral functions
(\ref{resPILR}), (\ref{F11}) and (\ref{resPISP}). The main 
numerical results
are the VEVs (\ref{numO1}),(\ref{numO2}) and the bag 
parameters (\ref{numbag}).
These results are exact in the chiral limit, so we have
the $\Delta I=3/2$ part of $\varepsilon_K'/ \varepsilon_K$
model independently at all orders in $1/N_c$.  
This has been possible because our 
results for $\im(e^2G_E)$ can be written in an exact way in terms 
of integrals of full two-point functions in the chiral limit. They can be 
related to spectral functions via dispersion relations and resummation 
of the effect of all higher dimensional operators in the OPE of the 
relevant two-point functions. The information about these spectral functions 
comes from known results on the scalar spectral functions and from 
experimental $\tau$-data, that are the most important source of uncertainty. 
The results can be then substantially improved by obtaining  
better experimental data from $\tau$ decays. Such improvement can be 
performed in the B-factories (BaBar,Belle).

In order to reach final values, all effects which vanish 
in the chiral limit,
as final state interactions and the rest of higher CHPT corrections, 
isospin violation and long-distance electromagnetic effects, must 
be included as explained in Section \ref{secq7q8} of Chapter \ref{CPviolation}.

This calculation is a good example of two facts. First, in some cases 
we can extract information about low energy couplings from experimental 
data, using dispersion relations and the relation between these couplings 
and integrals over appropriate Green's functions. Another point is that 
the $X$-boson method can account for the scheme and scale dependence, 
regardless of wether we are using the large $N_c$ approach or not.

\hspace{1 cm}

\noindent{\bf C. Ladder Resummation Approximation to QCD Green's functions}

We have constructed a new approximation to low and intermediate energy 
hadronic quantities \cite{BGLP03}, 
described in Chapter \ref{matrixelements}. Our approach naturally fits 
in the large $N_c$ limit and incorporates chiral symmetry 
constraints by construction. It keeps the good features of the ENJL model
(CHPT at NLO, contains some short-distance QCD constraints, 
good phenomenology \dots) 
and improves adding more short-distance QCD constraints
and  the analytic structure of large $N_c$. 

We have shown that many short-distance constraints can be easily incorporated
but pointed out that our model, but also those with 
general Green's functions saturated by hadrons approaches, 
cannot reconcile all short-distance constraints due to 
a general conflict between short distance constraints on Green's functions
and those on form factors and cross-sections
that can be obtained from those Green's functions via LSZ reduction 
--see Section \ref{SDProblem}.

Our approach incorporates the gap equation and
the concept of a constituent quark mass following directly from the
Ward Identities and the resummation assumption.

We have also compared our results with experimental results for hadronic
observables and found reasonable agreement.

The calculation of the four-point functions that are needed  
in the calculation of some of the CP-violating parameters 
we have discussed in this Thesis is in progress \cite{BGLP03b}.

\hspace{1 cm}

\noindent{\bf D. Future Plans and Applications}

The next step we are already finishing is to use the Green's functions obtained
within the hadronic model introduced above  
in the calculation of hadronic matrix elements relevant for $\varepsilon_K'$ 
and the $\hat B_K$ parameter. We will   
analyze the relevant internal cancellations and dominant hadronic
parameters in the quantities we calculate with them. 

Work in this direction where we will study the
origin of the large chiral corrections to $\hat B_K$ 
found in \cite{BP95bk} is in progress \cite{BGLP03b}. This parameter 
can be determined by an integration over all the range of energies 
of the four-point function $\Pi_{PLPL}^{\alpha \alpha}$ \cite{BP95bk}, 
where P denotes a pseudoscalar density and 
L denotes a left-handed (V-A) current. The method we will follow is 
briefly described in Section \ref{scbk}.

This program will also be  extended
to study  the origin of the large value for $\im G_8$ 
which was obtained in \cite{epsprime}.
Large values of $\im G_8$ were previously pointed out
in \cite{HKPS00} and more recently in \cite{HPR03}. Another source of 
information about the imaginary part of this coupling, as discussed in 
the first part of this Chapter, is the measurement of the different 
CP-violating asymmetries in the charged kaon decays $K\to 3\pi$ since 
they depend in a determinant way on $\im G_8$.

Another more ambitious program is to use this hadronic model 
to construct the $\Delta S=1$ Green's functions that let us 
get systematically all the $\Delta S=1$ counterterms at NLO.

\appendix

\chapter{OPE of $\Pi_{LR}^T(Q^2)$ and  
$\Pi_{SS+PP}^{(0-3)\, \rm conn}(Q^2)$}

\label{appendixq7q8}

\section{Calculation of the Corrections of $\mathcal{O}(a^2)$ to the
Dimension Six Contribution to $\Pi_{LR}^T(Q^2)$}
\label{AppA}
\subsection{Renormalization Group Analysis}

We have the two-point function
\begin{eqnarray}
\Pi^{\mu \nu}_{LR} (q)&\equiv& \frac{1}{2}\,
 i\int d^Dy\, e^{iq \cdot y}\langle 0\vert
T(L^{\mu}(y)R^{\nu}(0)^{\dag} )\vert 0\rangle \equiv
(q^{\mu}q^{\nu}-g^{\mu \nu} q^2)\Pi_{LR}^{T} (q^2) \nonumber \\ 
&+&q^{\mu} q^{\nu}\Pi_{LR}^{L}(q^2)
\end{eqnarray}
The contribution of dimension six operators to $\Pi_{LR}^{T}(Q^2)$ (where
$Q^2=-q^2$) can be written in $D=4-2\epsilon$ dimensions as
\begin{equation}\label{dimension6}
Q^6\, \Pi_{LR}^{T} (Q^2)\Big \vert_{D=6}
\,\equiv \, \nu^{2\epsilon} \, \sum_{i=1,2} C_i (\nu,Q^2) <O_i>(\nu)
\end{equation}
with
\begin{eqnarray}\label {operadores}
<O_1>&\equiv& \langle 0 | O_6^{(1)} | 0 \rangle =
\frac{1}{4} \,
 <0\vert(\overline s \gamma ^{\nu} d)_L (\overline
d\gamma_{\nu} s)_R\vert 0> \nonumber \\
<O_2>&\equiv& \langle 0 | O_6^{(2)} | 0 \rangle =
3 \, <0\vert(\overline d d)_L (\overline s s)_R\vert 0>
\end{eqnarray}
and
\begin{equation}
\label{Wilson}
C_i (\nu,s)=a(\nu)\sum_{k=0} a(\nu)^k\,  C_i ^{(k)} (\nu,s)
\end{equation}
where the dependence in $\nu$ and $s$ of $C_i^{(k)}$ is only
logarithmic. Everything here we define in the $\overline{MS}$ scheme.

In absence of electromagnetic interactions the matrix elements
(\ref{operadores}) only mix between themselves. The 
renormalization group equations (RGE) they satisfy are
\begin{eqnarray}\label{EGR}
\nu \frac{d<O_1> (\nu)}{d\nu} &=&-\gamma_{77}(\nu)<O_1>(\nu)+\frac{1}{6}\,
\gamma_{87}(\nu)<O_2>(\nu) \nonumber\\
\nu \frac{d<O_2> (\nu)}{d\nu} &=&-\gamma_{88}(\nu)<O_2>(\nu)+6\,
\gamma_{78}(\nu)<O_1>(\nu)
\end{eqnarray}
With $\gamma(\nu)$ the QCD anomalous dimension matrix
defined in (\ref{gamma}).
In the $NDR$ scheme \cite{NLOWilscoef,CFMR94,BBL96}
\footnote{For these operators the Fierzed
version and the $Q_7$-$Q_8$ version have the same anomalous 
dimension matrix.} for $n_f=3$ flavours\footnote{We will use
along this work $n_f=3$ since this is the number of active flavours
of the QCD effective theory where $Q_7$ and $Q_8$ appear.}, 
\ba
\label{NDR}
\gamma(\nu)&=& {\dis  \sum_{n=1}} \gamma^{(n)} 
 a(\nu)^n  \, \nonumber \\
\gamma^{(1)}&=& -\frac{3}{2 \, N_c} \, \left( 
\begin{array}{c|c}
-1 & 0 \nonumber \\ N_c & N_c^2 -1 \nonumber \\ 
\end{array} \right) ;  \nonumber \\ 
\gamma^{NDR (2)} &=&   - \frac{1}{96 N_c^2} \left( 
\begin{array}{c|c}
-137 N_c^2 + 132 N_c -45       &  
213 N_c^3 - 72  N_c^2  +  108 N_c \nonumber \\ 
200 N_c^3  - 132  N_c^2 -18 N_c & 
203 N_c^4 - 60 N_c^3  - 479 N_c^2 + 132  N_c  - 45  
\end{array} \right)  \, . \nonumber \\ 
\ea
In the HV scheme of \cite{NLOWilscoef,BBL96} \footnote{I.e. without
the $\beta_1 \, C_F  $ terms from renormalizing the axial current
in the diagonal coefficients \cite{BBL96}.}
\ba
\label{HV}
\gamma^{HV (2)} &=&   - \frac{1}{96 N_c^2} \left( 
\begin{array}{c|c}
 - 17 N_c^2 - 12 N_c - 45 &
-107 N_c^3 + 24 N_c^2 + 108 N_c \nonumber \\ 
80 N_c^3 + 12 N_c^2  - 18 N_c & 
115 N_c^4 - 12  N_c^3 - 71 N_c^2 - 12 N_c - 45 
\end{array} \right)  \, . \nonumber \\ 
\ea

We also need the quark mass anomalous dimension
in the $\overline{MS}$ scheme, 
\ba
\gamma_m(a)\equiv -\frac{\nu}{m} \, \frac{{\rm d} m}{{\rm d} \nu}
= {\dis \sum_{k=1}} \, \gamma_m^{(k)} a(\nu)^k \, 
\ea
where $m$ is a quark mass. 
The first coefficient is scheme independent
\be
\gamma_m^{(1)}= \frac{3}{2} C_F \, . 
\ee
Notice that $\gamma_{88}^{(1)}=- 2 \gamma_m^{(1)}$
to {\em all} orders in $1/N_c$ \cite{BG86,deR89},
this is the reason why $B_8$ in the chiral limit
is very near to 1 \cite{epsprime}.  The large $N_c$ result absorbs
{\em all} the one-loop scale dependence.
This exact scale cancellation {\em does} not occur
for $Q_6$ even at leading order
in $\alpha_S$. There is  a remnant diagonal anomalous dimension at one-loop
of order one  in $1/N_c$ which is not taken into account by
the large $N_c$ matrix element.
There is therefore no reason to expect $B_6$ around 1
as sometimes is claimed in the literature.

$\gamma_m^{(2)}$
is the same for both the NDR and HV schemes\cite{TAR81}, 
\ba
\gamma^{\overline {MS} (2)}_m&=&
\frac{C_F}{96 N_c} \left[203 N_c^2 - 60  N_c - 9\right] \, .
\ea
The relation $\gamma_{88}^{(2)}=- 2 \gamma_m^{(2)}$ is  not valid:
\ba
\gamma_{88}^{NDR (2)} &=& - 2 \gamma_m^{\overline{MS}(2)}
 +\frac{1}{32 N_c^2}
\left[ 89 N_c^2 - 24 N_c + 18 \right]  \,.
\ea 

The two-point function $\Pi_{LR}^T(Q^2)$ is  independent of
the scale $\nu$ in $D=4$
\begin{equation}\label{A7}
\frac{d}{d\nu}\Big(Q^6\, \Pi_{LR}^T (Q^2)\Big \vert_{D=6}\Big )=0 \, .
\end{equation}
This is also true in $D$ dimensions if 
$\gamma_5$ is anti-commuting like in the NDR scheme. 
The HV results are obtained from the NDR ones using 
the published results in  \cite{CFMR94}.

In $D=4-2\epsilon$ (\ref{A7})  yields the general condition
\ba
0&=&\dis \sum_{k=0} a^k(\nu) 
\left(  \beta(a) (k+1) - 2 \epsilon k) C_i^{(k)}(\nu,Q^2)\, <O_i>(\nu)
\right. \nonumber \\ 
&+& \left. \nu \frac{d C_i^{(k)}(\nu,Q^2)}{d \nu} <O_i>(\nu) 
+ C_i^{(k)}(\nu,Q^2) \, \nu \frac{d<O_i>(\nu)}{d\nu} \right)
\ea
with
\be
\nu \frac{da(\nu)}{d\nu} = a \, (\beta(a) - 2\epsilon)
\ee
and
$\beta(a)= \sum_{k=1} \, \beta_{k} a(\nu)^{k} $
with  first coefficient $\beta_1=1- 11 N_c/6$ for $n_f=3$.

To order $a(\nu)^0$,  one gets 
\begin{equation}
\frac{dC_i^{(0)}(\nu,Q^2)}{d\nu}=0
\end{equation}
so the $C_i^{(0)}$ are constants.

To order $a(\nu)$
\begin{eqnarray}
&& \beta_1 C_1^{(0)}+\nu
\frac{dC_1^{(1)}(\nu,Q^2)}{d\nu}- \gamma_{77}^{(1)} C_1^{(0)}
-2 \epsilon C_1^{(1)}=0 \, , \nonumber \\
&&\beta_1 C_2^{(0)}+  \nu
\frac{dC_2^{(1)}(\nu,Q^2)}{d\nu}- \gamma_{88}^{(1)}C_2^{(0)}
+\frac{1}{6} \gamma_{87}^{(1)} C_1^{(0)} 
-2 \epsilon C_2^{(1)}=0\, .
\end{eqnarray}
Integrating these two equations we obtain
\begin{eqnarray}
\label{C11}
C_1^{(1)} (\nu,Q^2)&=&
\frac{D_1^{(1)}}{2 \epsilon}+
\left(\frac{Q^2}{\nu^2}\right)^{- \epsilon} 
\left[ - \frac{{D_1^{(1)}}}{2\epsilon} + F_1^{(1)}\right]\, ,
\nonumber \\ 
C_2^{(1)} (\nu,Q^2)&=&
\frac{D_2^{(1)}}{2 \epsilon}+
\left(\frac{Q^2}{\nu^2}\right)^{- \epsilon} 
\left[ - \frac{{D_2^{(1)}}}{2\epsilon} + F_2^{(1)}\right]
\end{eqnarray}
with
\begin{equation}
\label{D11}
D_1^{(1)}=C_1^{(0)} \left[
\beta_1 - \gamma_{77}^{(1)} \right] \, ; \hspace{1 cm}
D_2^{(1)}=C_2^{(0)} \left[ 
\beta_1 -\gamma_{88}^{(1)} \right] +
\frac{1}{6} C_1^{(0)} \, \gamma_{87}^{(1)}
\end{equation}
which are valid in $D=4-2\epsilon$. 
The coefficients $C_i^{(0)}$,
$D_i^{(1)}$, and $F_i^{(1)}$ depend on $\epsilon$.
The anomalous dimensions $\beta_1$
and $\gamma_{ij}$ do not depend on $\epsilon$ in $MS$,
and in $\overline{MS}$ schemes in a known fashion.

\subsection{Calculation of the Constants $C_i^{(0)}$ and $F_i^{(1)}$}
\label{appCi}

The bare vacuum expectation value of $<O_1>$ can be expressed as an
integral as follows
\begin{equation} 
\label{O1integral}
<O_1>^{\rm bare}=-\frac{i}{2}g_{\mu \nu}\int
\frac{d^Dq}{(2\pi)^D}\Pi^{\mu\nu}_{LR}(q)=
\frac{D-1}{2}\int\frac{d^DQ}{(2\pi)^D} (Q^2\Pi_{LR}^T(Q^2)) \, . 
\end{equation}
The scheme used here to regularize this integral is 
the  $\overline{MS}$ scheme with $D=4-2\epsilon$, 
\begin{equation}\label{A2}
<O_1>^{\rm bare}=\frac{3-2\epsilon}
{32\pi^2}\frac{(4\pi)^{\epsilon}}
{\Gamma(2-\epsilon)}\int _0^{\infty} dQ^2 (Q^2)^{1-\epsilon}
(Q^2 \Pi_{LR}^T(Q^2))
\end{equation}
Notice that $<O_1>^{\rm bare}$  is scale independent.
The integral (\ref{A2}) diverges due to the high energy behaviour of
$\Pi_{LR}^T(Q^2)$. It is enough then to use the large $Q^2$
expansion of $\Pi_{LR}^T(Q^2)$ in $D$ dimensions.
This is a series in $\left(1/Q^2\right)^n$ starting at
$n=3$ in the chiral limit, Eq. (\ref{dimension6}).
Each coefficient of this series is finite
and can be written as a Wilson coefficient times the vacuum 
expectation value of some operator.
We now put (\ref{dimension6}) and (\ref{C11})
in (\ref{A2}) and perform the integral to find the divergent part.
For that we need the integral,
\begin{eqnarray}
\int_{\mu^2}^{\infty}
dQ^2\frac{1}{(Q^2)^{1+\epsilon}}&=&\frac{1}
{\epsilon}\mu^{-2\epsilon}\,.
\end{eqnarray}
We will set $\mu=\nu$ afterwards.

The $\overline{MS}$ subtraction needed then gives the full 
dependence on $\nu$.
\begin{eqnarray}
\label{o1run}
\nu\frac{d<O_1>^{\overline{MS}}(\nu)}{d\nu}
&=&\frac{3}{16\pi^2}a(\nu)\left\lbrace
\overline C_1^{(0)} + a(\nu)\left\lbrace  
\frac{G_1^{(1)}}{2} + \frac{\overline D_1^{(1)}}{6} + \overline F_1^{(1)} 
\right\rbrace\right\rbrace<O_1>^{\overline{MS}}(\nu)\nonumber \\
&+&\frac{3}{16\pi^2}a(\nu) \left\lbrace \overline C_2^{(0)} +a(\nu) 
\left\lbrace 
\frac{G_2^{(1)}}{2} + \frac{\overline D_2^{(1)}}{6}+\overline F_2^{(1)}
\right\rbrace\right\rbrace<O_2>^{\overline{MS}}(\nu) \, .\nonumber \\
\end{eqnarray}
Overlined quantities are in four dimensions and
\ba
G_i^{(1)} &=& \lim_{\epsilon\to 0}\frac{1}{\epsilon}
 \left(D_i^{(1)}-\overline D_i^{(1)}\right)\,.
\ea
Comparing (\ref{o1run}) and  (\ref{EGR}) 
order by order in $a$ and using (\ref{D11}), we get
up to the needed order in $\epsilon$
\ba
\label{resPILR}
C_1^{(0)}&=& -\frac{16\pi^2}{3} 
\left[ \gamma_{77}^{(1)}+ p_{77} \, \epsilon \right]
\,;
 \nonumber \\
C_2^{(0)}&=&\frac{8 \pi^2}{9}
\left[ \gamma_{87}^{(1)} + p_{87} \, \epsilon \right];
\ea
$\overline{D_1^{(1)}}$, $\overline{D_2^{(1)}}$, $G_1^{(1)}$
and   $G_2^{(1)}$ are then determined up to the $p_{ij}$ from Eq. (\ref{D11}).
We also get
\ba
\label{F11}
\overline{F}_1^{(1)}&=& - \frac{16 \pi^2}{3} \gamma_{77}^{(2)}-
\frac{1}{6}\overline{D}_1^{(1)}-\frac{1}{2}G_1^{(1)}\,;
\nonumber\\
\overline{F}_2^{(1)}&=& \frac{ 8 \pi^2}{9 } \gamma_{87}^{(2)}
-\frac{1}{6}\overline{D}_2^{(1)}-\frac{1}{2}G_2^{(1)}\,.
\ea
The constants $p_{ij}$ we determine below.

\subsection{The constants $p_{ij}$}

We now evaluate Eq. (\ref{A2}) to ${\cal O}(a)$ fully with its subtraction
in dimensional regularization using the same split in the integral
at $\mu^2$ as we used in the main text. The short-distance dimension
six part is the only divergent part, now regulated by dimensional
regularization rather than the $X$-boson propagator as in the main text.
The result is
\ba
<O_1>^{\overline{MS}}(\nu) &=&
\frac{3}{32\pi^2}a(\nu)\Bigg[\left(\frac{1}{3}\overline{C}_1^{(0)}
-\frac{16\pi^2}{3}p_{77}\right)<O_1>^{\overline{MS}}(\nu)
\nonumber\\
&+&\left(\frac{1}{3}\overline{C}_2^{(0)}
+\frac{8\pi^2}{9}p_{87}\right)<O_2>^{\overline{MS}}(\nu)\Bigg]
-\frac{3}{32\pi^2}{\cal A}_{LR}(\nu).
\ea
Comparison with Eq. (\ref{numresult}) and (\ref{numresultHV})
allows to determine $p_{77}$ and $p_{87}$. The finite coefficients
there are basically the $\Delta r_{ij}$ that corrected for the
dimensional regularization to the $X$-boson scheme. If one works
fully in dimensional regularization, it is here that these finite parts
surface.

The result is
\ba
p_{77}^{NDR}&=&-\frac{3}{4 N_c}\;; \quad 
p_{87}^{NDR}=\frac{3}{4}\;;
\nonumber\\
p_{77}^{HV}&=&-\frac{9}{4 N_c}\;; \quad 
p_{87}^{HV}=\frac{9}{4} \, .
\ea 
The transition between both agrees with the results in \cite{CFMR94}.

Putting in (\ref{NDR}) and (\ref{HV}) to obtain numerical values
\ba
-\frac{3}{16\pi^2}\overline{C}_1^{(0)}&=&
%\frac{3}{2 N_c} =
 \frac{1}{2}\, ;
\quad\quad
-\frac{3}{16\pi^2}\overline{D}_1^{(1)}
 =  -\frac{5}{2} \, ;
\nonumber \\
\frac{9}{8\pi^2}\overline{C}_2^{(0)}&=-&\frac{3}{2}  ;
\quad\quad
\frac{9}{8\pi^2} \, 
\overline{D}_2^{(1)}
%&=& \frac{1}{2 N_c}[N_c^2- 3 N_c+9]
=\frac{3}{2} \, .
\ea
For (\ref{F11}), in the NDR case we get
\ba
-\frac{3}{16\pi^2}\overline{F}_1^{NDR(1)} &=&
% \frac{1}{96 N_c^2} \left[ 181 N_c^2 
% - 156  N_c + 81 \right] = 
\frac{13}{16}\, \quad\quad
 \frac{9}{8\pi^2}\overline{F}_2^{NDR(1)} 
% &=& \frac{-1}{48 N_c} \left[ 104 N_c^2 
% - 78 N_c + 27 \right]
  = -\frac{75}{16} 
\ea
and in the HV scheme
\ba
-\frac{3}{16\pi^2} \overline{F}_1^{HV(1)} &=& 
% -\frac{3}{16\pi^2} F_1^{NDR(1)} + \frac{3}{2 N_c } \beta_1 
% + \frac{3}{2}  =
 \frac{41}{16}\,; \quad\quad
\frac{9}{8\pi^2}\overline{F}_2^{HV(1)} 
% &=&  \frac{9}{8\pi^2} F_2^{NDR(1)} - \frac{3}{2} \beta_1 
%- \frac{3}{2} N_c   
= - \frac{63}{16}\,.
\ea
All the expressions above are for $n_f=3$ flavours.

\section{Calculation of the Corrections of $\mathcal{O}(a^2)$ to the
Dimension Six Contribution to $\Pi_{SS+PP}^{(0-3)\, \rm conn}(Q^2)$}
\label{AppB}

\subsection{Renormalization Group Analysis}

The function we have to study here is 
\be
\label{B1p}
\Pi_{SS+PP}^{(0-3)}(q)\equiv i\int\,d^Dy \, e^{iy \cdot q}\,
\langle 0\vert T[(S+iP)^{(0-3)}(y)(S-iP)^{(0-3)}(0)]\vert 0 \rangle
\ee
with the definitions appearing in Section \ref{Q7Q8}.

The contribution of dimension six to the connected part of 
$\Pi_{SS+PP}^{(0-3)}(Q^2)$ can be written as (\ref{SVZSP})
\be\label{expsspp}
Q^4\,\Pi_{SS+PP}^{(0-3)\, \rm conn}(Q^2)\Big \vert _{D=6}
= \nu^{2 \epsilon}
\sum_{i=1,2} \tilde C_i(\nu,Q^2)  \langle O_i \rangle(\nu) \, 
\ee
with
\begin{equation}
\label{SSWilson}
\tilde C_i (\nu,s)=a(\nu)\sum_{k=0} a^{k}(\nu) \tilde C_i ^{(k)} (\nu,s)
\end{equation}
and the operators $O_1$  and $O_2$ were 
defined in (\ref{operadores}). 

From (\ref{scaleSS}) and (\ref{dispersionSS}),
we have now  in $D=4-2\epsilon$, 
\be
\nu \frac{d}{d\nu}\Pi_{SS+PP}^{(0-3)}(Q^2)=2\gamma_m(\nu)
\, \Pi_{SS+PP}^{(0-3)}(Q^2) \, .
\ee
Using this relation and the renormalization group equations, 
we get
\begin{eqnarray}
\tilde C_1^{(1)} (\nu,Q^2)&=&
\frac{\tilde D_1^{(1)}}{2 \epsilon}+
\left(\frac{Q^2}{\nu^2}\right)^{- \epsilon} 
\left[ - \frac{{\tilde D_1^{(1)}}}{2\epsilon} 
+ \tilde F_1^{(1)}\right]\, ,
\nonumber \\ 
\tilde C_2^{(1)} (\nu,Q^2)&=&
\frac{\tilde D_2^{(1)}}{2 \epsilon}+
\left(\frac{Q^2}{\nu^2}\right)^{- \epsilon} 
\left[ - \frac{{\tilde D_2^{(1)}}}{2\epsilon} 
+ \tilde F_2^{(1)}\right]
\end{eqnarray}
with
\ba
\label{D11tilde}
\tilde D_1^{(1)}&=&\tilde C_1^{(0)}\left[ 
\beta_1- 2\gamma_m^{(1)} - \gamma_{77}^{(1)} \right] 
 \, ,\nonumber\\
\tilde D_2^{(1)}&=&\tilde C_2^{(0)} 
\left[ \beta_1 - 2\gamma_m^{(1)}- \gamma_{88}^{(1)} \right]
+ \frac{1}{6} \tilde C_1^{(0)} \, \gamma_{87}^{(1)} \, .
\ea

In the next Section we determine the values of the constants
 $\tilde C_i^{(0)}$ and  $\tilde F_i^{(1)}$, 
which depend on $\epsilon$.

\subsection{Calculation of the Constants 
$\tilde C_i^{(0)}$ and $\tilde F_i^{(1)}$}
\label{AppA2}

The connected part of 
$\Pi_{SS+PP}^{(0-3)}(Q^2)$ can be related to the bare vacuum 
expectation value of the connected part of 
$<O_2>(\nu)$ through the relation \be\label{o2}
<O_2>^{\rm bare}_{\rm conn}(\nu)
=-i\,\int\frac{d^Dq}{(2\pi)^D}\,\Pi_{SS+PP}^{(0-3) \, \rm conn}(q)=
\int\frac{d^DQ}{(2\pi)^D}\,\Pi_{SS+PP}^{(0-3) \, \rm conn}(Q^2)
\ee
In the $\overline {MS}$ scheme with $D=4-2\epsilon$ and with renormalized 
$\Pi_{SS+PP}^{(0-3)}(Q^2)$
\be\label{O2MS}
<O_2>^{\rm bare}_{\rm conn}(\nu)
=\frac{(4\pi)^{\epsilon}}
{16 \pi^2 \, \Gamma(2-\epsilon)}\,\int _0 ^{\infty} \, dQ^2\,
(Q^2)^{1-\epsilon}\,  \Pi_{SS+PP}^{(0-3)\, \rm conn}(Q^2)
\ee

Proceeding analogously to the case of $\Pi_{LR}^T(Q^2)$
in Appendix \ref{appCi} and using that 
there is now a non vanishing  contribution coming 
from the anomalous dimensions of $\Pi_{SS+PP}^{(0-3)\, \rm conn}$, 
namely, 
\be
\int\frac{d^DQ}{(2\pi)^D}\,
\nu \, \frac{d \Pi_{SS+PP}^{(0-3)\, \rm conn}(Q^2)}{d\nu}
=2\gamma_m(\nu) <O_2>^{\rm bare}_{\rm conn}(\nu)
\ee
that we have to add to the one from the $\nu$-dependence
of the subtraction determined by the integration of $Q^2$ in 
(\ref{o2}).

The scale dependence  of  
the total $<O_2>$ can be obtained by adding both, we get 
in $D=4-2\epsilon$
\begin{eqnarray}
\nu\frac{d<O_2>^{\overline{MS}}(\nu)}{d\nu}
&=&\frac{1}{8\pi^2}a(\nu)\left\lbrace
\overline{\tilde C}_1^{(0)}+a(\nu)\left\lbrace 
\frac{\tilde G_1^{(1)}}{2} + \frac{\overline{\tilde D}_1^{(1)}}{2} +
\overline{\tilde F}_1^{(1)}
\right\rbrace \right\rbrace<O_1>^{\overline{MS}}(\nu)\nonumber \\
&+&\frac{1}{8\pi^2}a(\nu)\left\lbrace \overline{\tilde C}_2^{(0)}+
a(\nu)\left\lbrace  \frac{\tilde G_2^{(1)}}{2}
+ \frac{ \overline{\tilde D}_2^{(1)}}{2}
+ \overline{\tilde F}_2^{(1)}\right\rbrace \right \rbrace
<O_2>^{\overline{MS}}(\nu) \nonumber \\
&+&2\,\gamma_m <O_2>^{\overline{MS}}(\nu)
\end{eqnarray}
Again the barred quantities have to be taken at  $\epsilon=0$ and
\ba
\tilde G_i^{(1)} = \lim_{\epsilon\to0}
\frac{1}{\epsilon}\left(\tilde D_i^{(1)}-\overline{\tilde D}_i^{(1)}\right)\,.
\ea

Comparing this equation with (\ref{EGR}) order by order in $a(\nu)$ one 
obtains,
\begin{eqnarray}
\label{resPISP}
\tilde C_1^{(0)}&=& 48 \pi^2  p_{78}\, \epsilon \, ;
 \quad\quad\quad\quad
\tilde C_2^{(0)}= -8 \pi^2 \, 
\left[ \gamma_{88}^{(1)}+ 2 \gamma_m^{(1)}+  
p_{88} \, \epsilon \right] \, ;
\nonumber \\ 
\overline{\tilde F}_1^{(1)}&=& 48\pi^2 \gamma_{78}^{(2)}
-\frac{1}{2}\tilde G_1^{(1)}
-\frac{1}{2}\overline{\tilde D}_1^{(1)}
\, ;\nonumber \\
\overline{\tilde F}_2^{(1)}&=& -8 \pi^2
\left[ \gamma_{88}^{(2)}+2\gamma_m^{(2)}\right]
-\frac{1}{2}\tilde G_2^{(1)}
-\frac{1}{2}\overline{\tilde D}_2^{(1)}
\end{eqnarray}
Using Eq. (\ref{D11tilde}) everything can then be determined in terms of
the $p_{ij}$.

\subsection{Calculation of the $p_{ij}$.}

We now evaluate also the finite part from Eq. (\ref{B1p}) fully in dimensional
regularization to ${\cal O}(a)$ and obtain
\ba
<O_2>^{\overline{MS}}(\nu) &=&
\frac{1}{16\pi^2}a(\nu)\Bigg[\left(\overline{\tilde C}_1^{(0)}
+{48\pi^2}p_{78}\right)<O_1>^{\overline{MS}}(\nu)
\nonumber\\
&+&\left(\overline{\tilde C}_2^{(0)}
-{8\pi^2} p_{88}\right)<O_2>^{\overline{MS}}(\nu)\Bigg]
+3<0|\bar qq|0>^2(\nu)
\nonumber\\
&+&\frac{1}{16\pi^2}{\cal A}_{SP}(\nu)\,.
\ea
Comparison with Eq. (\ref{O6operator})
allows to determine $p_{78}$ and $p_{88}$. The finite coefficients
there are basically the $\Delta r_{ij}$ that corrected for the
dimensional regularization to the $X$-boson scheme. If one works
fully in dimensional regularization, it is here that these finite parts
surface for the $Q_8$ contribution.

The results are
\ba
p_{88}^{NDR}&=&-\frac{5}{4} N_c - \frac{1}{4 N_c}\;; \quad 
p_{78}^{NDR}=\frac{3}{2} \;;
\nonumber\\
p_{88}^{HV}&=&-\frac{9}{4} N_c +\frac{11}{4 N_c}\;; \quad 
p_{78}^{HV}=-\frac{1}{2} \, .
\ea 
again agreeing with the transition between both from 
\cite{CFMR94}.

Putting numbers, we get
\ba
\overline{\tilde C}_1^{(0)} &=& 
\overline{\tilde C}_2^{(0)} =
\overline{\tilde D}_1^{(1)} =
\overline{\tilde D}_2^{(1)} = 0\;
\nonumber\\
\frac{1}{48\pi^2} \overline{\tilde F}_1^{(1)} 
&=&  \frac{15}{32}\,; \quad\quad
 -\frac{1}{8\pi^2}\overline{\tilde F}_2^{(1)}
  = -\frac{211}{32}\;
\ea
which are scheme independent.
All the expressions above are for $n_f=3$ flavours.

\chapter{Analytical formulas for $K\to 3\pi$ decays}

\label{appendixK3pi}

Here we give some analytical formulas for the amplitudes and the 
CP-violating parameters in the $K\to 3\pi$, for which we give 
numerical results in Chapter \ref{chK3pi}. We also explain the method 
followed in the calculation of the dominant FSI at NLO for these processes.

\section{$K\to 3 \pi$ Amplitudes at NLO}
\label{Amplitudes}

A general way of writing the decay amplitude for $K^+\rightarrow 3\pi$ at 
NLO including FSI effects also at NLO is
\ba \label{genamplitude}
 A(K^+\rightarrow 3\pi)\,(s_1,s_2,s_3)= G_8\,a_{8}(s_1,s_2,s_3)\,
+\,G_{27}\,a_{27}(s_1,s_2,s_3)\,+\,e^2 G_E\,a_{E}(s_1,s_2,s_3)\,
\hspace{0.5 cm}&&\nonumber\\
+\,F^{(4)}(\widetilde K_i,Z_i,s_1,s_2,s_3)
\,+\,i\,F^{(6)}(\widetilde K_i,Z_i,s_1,s_2,s_3)\,.
\hspace{2 cm}&&
\ea
While for the corresponding CP conjugate the amplitude is
\ba
A(K^-\rightarrow 3\pi)\,(s_1,s_2,s_3)= G_8^*\,a_{8}(s_1,s_2,s_3)\,
+\,G_{27}\,a_{27}(s_1,s_2,s_3)\,+\, e^2 G_E^*\,a_{E}(s_1,s_2,s_3)\,
\hspace{0.5 cm}&&\nonumber\\
+\,F^{(4)}(\widetilde K_i^*,Z_i^*,s_1,s_2,s_3) 
\,+\,i\,F^{(6)}(\widetilde K_i^*,Z_i^*,s_1,s_2,s_3)\,.
\hspace{2 cm}&&
\ea
The energies $s_i$ are defined in Section \ref{cpK3pi},
 the $\widetilde K_i$  and $Z_i$ 
are counterterms appearing at  $\order(p^4)$
 and $\order(e^2 p^2)$ respectively,
 see Table \ref{tabKdef} and (\ref{EMdeltaS1}) for definitions. The functions 
$F^{(4)}(s_1,s_2,s_3)$ and $F^{(6)}(s_1,s_2,s_3)$ are
\ba \label{F4F6def}
F^{(4)}(\widetilde K_i,Z_i,s_1,s_2,s_3)&=&\sum_{i=1,11}H_i^{(4)}(s_1,s_2,s_3)
\widetilde K_i
+\sum_{i=1,14}J_i^{(4)}(s_1,s_2,s_3)Z_i \,,\nonumber\\
F^{(6)}(\widetilde K_i,Z_i,s_1,s_2,s_3)
&=& \sum_{i=1,11}H_i^{(6)}(s_1,s_2,s_3)
\widetilde K_i
+ \sum_{i=1,14}J_i^{(6)}(s_1,s_2,s_3)Z_i \,.
\ea

The complex functions $a_i$ can be written in terms of real functions 
as
\ba \label{functionsNLO}
a_i\,(s_1,s_2,s_3)\,&=&\,B_i(s_1,s_2,s_3)\,+\,i\,C_i(s_1,s_2,s_3)
\ea
$B_i(s_1,s_2,s_3)$ and $C_i(s_1,s_2,s_3)$ are real functions 
corresponding to the dispersive and absorptive amplitudes respectively and 
admit a CHPT expansion
\ba \label{amplitudesNLO}
B_i(s_1,s_2,s_3) &=& B_i^{(2)}(s_1,s_2,s_3) + B_i^{(4)}(s_1,s_2,s_3)
+\order(p^6)\,,\nonumber\\
C_i(s_1,s_2,s_3) &=& C_i^{(4)}(s_1,s_2,s_3) + C_i^{(6)}(s_1,s_2,s_3)
+\order(p^8)\, ,
\ea
where the superscript ${(2n)}$ indicates that the function is 
$\order (p^{2n})$ in CHPT. 

The functions $B_i^{(2)}$, $B_i^{(4)}$,  
$C_i^{(4)}$ and the part depending on $\widetilde K_i$ of $F^{(4)}$ in 
(\ref{genamplitude}) and (\ref{amplitudesNLO}) 
for $i=8,27$, which
 correspond to the CP-conserving amplitudes up to order 
$\order (p^4)$ and without electroweak corrections, that is,
\ba
A(K \to 3\pi ) &=& \re G_8 \left(B_8^{(2)}(s_1,s_2,s_3)+B_8^{(4)}(s_1,s_2,s_3)
+iC_8^{(4)}(s_1,s_2,s_3)\right)\nonumber\\
&&+ G_{27} \left(B_{27}^{(2)}(s_1,s_2,s_3)+B_{27}^{(4)}(s_1,s_2,s_3)
+iC_{27}^{(4)}(s_1,s_2,s_3)\right)\nonumber\\
&&+F^{(4)}(\re \widetilde K_i,s_1,s_2,s_3)\,
\ea
were obtained in \cite{BDP03}. 
We calculated these amplitudes for all the decays defined in (\ref{defdecays}) 
and got total agreement with \cite{BDP03}. 
The explicit expressions can be found there taking into account that the 
relation between the functions defined here and those used in \cite{BDP03} is, 
for the charged Kaon decays, 
\ba
&&M_{10}(s_3)+M_{11}(s_1)+M_{11}(s_2)+M_{12}(s_1)(s_2-s_3)+M_{12}(s_2)(s_1-s_3)
\nonumber\\
&&\hspace{1 cm}=\re G_8 \left(B_8^{(2)}+B_8^{(4)}+iC_8^{(4)}
\right)_{(++-)}\nonumber\\
&&\hspace{2 cm}+ G_{27} \left(B_{27}^{(2)}+B_{27}^{(4)}
+iC_{27}^{(4)}\right)_{(++-)}+F^{(4)}_{(++-)}\,,\nonumber\\
&&M_{7}(s_3)+M_{8}(s_1)+M_{8}(s_2)+M_{9}(s_1)(s_2-s_3)+M_{9}(s_2)(s_1-s_3)
\nonumber\\
&&\hspace{1 cm}=\re G_8 \left(B_8^{(2)}+B_8^{(4)}+iC_8^{(4)}
\right)_{(00+)}\nonumber\\
&&\hspace{2 cm}+ G_{27} \left(B_{27}^{(2)}+B_{27}^{(4)}
+iC_{27}^{(4)}\right)_{(00+)}+F^{(4)}_{(00+)}\, .
\ea

The functions $B_i^{(2)}$, $B_i^{(4)}$,  
$C_i^{(4)}$ and the part depending on $\widetilde K_i$ of $F^{(4)}$ in 
(\ref{genamplitude}) and (\ref{amplitudesNLO}) 
were calculated for $i=8,27$ in \cite{BDP03}. 
We calculated these quantities and got total agreement with \cite{BDP03},
 the explicit expressions can be found there. The functions 
$C_i^{(6)}(s_1,s_2,s_3)$ (for i=8,27) and $F^{(6)}$ are associated to FSI 
at NLO coming 
from two loops diagrams and are discussed in Appendix \ref{FSI6}. 

We have also calculated the contributions 
of order $e^2p^0$ and $e^2 p^2$ from the CHPT Lagrangian in (\ref{lagdS1}) 
and  (\ref{EMdeltaS1}) in presence of strong interactions for all the 
$K \to 3\pi$ transitions, that fix the functions $B_E^{(2)}$,  $B_E^{(4)}$, 
$C_E^{(4)}$ and $J_i^{(4)}$. The results are in the Appendix B.1 of 
reference \cite{GPS03}.

In order to calculate the asymmetries in the slope g defined in 
(\ref{gdefinition}) we need to expand these amplitudes in powers of the 
Dalitz plots variables $x$ and $y$, 
\ba
x\equiv \frac{s_1-s_2}{m_{\pi^+}^2} & {\rm and}&
\quad y \equiv \frac{s_3-s_0}{m_{\pi^+}^2}\,.
\ea
The notation we are going to use here is
\be \label{yexpansion}
G_i^{(2n)}(s_1,s_2,s_3)\,=\,G_{i,0}^{(2n)}\,+\,y\,G_{i,1}^{(2n)} 
\,+\,\order(x,y^2)\,;
\ee
where the function $G_i(s_1,s_2,s_3)$ can be any of the functions 
$B_i(s_1,s_2,s_3)$, $C_i(s_1,s_2,s_3)$ defined in (\ref{functionsNLO}) or 
$H_i(s_1,s_2,s_3)$, $J_i(s_1,s_2,s_3)$ in 
(\ref{F4F6def}). The coefficients $G_{i,0(1)}^{(2n)}$ are real quantities 
that depend on the masses $m_{\pi}^2$, $m_K^2$, 
the pion decay constant and the strong counterterm 
couplings of $\order (p^4)$, i.e., $L_i^r$.

\section{The Slope $g$ and $\Delta g$ at LO and NLO} 
\label{Adeltag}

We have checked that the following relations
\begin{itemize}
  \item $F_0^2\, \re \left(e^2G_E\right)<<m_\pi^2\re G_8$ 
  \item $\im G_8<<\re G_8$
  \item $\im (e^2 G_E)<<\re G_8$
\end{itemize}
can be used in this and the next sections to simplify the analytical 
expressions. To obtain the numerical results included in the
 text we use the full expressions, with no simplifications. We have also
checked that the terms 
disregarded with the application of these relations generate very small 
changes in the numbers.

Using the simplifications above, 
the value of $g$ at LO can be written trivially as 
\be \label{gLO}
g^{\rm LO}\,=\,2\frac{B_{8,1}^{(2)}\re G_8+B_{27,1}^{(2)}G_{27}
+B_{E,1}^{(2)}\re \left(e^2G_E\right)}
{B_{8,0}^{(2)}\re G_8 +B_{27,0}^{(2)}G_{27}
+B_{E,0}^{(2)}\re \left(e^2G_E\right)}\,.
\ee
The expressions for $B_{8,j}^{(2)}$, $B_{27,i}^{(2)}$,
and $B_{E,i}^{(2)}$ needed above can be obtained from
the expressions of the corresponding $B$'s for the
charged Kaon decays $++-$ and $00+$  in Appendix \ref{ANLO}
 and expanding them 
as in (\ref{yexpansion}). The results we  get 
are in (\ref{gLOCN}). 

We consider now the NLO corrections to the slope  $g$. 
Disregarding the tiny CP-violating we have $g[K^+\to 3\pi]=
g[K^-\to 3\pi]$ at NLO we get 
\ba \label{AgNLO}
A_0^{\rm NLO} &=& \left\lbrack \sum _{i=8,27}\left(B_{i,0}^{(2)}
+B_{i,0}^{(4)}\right)\re G_i 
\,+\,\sum_{j=1,11}H_{j,0}^{(4)}\re \widetilde K_j
\right\rbrack^2
\,+\,\left\lbrack \sum _{i=8,27}\left(C_{i,0}^{(4)}\re G_i\right)
\right\rbrack^2\, ,\nonumber\\
A_y^{\rm NLO} &=& 2\,\Bigg\lbrace
 \left\lbrack \sum _{i=8,27}\left(B_{i,0}^{(2)}
+B_{i,0}^{(4)}\right)\re G_i \,+\,\sum_{j=1,11}H_{j,0}^{(4)}\re \widetilde K_j
\right\rbrack\nonumber\\ &&\times
\left\lbrack \sum _{i=8,27}\left(B_{i,1}^{(2)}
+B_{i,1}^{(4)}\right)\re G_i \,+\,\sum_{j=1,11}H_{j,1}^{(4)}\re \widetilde K_j
\right\rbrack
\nonumber\\
&&+\left\lbrack \sum _{i=8,27}\left(C_{i,0}^{(4)}\re G_i\right)
\right\rbrack
\times\left\lbrack \sum _{i=8,27}\left(C_{i,1}^{(4)}\re G_i\right)
\right\rbrack
\Bigg \rbrace\,
\ea
for the coefficients defined in (\ref{amp2}).

One can get   $g_{C(N)}$ at NLO 
using (\ref{AmpDeltag}) and the results above
  substituting the coefficients $C_{i,j}^{(4)}$, $B_{i,j}^{(2n)}$ and 
$H_{i,j}^{(4)}$ by their values calculated expanding in the Dalitz variables 
the results in \cite{BDP03}.
 
 The slope $g$ asymmetry  in (\ref{defDeltag}) can be written at LO as 
in (\ref{defDeltag}). At NLO $\Delta g$ depend on the sum 
$A^+_yA^-_0 + A^+_0 A^-_y$ -see (\ref{amp2}) and (\ref{AmpDeltag}). It 
can be obtained directly from 
(\ref{AgNLO}) where we have neglected the small CP-violating
effects. 

For the difference  $A^+_yA^-_0 - A^+_0 A^-_y$, we get 
\ba 
\left(A^+_yA^-_0 - A^+_0 A^-_y \right)_{\rm NLO}
&=& 4 {\cal A}_I\left\lbrack
\left({\cal A}_R^2-{\cal C}_R^2\right){\cal D}_R-2{\cal A}_R
{\cal B}_R{\cal C}_R
\right\rbrack+4 {\cal C}_I\left\lbrack
\left({\cal A}_R^2-{\cal C}_R^2\right){\cal B}_R
\right. \nonumber\\  &&\left.
+2{\cal A}_R
{\cal C}_R{\cal D}_R
\right\rbrack+4\left({\cal B}_I{\cal C}_R-{\cal D}_I{\cal A}_R\right)
\left({\cal A}_R^2+{\cal C}_R^2\right)
\, , 
\ea
where ${\cal A}_R$, ${\cal B}_R$, ${\cal C}_R$ and ${\cal D}_R$ 
contain the contributions from the real parts 
of the counterterms 
\ba
{\cal A}_R &=& \sum_{i=8,27,E} 
\left(B_{i,0}^{(2)}+B_{i,0}^{(4)}\right)\, \re G_i\, 
+\, \sum_{i=1,11}H_{i,0}^{(4)} \, \re \widetilde K_i \, ,\nonumber\\
{\cal B}_R &=& \sum_{i=8,27,E} 
\left(B_{i,1}^{(2)}+ B_{i,1}^{(4)}\right)\,
\re G_i\, +\,
  \sum_{i=1,11}H_{i,1}^{(4)}\, \re \widetilde K_i \,  ,\nonumber\\
{\cal C}_R &=& \sum_{i=8,27,E} 
\left(C_{i,0}^{(4)} +C_{i,0}^{(6)}\right)  \,
\re G_i\, 
+\,\sum_{i=1,11}H_{i,0}^{(6)} \, \re \widetilde K_i\, ,\nonumber\\
{\cal D}_R &=& \sum_{i=8,27,E} 
\left( C_{i,1}^{(4)} + C_{i,1}^{(6)}\right)  \,
\re G_i\, +\,\sum_{i=1,11}H_{i,1}^{(6)}\, \re \widetilde K_i\, .
\ea
While  ${\cal A}_I$, ${\cal B}_I$, ${\cal C}_I$ 
are the same expressions but substituting 
the real parts of the counterterms by their imaginary parts.

The coefficients  $B_{i,0(1)}^{(2n)}$, $C_{i,0(1)}^{(2n)}$ and 
$H_{i,0(1)}^{(2n)}$  defined in (\ref{yexpansion}) are real.

\section{The Quantities  $|A|^2$ and $\Delta |A|^2$ at LO and NLO} 
\label{ANLO}

Here we give the results for the quantities $A$ and $\Delta A$
defined in (\ref{defACN}) and (\ref{Deltapm}), respectively. 
They enter in the integrands of the decay rates 
$\Gamma$  in (\ref{eq:extrs})  and the CP-violating asymmetries
$\Delta \Gamma$, see (\ref{eq:dGLO}).

To simplify the analytical expressions, 
we have made use
 of the fact that  the imaginary  part of the counterterms 
is much smaller than their  real parts. 
The $|A_{C(N)}|^2$  which give the asymmetries 
$\Delta \Gamma$ at LO are in  (\ref{GammaLO}).  

The result for $\Delta |A_{C}|^2$ at LO
can be obtained  substituting  in (\ref{DGammaLO}) 
the functions $B_i^{(2)}(s_1,s_2,s_3)$ and $C_i^{(4)}(s_1,s_2,s_3)$  
for $i=8$ and $i=27$  by
\ba \label{DeltapmLOppm}
B_{8\,++-}^{(2)}(s_1,s_2,s_3) &=& i\,\frac{C\,F_0^4}{f_{\pi}^3f_K}
\,(s_3-\mkd-\mpd)\, ,\nonumber\\
B_{27\,++-}^{(2)}(s_1,s_2,s_3) &=& i\,\frac{C\,F_0^4}{f_{\pi}^3f_K}
\,\frac{1}{3}(13\mpd+3\mkd-13s_3)\, ,\nonumber\\
C_{8\,++-}^{(4)}(s_1,s_2,s_3) &=&  i\,\frac{C\,F_0^4}{f_{\pi}^3f_K}\,
\left(-\frac{1}{16\pi f_\pi^2} \right)\nonumber\\
&&\times\Bigg \lbrace \frac{1}{2}
\left\lbrack s_3^2-s_3(3m_{\pi}^2+m_K^2)+2m_{\pi}^2(m_K^2+m_{\pi}^2)
\right\rbrack\sigma(s_3)\nonumber\\
&+&
\frac{1}{6}\left\lbrack 4s_2^2+s_2(-4m_{\pi}^2+2m_K^2-s_3) 
+m_{\pi}^2(4s_3-2m_K^2-3m_{\pi}^2)\right\rbrack
\sigma(s_2)\nonumber\\
&+&({\rm exchange\,\, s_1 \,\,  and\,\, s_2\,\,  in\,\, the\,\, 
 second\,\, term})\Bigg\rbrace\,,\nonumber\\
C_{27\,++-}^{(4)}(s_1,s_2,s_3) &= &i\,\frac{C\,F_0^4}{f_{\pi}^3f_K}\,
\left(-\frac{1}{16\pi f_\pi^2} \right)\nonumber\\
&&\times\Bigg \lbrace \frac{1}{6}\left\lbrack-13s_3^2+3s_3(13m_{\pi}^2
+m_K^2)-2m_{\pi}^2(3m_K^2+13m_{\pi}^2)\right\rbrack
\sigma(s_3)\nonumber\\
&&+\frac{1}{36(m_K^2-m_{\pi}^2)} \Big\lbrack s_2^2(14m_{\pi}^2+31m_K^2)
+s_2(26s_3(m_K^2-m_{\pi}^2)\nonumber\\
&&-174m_K^2m_{\pi}^2-7m_K^4+76m_{\pi}^4)+(104m_{\pi}^2s_3(m_{\pi}^2-m_K^2)
\nonumber\\
&&-168m_{\pi}^6+161m_K^2m_{\pi}^4+67m_K^4m_{\pi}^2) \Big\rbrack \,
\sigma(s_2)\nonumber\\
&+&({\rm exchange\,\, s_1 \,\,  and\,\, s_2\,\,  in\,\, the\,\, 
 second\,\, term}) 
\Bigg \rbrace \,,
\ea
and $B_E^{(2)}$, $C_E^{(4)}$ by
\ba \label{DeltapmLOppmEM}
B_{E\,++-}^{(2)} &=& i\,\frac{C\,F_0^4}{f_{\pi}^3f_K}
\, \left(-2F_0^2\right)\,,\nonumber\\
C_{E\,++-}^{(4)} &=& i\,\frac{C\,F_0^4}{f_{\pi}^3f_K}\, 
\left(\frac{-F_0^2}{16\pi f_\pi^2}\right) 
\,\Bigg \lbrace (2m_{\pi}^2-s_3)\sigma(s_3)
\nonumber\\
&&+\frac{1}{4(m_K^2-m_{\pi}^2)} \left \lbrack 3s_2^2+s_2
(5m_K^2-12m_{\pi}^2)+m_{\pi}^2(5m_{\pi}^2-m_K^2)\right\rbrack
\sigma(s_2)\nonumber\\
&&+ ({\rm exchange\,\, s_1 \,\,  and\,\, s_2\,\,  in\,\, the\,\, 
 second\,\, term}) \Bigg \rbrace \, . 
\ea

One can get $\Delta |A_N|^2$ at LO substituting  in (\ref{DGammaLO}) 
the functions $B_i^{(2)}(s_1,s_2,s_3)$ and $C_i^{(4)}(s_1,s_2,s_3)$  
for $i=8$  and $i=27$ by  
\ba\label{DeltapmLO00p}
B_{8\,00+}^{(2)}(s_1,s_2,s_3) &=& i\,\frac{C\,F_0^4}{f_{\pi}^3f_K}
\,(\mpd-s_3)\, ,\nonumber\\
B_{27\,00+}^{(2)}(s_1,s_2,s_3) &=& i\,\frac{C\,F_0^4}{f_{\pi}^3f_K}
\,\frac{1}{6(\mkd-\mpd)}\left\lbrack 5m_K^4+19\mpd\mkd-4m_{\pi}^4
+s_3(4\mpd-19\mkd)\right\rbrack\, ,\nonumber\\
C_{8\,00+}^{(4)}(s_1,s_2,s_3) &=&  i\,\frac{C\,F_0^4}{f_{\pi}^3f_K}\,
\left(-\frac{1}{16\pi f_\pi^2} \right)\nonumber\\
&&\times\Bigg \lbrace \frac{1}{2}
\left\lbrack s_3^2+s_3(m_K^2-m_{\pi}^2)-m_{\pi}^2m_K^2
\right\rbrack\sigma(s_3)\nonumber\\
&+&
\frac{1}{6}\left\lbrack 2s_2^2+s_2(s_3-2(4m_{\pi}^2+m_K^2)) 
+m_{\pi}^2(-4s_3+5m_K^2+9m_{\pi}^2)\right\rbrack
\sigma(s_2)\nonumber\\
&+& ({\rm exchange\,\, s_1 \,\,  and\,\, s_2\,\,  in\,\, the\,\, 
 second\,\, term}) \Bigg\rbrace\,,\nonumber\\
C_{27\,00+}^{(4)}(s_1,s_2,s_3) &= &i\,\frac{C\,F_0^4}{f_{\pi}^3f_K}\,
\left(-\frac{1}{16\pi f_\pi^2} \right)\frac{1}{(\mkd-\mpd)}\nonumber\\
&&\times\Bigg \lbrace \frac{-1}{12}\Big\lbrack 26s_3^2(m_K^2-m_{\pi}^2)
+s_3(56m_{\pi}^4-57m_K^2m_{\pi}^2-14m_K^4)\nonumber\\
&&\hspace{0.5cm}+m_{\pi}^2(31m_K^2m_{\pi}^2-30m_{\pi}^4+19m_K^4)\Big\rbrack
\sigma(s_3)\nonumber\\
&&+\frac{1}{36} \Big\lbrack s_2^2(-8m_{\pi}^2+38m_K^2)
+s_2(s_3(19m_K^2-4m_{\pi}^2)\nonumber\\
&&\hspace{0.5 cm} -144m_K^2m_{\pi}^2-23m_K^4+32m_{\pi}^4)
+s_3(16m_{\pi}^2-76m_K^2)m_{\pi}^2\nonumber\\
&&\hspace{0.5 cm}-36m_{\pi}^6+151m_K^2m_{\pi}^4+65m_K^4m_{\pi}^2) \Big\rbrack 
 \sigma(s_2)\nonumber\\
&&+ ({\rm exchange\,\,  s_1\,\,   and \,\, s_2 \,\, in\,\, the \,\, 
second\,\, term}) \Bigg \rbrace \,,
\ea
and 
\ba \label{DeltapmLO00pEM}
B_{E\,00+}^{(2)} &=& i\,\frac{C\,F_0^4}{f_{\pi}^3f_K}
\, \frac{F_0^2}{2(\mkd-\mpd)}\left(5\mpd-\mkd-3s_3\right)\,,\nonumber\\
 C_{E\,00+}^{(4)} &=& i\,\frac{C\,F_0^4}{f_{\pi}^3f_K}\, 
\left(\frac{-1}{16\pi(\mkd-\mpd)} \right)\nonumber\\
&&\times\Bigg \lbrace \frac{1}{4}\left\lbrack s_3(8m_K^2-5m_{\pi}^2)
+m_{\pi}^2(3m_{\pi}^2-7m_K^2)\right\rbrack\sigma(s_3)
\nonumber\\
&&+\frac{1}{4} \left \lbrack 2s_2^2+s_2
(s_3-3(m_K^2+2m_{\pi}^2))+m_{\pi}^2(-4s_3+5m_{\pi}^2+7m_K^2)\right\rbrack
\sigma(s_2)\nonumber\\
&&+\frac{1}{4} \left \lbrack 2s_1^2+s_1
(s_3-3(m_K^2+2m_{\pi}^2))+m_{\pi}^2(-4s_3+5m_{\pi}^2+7m_K^2\right\rbrack
\sigma(s_1) \Bigg \rbrace\,. \, 
\ea

The function $\sigma(s)$ appearing in all the formulas
above is
\be
\label{sigma}
\sigma(s)\,=\,\sqrt{1-\frac{4\mpd}{s}}\, .
\ee
In all the expressions at LO
we  use $f_K=f_\pi=F_0$.

At NLO,  we get
\ba
\label{eqANLO}
|A_{NLO}|^2 &=&  
\left\lbrack\left(B_8^{(2)}+B_8^{(4)}\right)\re G_8
+\left(B_{27}^{(2)}+B_{27}^{(4)}\right)G_{27}+\sum_{i=1,11}H_i^{(4)}\re 
\widetilde K_i
\right\rbrack^2\nonumber\\
&&+\left\lbrack C_8^{(4)}\re G_8 +C_{27}^{(4)}G_{27}\right\rbrack^2
\, ,\nonumber\\
\Delta |A_{NLO}|^2 &=&  
-2 \im G_8 \Bigg \lbrace \left\lbrack G_{27}\left(B_{27}^{(2)}+B_{27}^{(4)}
\right)+\sum_{i=1,11}H_i^{(4)}\re 
\widetilde K_i\right\rbrack\left(C_8^{(4)}+C_8^{(6)}\right)
\nonumber\\
&&\hspace{1. cm}-\left(B_{8}^{(2)}+B_{8}^{(4)}
\right)\left\lbrack G_{27}\left(C_{27}^{(4)}+C_{27}^{(6)}\right)
+H_i^{(6)}\re \widetilde K_i\right\rbrack
\Bigg\rbrace\nonumber\\
&&-2\im \left(e^2G_E\right)\Bigg \lbrace \left( \sum_{i=8,27}\left\lbrack
\re G_i\left(B_i^{(2)}+B_i^{(4)}
\right)\right\rbrack+\sum_{i=1,11}H_i^{(4)}\re 
\widetilde K_i\right)\,C_E^{(4)}\nonumber\\
&&\hspace{1. cm}-\left(B_{E}^{(2)}+B_{E}^{(4)}\right)\left( 
\sum_{i=8,27}\left\lbrack\re G_{i}\left(C_{i}^{(4)}+C_{i}^{(6)}\right)
\right\rbrack+\sum_{i=1,11}H_i^{(6)}\re \widetilde K_i\right)
\Bigg\rbrace\nonumber\\
&&+\left(2\sum_{i=1,11}H_i^{(4)}\im 
\widetilde K_i \right) \Bigg\lbrace
\sum_{i=8,27}\left\lbrack\re G_i \,\left(C_i^{(4)}+C_i^{(6)}
\right)\right\rbrack \nonumber\\
&&\hspace{7 cm}
+\sum_{i=1,11}H_i^{(6)}\re \widetilde K_i\Bigg\rbrace\nonumber\\
&&-2\left(\sum_{i=1,11}H_i^{(6)}\im \widetilde K_i\right)\Bigg\lbrace
\sum_{i=8,27}\left\lbrack\re G_i \,\left(B_i^{(2)}+B_i^{(4)}
\right)\right\rbrack \nonumber\\
&&\hspace{7 cm}+\sum_{i=1,11}H_i^{(4)}\re \widetilde K_i\Bigg\rbrace
\,.
\ea
Again, we disregarded the $s_i$ dependence of the functions $B_i^{(2n)}$, 
$C_i^{(2n)}$ and $H_i^{(2n)}$. The functions $B_{8(27)}^{(4)}$ and 
$H_i^{(4)}$ can be deduced from the results in \cite{BDP03} and 
the functions 
$B_{E}^{(4)}$ from Appendix B.1 of reference \cite{GPS03}. Finally, the functions 
$C_i^{(6)}$ and $H_i^{(6)}$ are discussed in Appendix \ref{FSI6}.

\section{Final State Interactions at NLO} 
\label{FSI6}

\begin{figure}
\begin{center}
\epsfig{file=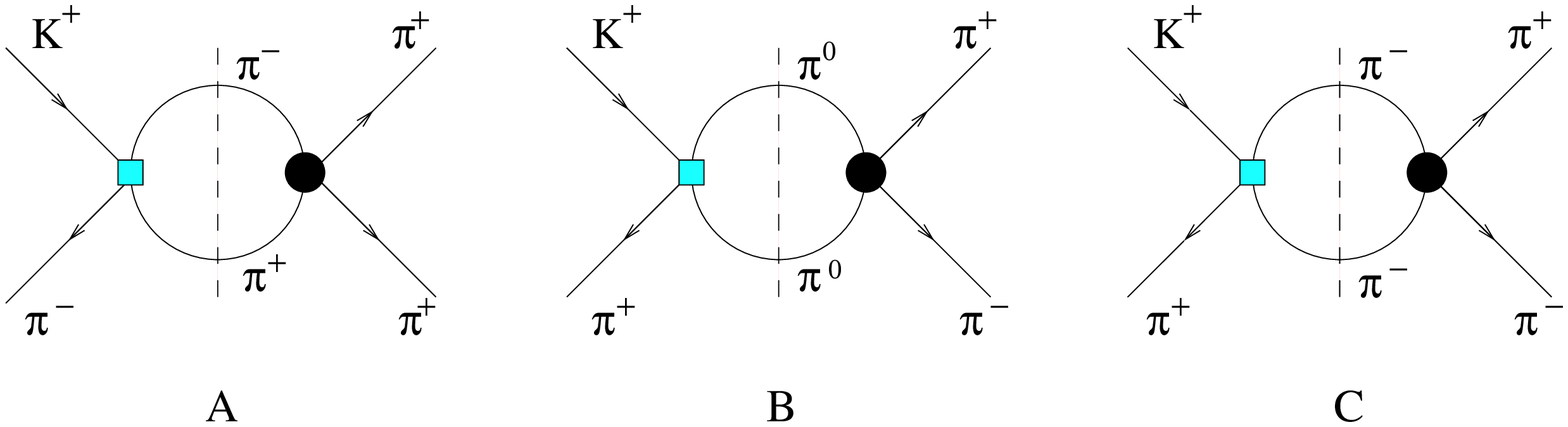,width=10cm}
\end{center}
\caption{\label{fig:imp6ppm} Relevant diagrams for the calculation of FSI for
 $K^+\rightarrow\pi^+\pi^+\pi^-$. 
The square vertex is the weak vertex and the
 round one is  the strong vertex}
\end{figure}

\begin{figure}
\begin{center}
\epsfig{file=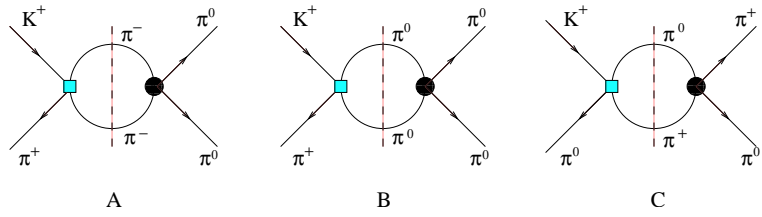,width=10cm}
\end{center}
\caption{\label{fig:imp6p00} Relevant diagrams for the calculation of FSI for
 $K^+\rightarrow\pi^0\pi^0\pi^+$. The square vertex is the weak 
vertex and the
 round one is  the strong vertex.}
\end{figure}

In this Appendix we provide some details of the calculation 
of the FSI using
the optical theorem in the framework of CHPT. We compute the imaginary
part of the amplitudes at ${\cal O}(p^6)$. 
The calculation corresponds to the diagrams shown 
in Figures \ref{fig:imp6ppm} and \ref{fig:imp6p00}.
We can distinguish the cases in which the weak vertex 
is of $ {\cal O}(p^4)$
and the strong vertex of order  $ {\cal O}(p^2)$ and the inverse 
case in which
the weak vertex is of order  $ {\cal O}(p^2)$ and the strong 
vertex of order 
 $ {\cal O}(p^4)$. In this paper we will not consider the weak 
vertices  generated by the electroweak penguin.
In  Subsection \ref{sec:not6} we provide  some notation. In Subsections 
\ref{sec:ppm6} and \ref{sec:p006}
we report the  calculation for the  charged Kaon decays.
An example of  the calculation of the integrals 
that must be performed is given in Subsection 
\ref{sec:integ6}. Finally, in 
Subsection \ref{phasesNLO} we give analytical results for 
the strong phases at NLO.

\subsection{Notation}
\label{sec:not6}
In order to be concise we use  the functions $M_i$  for the weak 
amplitudes given in \cite{BDP03}. We define
\eq\label{asbs}
\widetilde{M}_i(s)&=& \int_{-1}^1 d\cos\theta\;  \left. M_i(a(s)+b(s) \cos\theta)\right|_{p^4}\, ,\\
\widetilde{M}_i^s(s)&=& \int_{-1}^1 d\cos\theta\;  \left.  (a(s)+b(s) \cos\theta)
M_i(a(s)+b(s) \cos\theta)\right|_{p^4}\,,\\
%t&=& a+ b \cos\theta\nn \\
\widetilde{M}_i^{ss}(s)&=& \int_{-1}^1 d\cos\theta\; \left.  (a(s)+b(s) \cos\theta)^2 M_i(a(s)+b(s) \cos\theta)\right|_{p^4}\,,\\
a(s)&=& \fr{1}{2}( m_K^2+3m_\pi^2-s)\,,\nn\\
b(s)&=& \fr{1}{2}\sqrt{(s-4 m_\pi^2)\left(s-2  (m_K^2+m_\pi^2)+\fr{(m_K^2-m_\pi^2)^2}{s}\right)}\,.
\en

The amplitudes at ${\cal O}(p^4)$ for the $\pi\pi\rightarrow \pi\pi$ 
scattering  in a theory with three flavors can be found  in \cite{BKM91}.
We decompose the amplitudes in the various cases as follows.
For the case $\pi^+\pi^+\rightarrow \pi^+\pi^+$ the amplitude at ${\cal O}(p^4)$
is
\eq
\Pi_1&=& P_1(s)+P_2(s,t)+P_2(s,u) .
\en
For the case $\pi^0\pi^0\rightarrow \pi^+\pi^-$ the amplitude at ${\cal O}(p^4)$
is
\eq
\Pi_2&=& P_3(s)+P_4(s,t)+P_4(s,u) .
\en
For the case $\pi^+\pi^-\rightarrow \pi^+\pi^-$ the amplitude at ${\cal O}(p^4)$
is
\eq
\Pi_3&=& P_5(s)+P_6(s,t)+P_6(s,u)+P_7(s,t)-P_7(s,u) .
\en
Finally the amplitude
$\pi^0\pi^0\rightarrow \pi^0\pi^0$ at ${\cal O}(p^4)$
is
\eq
\Pi_4&=& P_8(s)+P_8(t)+P_8(u) .
\en
The value for the various $P_i$ can be deduced from \cite{BKM91}.
In the following we use
\eq\!\!\!\!\!\!\!\!\!
\widetilde{P}^{(n,m)}_i(s)&=& \int_{-1}^1 d\cos\theta\;s^n  
(c(s) (1- \cos\theta))^m 
P_i(s,c(s) (1- \cos\theta))\,,\\
\!\!\!\!\!\!\!\!\!
\widehat{P}^{(n)}_{1,i}(s)&=&\int_{-1}^1 d\cos\theta\;(c(s)(1-\cos\theta))^n
P_i(c(s)(1+\cos\theta),c(s)(1-\cos\theta))\,, \\
\!\!\!\!\!\!\!\!\!
\widehat{P}^{(n)}_{2,i}(s)&=&\int_{-1}^1 d\cos\theta\;(c(s)(1-\cos\theta))^n
P_i(c(s)(1-\cos\theta),s)\,, \\
\!\!\!\!\!\!\!\!\!
c(s)&=&-\fr{1}{2} (s-4 m_\pi^2)\,.
\en
Another function we use in the next subsections  is
$\sigma(s)$ which was defined already in (\ref{sigma}).

\subsection{Final State Interactions for $K^+\rightarrow\pi^+\pi^+\pi^-$}
\label{sec:ppm6}

We first  compute the  contributions
depicted in Figure  \ref{fig:imp6ppm} in which the weak vertex is of
${\cal O}(p^4) $ and the strong vertex of ${\cal O}(p^2) $.
 The results for the diagrams A and B are
\eq
{\rm Im}A^{(6,1)}_W&=& \fr{\sig(s_3)}{32\pi}
\frac{(2 m_\pi^2-s_3)}{f_{\pi}^2}\left[
  \left. M_{10}(s_3)\right|_{p^4}+ \widetilde{M}_{11}(s_3)+ 
\widetilde{M}_{12}(s_3) (m_K^2\right. \nonumber \\
&&+3 m_\pi^2-2
  s_3)\left. - \widetilde{M}_{12}^s(s_3)\right]
\,,\\
{\rm Im}A^{(6,2)}_W&=& \fr{\sig(s_1)}{32\pi}\frac{( s_1-m_\pi^2)}{f_{\pi}^2}\left[
  \left.M_{7}(s_1)\right|_{p^4}+\widetilde{M}_{8}(s_1)+\widetilde{M}_{9}(s_1) (m_K^2+3 m_\pi^2-2
  s_1)\right.\nonumber \\
&&
\left. -\widetilde{M}_{9}^s(s_1)\right] 
 +(s_1\leftrightarrow s_2)
\,,
\en
respectively. For diagram C we have both $S$-wave  and 
$P$-wave contributions. We get for them  
\eq 
\label{eq:im6,3ws}
{\rm Im}A^{(6,3)}_{W,S}&=& \fr{\sig(s_1)}{64\pi} \frac{s_1}{f_{\pi}^2}\left[
  2
  \left. M_{11}(s_1)\right|_{p^4}+\widetilde{M}_{11}(s_1)+\widetilde{M}_{10}(s_1)+\widetilde{M}_{12}^s(s_1) \right. \nonumber \\ &&
\left.
-\widetilde{M}_{12}(s_1) (m_K^2+3 m_\pi^2-2
  s_1) \right]  +(s_1\leftrightarrow s_2)
\,,  \\ 
\label{eq:im6,3wp}
{\rm Im}A^{(6,3)}_{W,P}&=& \fr{\sig(s_1)}{64\pi}\frac{1}{f_{\pi}^2} \fr{s_1 (s_3-s_2)}{s_1^2-2 (m_K^2+m_\pi^2) s_1+(m_K^2-m_\pi^2)^2}
\left[(s_1-(m_K^2+3 m_\pi^2))\right. \nonumber \\
\et
\times(\widetilde{M}_{11}(s_1)-\widetilde{M}_{10}(s_1)+\widetilde{M}_{12}(s_1)(2
s_1-m_K^2-3  m_\pi^2)) 
+2 \widetilde{M}^s_{11}(s_1) \nonumber \\
\et-2 \widetilde{M}^s_{10}(s_1)+\widetilde{M}^s_{12}(s_1)(5 s_1-3(m_K^2+3
  m_\pi^2))+2 \widetilde{M}^{ss}_{12}(s_1) \nonumber \\
\et
+\fr{8}{3} b^2(s_1) \left. M_{12}(s_1)\right|_{p^4}
  \Big] 
 +(s_1\leftrightarrow s_2)\, , 
\en
respectively.

Secondly, 
 we report the calculation of  the  case in which the strong vertex
 is of ${\cal O}(p^4)$ and the weak vertex is of ${\cal O}(p^2)$.
With analogous notation as above, we get
\eq
{\rm Im}A^{(6,1)}_\pi&=& \fr{\sig(s_3)}{32\pi} 
 \left.M_{10}(s_3)\right|_{p^2}\ 
(P_1(s_3)+\widetilde{P}^{(0,0)}_2(s_3))\,,
\\ 
{\rm Im}A^{(6,2)}_\pi&=& \fr{\sig(s_1)}{32\pi} \left. (M_{7}(s_1)+M_{8}(s_2)+M_{8}(s_3))\right|_{p^2}\, 
(P_3(s_1)+\widetilde{P}^{(0,0)}_4(s_1))\nonumber
\\ \et
+(s_1\leftrightarrow s_2) \,,
\\ 
{\rm Im}A^{(6,3)}_{\pi,S}&=& \fr{\sig(s_1)}{32\pi}\left(\left.M_{10}(s_3)\right|_{p^2}+\left.M_{10}(s_2)\right|_{p^2}\right)
(P_5(s_1)+\widetilde{P}^{(0,0)}_6(s_1))
\nonumber \\ \et
+(s_{1}\leftrightarrow s_{2}) \,,
\\
{\rm Im}A^{(6,3)}_{\pi,P}&=&
\fr{\sig(s_1)}{32\pi}\left(\left.M_{10}(s_3)\right|_{p^2}-\left.M_{10}(s_2)\right|_{p^2}\right)\fr{1}{s_1-4 m_\pi^2}
\Big( (s_1-4 m_\pi^2)\widetilde{P}_7^{(0,0)}(s_1)
\nonumber \\ \et
+2\widetilde{P}_7^{(0,1)}(s_1)\Big) +(s_{1}\leftrightarrow s_{2})\, .
\en
The final result for the Im$A^{(6)}$ is given by the sum
\eq
\label{eq:sumFSI}
 {\rm Im} A^{(6)}&=& \sum_{i=1,2;\ j=W,\pi}{\rm Im} A^{(6,i)}_{j}+\sum_{j=W,\pi;\
 k=S,P}{\rm Im} A^{(6,3)}_{j,k}\,.
\en

The relation between 
this imaginary amplitude and the functions defined in Appendix 
\ref{Amplitudes} is
\begin{equation}
\im A^{(6)} = \sum _{i=8,27} G_i\,C_i^{(6)}(s_1,s_2,s_3)\,+\,
\sum_{i=1,11}H_i^{(6)}(s_1,s_2,s_3)\widetilde K_i \, .
\end{equation}
This relation is also valid for $K^+\to \pi^0\pi^0\pi^+$.

\subsection{Final State Interactions for  $K^+\rightarrow\pi^0\pi^0\pi^+$}
\label{sec:p006}

The calculation is analogous to the one for 
$K^+\rightarrow\pi^+\pi^+\pi^-$.
The relevant graphs are depicted in Figure \ref{fig:imp6p00}. 
In the case  in which the weak vertex is of ${\cal O}(p^4)$, we get 
\eq
{\rm Im}A^{(6,1)}_W&=& \fr{\sig(s_3)}{32\pi}\frac{(s_3- m_\pi^2)}
{f_{\pi}^2}\left[
 2 M_{11}(s_3)+ \widetilde{M}_{11}(s_3)+ \widetilde{M}_{10}(s_3)- 
\widetilde{M}_{12}(s_3) (m_K^2\right. \nonumber \\
&&\left.
+3 m_\pi^2-2
  s_3) + \widetilde{M}_{12}^s(s_3)\right]
\,,\\
{\rm Im}A^{(6,2)}_W&=& \fr{\sig(s_3)}{32\pi}\frac{m_\pi^2}{f_{\pi}^2}\left[
  M_{7}(s_3)+\widetilde{M}_{8}(s_3)+\widetilde{M}_{9}(s_3) (m_K^2+3 m_\pi^2-2
  s_3)\right.\nonumber \\
&&
\left. -\widetilde{M}_{9}^s(s_3)\right] \,,
\\
{\rm Im}A^{(6,3)}_{W,S}&=& \fr{\sig(s_1)}{64\pi} \frac{(2 m_\pi^2-s_1)}{f_{\pi}^2}\left[
   2 M_{8}(s_1)+\widetilde{M}_{8}(s_1)+\widetilde{M}_{7}(s_1)-\widetilde{M}_{9}(s_1) (m_K^2\right. \nonumber \\ &&+3 m_\pi^2-2
  s_1)
\left.+\widetilde{M}_{9}^s(s_1)\right]
 +(s_1\leftrightarrow s_2)\,.
\en

Also in this case diagram  C generates both 
 $S$-wave and  $P$-wave contributions.
The P-wave contribution due to the diagram  C in \ref{fig:imp6p00} 
 is
\eq
{\rm Im}A^{(6,3)}_{W,P}&=& \fr{\sig(s_1)}{64\pi}\frac{1}{f_{\pi}^2} 
\fr{s_1 (s_3-s_2)}{s_1^2-2 (m_K^2+m_\pi^2) s_1+(m_K^2-m_\pi^2)^2}
\left[(s_1-(m_K^2+3 m_\pi^2))\right. \nonumber \\
\et
\times(\widetilde{M}_{8}(s_1)-\widetilde{M}_{7}(s_1)+\widetilde{M}_{9}(s_1)(2
s_1-m_K^2-3  m_\pi^2)) 
+2 \widetilde{M}^s_{8}(s_1) \nonumber \\
\et-2 \widetilde{M}^s_{7}(s_1)+\widetilde{M}^s_{9}(s_1)(5 s_1-3(m_K^2+3
  m_\pi^2))+2 \widetilde{M}^{ss}_{9}(s_1) \nonumber \\
\et
+\fr{8}{3} b^2(s_1) M_{9}(s_1)
  \Big] 
 +(s_1\leftrightarrow s_2) \,.
\en

If the strong  vertex  is ${\cal O}(p^4)$ and the weak
vertex is  order ${\cal O}(p^2)$, we get
\eq
{\rm Im}A^{(6,1)}_\pi&=& \fr{\sig(s_3)}{32\pi} 
 \left.(M_{10}(s_1)+M_{10}(s_2))\right|_{p^2}\ 
(P_3(s_3)+\widetilde{P}^{(0,0)}_4(s_3)) \,,
\\
{\rm Im}A^{(6,2)}_\pi&=& \fr{\sig(s_3)}{32\pi} \left. (M_{7}(s_3)+M_{8}(s_1)+M_{8}(s_2))\right|_{p^2}\, 
(P_8(s_3)+\widetilde{P}^{(0,0)}_8(s_3)) \,,\nonumber 
\\ \\
{\rm Im}A^{(6,3)}_{\pi ,S}&=& \fr{\sig(s_1)}{64\pi}\left.\left(2 M_{8}(s_1)+M_{8}(s_2)+M_{8}(s_3)+M_{7}(s_2)+M_{7}(s_3)\right)\right|_{p^2}\nn\\ \et
\times \Big(\widetilde{P}^{(0,0)}_3(s_1)
+\widehat{P}^{(0)}_{1,4}(s_1)+\widetilde{P}^{(0)}_{2,4}(s_1)\Big)
+(s_{1}\leftrightarrow s_{2}) \,,
\\
{\rm Im}A^{(6,3)}_{\pi ,P}&=&
\fr{\sig(s_1)}{64\pi}\left.\left(M_{7}(s_3)-M_{8}(s_3)-M_{7}(s_2)+M_{8}(s_2)\right)\right|_{p^2}\fr{1}{s_1-4 m_\pi^2}\nonumber \\ \et
\times \Big( (s_1-4 m_\pi^2)
(\widetilde{P}^{(0,0)}_3(s_1)-\widehat{P}^{(0)}_{1,4}(s_1)+\widetilde{P}^{(0)}_{2,4}(s_1))+2 \widetilde{P}^{(1,0)}_3(s_1)
\nonumber \\ \et
- 2 \widetilde{P}^{(1)}_{1,4}(s_1) +2\widehat{P}^{(1)}_{2,4}(s_1)\Big)
 +(s_{1}\leftrightarrow s_{2})
\,.
\en
The total contribution is given by the sum  
of (\ref{eq:sumFSI}) with the proper right-hand side terms.

\subsection{Integrals}
\label{sec:integ6}
The integrals necessary to compute the two-bubble 
FSI we discussed in the previous subsection can be calculated 
generalizing the method outlined in \cite{BCEGS97}.
As an example  we show the integration of the function
\eq
\label{eq:Bdefinition}
32 \pi^2 B(m_1,m_2,t)&=& C_B +\left\{\fr{2
    \eta\delta}{t}\ln\fr{\eta-\delta}{\eta+\delta}
+\fr{\la}{t}\ln\fr{(\la-t)^2-\eta^2\de^2}{(\la+t)^2-\eta^2\de^2}  \right\}
\en
where $C_B$ is a  term which does not depend on $t$, 
\eq
C_B &=&
2 \left(1-\ln\fr{\eta^2-\delta^2}{4\nu^2}\right)
\label{eq:cdef}
\en
and
\eq
\eta&=& m_1+m_2\nn\\
\de&=&  m_1-m_2\nn \\
\la&=& \sqrt{ \left[\left( t-\eta^2\right)\left( t-\de^2\right)\right]}\,.
\label{eq:etadelta}
\en
In the center of mass frame one  can define
\eq
Q&=& p_K+p_\pi=(\sqrt{s},0,0,0)\nn \\
p_\pi&=&\fr{\sqrt{s}}{2}\left(\left(1- \fr{m_K^2-m_\pi^2}{s}\right),0,0,
\sqrt{1-\fr{2 (m_K^2+m_\pi^2)}{s}+\fr{(m_K^2-m_\pi^2)^2}{s^2}}\right)\nonumber
\\
\en
where $p_\pi$ is the momentum of the external 
pion entering in the same vertex of the Kaon.
The functions $B$ can also be generated  in the strong vertex.
  In this case $p_k$ is the momentum of an external pion. 
The contribution to the imaginary part of the amplitude $A$ is
\eq
{\rm Im} \, A=\fr{1}{32\pi}\sig(s)\int_{-1}^1 
{\rm d} \,\cos\theta\, B[m_1,m_2,t]
\en
with 
\eq
t&=& a+ b \cos\theta
\en
and $a\equiv a(s)$, $b\equiv b(s)$ in (\ref{asbs}).
In order  to solve the difficult part of the integral one can put
\eq
\label{eq:tdef}
t=\fr{1}{2}\left[\eta^2+\de^2-(\eta^2-\de^2)\fr{1+x^2}{2 x}\right]\,.
\en
In this way
%\eq
% x^2&=& \fr{(\la+t)^2-\eta^2\de^2}{(\la-t)^2-\eta^2\de^2}\\
%\fr{\la}{t}&=&\fr{-1+x^2}{1+x^2-2 x (\eta^2+\delta^2)/(\eta^2-\delta^2)}
%\en
%and
\eq
\int_{-1}^1  d\!\cos\theta 
\fr{\la}{t}\ln\fr{(\la-t)^2-\eta^2\de^2}{(\la+t)^2-\eta^2\de^2}&=&
\fr{\eta^2-\delta^2}{2 b}\int_{x_{\rm min}}^{x_{\rm max}} dx\;
  \fr{(1-x^2)^2\ln x}{x^2\left( x^2+1-2 x\alp \right)} 
\\
&=& \fr{\eta^2-\delta^2}{2 b x}\biggr\{-1-x^2-(1-x^2)\ln x+\alp x \ln^2 x
\nn\\
\et
+2
  x\sqrt{\alp^2-1}\biggl(\ln x \ln\fr{1-\alp+x\sqrt{a^2-1}}{1-\alp-x\sqrt{a^2-1}}
\biggr.
\nn\\
\et
+Li_2\left(\fr{x}{\alp+\sqrt{\alp^2-1}}\right)\nn\\ \et
\left.
\biggl.
+Li_2\left(x(\alp+\sqrt{\alp^2-1})\right)\biggr)\biggr\}\right|_{x_{\rm min}}^{x_{\rm max}}
\en
where
\eq
\alp&=& {(\eta^2+\delta^2)\over (\eta^2-\delta^2)}\nonumber\\
x_{\rm max}&=&\fr{2}{\eta^2-\delta^2}\left\{
\fr{\eta^2+\delta^2}{2}-a-b+\fr{1}{2}\sqrt{(2
  (a+b)-(\eta^2+\delta^2))^2-(\eta^2-\delta^2)^2}\right\}\nn \\
x_{\rm min}&=&\fr{2}{\eta^2-\delta^2}\left\{
\fr{\eta^2+\delta^2}{2}-a+b+\fr{1}{2}\sqrt{(2
  (a-b)-(\eta^2+\delta^2))^2-(\eta^2-\delta^2)^2}\right\}\,.\nn \\
\label{eq:xdef}
\en
In the case $m_K=m_\pi$, $a+b=0$ and one recovers the 
formulas of \cite{BCEGS97}.

\subsection{Analytical Results for the Dominant FSI Phases at NLO}
\label{phasesNLO}

The elements of the matrices defined in (\ref{Rdelta2def}) have the next 
analytical expressions at NLO
\ba
\mathbb{R}^{\rm LO} &=& \left(\begin{array}{cc}R_{11}&R_{12}
\\R_{21}&R_{22}\end{array}\right)\,,
\nonumber\\
\delta_2^{\rm NLO}&=&\frac{\dis \sum_{i=8,27,E} G_i \left(C_{i,1}^{(++-)}
+C_{i,1}^{(00+)}\right)
+\sum_{i=1,11} \left(H_{i,1}^{(6)(++-)}+ H_{i,1}^{(6)(00+)}\right)
\widetilde K_i }
{\dis \sum_{i=8,27,E} G_i  \left(B_{i,1}^{(++-)}+B_{i,1}^{(00+)}\right)
+\sum_{i=1,11} \left(H_{i,1}^{(4)(++-)}+ H_{i,1}^{(4)(00+)}\right)
\widetilde K_i }\,,
\ea
with 
\ba
R_{11} &=& \frac{\left(-\beta_1+\frac{1}{2}\beta_3\right)^{\rm NR}
\left(\alpha_1+\alpha_3\right)^{\rm R-NR}-
\left(\beta_1+\beta_3\right)^{\rm NR}
\left(-\alpha_1+\frac{1}{2}\alpha_3\right)^{\rm R-NR}}
{\left(-\beta_1+\frac{1}{2}\beta_3\right)^{\rm NR}
\left(\alpha_1+\alpha_3\right)^{\rm NR}-
\left(\beta_1+\beta_3\right)^{\rm NR}
\left(-\alpha_1+\frac{1}{2}\alpha_3\right)^{\rm NR}}
\, ,\nonumber\\
R_{21} &=& \frac{\left(-\beta_1+\frac{1}{2}\beta_3\right)^{\rm NR}
\left(\beta_1+\beta_3\right)^{\rm R-NR}-
\left(\beta_1+\beta_3\right)^{\rm NR}
\left(-\beta_1+\frac{1}{2}\beta_3\right)^{\rm R-NR}}
{\left(-\beta_1+\frac{1}{2}\beta_3\right)^{\rm NR}
\left(\alpha_1+\alpha_3\right)^{\rm NR}-
\left(\beta_1+\beta_3\right)^{\rm NR}
\left(-\alpha_1+\frac{1}{2}\alpha_3\right)^{\rm NR}}
\ ,\nonumber\\
R_{12} &=& -\frac{\left(-\alpha_1+\frac{1}{2}\alpha_3\right)^{\rm NR}
\left(\alpha_1+\alpha_3\right)^{\rm R-NR}-
\left(\alpha_1+\alpha_3\right)^{\rm NR}
\left(-\alpha_1+\frac{1}{2}\alpha_3\right)^{\rm R-NR}}
{\left(-\beta_1+\frac{1}{2}\beta_3\right)^{\rm NR}
\left(\alpha_1+\alpha_3\right)^{\rm NR}-
\left(\beta_1+\beta_3\right)^{\rm NR}
\left(-\alpha_1+\frac{1}{2}\alpha_3\right)^{\rm NR}}
\, ,\nonumber\\
R_{22} &=& -\frac{\left(-\alpha_1+\frac{1}{2}\alpha_3\right)^{\rm NR}
\left(\beta_1+\beta_3\right)^{\rm R-NR}-
\left(\alpha_1+\alpha_3\right)^{\rm NR}
\left(-\beta_1+\frac{1}{2}\beta_3\right)^{\rm R-NR}}
{\left(-\beta_1+\frac{1}{2}\beta_3\right)^{\rm NR}
\left(\alpha_1+\alpha_3\right)^{\rm NR}-
\left(\beta_1+\beta_3\right)^{\rm NR}
\left(-\alpha_1+\frac{1}{2}\alpha_3\right)^{\rm NR}}\,,
\ea
The definitions of $\alpha_1$, $\alpha_3$, $\beta_1$ and $\beta_3$ are in 
(\ref{amp1}) and the values of their relevant combinations are
\ba
\left(-\alpha_1+\frac{1}{2}\alpha_3\right)^{\rm NR} &=&
\sum_{i=8,27,E} G_i B_{i,0}^{(00+)}
+\sum_{i=1,11} H_{i,0}^{(4)(00+)} \widetilde K_i \, ,\nonumber\\
\left(-\alpha_1+\frac{1}{2}\alpha_3\right)^{\rm R-NR} &=&
\sum_{i=8,27,E} G_i C_{i,0}^{(00+)}
+\sum_{i=1,11} H_{i,0}^{(6)(00+)} \widetilde K_i\, ,\nonumber\\
\left(-\beta_1+\frac{1}{2}\beta_3\right)^{\rm NR} &=& \frac{1}{2}
\left\lbrack \sum_{i=8,27,E} G_i \left(B_{i,1}^{(++-)}-B_{i,1}^{(00+)}
\right)\right.\nonumber\\
&&\left.+\sum_{i=1,11} \left(H_{i,1}^{(4)(++-)}- H_{i,1}^{(4)(00+)}\right)
\widetilde K_i 
\right\rbrack\, ,\nonumber\\
\left(-\beta_1+\frac{1}{2}\beta_3\right)^{\rm R-NR} &=& \frac{1}{2}
\left\lbrack \sum_{i=8,27,E} G_i \left(C_{i,1}^{(++-)}-
C_{i,1}^{(00+)}\right)\right.\nonumber\\
&&\left.+\sum_{i=1,11} \left(H_{i,1}^{(6)(++-)}-H_{i,1}^{(6)(00+)}\right)
 \widetilde K_i \right\rbrack\, ,\nonumber
\ea
\ba
\left(\alpha_1+\alpha_3\right)^{\rm NR} &=&
\sum_{i=8,27,E} G_i B_{i,0}^{(+-0)}
+\sum_{i=1,11} H_{i,0}^{(4)(+-0)} \widetilde K_i \, ,\nonumber\\
\left(\alpha_1+\alpha_3\right)^{\rm R-NR} &=&
\sum_{i=8,27,E} G_i C_{i,0}^{(+-0)}
+\sum_{i=1,11} H_{i,0}^{(6)(+-0)} \widetilde K_i \, ,\nonumber\\
\left(\beta_1+\beta_3\right)^{\rm NR} &=& -\frac{1}{2}\left\lbrack
\sum_{i=8,27,E} G_i B_{i,1}^{(+-0)}
+\sum_{i=1,11} H_{i,1}^{(4)(+-0)} \widetilde K_i\right\rbrack
\, ,\nonumber\\
\left(\beta_1+\beta_3\right)^{\rm R-NR} &=& \frac{1}{2}\left\lbrack
\sum_{i=8,27,E} G_i C_{i,1}^{(+-0)}
+\sum_{i=1,11} H_{i,1}^{(6)(+-0)} \widetilde K_i\right\rbrack\, .
\ea
where the functions $B_{i,0(1)}$, $C_{i,0(1)}$ and  $H_{i,0(1)}$  
are those obtained form the expansion in  (\ref{yexpansion})
of the corresponding full quantities that can be found
in Appendix \ref{ANLO}.

Disregarding the tiny CP-violating (less than 1\%)
and the effects of order $e^2p^2$ (the loop contribution
 is less than 2\%), 
we obtain the numbers in (\ref{RmatrixNLO}).

\chapter{Some Short-Distance Relations for Three-point Functions}

\label{apendixshortdist}

%Following the same lines we have just explained for obtaining  
%the OPE of the four point function $\langle 0|PL ^\mu PL_\mu|0\rangle$, 
We calculated or recalculated several short-distance behaviours of three-point
functions in \cite{BGLP03}. The results are

\be
\label{PSPSD}
\lim_{\lambda\to \infty}\Pi^{PSP}(\lambda p_1,\lambda p_2)^{\chi}
=
\frac{\condc}{2\lambda^2}
\left\{\frac{p_2^2}{q^2 p_1^2}+\frac{q^2}{p_1^2 p_2^2}
-\frac{p_1^2}{q^2 p_2^2}-\frac{2}{p_1^2}\right\}
\ee
\ba
\label{SVVSD}
\lefteqn{
\lim_{\lambda\to\infty}\Pi^{SVV}(\lambda p_1,\lambda p_2)^\chi
= \frac{\condc}{2\lambda^2 q^2 p_1^2 p_2^2}
\Bigg\{ -4 p_2^2\,p_{1\mu} p_{1\nu} - 2 (p_1^2+p_2^2-q^2) \,p_{1\mu} p_{2\nu}
}
&&\nonumber\\&&
-2 (p_1^2 + p_2^2 + q^2) \,p_{2\mu} p_{1\nu} -4 p_1^2\,p_{2\mu} p_{2\nu} 
+ \left( q^4 -(p_1^2-p_2^2)^2 \right)\,g_{\mu\nu}\Bigg\}
\ea
\ba
\label{PVASD}
\lefteqn{
\lim_{\lambda\to\infty}\Pi^{PVA}(\lambda p_1,\lambda p_2)^\chi
 = \frac{i\condc}{2\lambda^2 q^2 p_1^2 p_2^2}
\Bigg\{ 4 p_2^2\,p_{1\mu} p_{1\nu} - 2 (q^2+p_1^2-p_2^2) \,p_{1\mu} p_{2\nu}
}
&&\nonumber\\&&
+2 (q^2 + p_2^2 - p_1^2) \,p_{2\mu} p_{1\nu}  -4 p_1^2\,p_{2\mu} p_{2\nu}
+ \left( p_2^4 -(p_1^2-q^2)^2 \right)\,g_{\mu\nu}\Bigg\}
\ea
Some of these have been mentioned in Refs.~\cite{KN01,Moussallam}.

The following were first calculated to our knowledge in \cite{BGLP03} 
\ba
\label{PASSD}
\lefteqn{
\lim_{\lambda\to\infty}\left(\Pi^{PAS}(\lambda p_1,\lambda p_2)^\chi
+\Pi^{SAP}(\lambda p_1,\lambda p_2)^\chi\right)
\,=\,
i 4 \pi\alpha_S\frac{N_c^2-1}{N_c^2}\frac{\condc^2}{\lambda^5}}
 \nonumber\\
&&\times\left\lbrace
\frac{2 p_{1\mu}}{p_1^2}
\left(\frac{1}{q^4}+\frac{1}{p_2^4}+\frac{1}{p_2^2 q^2}\right)
+\frac{p_{2\mu}}{p_2^2}
\left(\frac{-1}{q^4}+\frac{1}{p_2^2 q^2}\right)
+\frac{q_{\mu}}{q^2}
\left(\frac{1}{p_2^4}-\frac{1}{p_2^2 q^2}\right)
\right\rbrace
\ea

\ba
\label{PASSDi}
\lim_{\lambda\to\infty}\lim_{m_q\to 0}\frac{\partial}{\partial m_i}
\left(\Pi^{PAS}(\lambda p_1,\lambda p_2)^{ijk}
+\Pi^{SAP}(\lambda p_1,\lambda p_2)^{ijk}\right)
&=&
-2 i  \condc \frac{p_{2\mu}}{\lambda^3 p_2^2 q^2}\,,
\nonumber\\
\lim_{\lambda\to\infty}\lim_{m_q\to 0}\frac{\partial}{\partial m_j}
\left(\Pi^{PAS}(\lambda p_1,\lambda p_2)^{ijk}
+\Pi^{SAP}(\lambda p_1,\lambda p_2)^{ijk}\right)
&=&
2 i  \condc \frac{p_{1\mu}}{\lambda^3 p_1^2}
\left(\frac{1}{q^2}+\frac{1}{p_2^2}\right)\,,
\nonumber\\
\lim_{\lambda\to\infty}\lim_{m_q\to 0}\frac{\partial}{\partial m_k}
\left(\Pi^{PAS}(\lambda p_1,\lambda p_2)^{ijk}
+\Pi^{SAP}(\lambda p_1,\lambda p_2)^{ijk}\right)
&=&
2 i  \condc \frac{q_{\mu}}{\lambda^3 p_2^2 q^2}\,.
\ea

\newpage
\listoffigures

\newpage
\listoftables

\end{document}